\documentclass[11pt,twoside]{book}
\usepackage{ml-books}
\usepackage{wrapfig}
\usepackage{enumerate}
\usepackage[export]{adjustbox}
\usepackage{float}
\usepackage{mathtools}

\usepackage{makeidx}                       % Indice analitico
\makeindex

\newcommand{\Id}{\hat{\mathbb{1}}}
\newcommand{\PauliX}{\hat{X}}
\newcommand{\PauliY}{\hat{Y}}
\newcommand{\PauliZ}{\hat{Z}}

\newcommand{\Hgate}{\hat{H}}
\newcommand{\Sgate}{\hat{S}}
\newcommand{\Tgate}{\hat{T}}

\newcommand{\St}{\text{Stab}}

\newcommand{\dbraket}[2]{\langle\!\langle#1|#2\rangle\!\rangle}

\definecolor{covercolor}{RGB}{24,27,35}

\begin{document}

%%%%%%%%%%%%%%%%%%%%%%% TITLE PAGE %%%%%%%%%%%%%%%%%%%%%%%%%

\begin{titlepage}
\newgeometry{top=1in,bottom=1in,right=0in,left=0in}
\thispagestyle{empty}

\pagecolor{covercolor}
\color{white}

\vspace{2cm}
{\fontsize{18}{0}\selectfont  Mario Collura  $\cdot$ Guglielmo Lami\par
Nishan Ranabhat $\cdot$ Alessandro Santini} \par
\rule{5cm}{1pt}\par
\vspace{1cm}

{\fontsize{32}{0}\selectfont  
Tensor Network Techniques \par for Quantum Computation}
%\vspace{0.2cm}
\begin{center}
\vskip\baselineskip
\adjincludegraphics[width=\paperwidth]{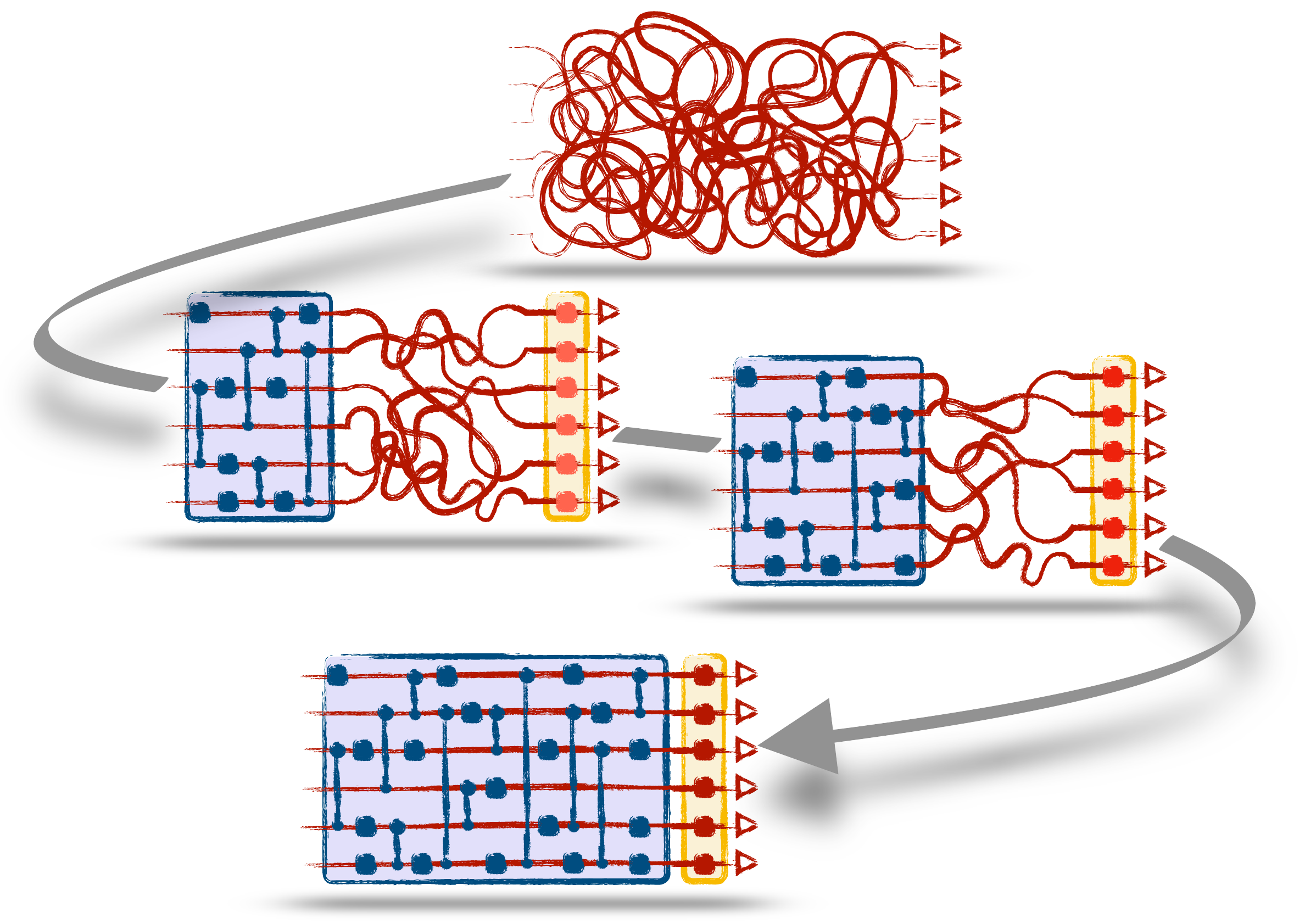}
\vskip\baselineskip
{\fontsize{16}{0}\selectfont Trieste, 2025}

\end{center}
\end{titlepage}
\pagecolor{white}

\epigraph{I am withdrawing to write a book. And another time: I am withdrawing to construct a labyrinth
.}{``Ts'ui P\^en'' in The Garden of Forking Paths by Jorge Luis Borges}

%%%%%%%%%%%%%%%%%%%%%%%%%%%%%%%%%%%%%%%%%%%%%%%%%%%%%%%%%%%%

%%%%%%%%%%%%%%%%%%%%%%PREFACE%%%%%%%%%%%%%%%%%%%%%%%%
\chapter*{Preface}\label{preface}
The last decade has seen a revolutionary shift in the overlap between quantum information science and the study of quantum many-body systems, two fields that traditionally developed along separate tracks. This book, designed for undergraduate students and early researchers, builds a bridge between these disciplines by exploiting tensor networks—a powerful computational framework with deep relevance for both many-body physics and quantum computation.

Tensor networks emerged initially as a way to represent complex quantum states compactly in many-body physics. However, as quantum computing has advanced, tensor networks have also become indispensable tools for understanding and simulating quantum systems. They allow us to visualize and capture intricate entanglement structures and to represent states in a way that is both computationally manageable and mathematically rich. As a result, tensor networks offer a critical foundation for tackling challenges in quantum computation, where state complexity, entanglement, and even more elusive aspects of quantum mechanics—like quantum magic—play key roles.

The book begins with an introduction to fundamental concepts, presenting tensor networks as efficient tools for representing and manipulating quantum states. We cover the basics of matrix product states (MPS), the foundational building blocks of tensor network theory quantum circuits, which allow us to approximate low-entanglement quantum states with minimal computational resources.
We examine how they allow us to quantify and manipulate entanglement and delve into topics that go beyond entanglement alone, including recent advances in quantifying non-stabilizer resources—referred to as quantum magic—which are vital for achieving quantum speedup. This book discusses the implications of these non-stabilizer resources in the context of stabilizer circuits and Clifford transformations, offering insight into how tensor networks can represent both stabilizer and non-stabilizer states within the same framework. Beside that, we revisit well known techniques like the Time-Dependent Variational Principle (TDVP) and Matrix Product Operator (MPO) based Lindblad dynamics, in order to give the proper background to advance into new methods which combine tensor networks with Clifford stabiliser formalism. 

Furthermore, the book explores the connection between tensor networks and classical simulations of quantum circuits, highlighting the power of tensor networks to simulate complex quantum algorithms. We delve into the current limitations and challenges associated with scaling these simulations and discuss how advances in tensor network methods might overcome these hurdles in the future.

Our goal with this book is not to provide an exhaustive review of tensor networks or quantum computing, but rather to present these ideas in a pedagogical, hands-on manner, using accessible language and representative examples to illuminate complex topics. By emphasizing recent developments in tensor network methods for studying quantum complexity, quantum magic, and non-stabilizerness, we aim to provide readers with a toolkit for exploring the cutting-edge intersections of quantum computing and quantum many-body physics.

It is our hope that this book will inspire readers to further explore the interface between quantum information science and many-body physics, equipping them with the theoretical insights and practical tools needed to contribute to this rapidly evolving field.\\

The Authors
%%%%%%%%%%%%%%%%%%%%%%%%%%%%%%%%%%%%%%%%%%%%%%%%%%%%%%

%%%%%%%%%%%%%%%Acknowledgemts%%%%%%%%%%%%%%%%%%%%%%
\chapter*{Acknowledgments}
We are beholden to numerous colleagues and students for their invaluable insights and suggestions.
In particular, we gratefully acknowledge Pasquale Calabrese and Marcello Dalmonte for their critical reading of parts of the manuscript.
Obviously, we take full responsibility for any remaining errors or oversights, and no blame should fall on the individuals mentioned above.
%%%%%%%%%%%%%%%%%%%%%%%%%%%%%%%%%%%%%%%%%%%%%%%%%%%

\tableofcontents
% only 2 Algorithms
%\newpage
%\listof{boxedalgorithm}{List of Algorithms}
\newpage
\tcblistof[\chapter*]{definition}{List of Definitions}
\newpage
\tcblistof[\chapter*]{example}{List of Examples}
% only 2 Lemmas
%\newpage
%\tcblistof[\chapter*]{lemma}{List of Lemmas}
% only Theorem
%\newpage
%\tcblistof[\chapter*]{theorem}{List of Theorems}

\newpage

\part{Preliminaries}
Tensor network methods encompass a set of strategies designed to comprehend and analyze multi-linear maps, proving particularly valuable in the realm of quantum computation and information processing, as well as in quantum many-body physics. These techniques serve as the foundation for tensor network contraction algorithms, essential for modeling physical systems. Employed within abstract graphical frameworks, these methods proficiently depict channels, maps, states, and algorithms applicable across various domains within the field of quantum physics.

While tensor networks span various subjects, existing literature tends to be highly specific, often catering to a narrow community.
Here we take the typical approach of tensor network for many-body physics - by offering a instructive exploration of the foundational principles of tensor network theory - exploiting its broad applicability in the realm of quantum computation and information.

The chapters presented in this section serves as the essential foundation for understanding the core concepts and mathematical tools required to explore the applications of tensor networks in quantum physics and quantum computing. This part introduces the reader to the building blocks of tensor network theory, starting with the basic concept of tensors and expanding into more complex tensor network structures that form the backbone of modern computational methods in many-body quantum systems.

\paragraph{Tensors and Tensor Networks ---}
At the heart of this discussion are \emph{tensors}, which generalize matrices to higher dimensions. Tensors represent multi-indexed arrays that serve as the fundamental objects for encoding and manipulating data in various dimensions. In quantum physics, they play a critical role in simplifying and representing complex multi-particle quantum states. The section on ``Special Tensors'' explores key forms such as \emph{identity tensors} and \emph{Kronecker delta tensors}, which have specific properties essential for simplifying tensor operations.

Building on this, the book introduces \emph{tensor networks}, a graphical formalism to represent complex tensors by breaking them down into smaller components. This decomposition of large tensors into interconnected networks enables computational efficiency, especially for simulating large quantum systems. A deep dive into specific tensor network architectures, such as \emph{Matrix Product States (MPS)} and \emph{Tree Tensor Networks (TTN)}, demonstrates their wide-ranging applicability to problems in quantum physics and quantum computation.

\paragraph{Quantum Physics with Tensors ---}
To establish the relevance of tensor networks to quantum mechanics, the next section provides an overview of quantum physics, including its foundational concepts like \emph{states}, \emph{observables}, and \emph{measurements}. The representation of quantum systems in terms of tensors naturally arises here. For instance, \emph{quantum states}, especially in many-body systems, are often expressed using tensor networks, which significantly reduce the complexity of handling these states.

\emph{Matrix Product States (MPS)} and \emph{Matrix Product Operators (MPO)} are key methods discussed in this section. They are shown to efficiently represent quantum states and operations, even for systems with large numbers of qubits. Moreover, \emph{entanglement} --- a hallmark feature of quantum systems --- naturally emerges in tensor network representations. In this section we exploit the \emph{Schmidt decomposition} and other key tools for understanding and quantifying entanglement in many-body systems.

\paragraph{Quantum Computing with Tensor Networks ---}
The preliminaries culminate with an exploration of how tensor networks can be leveraged for \emph{quantum computing}. As quantum circuits grow more complex, traditional methods become computationally expensive. Tensor networks, particularly in their MPS form, offer a powerful way to simulate quantum circuits and manipulate quantum data. Topics like \emph{state preparation}, \emph{quantum gates}, and \emph{measurements} are discussed in detail, demonstrating how tensor networks streamline these processes, particularly in noisy intermediate-scale quantum (NISQ) devices.

Additionally, the section highlights real-world implementations of these techniques across various \emph{quantum platforms}, such as \emph{superconducting qubits}, \emph{neutral-atom qubits}, and \emph{trapped-ion qubits}. Each of these platforms provides distinct advantages and challenges, and tensor network methods are presented as valuable tools for overcoming the computational challenges associated with simulating these systems.

In conclusion, the ``Preliminaries'' section lays the groundwork for the more advanced applications that follow, by equipping the reader with both the theoretical knowledge and practical tools needed to navigate the landscape of quantum physics and quantum computation using tensor networks.

%% \include{chapters/chapt_1}
%%%%%%%%%%%%%%%%%%%%% TENSOR NETWORK BASICS %%%%%%%%%%%%%%%%%%%%%%%%%%%%

\chapter{Tensor Network Basics}\label{chap1}
\epigraph{Everything should be made as simple as possible, but not simpler.}{Albert Einstein}

\emph{Tensor Network (TN) methods} often refer to a comprehensive set of tools commonly utilized in contemporary quantum information science, condensed matter physics, mathematics, and computer science.
Essentially, these methods constitute a set of techniques employed to systematically organize and manipulate vast numerical datasets arranged in multidimensional arrays, also known as tensors, interconnected to form a network~\cite{biamonte2020,Ran_2020,Evenbly_2022}. These methods allow for a simple and appealing diagrammatic notation that facilitates the understanding of complex linear algebra operations in a succinct fashion, making them well suited for modern computational devices. In this section, we present a formal introduction to tensors, tensor networks, and essential operations associated with tensor networks. Let us start from the fundamental building block of any tensor network, namely, a tensor.

\section{Tensors}
In the realm of mathematics, tensors serve as a powerful algebraic entity that encodes a multi-linear relationship among sets of algebraic objects associated with a vector space. In practice, a generic tensor is a scalar-valued function of multiple parameters which are linear with respect to each other. Remarkably, tensors transcend specific bases, asserting their definition independent of any particular basis.
Formally we have the following definition:

\begin{definition}{Tensor}{tensor}
 Let $V$ is a vectorial space with dimension $d$ over the complex numbers $\mathbb{C}$, i.e.\ $V \simeq \mathbb{C}^d$. The dual space $V^{*}$ is the vector space defined as the set of all linear map $\varphi: V  \to \mathbb{C}$. It has dimension $d$ as well. Its element are called \emph{covectors}. A type $(p, q)$ {\bf tensor} is therefore a multi-linear map
$$T: \underbrace{V^{*} \times \dots \times V^{*}}_{p} \times
\underbrace{V \times \dots \times V}_{q} \to \mathbb{C}$$
Therefore, a tensor $ T$ associates $q$ vectors $\{v_1,\dots v_q\}$ and $p$ covectors
$\{w_1,\dots w_p\}$ a scalar $T(w_1,\dots w_p,v_1,\dots v_q)$.
%\Guglielmo{I guess at this point we should introduce a basis for $V$ and $V^*$, so that later we can introduce coordinates (i.e.\ with tensors of numbers)}
\end{definition}

However, in practical applications, tensors often find representation through their components within a basis tied to a specific coordinate system. These components assemble into an array, akin to a high-dimensional matrix, facilitating an easier understanding.

For example, following the Dirac convection of modern quantum physics,
we may choose the Hilbert space $\mathcal{H}\simeq \mathbb{C}^{d}$ as the typical vector space wherein tensors act as multi-linear map,\footnote{We refer the reader to the next chapter to a tighter connection with quantum physics and quantum computing.} with standard computational basis $\{\ket{j}:j=0,\dots,d-1\}$.
In this context, a vector $\ket{\psi}\in\mathcal{H}$ is essentially an order-$1$ tensor, which we may express in terms of its tensor components
$\psi_j =\braket{j}{\psi}$ with respect to the computational basis, such that
$\ket{\psi}=\sum_j \psi_j \ket{j}$.
In strict analogy, we introduce the standard covectors of the dual space $\mathcal{H}^*$ whose basis is $\{\bra{j}:j=0,\dots,d-1\}$.
Linear operators $\hat O \in \mathcal{H}\times \mathcal{H}^*$ can be represented as order-2 tensors with components $O_{ij} = \mel{i}{\hat O}{j}$, such that $ \hat O = \sum_{i,j} O_{ij}  \dyad{i}{j}$.
Here we implicitly introduced the concept of \emph{order} of a tensor as

\begin{definition}{Tensor order}{order}
 We define the {\bf order of a tensor} as the number of indices it possesses, no matter whether the indices are spanning over the vectors or the covectors.
\end{definition}

With this respect, and after having introduced a suitable basis, tensor serves as a generalization encompassing both vectors and matrices. Specifically, a generic order-$k$ tensor is portrayed as a complex multidimensional array denoted by
$T_{j_1, \ldots, j_k} \in \mathbb{C}^{d_1}\times \dots \times\mathbb{C}^{d_k}$. Each index $j_i$ for $i\in \{1,\dots k\}$, colloquially referred to as a \emph{leg} of the tensor, takes values from the corresponding set $\{1,2,\ldots,d_i\}$. In this context, $d_i$ represents the dimension associated with the index $j_i$, and the overall dimension of the entire tensor is expressed as the product $\prod_{i=1}^k d_i$. Consequently, each individual element within the multi-dimensional array $T_{j_1, \ldots, j_k}$ can be unequivocally identified by a the $k$-tuple $\{j_1,\dots, j_k\}$.
As passing by we introduce the following:

\begin{definition}{Tensor dimension}{dimension}
 The number of values a tensor index can take is referred to as the {\bf dimension of an index}, which is most often denoted by $\chi$, but can also be denoted by $d$, especially when it spans a local physical vector space (or its dual). The product of all those dimensions defines the total dimension of a tensor.
\end{definition}

A scalar is an order-$0$ tensor, a vector is an order-$1$ tensor, and a matrix an order-$2$ tensor and so on. When dealing with tensors of higher order that feature multiple legs, tensor diagrams offer a powerful tool for representing them and eventually executing intricate operations. These diagrams serve as a visual and intuitive representation, simplifying complex computations and enhancing comprehension of tensor behavior.

In the following we illustrate the tensor network diagram for a scalar $S$, a vector $V$, a matrix $M$, and a generic order-$k$ tensor $T$:\index{Tensors!network diagram}
$$
\includegraphics[width=\textwidth]{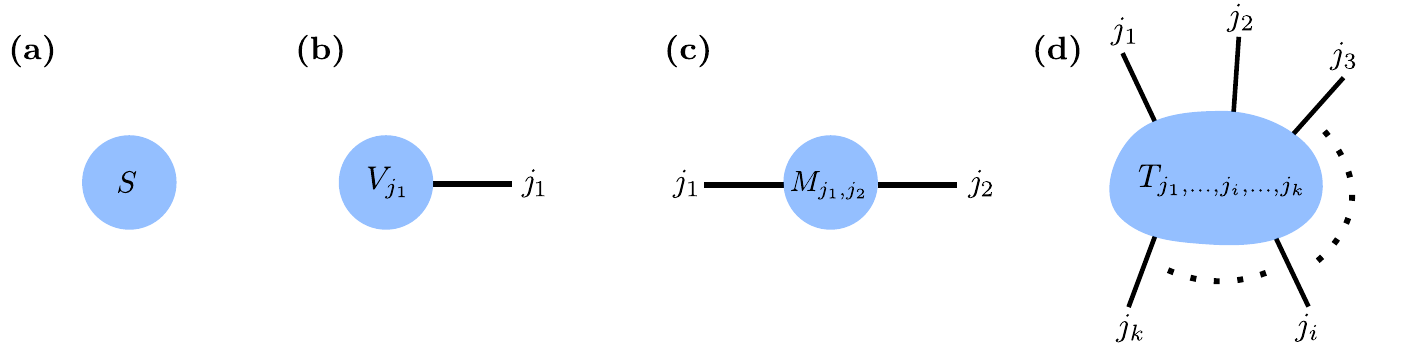}
$$

\paragraph{Tensor reshaping by index fusion ---}\index{Tensors!index fusion}
Given a generic order-$k$ tensor, one can always reshape the tensor into an order-$k'$ tensor with $ k' < k$. This fundamental operation is called index fusion. We first identify the $q\leq k$ indices we want to fuse together, for example $\{j_1,\dots, j_q\}$. Each of those indices is labeling a local Hilbert space vector $\{\ket{j_i}:j_i = 1, \dots, d_i\}$. We then consider the tensor product
of those Hilbert spaces $\mathcal{H}_1\otimes\dots\otimes\mathcal{H}_q$ and its full computational basis $\{\ket{j_1,\dots,j_q}\equiv\ket{\alpha}:\alpha=1,\dots,\prod_{i=1}^{q} d_i\}$.
In the end we get a tensor with order $ k' = k-q+1$.
Let us mention that, a tensor can be always reshaped back to its original form (or just partially reshaped)
provided that we keep track of how the new index $\alpha$ is targeting the tensor product states $\ket{\alpha} \equiv \ket{j_1,\dots,j_q}$ associated with it, or in other words how it has been computed starting from the original indices $\{j_1,\dots,j_q\}$. The following tensor network diagram illustrates the reshaping of a generic tensor with index fusion,
$$
\includegraphics[width=0.7\textwidth]{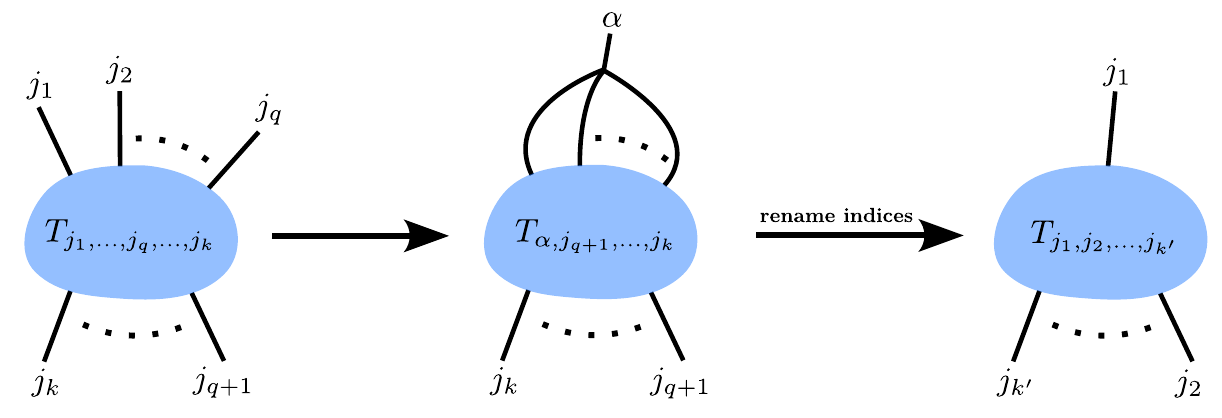}
$$
\paragraph{Tensor reshaping by index splitting ---}\index{Tensors!index splitting} Alternatively, we can reshape a generic order-$k$ tensor into an order-$k'$ tensor with $k'>k$ by splitting one or several indices. This operation is the inverse of index fusion and involves splitting an index into several indices of smaller dimension, $\{\ket{j_i}: j_i = 1,\dots,d_i\} \rightarrow \ket{j_i} \equiv \{\ket{\alpha_1,\dots,\alpha_q} : \alpha_l = 1,\dots,d'_l \hspace{0.1cm}\text{where,} \hspace{0.1cm} d_i = \prod_{l=1}^q d'_l \}$. The result is a tensor of order $k' = k+q-1$. Notably, reshaping a tensor by splitting an index is not a unique operation as an index of generic dimension $d$ can be split into $q$ indices in more than one ways. Therefore, index splitting is an inverse of index fusion only if we keep track of the way indices were fused. The following tensor network diagram illustrates the reshaping of a generic tensor with index splitting,
$$
\includegraphics[width=0.7\textwidth]{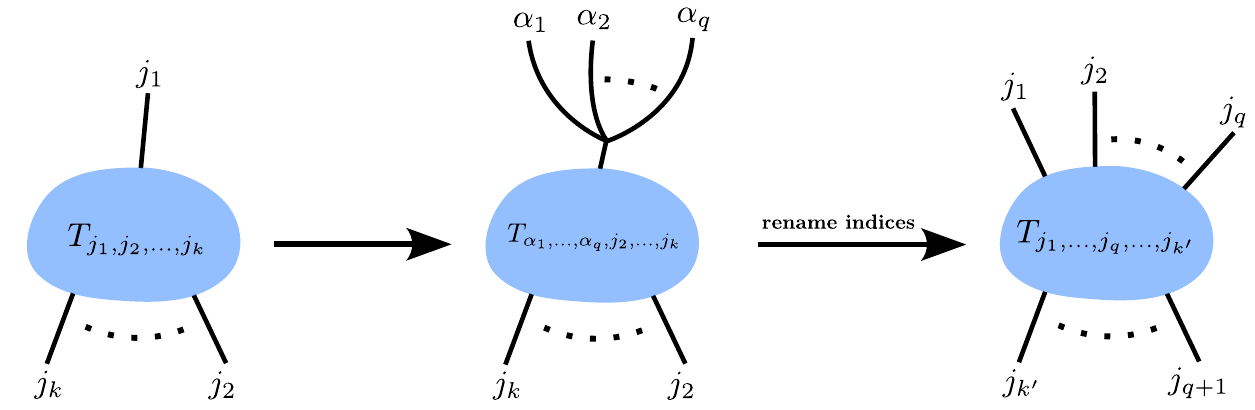}
$$

\paragraph{Tensor reshaping by outer product ---}\index{Tensors!outer product}
When multiple disconnected tensors appear within the same diagram, they are combined through the tensor product operation. In quantum physics notation, this operation is denoted by the symbol $\otimes$. However, in abstract index notation, the tensor product sign is not explicitly indicated. Notice that, a tensor by itself is not carrying any information about the algebra of the vector spaces (and dual) connected to its legs. Therefore, when tensors are drawn near each other, they can freely move through each others as
$$
\includegraphics[width=0.6\textwidth]{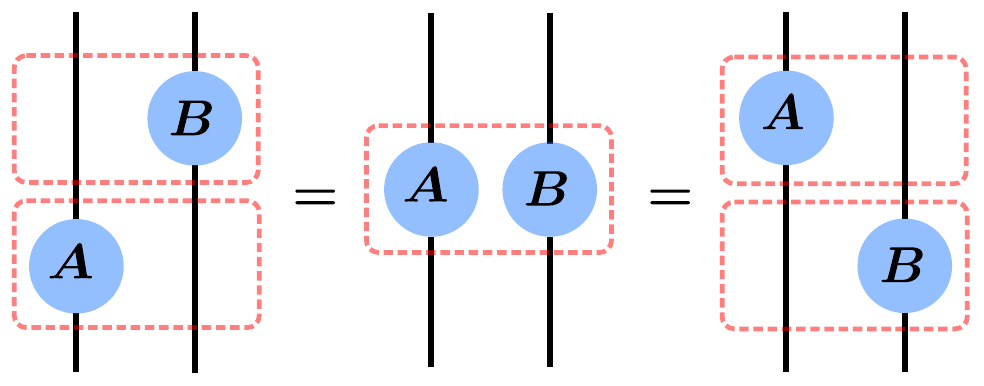}
$$
which, assuming applying tensors from bottom to up, basically means
\begin{equation}
(\mathbb{1}\otimes B)(A\otimes \mathbb{1})
=
A\otimes B
=
(A\otimes \mathbb{1}) (\mathbb{1}\otimes B),
\end{equation}
where $\mathbb{I}$ is the identity matrix and it has a very special representation as a tensor diagram: just a simple wire.

Notice that, the tensor product of two or more tensors may results in a final tensor with higher order. Let us formalise this idea by considering the elementary outer product operation: $A$ and $B$ are two tensors, such that indices
$\{\alpha_1,\dots \alpha_p\}$ of tensor $A$ need to be ``tensorised'' with indices $\{\beta_1,\dots \beta_q\}$ of tensor $B$ no matter the other indices of both tensors.
We can do that in two steps: (i) we first fuse the indices $\{\alpha_1,\dots \alpha_p\}$
and $\{\beta_1,\dots \beta_q\}$ respectively into a single index $\boldsymbol{\alpha}$ and
$\boldsymbol{\beta}$; (ii) we then fuse the new bold indices into one single index
$\boldsymbol{\gamma} = (\boldsymbol{\alpha},\boldsymbol{\beta})$ such that
\begin{equation}
A_{(i_1,\dots,i_n),\boldsymbol{\alpha}} \, B_{(j_1,\dots,j_m),\boldsymbol{\beta}}
\equiv C_{(i_1,\dots, i_n, j_1,\dots,j_m), \boldsymbol{\gamma}},
\end{equation}
which defines the new tensor $C$.

\subsection{Special Tensors}
Here, we gather some special tensors that we will encounter frequently during the subsequent sections and chapters.

\paragraph{Identity tensor ---}
In the graphical notation, there is one class of tensors which play a crucial role in many
different shape manipulation at it always plays the role of the \emph{Kronecker delta}: they are the so called \emph{identity tensors}\index{Identity tensor}:
$$
\includegraphics[width=0.15\textwidth]{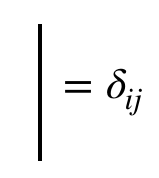}
$$
These tensors enable (i) the contraction of tensor indices through diagrammatic connection (see next section), (ii) the raising and lowering of indices (when index position may have a meaning related to the vector space associate to the specific leg), and (iii) they establish a duality between maps, states, and linear maps in general (see next chapter).

However, as we have already stressed, a tensor by itself does not carry any information about the algebra of the vector spaces associated to the specific legs that are going to be contracted; the same consideration are valid for the identity tensor, which therefore act as simple connecting wire. Notwithstanding, as we will briefly anticipate in Example~\ref{exmp:quantum_computing}, and we will thoroughly use in the next sections, when identities are promoted to ``operators'', then which states are contracted do matter.

\begin{example}{Glimpse on Quantum Computing}{quantum_computing}
As previously mentioned, tensors serve as multilinear maps. They can be decomposed in any given basis and expressed in terms of their components. In the realm of quantum information science, it is common to introduce a computational basis $\{\ket{0},\ket{1}\}$ for each Hilbert space and expand the tensors within it, utilizing kets $(\ket{})$ for vectors and bras $(\bra{})$ for dual vectors. This expansion is represented as follows:
$$
\includegraphics[width=0.35\textwidth]{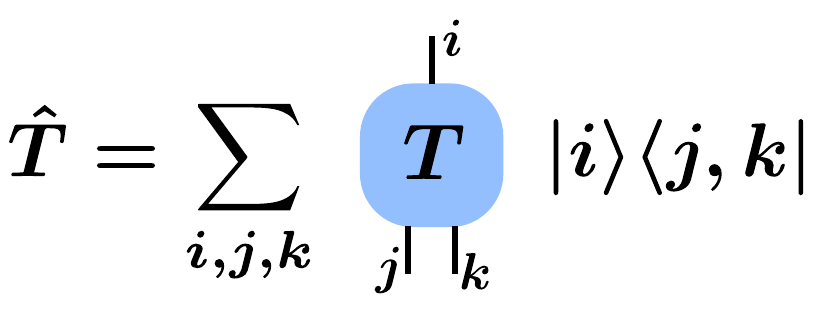}
$$
Here, $T^{i}_{jk}$ are the actual components of the tensor in the computational basis. Notice that, drawing the leg $i$ pointing up, and the legs $(j,k)$ pointing down is meaningless by itself (we could have rearranged the indices into any form without changing the meaning of the numeric tensor $T$); what is meaningful is the action of the \textbf{operator}
$\hat T$ once different Hilbert spaces have been linked to each leg of the tensor box $T$.\\

With this in mind, we can revisit the questions of how the Kronecker delta tensor may be used to construct different quantum operators. Let us consider two copies of a two-level quantum system, and let's introduce the computational basis $\{\ket{0},\ket{1}\}$.  We may thus construct the following tensor operators:
\begin{enumerate}
\item
$$
\includegraphics[width=0.6\textwidth]{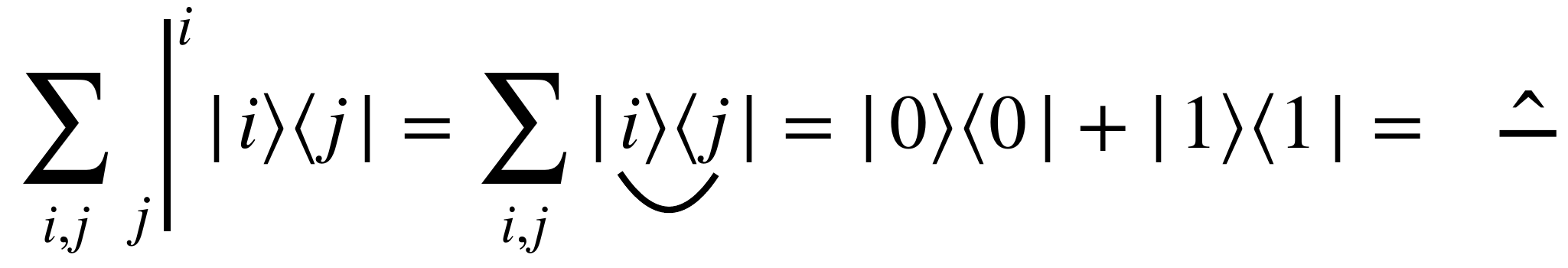}
$$
\item
$$
\includegraphics[width=0.6\textwidth]{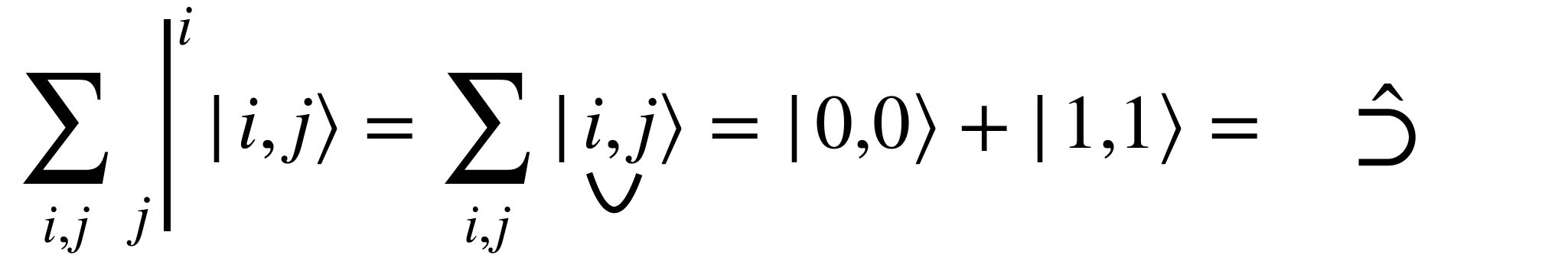}
$$
\item
$$
\includegraphics[width=0.6\textwidth]{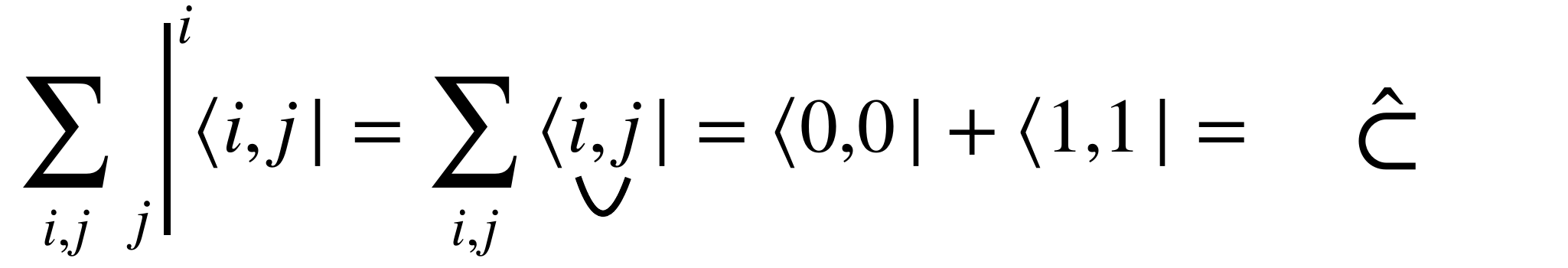}
$$
\end{enumerate}
In this context, the map labeled as 1. represents the identity operator, whereas the tensor operator labeled as 2. (or 3.) operates on a two-bra (or ket) system, yielding a scalar as the outcome. The connection between these three equations is established through the bending of wires. In a given basis, the act of bending a wire signifies the conversion of a bra to a ket, and vice versa.

In fact, with an abuse of notation, we can easily combine operators 1., 2. and 3. as they where
simple wires. For example, let us consider three local Hilbert spaces with each a two-level system, and let us label those systems as $\{a,b,c\}$. Operators $\hat\subset^{b}_{a}$ and
${}^{c}_{b}\hat\supset$ can be combined together giving
$$
\includegraphics[width=0.8\textwidth]{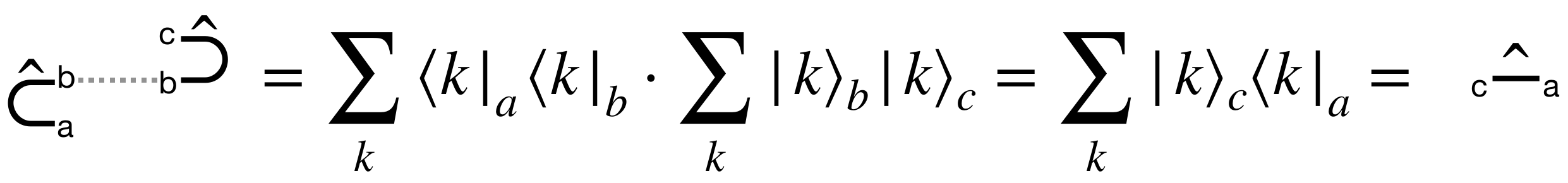}
$$

\end{example}

\paragraph{Copy tensor ---}
A very useful tensor is the multidimensional generalisation of the Kronecker delta\index{Copy tensor}, where many legs are forced to be equal. In graphic notation one has
$$
\includegraphics[width=0.4\textwidth]{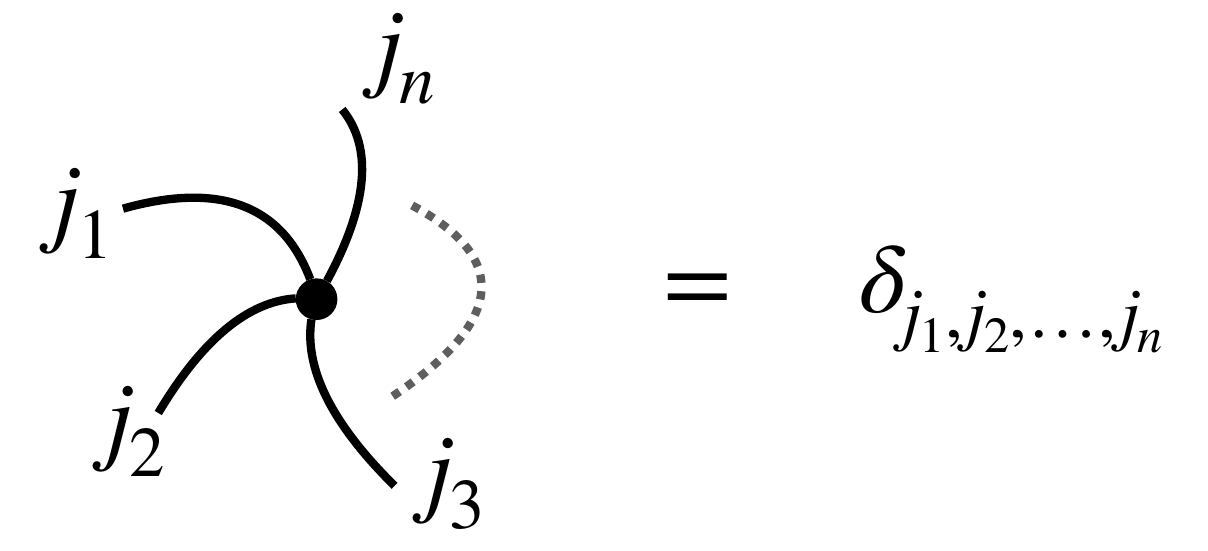}
$$
Interestingly, the natural decomposition of the multidimensional Kronecker delta
$\delta_{j_1,j_2,\dots,j_n} = \delta_{j_1,j_2}\delta_{j_2,j_3}\cdots\delta_{j_{n-1},j_n}$
(or for any other indices permutations)
induces the following graphical \emph{tensor network decomposition}
$$
\includegraphics[width=0.6\textwidth]{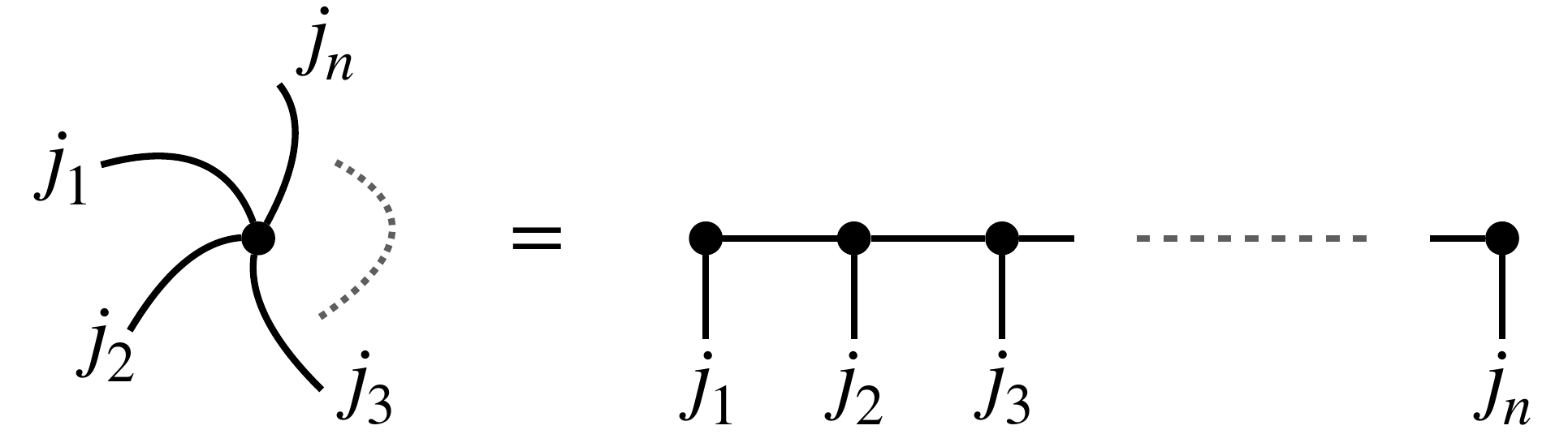}
$$

\paragraph{Swap tensor ---}
Let's examine the tensor product of $n$ equivalent Hilbert spaces denoted as $\mathcal{H}^{\otimes n}$. We can use two \emph{Identity tensors}\index{Swap tensor} to formally construct the \emph{Swap tensor}, which switches the positions of two Hilbert spaces within a composite system. Specifically, it exchanges two consecutive local Hilbert spaces $i$th and $(i+1)$th. The diagrammatic representation is
$$
\includegraphics[width=0.4\textwidth]{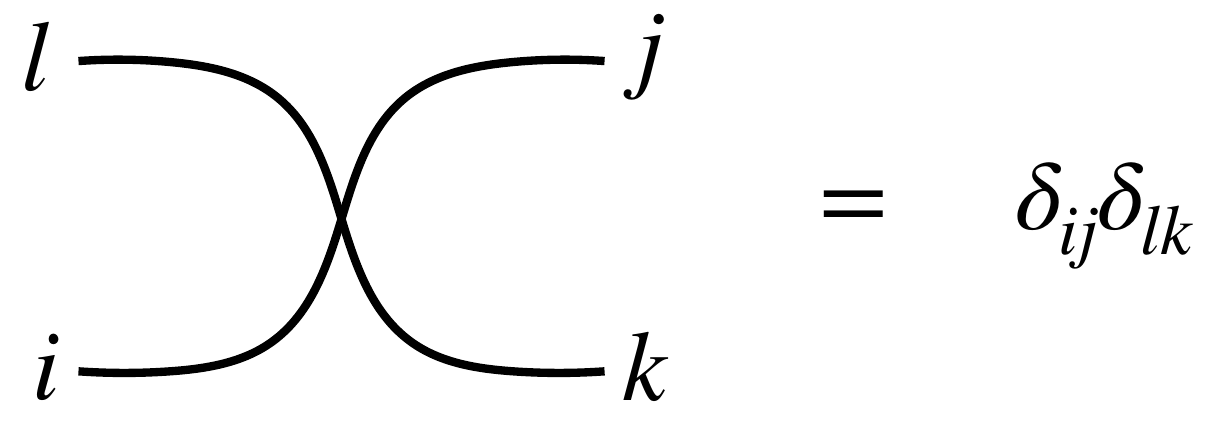}
$$
which, as we have consistently emphasized, it lacks any algebraic significance unless it is viewed in conjunction with the Hilbert spaces upon which it operates. In fact, from the operatorial point of view,
the \emph{Swap Operator} reads
$$
\includegraphics[width=0.6\textwidth]{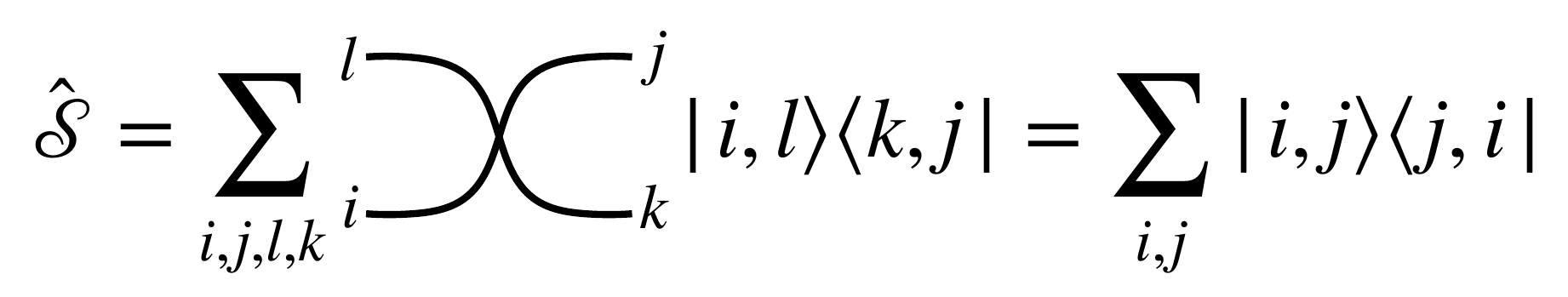}
$$
Notice that, this single operation it is sufficient to generate the all permutation group of the $n$ local spaces.

\begin{example}{Non-local swapping}{swap_example}
The Swap tensor serves as a generator for the permutation group. In this capacity, it enables the repositioning of two initially distant local systems, bringing them into proximity to facilitate interaction (cfr.\ Quantum Platform). This process can then be reversed to restore the entire system to its original configuration.

For clarity, let us consider a scenario involving four local Hilbert spaces denoted by $\{i_1,i_2,i_3, i_4\}$, where the objective is to swap the positions of the first and fourth systems. To achieve this, we can employ five local swap operators in combination; for example in this way
$$
\includegraphics[width=0.55\textwidth]{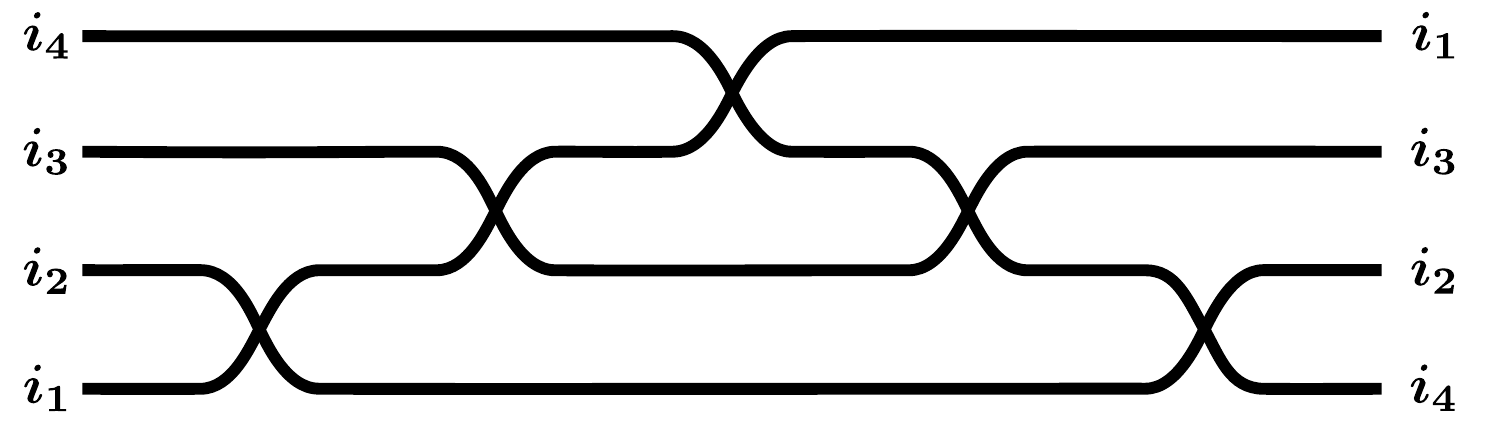}
$$
where the swap operations need to be performed sequentially. Certainly, there exist implementation options where we can parallelize multiple permutations simultaneously within the specific platform. This strategy can significantly accelerate the procedure, nearly doubling its speed. For instance, one such approach involves the following:
$$
\includegraphics[width=0.55\textwidth]{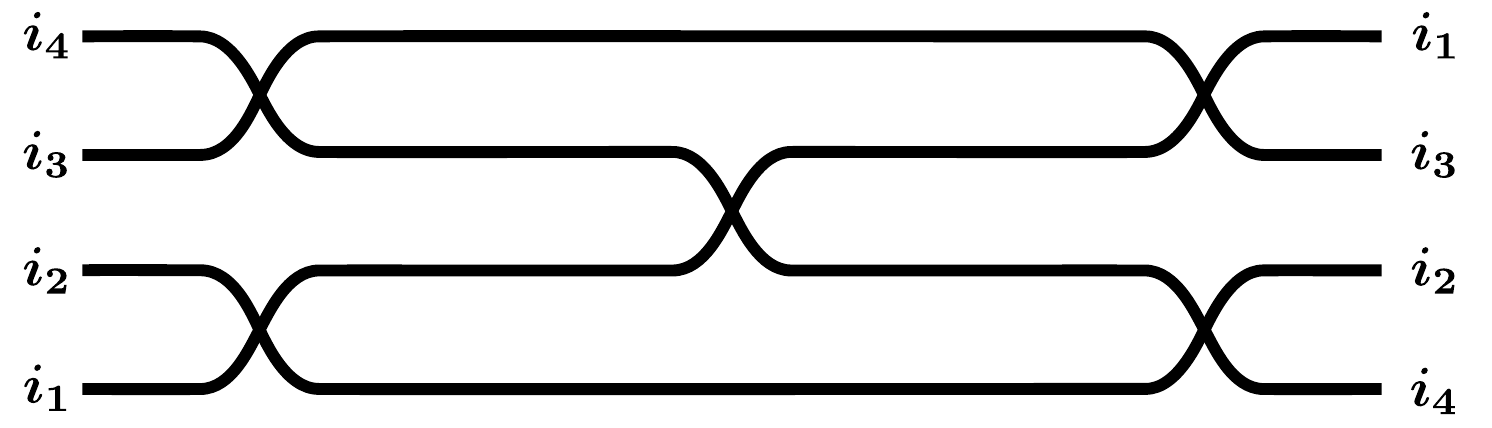}
$$
where, in this approach, the first swapping procedures involving $1\leftrightarrow 2$ and $3\leftrightarrow 4$ can be executed concurrently, as can the final $4\leftrightarrow 2$ and $3\leftrightarrow 1$ swaps.
\end{example}

\paragraph{XOR tensor ---}
The XOR (or Parity) tensor \index{Parity tensor} is a multi-dimensional array that operates on qubits or qubit-like systems in quantum computing. It represents a gate or operation that computes the parity (or XOR) of the qubits it acts upon. The XOR tensor outputs $1$ if the number of qubits in the state with value $1$ is even, and $0$ if it is odd.
Notice that, in this definition, there is an overall logical NOT operation if compared with the exclusive disjunction $\oplus$.
In graphic notation one has
$$
\includegraphics[width=0.5\textwidth]{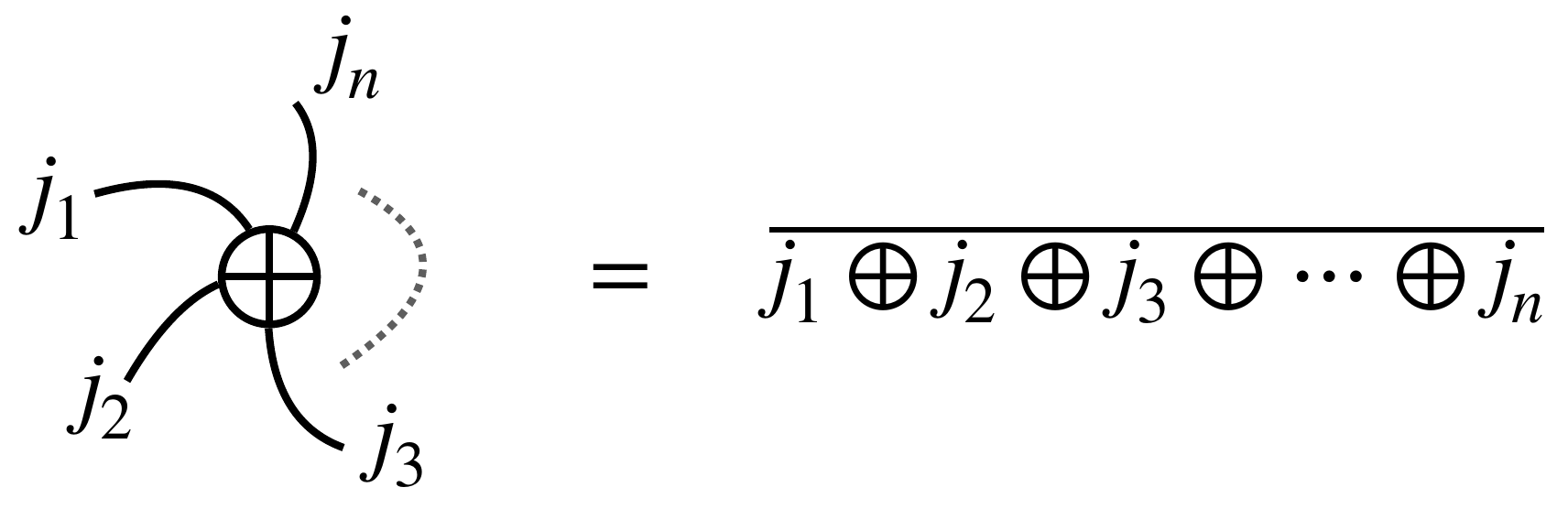}
$$
Also in this case, by using the associative property of the logical disjunction, we can easily decompose a $n$-legs XOR tensor into the following XOR chain
$$
\includegraphics[width=0.55\textwidth]{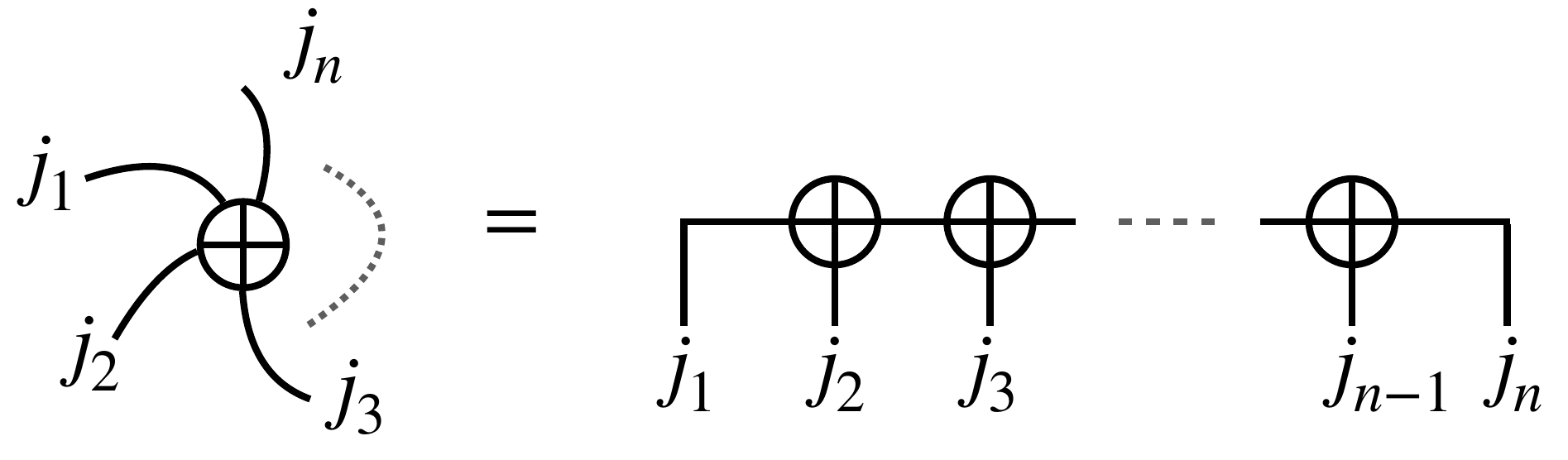}
$$
where we broke down the operation into $3$-qubits XOR operations,
such that
$$
\includegraphics[width=0.6\textwidth]{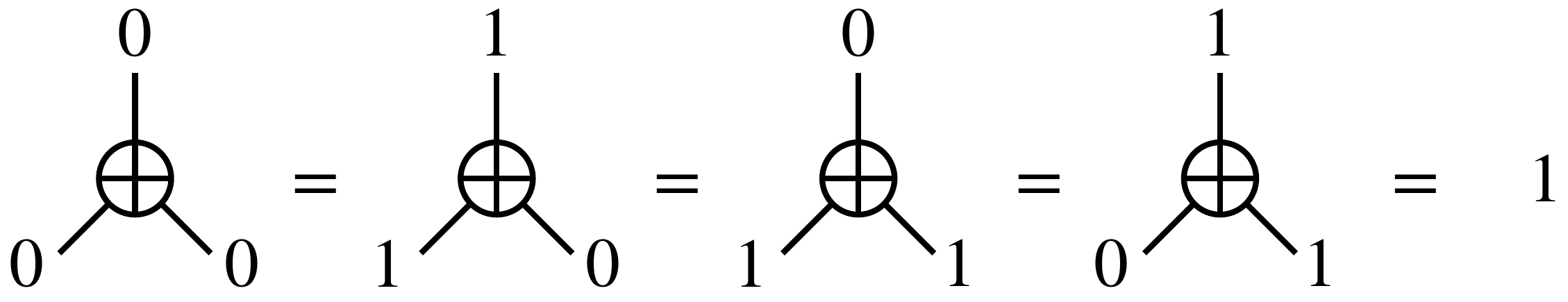}
$$

Let us mention the fact that, XOR and COPY tensors are related by local basis transformation induced by the Hadamard tensor $H$
(cfr.\ Chapter~\ref{chap2}), such that one has
$$
\includegraphics[width=0.5\textwidth]{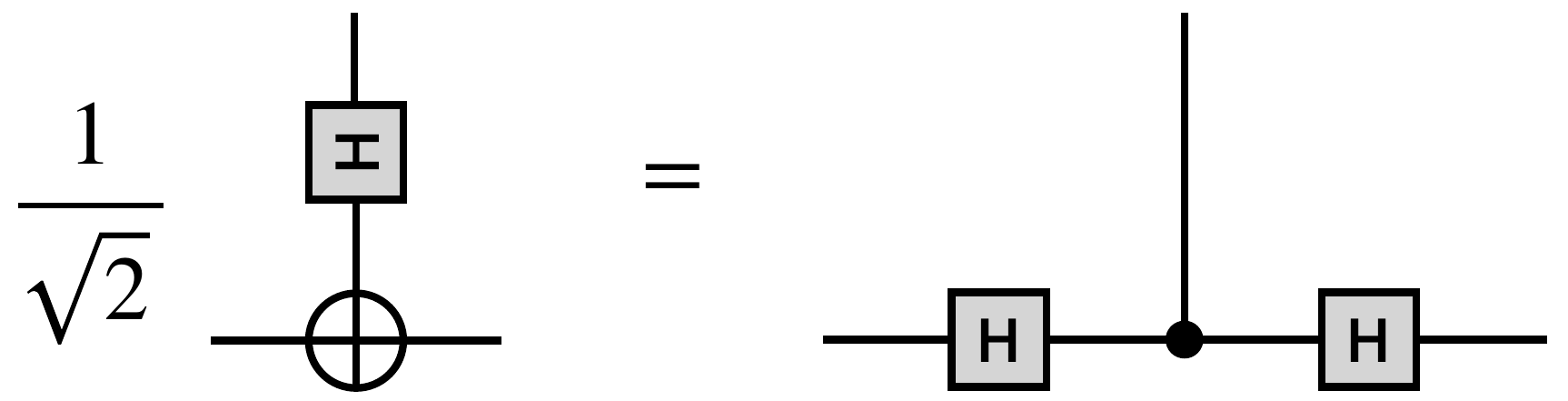}
$$

\paragraph{CNOT tensor ---}
The Control-NOT (CNOT) tensor\index{CNOT}, is a $4$-order tensor which can be constructed by attaching a copy tensor with a xor tensor.
It is of fundamental importance in quantum computation since it is used as entries of the homonym $2$-qubit gate (cfr.\ Chapter~\ref{chap2}).
It is diagrammatic represented~as
$$
\includegraphics[width=0.4\textwidth]{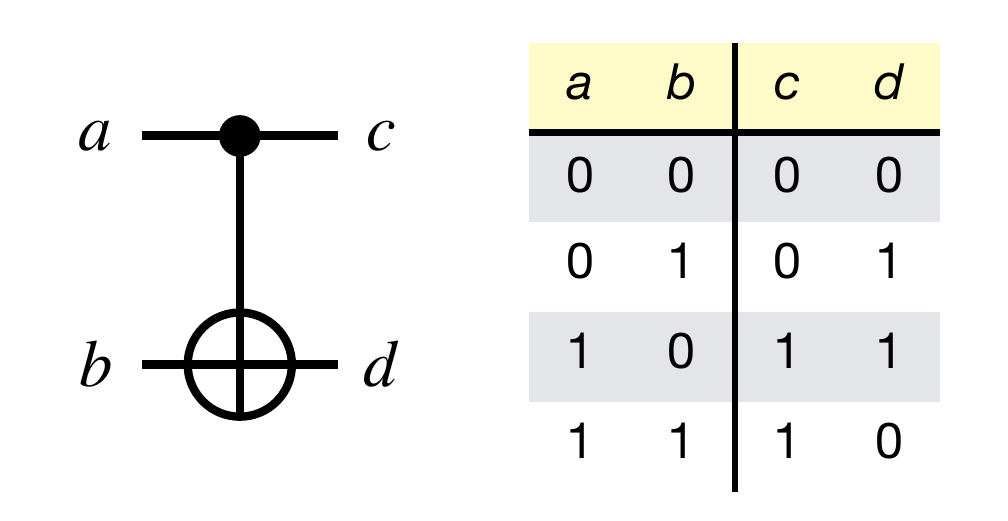}
$$
Basically, the copy tensor is attached to the control line, which let the input ``a'' to be copied to the output ``c''; in the meanwhile this control line is controlling indeed the xor tensor, whose input ``b'' is then transformed to the output ``d'' according to the value of the control line (see logic table above).

\section{Tensor Networks}
A tensor network is a structured assemblage of tensors, where a specific subset (up to the entire set) of their total indices is systematically contracted. The foundational operation within tensor networks involves the contraction of indices between two tensors, representing a fundamental and versatile generalization of traditional matrix multiplication. This operation lies at the core of tensor network methodologies, providing a framework for intricate manipulations and computations involving higher-order tensors.

\paragraph{Tensor index contraction ---}\index{Tensors!index contraction}
Diagrammatically, contracting two tensor indices corresponds to connect those indices with a single wire. Let's consider two tensors $A$ and $B$ of order respectively $k$ and $q$, and dimensions $\prod_{i=1}^k d^{(A)}_{i}$ and $\prod_{i=1}^{q} d^{(B)}_{i}$.
They can be \emph{contracted} along shared indices: for example, the rightmost $l$ indices of $A$ with the leftmost $l$ indices of $B$ by just connecting the corresponding legs as below
$$
\includegraphics[width=0.6\textwidth]{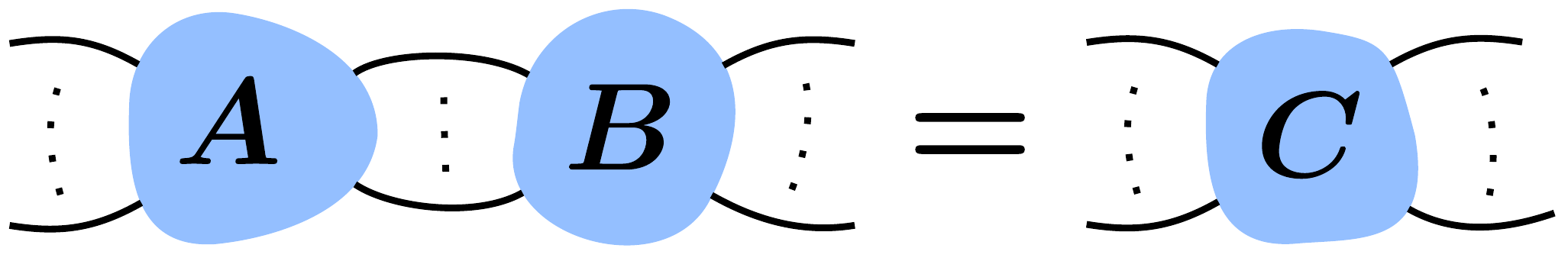}
$$
which correspond in summing over the spaces of the shared indices, namely
\begin{equation}\label{chapt1_eq:tensor_contraction}
    \sum_{j_1,\ldots,j_l}
    A_{m_1,\ldots,m_{k-l},j_1,\ldots,j_l} B_{j_1,\ldots,j_l,n_{1},\ldots,n_{q-l}} = C_{m_1,\ldots,m_{k-l},n_{1},\ldots,n_{q-l}}.
\end{equation}
The result is a order-$(k+q-2l)$ tensor.
Notice that the legs $\{j_1,\ldots,j_l\}$ can be contracted provided that they are referring to the same vector (or dual) spaces in both tensor $A$ and tensor $B$.

Alternatively, one can always achieve tensor contractions of arbitrary order and over arbitrary number of legs by resorting the operation to a vector-vector,  matrix-vector or matrix-matrix multiplication. This involves initially \emph{reshaping the tensors} into lower order tensors, by collecting via \emph{index fusion} all non-contracted indices into a single index and  all shared indices into another single index.
In our previous example,  we have considered the special case where the last $l$ indices of $A$ and the first $l$ indices of $B$ are the shared indices; this will correspond to reshape both tensors as
$A_{m_1,\ldots,m_{k-l},j_1,\ldots,j_l} \rightarrow A_{m,j}$
and
$B_{j_1,\ldots,j_l,n_1,\ldots,n_{q-l}} \rightarrow B_{j,n}$.
This is followed by regular a matrix-matrix multiplication along the shared index $j$, $\sum_{j} A_{m,j} B_{j,n} = C_{n,m}$. Finally, the free indices are reshaped back to the original indices $C_{n,m} \rightarrow C_{m_1,\ldots,m_{k-l},n_{1},\ldots,n_{q-l}}$. The modern scientific computing provides several options for performing efficiently algebraic manipulation of tensors.

\begin{example}{Contraction order do matter}{contractions}
At the core of all tensor network procedures lies the contraction of a network comprising multiple tensors, consolidating them into a single tensor. Here we illustrate an example of the problem where the goal is to contract the tensor network $\{A, B, C\}$ as depicted below
$$
\includegraphics[width=0.7\textwidth]{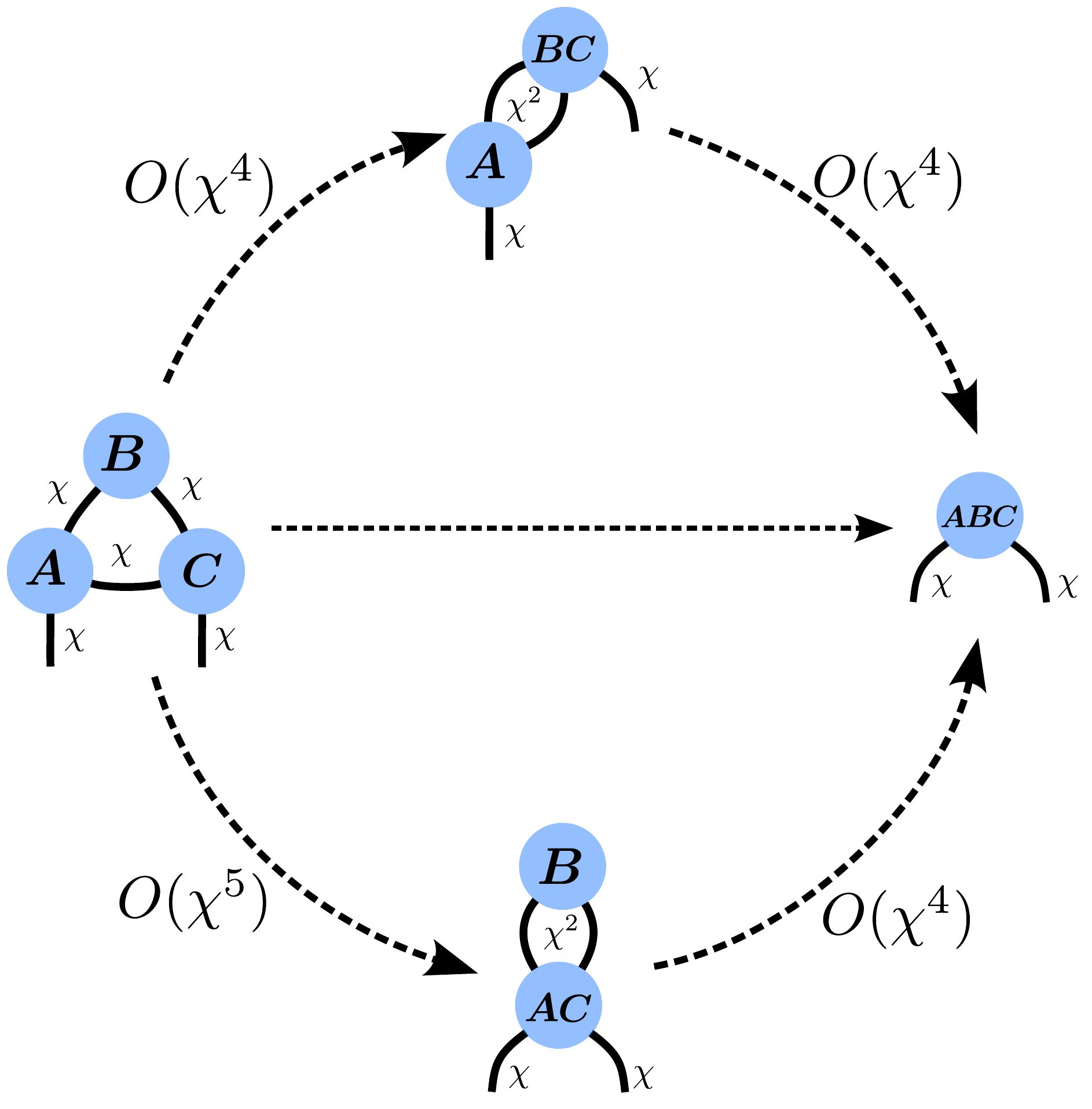}
$$
One straightforward approach to numerically evaluate this contraction would involve a direct summation over all the three shared indices, achievable through the implementation of a series of nested \texttt{for} loops.
Although this approach yields the correct result, it is generally not the preferred method due to computational inefficiency. Specifically, for this example, assuming all indices have bond dimension $\chi$, the direct summation over the internal legs requires $O(\chi^3)$ operations, for each entry of the tensor $ABC$, thus yielding to an overall computational cost~$O(\chi^5)$.

We can opt for an alternative approach and break down the assessment of the tensor ABC into two distinct steps. But also in this case, we may follow different paths; for example:
\begin{enumerate}
\item $A,B,C \to AC, B \to ABC$
\item $A,B,C \to A, BC \to ABC$
\end{enumerate}
Applying analogous reasoning as previously, it can be deduced that the computational cost for evaluating the intermediate tensor $AC$ in the path $1$ scales as $O(\chi^5)$, and only the subsequent contraction to get $ABC$ needs for a smaller cost of $O(\chi^4)$.
However, by following the path $2$ we only need an overall computational cost of $O(\chi^4)$.

Therefore, decomposing the network contraction into a series of smaller contractions, each involving only a pair of tensors (referred to as pairwise tensor contraction), proves to be computationally economical or even more cost-effective for any non-trivial bond dimension, provided that the chosen sequence of pairwise contractions have been properly selected.
\end{example}

\paragraph{Tensor contraction cost ---}\index{Tensors!contraction cost}
We have just considered the problem of efficiently contracting tensors, and we have realized from the Example~\ref{exmp:contractions}, that the elementary contraction within a tensor network is a \emph{pairwise contraction}.
This contraction is denoted as $(A \times B)$, and involves tensors $A$ and $B$ connected by one or more common indices. A straightforward method for executing such contractions, as exemplified in Equation~\eqref{chapt1_eq:tensor_contraction}, entails employing nested \texttt{for} loops to sum over the shared indices. The computational cost associated with this evaluation, measured in terms of the required number of scalar multiplications, can be precisely expressed as:
\begin{equation}\label{chapt1_eq:cost_contract}
    \text{cost}(A\times B) = \frac{\text{dim}(A)\times \text{dim}(B)}{\text{dim}(A \cap B)}
\end{equation}
where $\text{dim}(A)$ ($\text{dim}(B)$) represents the total dimension of $A$ ($B$) (i.e., the product of its index dimensions), and $\text{dim}(A\cap B)$ denotes the total dimension of the shared indices.

Alternatively, one can reinterpret a pairwise contraction as a matrix multiplication by following these steps: firstly, rearrange the free and contracted indices on each tensor so that they appear sequentially. Next, merge the free indices and the contracted indices into single separate indices. Subsequently, the contraction is performed through a single matrix-matrix multiplication operation, although the resulting product may need to be reshaped back into tensor form. This approach of recasting as matrix multiplication does not alter the formal computational cost outlined in Equation~\eqref{chapt1_eq:cost_contract}. However, in modern computing, utilizing highly optimized {\texttt BLAS} routines, matrix multiplications are typically executed more efficiently than equivalent {\texttt for} loop summations. Consequently, when dealing with tensors of large total dimensions, recasting as matrix multiplication is often favored for evaluating pairwise tensor contractions, despite requiring additional computational overhead due to the necessity of rearranging tensor elements in memory using ``reshaping'' (i.e.\ index fusion). However, for practical uses there are several packages available in majority of modern programming languages designed for scientific computing that allows for a very efficient and direct contraction of a complex tensor network.

\section{Tensor Network Decomposition}
Tensor decomposition involves the process of taking a single tensor and breaking it down into two or more constituent tensors. The original tensor indices are distributed among these constituents, which are connected through shared internal indices. Conceptually, tensor decomposition can be seen as the inverse operation of tensor contraction. In this section, we explore two specific decompositions that play a pivotal role in essential tensor network algorithms for quantum physics, namely the \emph{eigendecomposition}, \emph{singular value decomposition} (SVD), and the \emph{QR decomposition}. These decompositions are essentially matrix decomposition, for tensors of higher order decomposition is usually achieved by first reshaping the tensor into an appropriate matrix with index fusion followed by a matrix decomposition, and finally reshaping back to the higher order tensor by index splitting.

\paragraph{Eigendecomposition ---}\index{Diagonalization} Matrices that are diagonalizable can be factorized in terms of its eigen values and eigenvectors called eigendecomposition. Here we restrict to Hermitian matrix which is a special case of diagonalizable matrix relevant to quantum operators. In this case it is also called spectral decomposition. Consider a $n \times n$ Hermitian matrix $M$, the eigendecomposition of $M$ is given by,
\begin{equation}\label{eq:eig_decomp}
    M_{\alpha,\beta} = U_{\alpha,\gamma} \Sigma_{\gamma,\gamma} U_{\gamma,\beta}^{\dagger},
\end{equation}
where $U$ is a $n\times n$ unitary matrix ($UU^{\dagger} = U^{\dagger}U = \mathbb{1}$) whose columns are orthogonal eigenvectors of $M$ and $\Sigma$ is a $n\times n$ diagonal matrix whose diagonal entries are the corresponding eigenvalues. The computational cost for an eigendecomposition scales as $O(n^3)$. The following tensor network diagram shows the eigendecomposition of a generic Hermitian matrix $M$ and the associated properties. The vertical line represents the identity matrix $\mathbb{1}$.
$$
\includegraphics[width=1.0\textwidth]{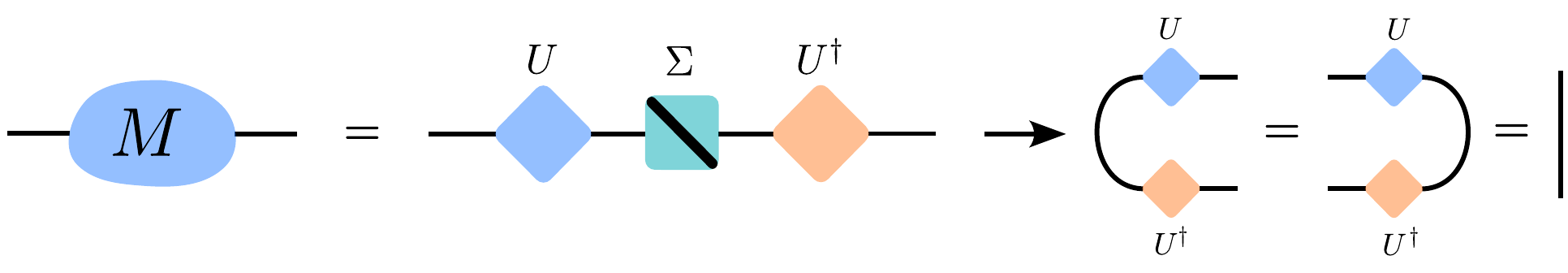}
$$

\paragraph{Singular Value Decomposition (SVD) ---}
Unlike eigendecomposition singular Value decomposition exists for any generic matrix. SVD\index{Singular Value Decomposition} factorizes a $m\times n$ rectangular matrix $M$ into three different components as follows,
\begin{equation}\label{eq:eig_decomp}
    M_{\alpha,\beta} = U_{\alpha,\gamma} S_{\gamma,\gamma} V_{\gamma,\beta}^{\dagger},
\end{equation}
where $U$ is a $m\times m$ unitary matrix, $S$ is a $m\times n$ rectangular diagonal matrix with positive entries, and $V^{\dagger}$ is a $n \times n$ unitary matrix. The diagonal entries of $S$ are known as the singular values and is usually written in descending order without loss of generality. The number of non-zero singular values $r = \text{min}(m,n)$ is known as the rank of matrix $M$. This allows for a compact representation of SVD, $M_{\alpha,\beta} = \sum_{\gamma = 1} ^{r} U_{\alpha,\gamma} S_{\gamma,\gamma} V_{\gamma,\beta}^{\dagger}$. In compact representation $U$ has dimension $m \times r$ and has orthonormal columns, $U^{\dagger}U = \mathbb{1}$ and are called left normalized, whereas $V^{\dagger}$ has dimension $r \times n$ and has orthonormal rows, $VV^{\dagger} = \mathbb{1}$ and are called right normalized. The computational cost of compact SVD is $O(mn^2)$ (assuming $m>n$). The following tensor network diagram shows the SVD of a generic rectangular matrix $M$ and the associated properties.
$$
\includegraphics[width=1.0\textwidth]{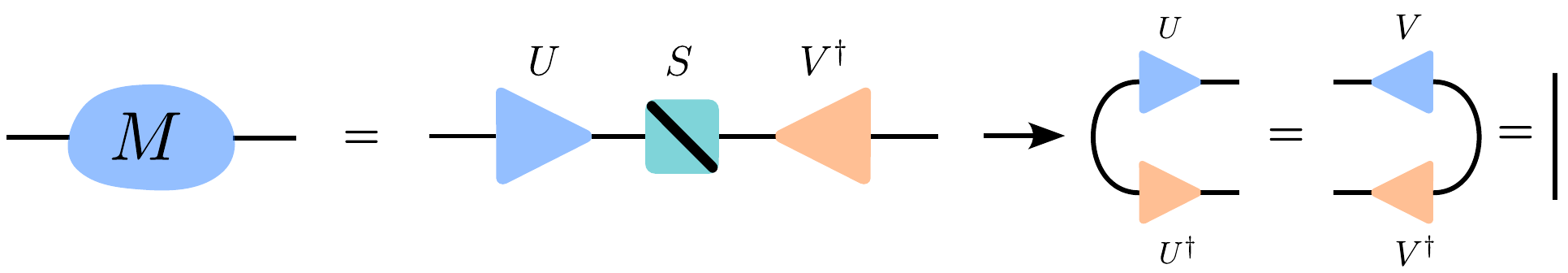}
$$

The eigendecomposition is related to SVD. The SVD of $M$ implies the following two expressions, $MM^{\dagger} = U(SS^{*})U^{\dagger}$ and $M^{\dagger}M = V(S^{*}S)V^{\dagger}$. Therefore the singular values are the square roots of the eigenvalues of $MM^{\dagger}$ (and $M^{\dagger}M$). If $M$ is Hermitian it has a unique eigendecomposition, $M = \overline{U} \Sigma \overline{U}^{\dagger}$ and the absolute eigenvalues of $M$ are the singular values. If $M$ is positive semi-definite the eigenvalues becomes equal to the singular values. In this scenario the SVD becomes equivalent to eigendecomposition.

\paragraph{QR decomposition ---} QR decomposition\index{QR decomposition} decomposes a matrix $M$ of dimension $m \times n$ into two constituent components as,
\begin{equation}\label{eq:qr_matrix}
M_{\alpha,\beta} =Q_{\alpha,\gamma}R_{\gamma,\beta}
\end{equation}
where, $Q$ is a $m \times m$ unitary matrix, $R$ is a $m \times n$ upper triangular matrix. The lowest $(m-n)$ rows of matrix $R$ are comprised of zeros, allowing for a more compact representation of the decomposition as follows: $M_{\alpha,\beta} = \sum_{\gamma} Q^1_{\alpha,\gamma}R^1_{\gamma,\beta}$, where $Q^1$ denotes a rectangular matrix with dimensions $m \times n$ and features orthonormal columns ($Q^{\dagger}Q = \mathbb{1}$), while $R^1$ is an upper triangular square matrix sized at $n \times n$. Although $Q^1$ shares the characteristic of left orthonormality with matrix $U$ in the SVD, it is important to note that these matrices are not identical. The following tensor network diagram shows the QR decomposition of a generic rectangular matrix $M$ and the associated properties.
$$
\includegraphics[width=1.0\textwidth]{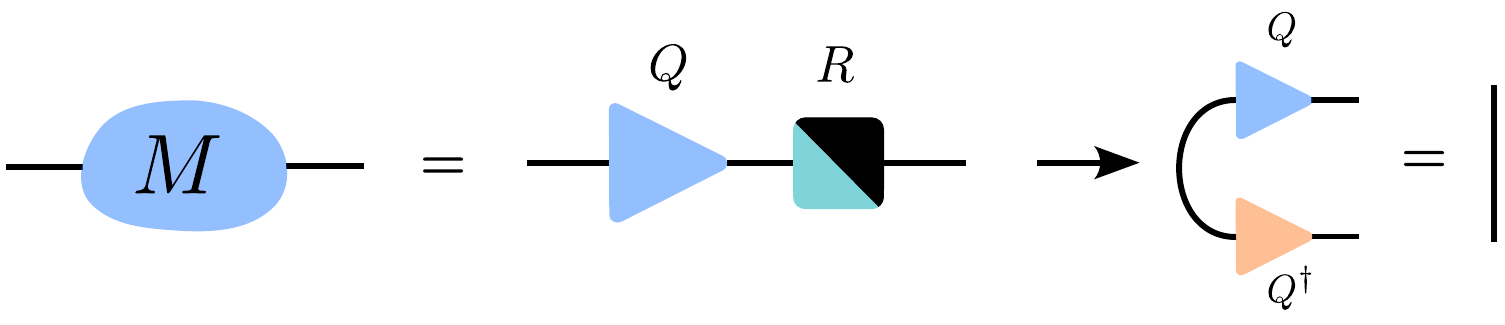}
$$

Analogously, we can define $LQ$ and $RQ$ decompositions, where $Q$ is now right normalized, $QQ^{\dagger} = \mathbb{1}$ and $L$ and $R$ are lower and upper triangular matrices respectively.

\subsection{Matrix product tensor network}

The matrix decompositions outlined above can be used to decompose a generic tensor into constituent tensors of lower order~\cite{TensorDecomposition_2009, TensorTrain_2011}. One such decomposition that has huge application in many body physics is matrix product decomposition where a generic tensor of order-$k$ is decomposed as the product of $k$ matrices (or more precisely order-3 tensors)\index{MPS}. This decomposition is obtained by a sequence of SVD or QR decompositions in a generic matrix. In the following example we will employ SVD to decompose a generic order-$k$ tensor into right canonical matrix product network.

\begin{example}{Left canonical matrix product decomposition}{MPT}
Consider a generic order-$k$ tensor $T_{j_1,j_2,\dots,j_k}$, we begin by reshaping the tensor into a matrix by index fusion, $T_{j_1,(j_2,\dots,j_k)}$followed by a SVD\index{Singular Value Decomposition},
\begin{equation*}\label{eq:first_svd}
T_{j_1,(j_2,\dots,j_k)} = \sum_{c_1} U_{j_1,c_1} S_{c_1,c_1} V^{\dagger}_{c_1,(j_2,\dots,j_k)} = \sum_{c_1} A_{j_1}^{c_1} T_{c_1,j_2,\dots,j_k}.
\end{equation*}

The left normalized matrix $U_{j_1,c_1}$ is renamed as $A^{c_1}_{j_1}$ and $S$ and $V^{\dagger}$ is contracted and reshaped into $T_{c_1,j_2,\dots,j_k}$. In the second step we reshape $T_{c_1,j_2,\dots,j_k}$ into a matrix $T_{(c_1,j_2),(j_3,\dots,j_k)}$ followed by a SVD,
\begin{align*}\label{eq:second_svd}
\sum_{c_1} A_{j_1}^{c_1} T_{(c_1,j_2),(j_3\dots,j_k)} &= \sum_{c_1,c_2} A_{j_1}^{c_1} U_{c_1,j_2,c_2}  S_{c_2,c_2} V^{\dagger}_{c_2,(j_3,\dots,j_k)}\\
&= \sum_{c_1,c_2} A_{j_1}^{c_1} A_{j_2}^{c_1,c_2}   S_{c_2,c_2} V^{\dagger}_{c_2,(j_3,\dots,j_k)}\\
&= \sum_{c_1,c_2} A_{j_1}^{c_1} A_{j_2}^{c_1,c_2} T_{c_2,j_3,\dots,j_k}.
\end{align*}

This procedure is sequentially repeated at each indices to obtain the full decomposition,
\begin{equation*}\label{eq:final_svd}
T_{j_1,j_2,\dots,j_k} = \sum_{c_1,c_2,\dots,c_{k-1}}  A_{j_1}^{c_1} A_{j_2}^{c_1,c_2} \dots A_{j_{k-1}}^{c_{k-2},c_{k-1}} A_{j_k}^{c_{k-1}}.
\end{equation*}

The full procedure is summarized in the following tensor network diagram,
$$
\includegraphics[width=1.0\textwidth]{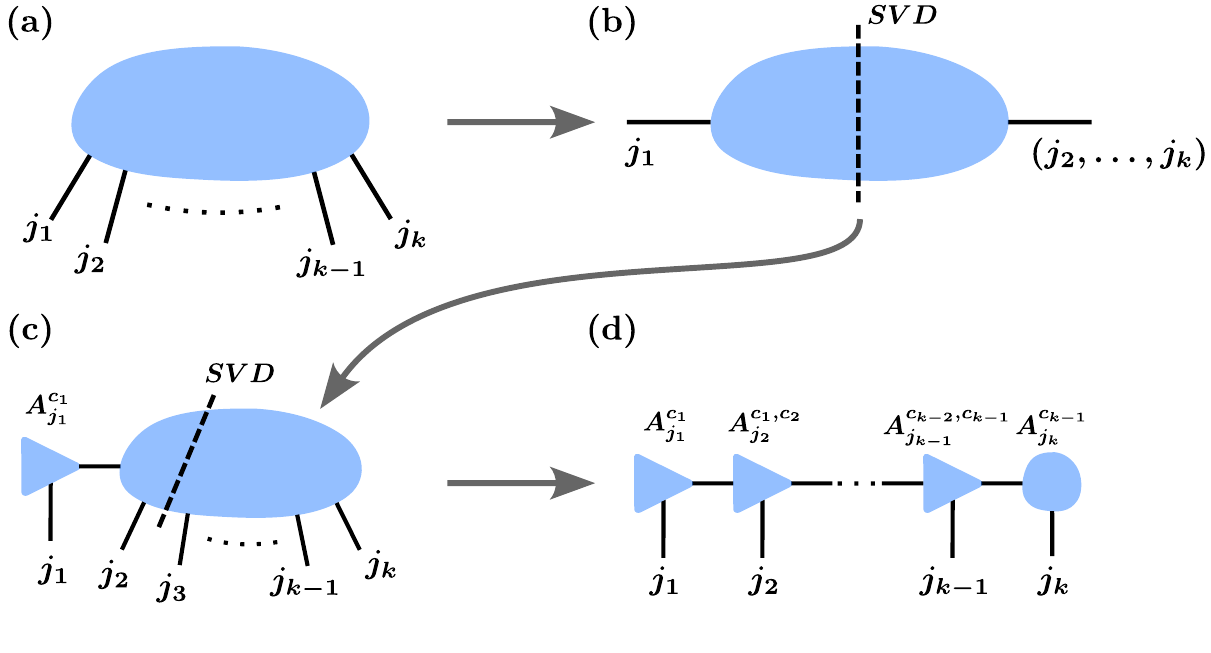}
$$

If we assume the dimension of the free indices to be $\chi$ then the dimension of the constituent tensors are $\chi \times \chi,\dots, \chi^{k/2-1} \times \chi \times \chi^{k/2}, \chi^{k/2} \times \chi \times \chi^{k/2-1},\dots, \chi \times \chi$. This is called a left canonical decomposition since the constituent tensors (apart from the rightmost tensor $A_{j_k}^{c_{k-1}}$) are left normalized, $\sum_{c_{l-1},j_l} A_{j_{l}}^{c_{l-1},c_{l}} \Big[A_{j_{l}}^{c_{l-1},c'_{l}}\Big]^{\dagger} = \delta^{c_l,c'_l} $.
\end{example}

The example presented above is not unique, alternatively we can start the SVD from the rightmost index $j_k$ and proceed from right to left to obtain the right canonical matrix product decomposition,
\begin{equation*}\label{eq:right_canonical_svd}
T_{j_1,j_2,\dots,j_k} = \sum_{c_1,c_2,\dots,c_{k-1}}  B_{j_1}^{c_1} B_{j_2}^{c_1,c_2} \dots B_{j_{k-1}}^{c_{k-2},c_{k-1}} B_{j_k}^{c_{k-1}}.
\end{equation*}

This is summarized by the following tensor network diagram,
$$
\includegraphics[width=1.0\textwidth]{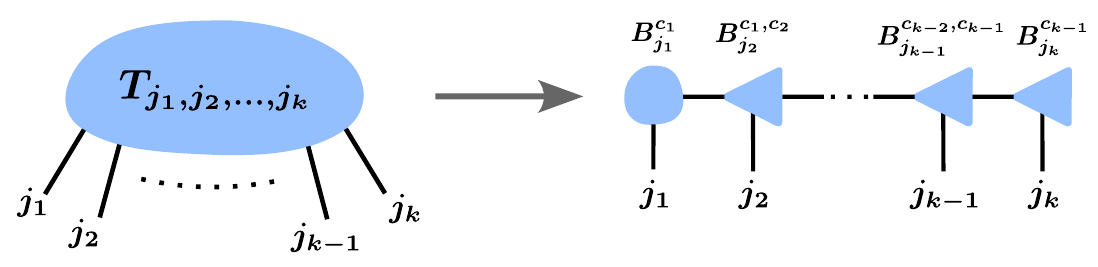}
$$
Each constituent tensors (apart from the leftmost tensor $B_{j_1}^{c_1}$) are right normalized, $\sum_{c_{l},j_l} B_{j_{l}}^{c_{l-1},c_{l}} \Big[B_{j_{l}}^{c'_{l-1},c_{l}}\Big]^{\dagger} = \delta^{c_{l},c'_{l}} $.

\subsection{Tree tensor network (TTN)}

Another relevant decomposition of a large tensor is the tree tensor network (TTN)\index{TTN} where the constituent tensors of the network lie on the nodes of a connected acyclic tree graph\cite{montangero2018tensor}. In the following examples we outline steps to decompose an order-8 tensor into a binary TTN (two tensors branching out of each tensor).

\begin{example}{Tree tensor network decomposition}{TTN1}
We start from a generic tensor of order-8. The tensor is split into two order-5 tensors via QR decomposition as shown in (a). Similar steps of consecutive QR and LQ decompositions are applied to get six order-3 tensors as shown in (b),(c), and (d) respectively. The underlying structure is a binary tree with an input branch and two output branches stemming out of each nodes. In the final form five of the six constituent tensors are right or left orthogonal as shown in (e). The non-orthogonal tensor (in orange full circle) is known as the orthogonality center. A tensor in a tensor network is called an orthogonality center if all the indices stemming out of the tensor annihilate to identity when contracted with its conjugate. We can also separate the orthogonality center into an order-2 tensor (in orange full triangle) lying at the geometrical center of the tensor network.
$$
\includegraphics[width=0.8\textwidth]{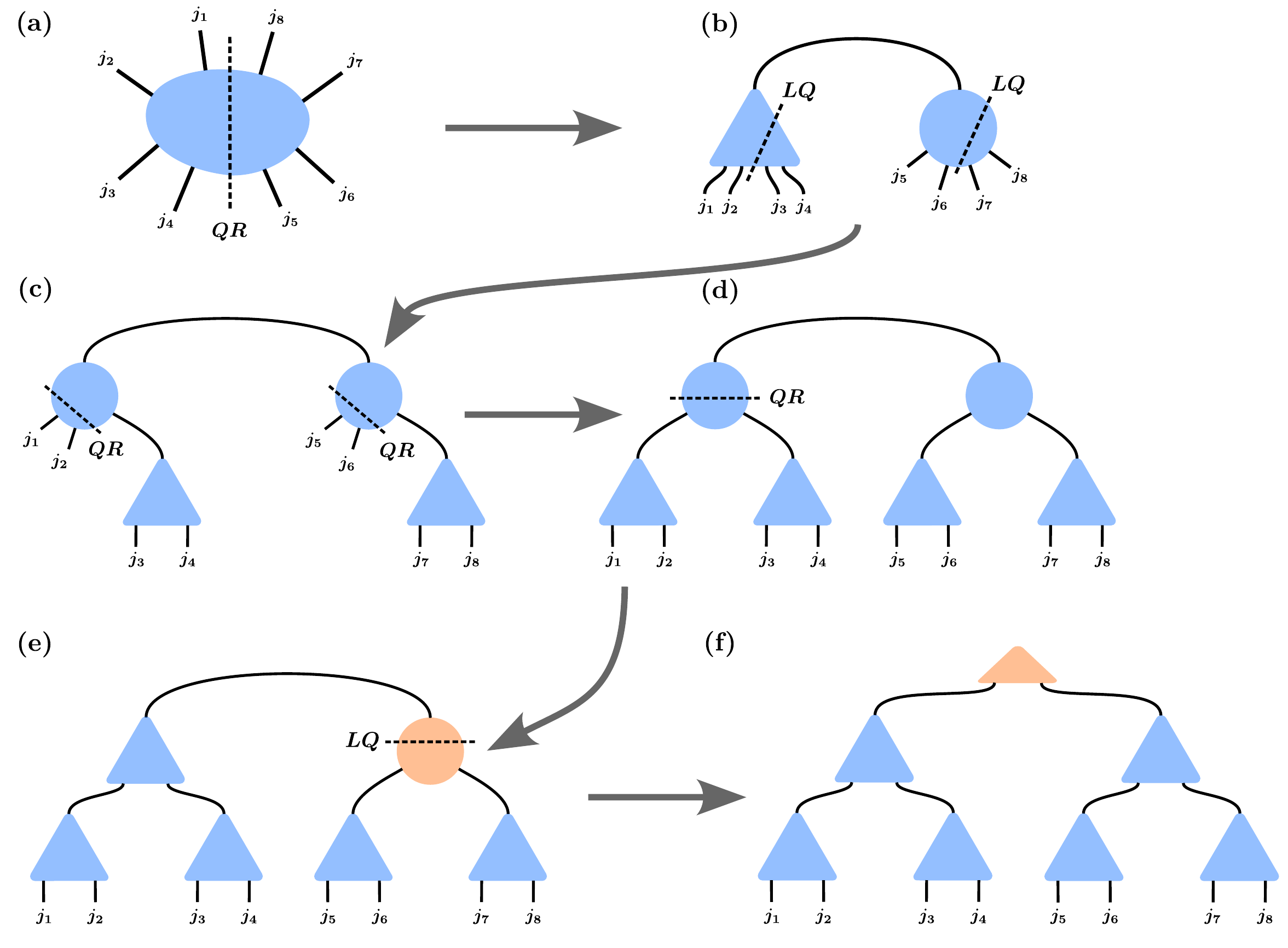}
$$

The position of orthogonality center within a tree tensor network is flexible. We can shift the orthogonality center to any node by a sequence of QR and LQ decompositions. The following figure depicts the shifting of orthogonality center in the tree tensor network from one node to another.
$$
\includegraphics[width=0.7\textwidth]{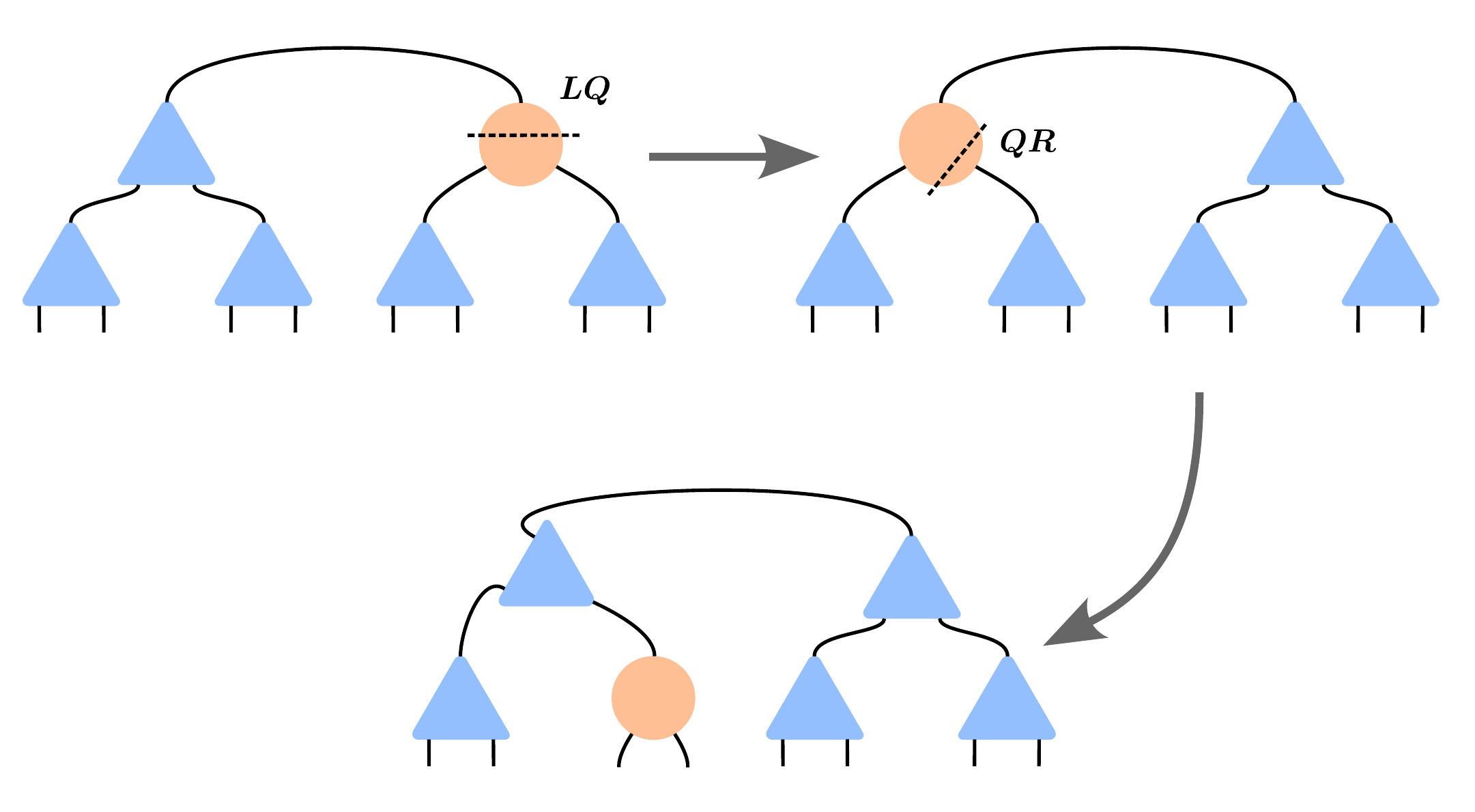}
$$
\end{example}

%\begin{example}{Shifting the orthogonality center}{TTN2}
%$$
%\includegraphics[width=0.8\textwidth]{figs/chapt_1/%TTN_gauge_centre.pdf}
%$$
%\end{example}

\biblio
\chapter{Quantum Physics with Tensors}\label{chap2}
\epigraph{Reality is not always probable,\\ or likely.}{Jorge Luis Borges}%,\\ The Lottery in Babylon, 1944}

\section{Introduction to quantum mechanics}

Quantum mechanics is universally recognized as one of the main achievement in humanity's understanding of the natural world~\cite{penrose2004road}. Despite its incredible generality and astonishing predictive power, this theory is based on a few simple foundational elements. In the following, we will shortly summarize these postulates, without explaining how they were originally derived by the founding fathers of quantum mechanics, about a hundred years ago.

\subsection{States and observables}
First, any quantum system is associated with an Hilbert space $\mathcal{H}$, i.e.\ a linear vector space over the field of complex numbers $\mathbb{C}$ with a metric induced by the inner product.\index{Hilbert space} Elements (vectors) of the Hilbert space represent possible states of the system, and are usually denoted with the ket notation $\ket{\psi}$. The inner scalar product between two states $\ket{\psi}$, $\ket{\phi}$ is denoted as $\braket{\psi}{\phi} \in \mathbb{C}$. Physical states have norm 1, meaning that $\norm{\ket{\psi}}^2 = \braket{\psi}{\psi}=1$. Since $\mathcal{H}$ is a linear space, in principle any linear combination $\alpha \ket{\psi} + \beta \ket{\phi}$ ($\alpha, \beta \in \mathbb{C}$) of two states $\ket{\psi}$, $\ket{\phi}$ is also a possible state (a part for an overall normalization constant).

Sometimes it is convenient to represent a state $\ket{\psi}$ in terms of its components with respect to a basis of the Hilbert space. Specifically, we consider an orthonormal basis, denoted as $\{ \ket{k} \}_{k=1}^D$ where $D$ is the Hilbert space dimension.
The vectors of this basis satisfies: $\braket{k}{l} = \delta_{kl}$ (orthogonality), and $\sum_{k=1}^D \ket{k} \bra{k} = \Id$ (completeness). The decomposition of $\ket{\psi}$ is the following
\begin{equation}
 \ket{\psi} = \sum_k \psi_k \ket{k} \, ,
\end{equation}
and the components are given by
\begin{equation}
\psi_k = \braket{\psi}{k} \, .
\end{equation}
Essentially, by choosing a specific basis, we can express all the states in the Hilbert space in terms of their components (coordinates). Consequently, we effectively establish an equivalence between $\mathcal{H}$ and $\mathbb{C}^D$. This is of great practical utility, as numbers can be easily handled for numerical calculations.

The second part of this first postulate asserts that any physical observable $O$ (such as energy, momentum, magnetization, etc.) is associated with a Hermitian operator $\hat{O}$ acting on $\mathcal{H}$ (i.e.\ $\hat{O}: \mathcal{H} \rightarrow \mathcal{H}$ and $\hat{O}^{\dag} = \hat{O}$, where $\dag$ indicates the Hermitian adjoint). Also operators can be expressed in coordinates as\index{Observables}
\begin{equation}
 \hat{O} = \sum_{k,l=1}^D O_{kl} \ket{k} \bra{l} \, ,
\end{equation}
with
\begin{equation}
O_{kl} = \matrixel{k}{\hat{O}}{l} \, .
\end{equation}
Notice that $\ket{k} \bra{l}$ is a convenient notation for the linear operator acting as $\ket{k} \braket{l}{k'} = \ket{k} \delta_{lk'}$ on a given vector of the basis $\ket{k'}$. The coordinates $O_{kl}$ of the operator are $D^2$ complex numbers. The fact that the operator must be Hermitian implies that the matrix of coefficients must satisfy the condition $O_{kl}^* = O_{lk}$. Even for operators, there exists a vector space structure, as Hermitian matrices of size $D \times D$ form a vector space over the field of real numbers $\mathbb{R}$ with dimension $D^2$.

As a final point in this section, we observe that in general one can contemplate a broader scenario where a quantum system exists in a certain statistical mixture $\{\pi_i,\ket{\psi_i} \}$ of possible states (i.e.\ the system can be in the state $\ket{\psi_i}$ with probability $\pi_i \geq 0$, and $\sum_i \pi_i = 1$).\index{Statistical mixture} In this context, the mathematical object that is convenient to consider for describing the system is the following operator\index{Density matrix}
\begin{equation}
    \hat{\rho} = \sum_i \pi_i \ket{\psi_i} \bra{\psi_i} \, ,
\end{equation}
which is known as density matrix. Here, $\ket{\psi_i} \bra{\psi_i}$ is the projector operator on the state $\ket{\psi_i}$. The density matrix is:
\begin{enumerate}[i.]
    \item Hermitian $\hat{\rho}^{\dag} = \hat{\rho}$;
    \item positive semi-definite, meaning that $\expval{\hat{\rho}}{\psi} \geq 0$, for any state $\ket{\psi}$;
    \item with trace $1$ since: $\Tr[\hat{\rho}] = \sum_i \pi_i \Tr[\ket{\psi_i} \bra{\psi_i}] = \sum_i \pi_i = 1$.
\end{enumerate}
 In general, any operator that meets these conditions i), ii), iii) is as a possible density matrix. Notice that since $\hat{\rho}$ is Hermitian, it can be diagonalized as:
 \begin{equation}
     \hat{\rho} = \sum_k p_k \ket{\varphi_k} \bra{\varphi_k} \, ,
 \end{equation}
 where $\ket{\varphi_k}$ are the eigenvectors of $\hat{\rho}$ and $p_k$ the associated eigenvalues. Conditions ii) and iii) can also be rephrased saying that a density matrix should have non negative eigenvalues $p_k \geq 0$ which sum to $1$, $\sum_k p_k =1$. This meas that eigenvalues of a density matrix can also be interpreted as a set of probabilities.
As for states, also density matrices can be linearly combined. However, since the trace of the sum has to be 1, only convex superposition are allowed, as for instance $p \hat{\rho}_1 + (1-p) \hat{\rho}_2$, with $p \in [0,1]$. Whenever a density matrix can be expressed as a projector into a single state $\hat{\rho} = \ket{\psi} \bra{\psi}$ (i.e.\ the statistical mixture is trivial), it is said that the system is in a pure state.

\subsection{Evolution}
The second postulate defines how a quantum mechanical system evolves. It states that any closed quantum system undergoes its evolution through a unitary transformation.\index{Unitary operator} This implies that in the absence of interaction with the environment, any physical process is mathematically represented by a unitary operator $\hat{U}$ acting on the Hilbert space and mapping a state $\ket{\psi}$ to a new state $U \ket{\psi}$. We recall that a linear operator is unitary when it preserves the inner product or equivalently when it satisfies $\hat{U}^{\dag} \hat{U} = \hat{U} \hat{U}^{\dag} =\Id$. This condition allows consistency with the first postulate, since the evolved state $U \ket{\psi}$ will still have a norm of 1 (since $\norm{\hat{U} \ket{\psi}}^2 = \expval{\hat{U}^{\dag}\hat{U}}{\psi} = \norm{\ket{\psi}}^2 = 1$). 

The most notable example of quantum evolution is given by the Schr\"odinger equation,\index{Schr\"odinger equation} which describes the evolution in time $t$ of a closed quantum system, relating the derivative with respect to time of the state $\ket{\psi(t)}$ to the action of a suitable Hermitian operator $\hat{H}$ called the Hamiltonian of the system~\cite{paffuti2009quantum}. The equation is:
\begin{equation}
    \frac{\partial}{\partial t} \ket{\psi(t)} = -i \hat{H} \ket{\psi(t)} \, .
\end{equation}
The Hermiticity property of the Hamiltonian, $\hat{H} = \hat{H}^{\dag}$, implies that
\begin{equation}
\frac{d}{dt} \braket{\phi(t)}{\psi(t)} = -i \langle \phi(t)|\hat{H}|\psi(t) \rangle + i \langle \phi(t)|\hat{H}^{\dag}|\psi(t) \rangle = 0 \,
\end{equation}
which implies at all time $t$
\begin{equation}
    \braket{\phi(t)}{\psi(t)} = \braket{\phi(0)}{\psi(0)} \, .
\end{equation}
Last equation signifies that the time evolution preserves the inner products between states. Thus, Schr\"odinger's time evolution is governed by a unitary operator $\hat{U}_t$ such that
\begin{equation}
   \ket{\psi(t)} = \hat{U}_t \ket{\psi(0)} \, .
\end{equation}
It is not difficult to realize that
\begin{equation}
    \hat{U}_t = \exp \left( -i \hat{H} t \right) \equiv \sum_{n=0}^{\infty} \frac{\big( -i \hat{H} t\big)^n}{n!} \, ,
\end{equation}
where we have defined the exponential of an operator through its Taylor expansion in powers~\cite{paffuti2009quantum}, i.e.\,
\begin{equation}
    \exp \left( \hat{A} \right) \equiv \sum_{n=0}^{\infty} \frac{\hat{A}^n}{n!} \, .
\end{equation}
In general, the Hamiltonian operator can depend on time $\hat{H}(t)$. This often occurs when the quantum system is coupled with external fields that change over time. In this case, the correct definition of the unitary operator describing the evolution is~\cite{paffuti2009quantum}
\begin{align}
    \begin{split}
    \hat{U}_t = & \mathcal{T} \left[ \exp( -i \int_0^t \hat{H}(s) ds ) \right] \equiv \\ \equiv& \sum_{n=0}^{\infty} \frac{(-i)^n}{n!} \int_0^t ds_1 \int_0^t ds_2 \ldots  \int_0^t ds_N \, \mathcal{T}\big[ \hat{H}(s_1) \hat{H}(s_2) \ldots  \hat{H}(s_N)\big] \, ,
    \end{split}
\end{align}
where we introduced the time-ordering operator $\mathcal{T}[\ldots ]$ which arranges operators in chronological order~\cite{paffuti2009quantum}. For instance:
\begin{equation}
\mathcal{T}[\hat{H}(s_1) \hat{H}(s_2)] = \begin{cases}
\hat{H}(s_1) \hat{H}(s_2) & \text{if } s_1 > s_2 \\
\hat{H}(s_2) \hat{H}(s_1) & \text{if } s_1 < s_2
\end{cases}
\end{equation}

This type of evolution (unitary dynamics) is not the only possible one. Indeed, in the density matrix formalism, it is possible to identify a much more general class of physically allowed transformations. These also include cases where the system is coupled with an ancillary system, evolved unitarily, and then the ancillary system is traced out. The resulting dynamics for the system's density matrix $\hat{\rho}$ is thus non-unitary. This type of generalized evolution is described by a set of Kraus operators $\hat{K}_i$ that must satisfy the completeness relation $\sum_i \hat{K}_i^{\dag} \hat{K}_i = \Id$.\index{Kraus operator} The evolution is given by the mapping
\begin{equation}\label{eq:kraus_eq}
    \hat{\rho} \,  \rightarrow \, \mathcal{E}(\hat{\rho}) = \sum_i \hat{K}_i^{\dag} \hat{\rho} \hat{K}_i \, .
\end{equation}
$\mathcal{E}$ is a super-operator, meaning that it maps operators to operators. In fact, the form given in Eq.~\eqref{eq:kraus_eq}
represents the most general form for a class of super-operators known as completely positive trace-preserving super-operators. It can be rigorously demonstrated that if $(\hat{\rho}$ is a well-defined density matrix, then $\mathcal{E}(\hat{\rho})$ will also be a well-defined density matrix. We refer the reader to Chapter~\ref{chap5} for a treatment of these topics.

\subsection{Measurements}
The third postulate of quantum mechanics concerns what happens when an external observer interacts with the quantum system to measure a physical quantity of interest. The possible outcomes of the measurement experiment are labeled by an index $k$, and for each of them we assume the existence of a certain operator $\hat{M}_k$ describing the evolution of the system just after the measurement. The postulate asserts that, with probability $p_k = \expval{\hat{M}_k^{\dag} \hat{M}_k}{\psi}$, the measurement results in $k$ and the system transitions (jumps) from its initial state $\ket{\psi}$ to\index{Measurement}
\begin{equation}\label{eq:jump_measurements}
    \frac{\hat{M}_k \ket{\psi}}{\sqrt{\expval{\hat{M}_k^{\dag} \hat{M}_k}{\psi}}} \, .
\end{equation}
Notice that the post-measurement state is again properly normalized to 1. The following completeness equation must be fulfilled by the measurement operators
\begin{equation}
    \sum_k \hat{M}_k^{\dag} \hat{M}_k = \Id \, ,
\end{equation}
so that the probabilities $p_k$ sum to $1$. In case of mixed state $\hat{\rho}$, Eq.~\eqref{eq:jump_measurements} is generalized to:
\begin{equation}
    \frac{\hat{M}_k \hat{\rho} \hat{M}_k^{\dag}}{\Tr[\hat{M}_k^{\dag} \hat{M}_k \hat{\rho}]} \, ,
\end{equation}
while the probability of outcome $k$ is $p_k = \Tr[\hat{M}_k^{\dag} \hat{M}_k \hat{\rho}]$.

The simplest example of measurement process is given by projective measurements. These are a sub-class of the general scenario described above. In this case, the observer measures a Hermitian operator $\hat{O}$, and the measurements operators are the projectors on the eigenstates of $\hat{O}$.\index{Observables} Since $\hat{O}$ is Hermitian, we can use its spectral decomposition $\hat{O} = \sum_k O_k \hat{P}_k$, where where $\hat{P}_k$ is the projector onto the eigenspace of $\hat{O}$ having eigenvalue $O_k$ and $\hat{P}_k \hat{P}_l = \delta_{kl} \hat{P}_k$. Since $\hat{P}_k$ are projectors, we have $p_k = \expval{\hat{P}_k}{\psi}$. Furthermore, it is particularly simple to compute the expectation value of $\hat{O}$ (namely, the value to which the empirical average of the outcomes converges as one repeats an increasing number of measurement experiments). Indeed,
\begin{equation}
    \overline{O} = \sum_k p_k O_k = \sum_k \expval{\hat{P}_k}{\psi} O_k = \expval{\sum_k O_k \hat{P}_k}{\psi} = \expval{\hat{O}}{\psi} \, ,
\end{equation}
where $\overline{\ldots }$ indicates average over possible outcomes.
The fluctuations around this value are governed by the variance of the observed outcomes, namely
\begin{equation}
    \overline{(O - \overline{O})^2} = \sum_k p_k \big( O_k - \overline{O} \big)^2  = \expval{\hat{O}^2}{\psi} - \expval{\hat{O}}{\psi}^2 \, .
\end{equation}

\subsection{Composite systems}\label{subsec:composite_systems}
It is possible for a quantum system to consist of multiple quantum constituents.\index{Many-body state} For instance, a molecule is composed of different atoms interacting with each other, and an atom consists of many electrons. Our last postulate specifies the relationship between the Hilbert space of the entire system $\mathcal{H}$ and the Hilbert space of the individual constituents $\mathcal{H}_i$. The statement is that\index{Hilbert space!many-body systems}
\begin{equation}
    \mathcal{H} \equiv \bigotimes_{i=1}^N \mathcal{H}_i \, ,
\end{equation}
i.e.\ the state space of a composite system is formed by the tensor product of the state spaces of the component physical systems. If we establish an orthonormal basis ${ \ket{k_i} }_{k_i=1}^{D_i}$ for each individual Hilbert space $\mathcal{H}_i$, it follows that the basis for the composite Hilbert space $\mathcal{H}$ is given by
\begin{equation}
\ket{k_1, k_2 \ldots  k_N} \equiv \ket{k_1} \ket{k_2} \ldots  \ket{k_N} \, .
\end{equation}
The orthogonality condition becomes
\begin{equation}
\braket{k_1, k_2 \ldots  k_N}{k_1', k_2' \ldots  k_N'} = \delta_{k_1' k_1} \delta_{k_2' k_2} \ldots  \delta_{k_N' k_N} \, .
\end{equation}
Therefore the dimension of $\mathcal{H}$ is the product of the dimension of individual Hilbert spaces, $D = \prod_{i=1}^N D_i$. \\

Now, suppose that our composite system is in a state $\hat{\rho}$. We divide the entire system into a specific subsystem $A$ and its complement $B$, such that $\mathcal{H} = \mathcal{H}_{A} \otimes \mathcal{H}_{B}$. If our focus is solely on describing the behavior of subsystem $A$, we have somehow to marginalize over $B$, i.e.\ to sum over over all potential states of $B$. Indeed for any density matrix $\hat{\rho}$ of the full system and any observable $\hat{O}_A: \mathcal{H}_{A} \rightarrow \mathcal{H}_{A}$ acting only on $A$, we have
\begin{equation}
\Tr[\hat{\rho} (\hat{O}_A \otimes \Id_{B})] = \Tr_A[\hat{\rho}_A \hat{O}_A ]
\end{equation}
where $\hat{\rho}_A$ is a density matrix for the subsystem only named reduced density matrix and defined as\index{Reduced density matrix}
\begin{equation}
\hat{\rho}_A \equiv \Tr_{B}[\hat{\rho}] \, .
\end{equation}
$\Tr_{B}[\ldots ]$ is a mathematical operation called partial trace, defined from the following equality\index{Partial trace}
\begin{equation}
\langle \alpha | \hat{\rho}_A | \alpha' \rangle = \sum_{\beta=1}^{D_{B}} \langle \alpha, \beta | \hat{\rho} | \alpha', \beta \rangle \, ,
\end{equation}
for any basis $\{ \ket{\alpha} \}_{\alpha=1}^{D_A}$ and $\{ \ket{\beta} \}_{\alpha=1}^{D_{B}}$ of $A$, $B$.
It is not difficult to show that if $\hat{\rho}$ is a well-defined density matrix, then so is $\hat{\rho}_A$, in the sense that it satisfies the three fundamental properties:\index{Density matrix} i) $\hat{\rho}_A^{\dag} = \hat{\rho}_A$; ii) $\hat{\rho}_A > 0$; iii) $\Tr[\hat{\rho}_A^{\dag}] = 1$. Besides, for any evolution operator $\hat{U}_A$ acting on $A$ only, we have: $\hat{U}_A^{\dag} \hat{\rho}_A \hat{U}_A \equiv \Tr_{B}[(\hat{U}_A \otimes \Id_{B})^{\dag} \hat{\rho} (\hat{U}_A \otimes \Id_{B})]$, meaning that the result is the same regardless if we first apply $\hat{U}_A \otimes \Id_{B}$ and then trace away $B$, or we first then trace away $B$ and then apply $\hat{U}_A$. Similar conclusions can be drawn in case of measurements performed on system $A$ only.

\begin{example}{Purification}{Purification}\index{Purification}
Purification refers to considering a mixed state (such as the infinite temperature state in the previous example) as derived from a pure state of a system and its environment by taking the partial trace over the environment. Specifically, consider a mixed state $\hat{\rho}_A$ for a system $A$ with Hilbert space $\mathcal{H}_A$. It is always possible to find a complementary system $B$ with Hilbert space $\mathcal{H}_{B}$ and a pure state $\ket{\psi} \in \mathcal{H}_A \otimes \mathcal{H}_{B}$ such that the reduced density matrix of $\ket{\psi}$ over $A$ looks like $\hat{\rho}_A$, i.e.\
\begin{equation}
    \hat{\rho}_A = \Tr_{B}[\hat{\rho}]
\end{equation}
with $\hat{\rho} = \ket{\psi} \bra{\psi}$. Indeed, to do this it is enough to write down $\hat{\rho}_A$ in its eigenbasis as: $\hat{\rho}_A = \sum_k p_k \ket{\varphi^a_k}\bra{\varphi^a_k}$, where $\{ \ket{\varphi^a_k} \}_k$ are the eigenvectors of  $\hat{\rho}_A$ and $p_k$ the associated eigenvalues. Afterwards, one can define $\ket{\psi} = \sum_k p_k^{1/2} \ket{\varphi^a_k} \ket{\varphi^b_k}$, where $\{ \ket{\varphi^b_k} \}_k$ is any other set of mutually orthogonal and normalized states defined in a suitable system $B$.
\end{example}

\enlargethispage*{\baselineskip}
\section{Quantum mechanics of a single qubit}\label{sec:single_qubit}
The simplest non-trivial quantum system is one in which the Hilbert space is spanned by only two states, which can be denoted as $\ket{0}$ and $\ket{1}$. Such a system is called a qubit (or quantum bit). A qubit is the fundamental unit of quantum information, analogous to the bit in classical computing. Unlike a classical bit, which can exist in only two distinct states $0$ or $1$, a qubit can be in a linear superposition of the two basis states, in accordance with the postulates of quantum mechanics. Any qubit state is therefore parameterized by two complex numbers $\alpha, \beta$ such that $\ket{\psi}=\alpha \ket{0} + \beta \ket{1}$ (for physical states $|\alpha|^2 + |\beta|^2 =1$). Thus, the Hilbert space has dimension $D=2$ ($\mathcal{H} \sim \mathbb{C}^2$). Any operator acting on $\mathcal{H}$ can be constructed using the identity matrix, plus the three Pauli matrices\index{Pauli matrices}, whose graphical tensor representation is given by
$$
\includegraphics[width=0.75\textwidth]{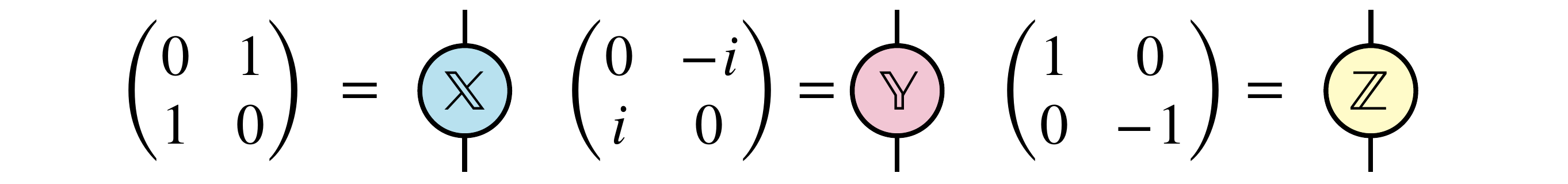}
$$
Formally, the operators associated to these tensors are constructed by exploiting the effect of these matrices in the tensor product space
$\mathcal{H}\otimes \mathcal{H}^*$.  This results in a similar construction to what was obtained for the identity in Example~\ref{exmp:quantum_computing},
$$
\includegraphics[width=0.45\textwidth]{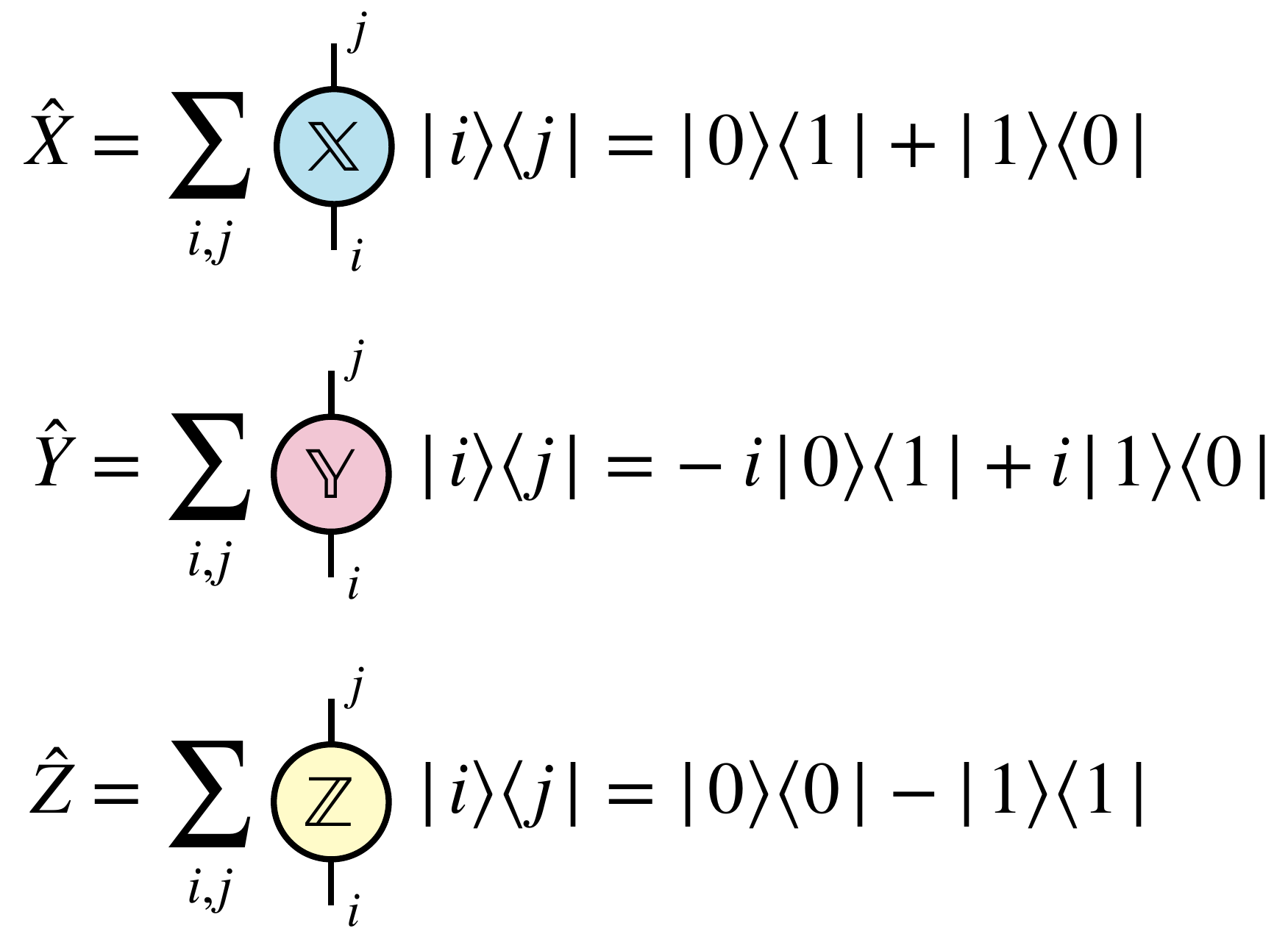}
$$
Observe that the representation of Pauli operators underscores the particular basis selected for their expression. Indeed, typically in quantum computation, $\ket{0}$ is designated as the $+1$ eigenvector of $\PauliZ$, and $\ket{1}$ as the $-1$ eigenvector of~$\PauliZ$. \\

%Notice that the Pauli matrices are also unitary.
Other relevant single qubit unitary transformations are the Hadamard gate $\Hgate$\index{Hadamard gate}, the phase gate $\Sgate$\index{Phase gate}
and the $\Tgate$ gate\index{Magic gate}, constructed in the similar way by exploiting the following tensors
$$
\includegraphics[width=0.8\textwidth]{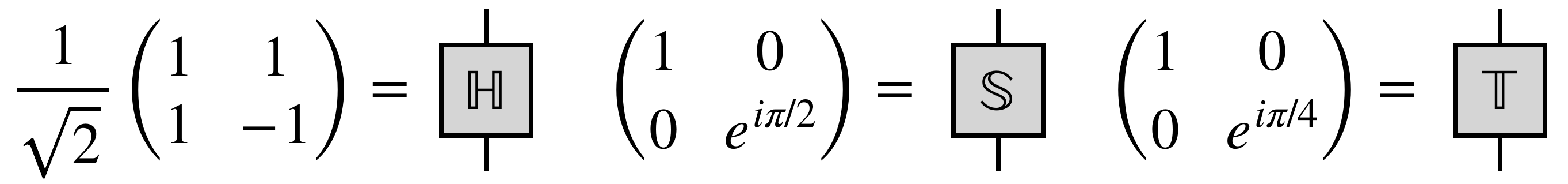}
$$

Note that we used the word `gate' as a synonym for unitary transformation, borrowing it from the language of classical computation, where operations are performed by logical gates acting on bits of information (0 or 1).

\subsection{Bloch sphere}
An useful reparametrization of the qubit state $\ket{\psi}=\alpha \ket{0} + \beta \ket{1}$ can be utilized to elucidate the effects of unitary transformations. Indeed, since $|\alpha|^2 + |\beta|^2 =1$, we can always find angles $\theta$ and $\phi$ such that\footnote{This expression holds true except for an overall global phase, which is entirely irrelevant as it cannot be detected by measurement experiments..}
\begin{equation}
    \ket{\psi}= \cos \frac{\theta}{2} \ket{0} +  e^{i \phi} \sin \frac{\theta}{2}\ket{1} \, .
\end{equation}
The fact that $\theta/2$ is used in this expression instead of $\theta$ can be clarified by calculating the expectation value of the Pauli matrices over the state. Indeed, we have:
\begin{align}
\begin{split}
    \expval{\PauliX}{\psi} &= \cos \frac{\theta}{2} \sin \frac{\theta}{2} e^{i \phi} \matrixel{0}{\PauliX}{1} + \cos \frac{\theta}{2} \sin \frac{\theta}{2} e^{-i \phi} \matrixel{1}{\PauliX}{0} = \\ &= 2 \cos \frac{\theta}{2} \sin \frac{\theta}{2} \cos \phi = \sin \theta \cos \phi \\
    \expval{\PauliY}{\psi} &= \cos \frac{\theta}{2} \sin \frac{\theta}{2} e^{i \phi} \matrixel{0}{\PauliY}{1} + \cos \frac{\theta}{2} \sin \frac{\theta}{2} e^{-i \phi} \matrixel{1}{\PauliY}{0} = \\ &= 2 \cos \frac{\theta}{2} \sin \frac{\theta}{2} \sin \phi = \sin \theta \sin \phi \\
    \expval{\PauliZ}{\psi}  &= \cos^2 \frac{\theta}{2}\expval{\PauliZ}{0} + \sin^2 \frac{\theta}{2}\expval{\PauliZ}{1} = \cos^2 \frac{\theta}{2} - \sin^2 \frac{\theta}{2} = \cos \theta \, .
\end{split}
\end{align}
Therefore, if we imagine placing the states $\ket{0}$ and $\ket{1}$ at the ``North Pole'' and ``South Pole'' of a sphere, respectively, we will find that the state $\ket{\psi}$ has polar coordinates $(\theta, \phi)$, with $\theta \in [0,\pi]$ and $\phi \in [0,2\pi]$. This sphere is known as Bloch sphere.\index{Bloch sphere} We will use the convention $\ket{\theta, \phi}$ to denote the state specified by the two angles $\theta, \phi$. \\

It should be noted that this is only an idealized framework and one should not think of the system as just a classic spin oriented in a certain direction. In fact, the Pauli operators exhibit non-zero (quantum) fluctuations from the expected value, since for instance $\expval{\PauliZ^2}{\psi} - \expval{\PauliZ}{\psi}^2 = 1 - \cos^2 \theta \geq 0$. This means that the state is effectively in a kind of probabilistic superposition around the point in the Bloch sphere. In fact, every time a (projective) measurement of $\PauliZ$ is performed, the system will always be found in one of the two eigenstates of $\PauliZ$, namely $\ket{0}$ or $\ket{1}$, never in rotated directions!\\

Now that we have an intuitive framework for single-qubit states, we can inquire about the unitary transformations that can implement rotations on the Bloch sphere. To answer, let us consider the following operator
\begin{equation}
    \hat{R}_Z(\alpha) = \exp\big( -i \frac{\alpha}{2} \PauliZ \big) \, .
\end{equation}
It is easy to realize that
\begin{equation}
    \hat{R}_Z(\alpha) = \begin{pmatrix}
        e^{-i \frac{\alpha}{2}} & 0 \\
        0 & e^{+i \frac{\alpha}{2}} \\
    \end{pmatrix} \, .
\end{equation}
Now we can apply this $\hat{R}_Z(\alpha)$ to the state $\ket{\theta, \phi}$, and result is
\begin{equation}
    \hat{R}_Z(\alpha) \ket{\theta, \phi} = \cos \frac{\theta}{2} e^{-i \frac{\alpha}{2}} \ket{0} +  e^{i \phi} \sin \frac{\theta}{2} e^{+i \frac{\alpha}{2}} \ket{1} = e^{-i \frac{\alpha}{2}} \ket{\theta, \phi + \alpha} \, .
\end{equation}
So, up to a global phase, the operator $\hat{R}_Z(\alpha)$ has rotated the state on the Bloch sphere by an angle $\alpha$ around the Z-axis. In general, it is easy to realize that the operator\index{Bloch sphere!rotation}
\begin{equation}
    \hat{R}_{\pmb{n}}(\alpha) = \exp\left(- i \frac{\alpha}{2} \pmb{n} \cdot \hat{\pmb{\sigma}} \right) \, ,
\end{equation}
where
\begin{equation}
\hat{\pmb{\sigma}} \equiv \big( \PauliX, \PauliY, \PauliZ \big)
\end{equation}
implements a rotation of an angle $\alpha$ around the axis specified by the unit vector $\pmb{n}$ ($|\pmb{n}|=1$). The rotation is in the counter-clockwise direction. It is also straightforward to demonstrate that
\begin{equation}
    \hat{R}_{\pmb{n}}(\alpha) = \cos (\frac{\alpha}{2}) \Id - i \, \pmb{n} \cdot \hat{\pmb{\sigma}} \sin (\frac{\alpha}{2}) \, .
\end{equation}
It is now easy to realize that the Hadamard gate corresponds to a rotation of $\pi$ around the axis $\frac{1}{\sqrt{2}}(1,0,1)$, while $\Sgate$ and $\Tgate$ are (apart for a global phase) rotations around the $Z-$axis of $\pi/2$ and $\pi/4$, respectively. In general, it is a well-known fact that every rotation can be decomposed into a composition of rotations around the axes $X$, $Y$, $Z$. It follows that every single qubit gate can be decomposed as:\index{Unitary operator!single qubit gate}
\begin{equation}
    \hat{U} = e^{i \delta} \hat{R}_{\pmb{x}}(\alpha) \hat{R}_{\pmb{y}}(\beta) \hat{R}_{\pmb{z}}(\gamma) \, ,
\end{equation}
where $\alpha, \beta, \gamma$ are known as Euler angles. In general, however, it is sufficient to implement rotations around two non-parallel generic axes $\pmb{n}$, $\pmb{m}$ in order to decompose a generic unitary as:
\begin{equation}
    \hat{U} = e^{i \delta} \hat{R}_{\pmb{n}}(\alpha) \hat{R}_{\pmb{m}}(\beta) \hat{R}_{\pmb{n}}(\gamma) \, ,
\end{equation}
% https://quantumcomputing.stackexchange.com/questions/17173/does-anyone-know-the-list-of-all-known-universal-sets-of-quantum-gates

\section{Many-body quantum systems and Entanglement}
Let us consider a system of $n$ qubits. According to the general postulates of quantum mechanics, a basis for the global Hilbert space $\mathcal{H}$ is given by the states
\begin{equation}
\ket{\pmb{x}} \equiv \ket{x_1} \ket{x_2} \dots \ket{x_{n}} \, ,
\end{equation}
where the indices $x_i \in \{0,1\}$ label the local computational basis (as before $\ket{0}$ is the $+1$ eigenvector of $\PauliZ$, and $\ket{1}$ is the $-1$ eigenvector of $\PauliZ$). Any state $\ket{\psi}$ can be therefore decomposed as\index{Many-body state}
\begin{equation}
\ket{\psi} = \sum_{\pmb{x}} \psi_{\pmb{x}} \ket{\pmb{x}} = \sum_{x_1} \sum_{x_2} \ldots  \sum_{x_n} \psi_{x_1 x_2 \ldots  x_n} \ket{x_1 x_2 \ldots  x_n} \, .
\end{equation}
Notice that $\psi_{x_1 x_2 \ldots  x_n}$ is a multi-dimensional tensor with $n$ indices. Using the tensor graphical representation introduced in Chapter~\ref{chap1} we can therefore represented this many-body wave function as:
\begin{equation}\label{chapt2_eq:psi_TN}
\includegraphics[width=0.5\textwidth,valign=c]{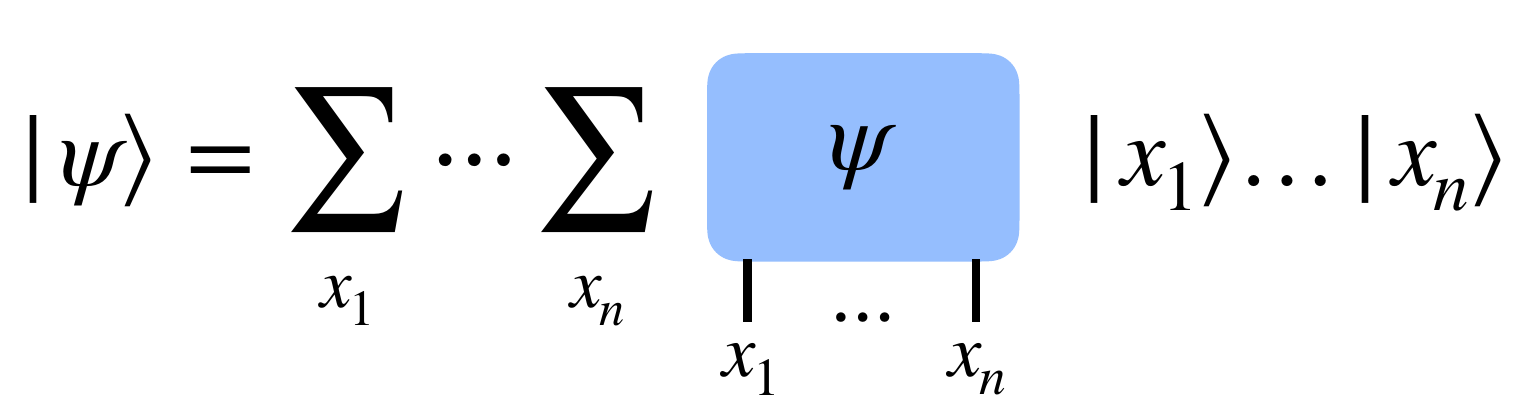}
\end{equation}
The tensor $\psi_{x_1 x_2 \ldots  x_n}$ contains $2^n$ complex numbers encoding all the information about the state of the system. It is usually an extraordinarily complex object. A primary source of this complexity is that typical many-body states $\ket{\psi}$ possess significant quantum correlations among their constituents (qubits). These correlations, known as entanglement, are crucial for understanding quantum behavior and will be further discussed in Section~\ref{sec:entanglement}.

\subsection{Universal set of gates }\label{chapter2_sec:universal_gates}

In addition to single-qubit gates discussed in Section~\ref{sec:single_qubit}, exploring the full complexity of the quantum many-body Hilbert space necessitates the use of unitary gates that can make the qubits interact. Obviously this requires the implementation of gates that act on at least two qubits. The fundamental two qubits gate is often considered to be the \textbf{CNOT gate}\index{CNOT}. In fact, we will see in Section~\ref{exmp:bell} that CNOT can be used to create prototypical entangled states, i.e.\ the Bell pairs. The CNOT gate is basically the operator associated to the CNOT tensor (cfr.\ Chapter~\ref{chap1})
$$
\includegraphics[width=0.85\textwidth]{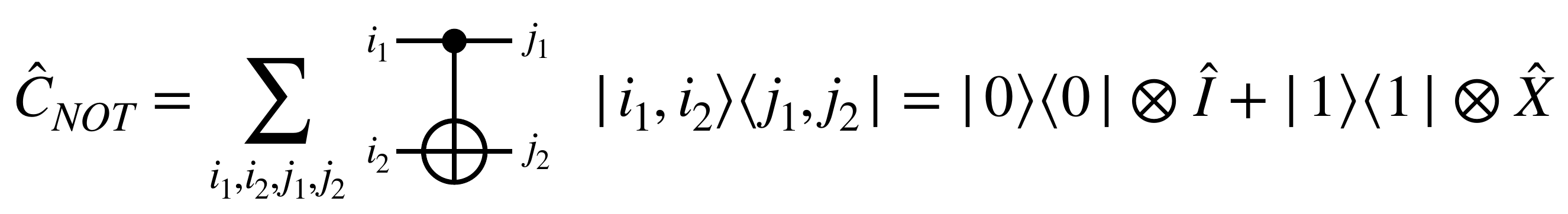}
$$
This means that the control bit remains unchanged, while the target bit is flipped if the control bit is 1.

At this point, it is appropriate to ask what is the minimal set of gates necessary to have available in order to implement every unitary transformation $\hat{U}$. In particular, given $\hat{U}$, one seeks a quantum circuit $\hat{\tilde{U}}$ composed of elementary gates that approximate it arbitrarily well, i.e.\ with error $||\hat{U} - \hat{\tilde{U}}|| < \epsilon$ (where $||\cdot||$ represents some appropriate matrix norm). It turns out that a set of gates satisfying this universality property is for instance given by:
\begin{equation}
    \{ \hat{\text{CNOT}}, \hat{H}, \hat{T} \}
\end{equation}
Indeed:

\begin{example}{Decomposition of $n$ qubits unitaries}{Decomposition of $n$ qubits unitaries}
Any $n$ qubits unitary $\hat{U}$ can
be decomposed into single qubit unitaries and CNOT gates, with a number $O(n^2 4^n)$ of elementary gates.
\end{example}

We emphasize that the decomposition described above is generally inefficient, requiring a number of basic operations that scale exponentially with the number of qubits $n$. Indeed, implementing a generic unitary transformation $\hat{U}$ involving $n$ qubits necessitates exponentially many elementary gates, as the matrix $\hat{U}$ is determined by $O(4^n)$ real parameters. A fundamental and still unresolved problem in quantum computation is identifying which special classes of unitary transformations can be computed in the quantum circuit model using a polynomial number of elementary gates.

Basically, in order to be universal a set of gates should be able to produce: complex wave function amplitudes, quantum superposition and quantum correlations (entanglement, see next sections). It is less obvious to observe that in addition to these properties, a set of gates must have an additional property, namely, not containing only unitaries from the Clifford group. These are defined as those unitaries that map Pauli matrices to Pauli matrices, i.e., they do not create superposition of Pauli operators. However, we refer the reader to Chapter \ldots  for a discussion of this aspect.

\subsection{Schmidt decomposition}\label{chapter2_sec:schmidt_decomposition}

We introduce a crucial tool in quantum mechanics known as the Schmidt decomposition~\cite{nielsen_chuang_2010}, which is fundamentally an application of the singular value decomposition (SVD)\index{Singular Value Decomposition} introduced in Chapter~\ref{chap1}. Let us consider a state $\ket{\psi}$ in a bipartite system, i.e.\ quantum mechanical system described by the tensor product of the two Hilbert spaces $\mathcal{H}_A \otimes \mathcal{H}_B$. Let us set a basis for the systems $A$ and $B$, i.e\ $\mathcal{H}_{A} = {\rm span} \{\ket{a_j}\}$ and $\mathcal{H}_{B} = {\rm span} \{\ket{b_j}\}$. The state $\ket{\psi}$ can now be written as
\begin{equation}
\ket{\psi}= \sum_{i,j} \Phi_{i j} \ket{a_i} \ket{b_j} \, \, ,
\end{equation}
or graphically as
\begin{equation}\label{chapt2_eq:psi_TN}
\includegraphics[width=0.7\textwidth,valign=c]{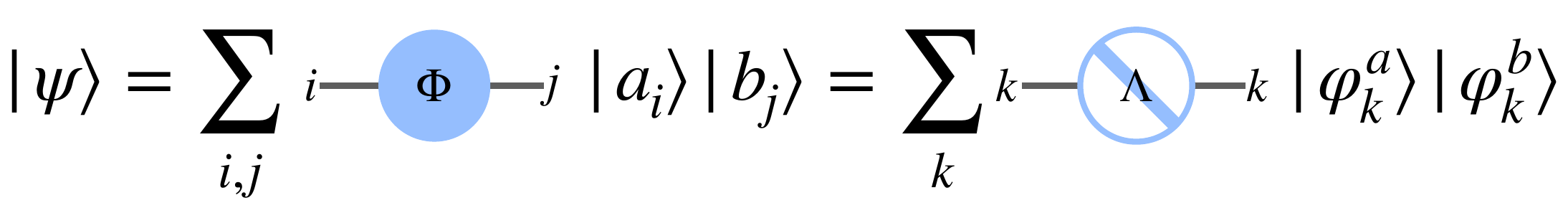}
\end{equation}\index{Entanglement!Schmidt decomposition}
In the last graphical equation we carried out a SVD of the matrix $\Phi = U \Lambda V^{\dag}$ and we defined new vectors $\ket{\varphi^a_k} = \sum_{i} U_{ik}\ket{a_i}$ and $\ket{\varphi^b_k} = \sum_{i} V^{*}_{ki}\ket{b_i}$. These are usually dubbed \emph{Schmidt vectors}. They are orthonormal, in fact
\begin{align}
\begin{split}
\braket{\varphi^a_{l}}{\varphi^a_{k}} &= \sum_{i,j} U_{jl}^* U_{ik} \braket{a_j}{a_i} = \sum_{i} (U^{\dag})_{li} U_{ik} = \delta_{lk} \\
\braket{\varphi^b_{l}}{\varphi^b_{k}} &= \sum_{i,j} V_{lj} V^{*}_{ki} \braket{b_j}{b_i} = \sum_{i} V_{li} (V^{\dag})_{ik} = \delta_{lk} \\
\end{split}
\end{align}
The real non-negative entries of the diagonal matrix $\Lambda_{kk}$ are the so called Schmidt values. The number of non vanishing Schmidt values is usually dubbed Schmidt rank.
Schmidt values (and vectors) are related to the eigenvalues (and eigenvectors) of the reduced density matrices of the subsystems A and B. In fact
\begin{equation}\label{chapt2_eq:psi_TN}
\includegraphics[width=0.85\textwidth,valign=c]{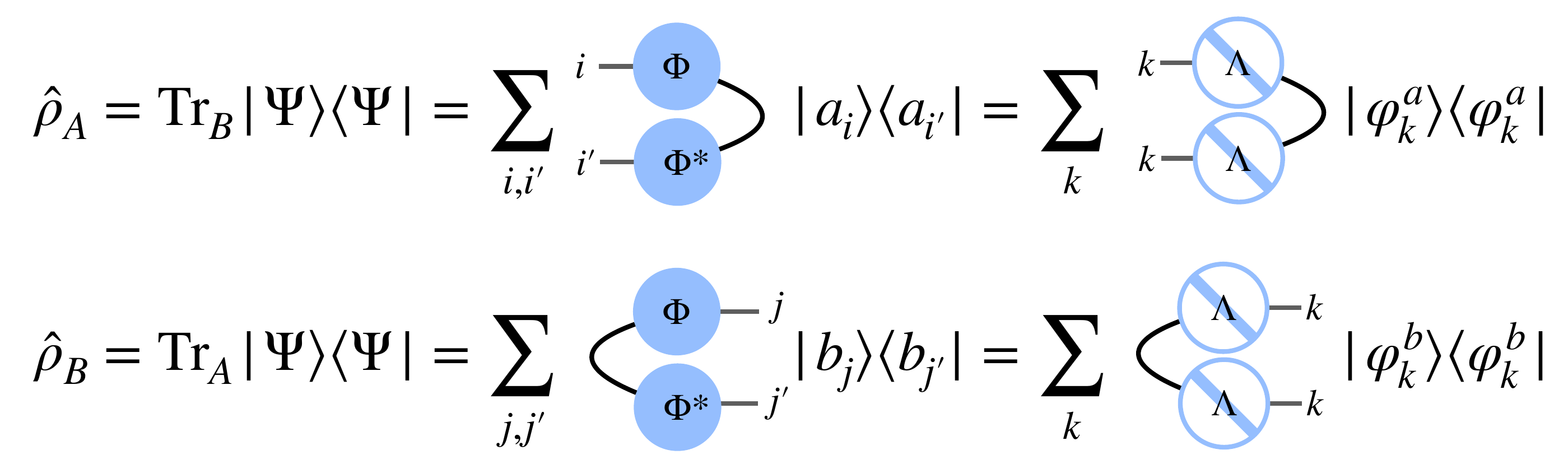}
\end{equation}
where last equality means
\begin{align}\label{eq:reduced_density_matrix_eig}
    \begin{split}
            \hat{\rho}_A & = \sum_k (\Lambda_{kk})^2 \ket{\varphi^a_k} \bra{\varphi^a_k} \\
            \hat{\rho}_B & = \sum_k (\Lambda_{kk})^2 \ket{\varphi^b_k} \bra{\varphi^b_k} \\
    \end{split}
\end{align}
Notice that if the initial state $\ket{\psi}$ is normalized, the reduced density matrices  $\hat{\rho}_A$, $\hat{\rho}_B$ are also normalized, in the sense that $\Tr[\hat{\rho}_A]=\Tr[\hat{\rho}_B]=1$. From this fact it follows that
\begin{equation}
\sum_k \Lambda^{2}_{kk} = 1 \, .
\end{equation}
Therefore the square of Schmidt coefficients can be also thought as a discrete probability distribution $p_k = \Lambda^{2}_{kk}$.

\subsection{Entanglement}\label{sec:entanglement}
Entanglement is the fundamental property of composite quantum systems, and accounts for the intrinsic quantum correlations among the constituencies (for instance qubits).\index{Entanglement}
Consider for instance a composite quantum system $\mathcal{H} = \mathcal{H}_A \otimes \mathcal{H}_B$ and a pure state $\ket{\psi} \in \mathcal{H}$. A natural question arises: can $\ket{\psi}$ be factorized as a product of states on $A$ and $B$, i.e.\ as $\ket{\psi_A} \otimes \ket{\psi_B}$?

\begin{definition}{Bipartite product state}{Bipartite product state}
Given a pure state  $\ket{\psi} \in \mathcal{H} = \mathcal{H}_A \otimes \mathcal{H}_B$ we define it to be a bipartite product state if $\ket{\psi} =\ket{\psi_A} \otimes \ket{\psi_{B}}$, otherwise we say the two partitions $A$ and $B$ of $\ket{\psi}$ are entangled.
\end{definition}

\begin{example}{Bell states, Bell measurements and quantum teleportation}{bell}
A prototypical example of entangled states is given by Bell pairs. Consider $N=2$ qubits and the following four states
\begin{align}
    \begin{split}
        \ket{\mathbb{1}} &= \frac{\ket{00} + \ket{11}}{\sqrt{2}} \\
        \ket{X} &= \frac{\ket{01} + \ket{10}}{\sqrt{2}} \\
        \ket{Y} &= i \frac{\ket{01} - \ket{10}}{\sqrt{2}} \\
        \ket{Z} &= \frac{\ket{00} - \ket{11}}{\sqrt{2}} \\
    \end{split}
\end{align}
The reason we adopted the names $\mathbb{1}, X, Y, Z$ becomes clear when one notices that these states are essentially the single qubit Pauli operators reshaped as vectors. Indeed we have for instance
\begin{equation}
    \frac{1}{\sqrt{2}} \big( \PauliX \big)_{x_0,x_1} = \braket{x_0,x_1}{X} \, ,
\end{equation}
meaning that components of state $\ket{X}$ match with matrix elements of $\PauliX$, a part for a normalization factor. The latter is chosen in order to make Bell states orthonormal. The reshaping can be graphically represent as follows
$$
\includegraphics[width=0.8\textwidth]{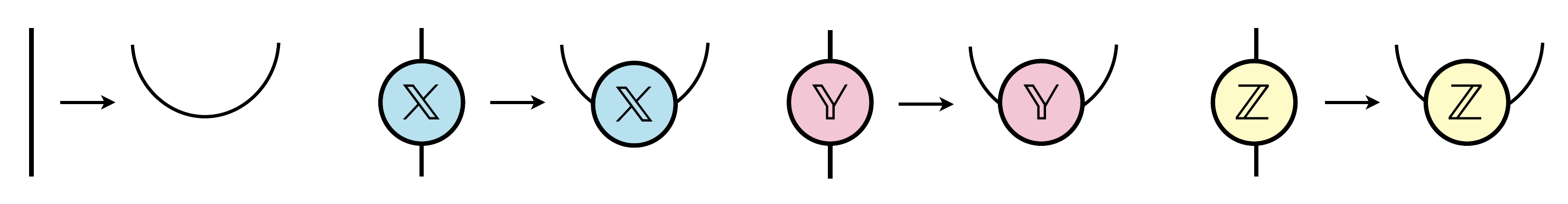}
$$
In this way, it is very easy to realize that inner scalar product between states is equal to Frobenius inner product of the corresponding Pauli matrices. For instance:
$$
\includegraphics[width=0.2\textwidth]{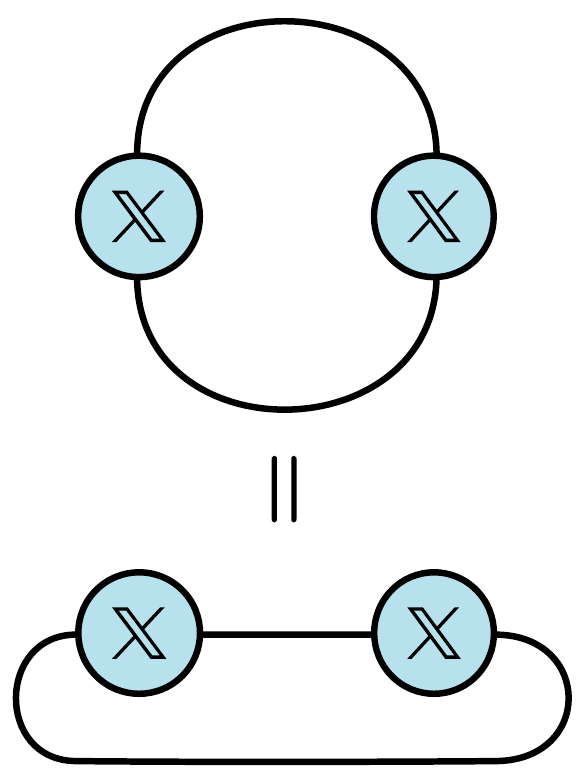}
$$
and therefore $\braket{X}{X}=\frac{1}{2} \Tr[X^{\dag} X]=\frac{1}{2} \Tr[\Id]=1$. \\

The set of Bell states form an orthonormal basis for the Hilbert space of a two qubits system.\index{Bell states} Consequently, they can be generated by applying a suitable unitary transformation $\hat{U}$ to the states of standard computational basis $\{ \ket{00}, \ket{01}, \ket{10}, \ket{11} \}$. In particular, it is easy to verify that one can realize $\hat{U}$ with a circuit consisting of one Hadamard gate, followed by a CNOT, i.e.\
$$\includegraphics[width=0.9\textwidth]{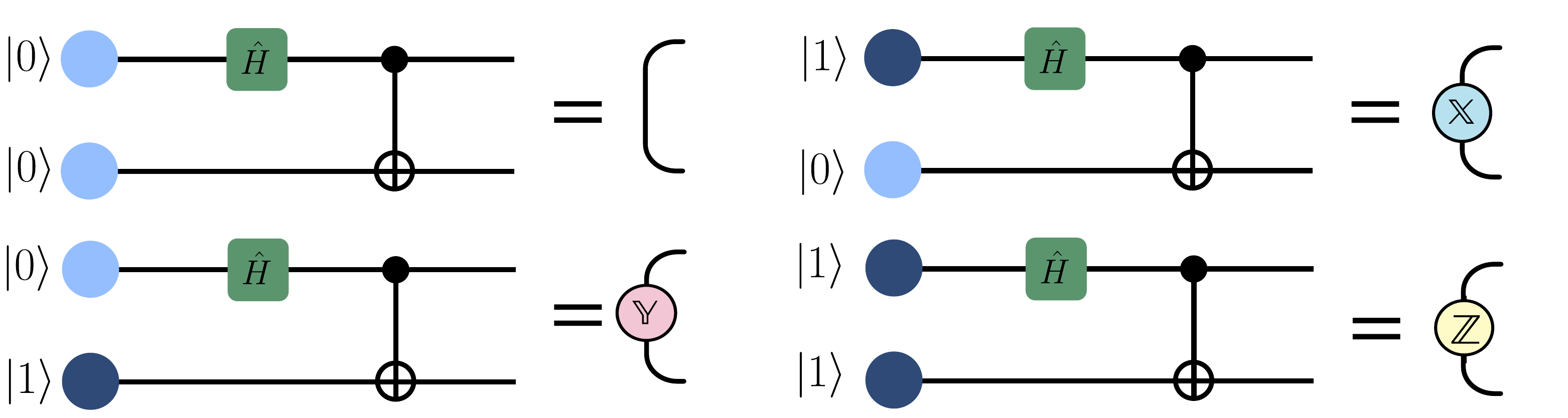}$$

Thus, if an unknown Bell state $\ket{\psi}$ is prepared one can determine which of the four Bell states the two qubits are in by first applying $\hat{U}^{\dag}$ as follows
$$\includegraphics[width=0.7\textwidth]{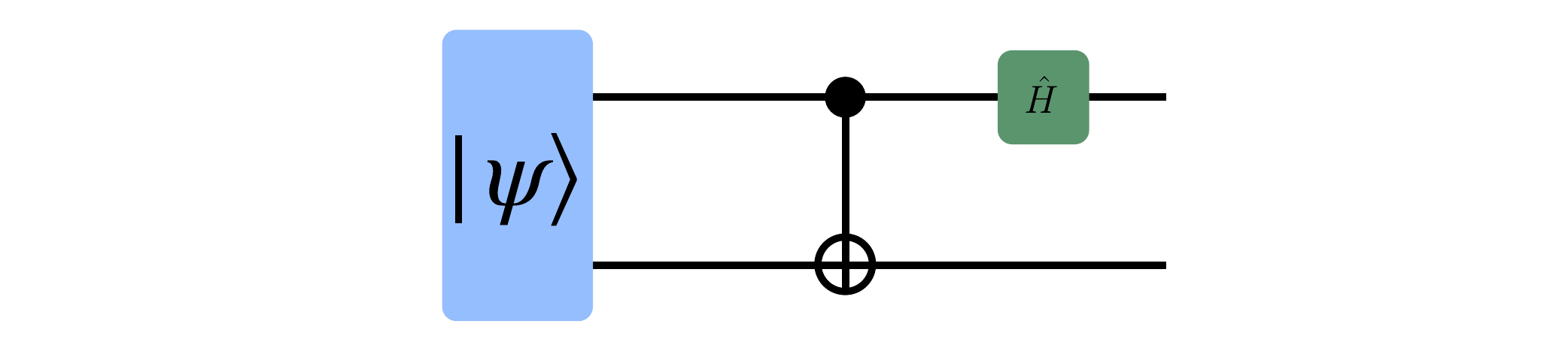}$$
and then perform a projective measurement of $\PauliZ$ on both qubits. If, for instance we find $\ket{0}$ and $\ket{0}$, that means the unknown state was $\ket{\mathbb{1}}$.

To conclude this paragraph, we illustrate an important use of entangled Bair pairs: the quantum teleportation.\index{Measurement!teleportation} Suppose Alice possesses a qubit, yet she is unaware of its state $\ket{\psi}$. Bob is in urgent need of this qubit. However, Alice is limited to sending only classical bits of information to Bob. A protocol to teleport the qubit state $\ket{\psi}$ can be realized by exploiting a shared Bell pair. Indeed, suppose Alice and Bob share each one qubit of the state $\ket{\mathbb{1}}$. Our initial set up can be therefore described by the following sketch
$$\includegraphics[width=0.61\textwidth]{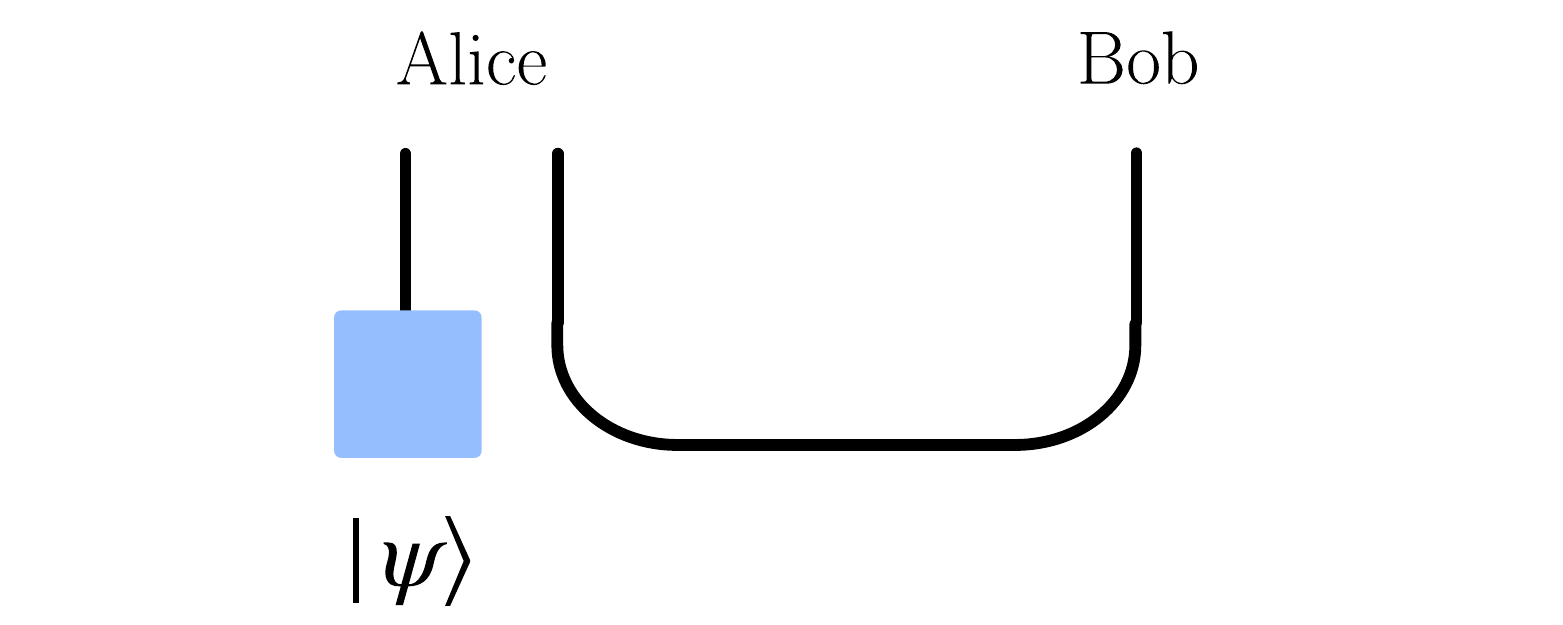}$$
Then suppose Alice perform a Bell measurement of the Bell state qubit and the qubit to be teleported. This yields one of four possible measurement outcomes and project the two qubits in $\ket{\mathbb{1}}, \ket{X}, \ket{Y}$ or $\ket{Z}$. Using a classical channel Alice can communicate the outcome to Bob, which can act with on its Bell qubit with the corresponding Pauli operator. The process can be summarized as follows:
$$\includegraphics[width=0.61\textwidth]{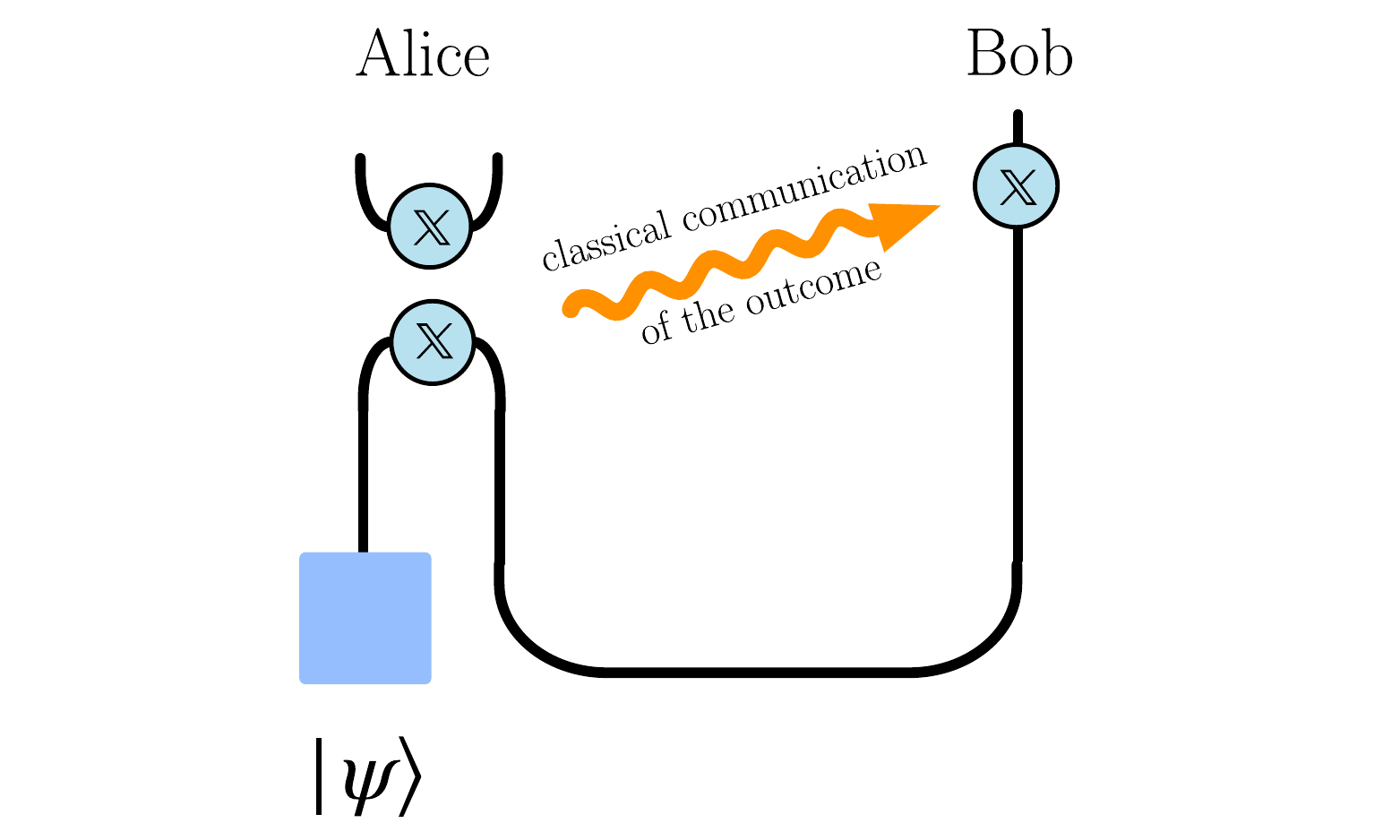}$$
in the case in which the outcome is $X$. Now notice the following graphical simplification
$$\includegraphics[width=0.61\textwidth]{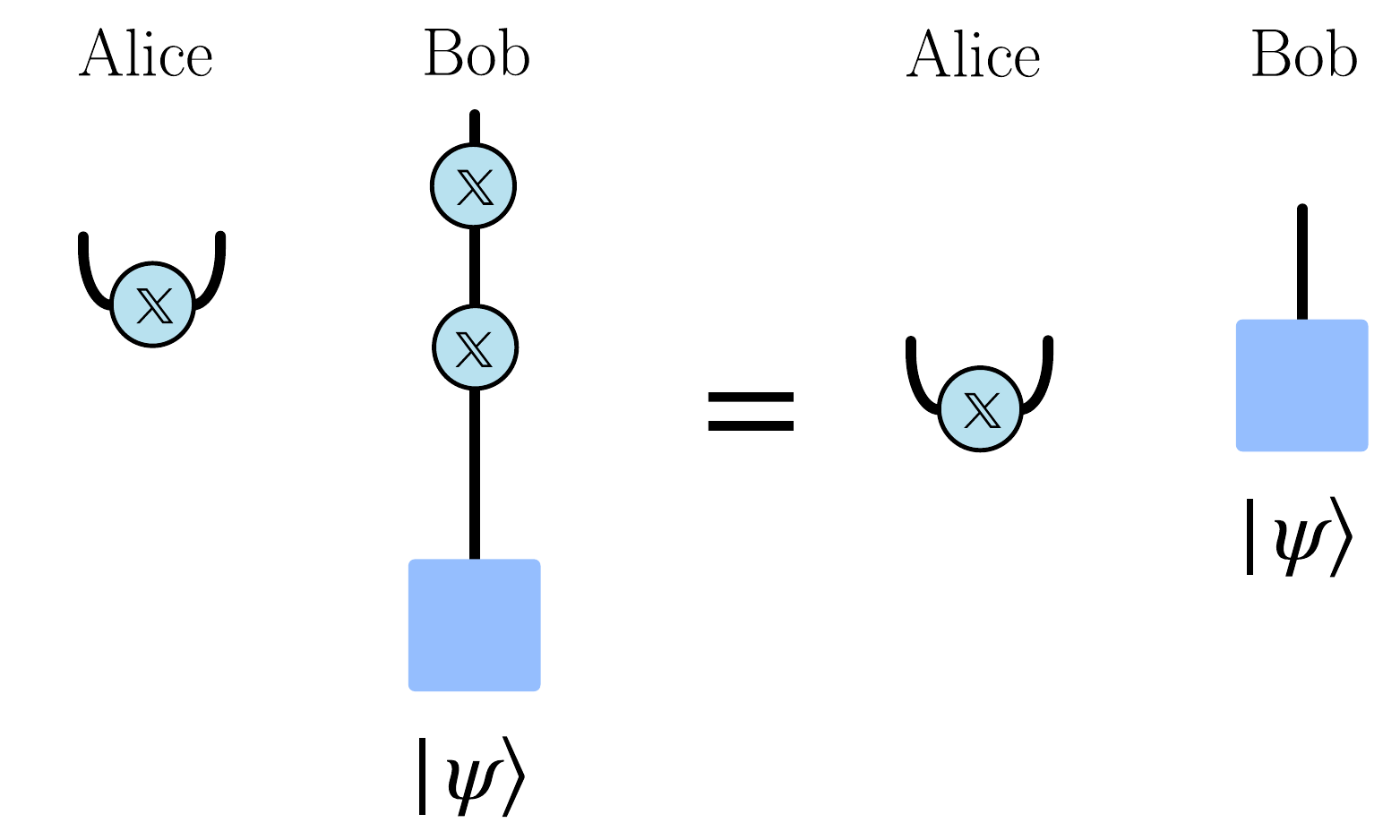}$$
where we used the fact that $\PauliX^2 = \Id$. Thus, Bob has successfully received the state $\ket{\psi}$, effectively teleporting it to his qubit!
\end{example}

Now given a pure state $\ket{\psi}$, what quantity can we use to discriminate whether it is a product state or an entangled state? To answer this question, we first provide a definition.

\begin{definition}{von Neumann entropy}{von Neumann entropy}
Given any density matrix $\hat{\rho}$, we define the von Neumann entropy as
\begin{equation}
    S(\hat{\rho}) = - \Tr[\hat{\rho} \log(\hat{\rho})] \, ,
\end{equation}
with $\log$ being matrix logarithm. This can be defined by using eigenstates $\{ \ket{\phi_k} \}$ of $\hat{\rho}$ as a basis. We have $\hat{\rho} = \sum_k p_k \ket{\phi_k} \bra{\phi_k}$, where $p_k$ are the corresponding eigenvalues (notice that $p_k \geq 0$ and $\sum_k p_k = 1$). In this basis, the von Neumann entropy is defined as
\begin{equation}\label{eq:vNinbasis}
    S(\hat{\rho}) = - \sum_k p_k \log p_k \, .
\end{equation}
\end{definition}

The von Neumann entropy\index{Entanglement!entropy} is a crucial quantity in quantum information, quantum computing and quantum many-body theory.
The following properties holds:
\begin{enumerate}
    \item invariance under unitary transformations: $S(\hat{U}^{\dag} \hat{\rho} \hat{U}) = S(\hat{\rho})$.
    \item non negativity: $S(\hat{\rho}) \geq 0$ and $S(\hat{\rho})=0$ if and only if the state is pure, i.e.\ $\hat{\rho}= \ket{\psi} \bra{\psi}$. Indeed in this case, $\hat{\rho}$ is a projector and has a single eigenvalue $1$, while all the others $p_k$ are $0$.
    \item upper bound: $S(\hat{\rho}) \leq \log d$, where $d$ is the dimension of the Hilbert space. The equality holds only for $\hat{\rho} = \frac{\mathbb{1}}{d}$, which is named maximally mixed state. Indeed, in this case all eigenvalues $p_k$ of $\hat{\rho}$ are $1/d$ and the sum Eq.~\eqref{eq:vNinbasis} is maximized.
    \item concavity: given a set of real positive numbers $\pi_i$ such that $\sum_i \pi_i =1$, then: $S(\sum_i \pi_i \hat{\rho}_i) \geq \sum_i \pi_i S(\hat{\rho}_i)$.
    \item sub-additivity: if we have a composite system $\mathcal{H} = \mathcal{H}_A \otimes \mathcal{H}_B$ then $S(\hat{\rho}_{AB}) \leq S(\hat{\rho}_A) + S(\hat{\rho}_B)$, where $\hat{\rho}_A = \Tr_{B}[\hat{\rho}_{AB}]$, $\hat{\rho}_B = \Tr_{A}[\hat{\rho}_{AB}]$ and $\hat{\rho}_{AB}$ is the global density operator of the two systems. Furthermore, equality holds when the systems are uncorrelated, i.e.\ $S(\hat{\rho}_A \otimes \hat{\rho}_B) = S(\hat{\rho}_A) + S(\hat{\rho}_B)$.
\end{enumerate}

Because of properties 2 and 3, we observe that the von Neumann entropy measures the intrinsic level of uncertainty encoded in the probability distribution of eigenvalues $p_k$ of $\hat{\rho}$. When there is no uncertainty, for instance, when $p_1 = 1$ and all other $p_k = 0$, we have $S(\hat{\rho}) = 0$. In contrast, when there are $d$ equally distributed possibilities, meaning $p_k = \frac{1}{d}$ for each $k$, the uncertainty is maximized and $S(\hat{\rho}) = \log d$.

\begin{definition}{Entanglement entropy}{entanglement_entropy}
The von Neumann entropy $S(\hat{\rho}_A)$ of the reduced density matrix $\hat{\rho}_A$ of a composite system is commonly referred to as the \textbf{entanglement entropy}.
\end{definition}

Notice that if we assume the system to be in a pure state $\ket{\psi}$, by using the Schmidt decomposition (Eq.~\eqref{eq:reduced_density_matrix_eig}) we find for $S(\hat{\rho}_A)$ an expression as Eq.~\eqref{eq:vNinbasis} in terms of the eigenvalues $p_k = \Lambda_{kk}^2$ of the reduced density matrix. We have therefore
\begin{align}
    \begin{split}
    S_{A/B} &= -{\rm Tr}_A\hat\rho_A\log\hat\rho_A
    = -{\rm Tr}_B\hat\rho_B\log\hat\rho_B \\ &= -\sum_{k} p_{k} \log p_{k}
    = -\sum_{k}\Lambda^{2}_{kk} \log \Lambda^{2}_{kk},
    \end{split}
\end{align}
Notice that a bound $\chi$ in the Schmidt rank of the state $\ket{\psi}$ reflects in a maximum value of the entanglement entropy. Indeed if $p_k  > 0$ only for $k \geq \chi$ it easy to show that the entanglement is upper bounded as $S_{A/B} \leq \log\chi$. In the case of product states $\ket{\psi} = \ket{\psi_A} \otimes \ket{\psi_B}$, one has a single positive Schmidt value, and therefore $\chi=1$. From this we find that $S_{A/B}=0$. Thus, entanglement entropy is zero for product states and positive for entangled states.

Reflecting on property 5 of the von Neumann, we observe that we found a measure that is non-negative and reaches zero precisely when the two systems are uncorrelated ($\rho_{AB} = \rho_A \otimes \rho_B$). Hence, we can employ this as a metric for the overall level of correlations.

\begin{definition}{Mutual information}{mutual_info}
Given a composite quantum systems we define the mutual information~as
\begin{equation}
    I(\hat{\rho}_{AB}) =  S(\hat{\rho}_{A}) + S(\hat{\rho}_{B}) - S(\hat{\rho}_{AB}) \, .
\end{equation}
\end{definition}

It is not difficult to show that if $\hat{\rho}_{AB}$ is a pure state then $S(\hat{\rho}_{A}) = S(\hat{\rho}_{B})$. Consequently, since $S(\hat{\rho}_{AB})=0$, the mutual information is $I(\hat{\rho}_{AB})=2S(\hat{\rho}_{A}) = 2S(\hat{\rho}_{B})$.

\enlargethispage*{\baselineskip}
We have so far provided definitions of entanglement based on rather simple arguments. However, it is possible to give a much deeper characterization of entanglement within the framework of the so-called quantum resource theories. In this context, entanglement is defined through measures known as entanglement monotones. An entanglement monotone is a non-negative function whose value does not increase under a certain class of quantum transformations that are axiomatically assumed not to create entanglement. These transformations are known as Local Operations and Classical Communication.

\begin{definition}{Local Operations and Classical Communication}{locc}
Local Operations and Classical Communication (LOCC) is a central concept in quantum information theory, especially in the study of quantum entanglement and quantum state manipulation~\cite{nielsen_chuang_2000,Chitambar_2014}. The operations allowed under LOCC are:
\begin{enumerate}
    \item \textbf{Local Operations:} each party can perform any quantum operation on their own subsystem. These operations include local unitary transformations, measurements, and the addition or removal of ancillary qubits.
    \item \textbf{Classical Communication:} the parties can communicate with each other using classical channels. This communication can be used for instance to coordinate their local operations based on the measurement outcomes.
\end{enumerate}

Importantly, LOCC do not allow for global unitary transformations and therefore the creation of genuine quantum correlations between parties, i.e.\ the generation of entanglement. As a result, LOCC cannot increase the amount of entanglement present in the system. However, they enable the manipulation and utilization of existing entanglement. For instance, LOCC can be used to concentrate entanglement, convert one form of entanglement into another, and distribute entanglement across different parties~\cite{wootters_zurek_1982,bennett_concentrating_1996,bennett_mixed-state_1996,horodecki_general_1999}.

\end{definition}

Essentially, entanglement can be defined as a kind of quantum correlation that cannot be created by LOCC alone. For pure states $\ket{\psi}$, the von Neumann entropy $S(\hat{\rho}_A)$ is recognized as a good measure of entanglement in this sense. In contrast, various functions have been developed to quantitatively measure entanglement in mixed states. Although a discussion of these aspects goes beyond the scope of this book, we can mention another highly relevant measure.

\begin{definition}{Negativity}{Negativity}
The negativity of a subsystem
$A$ of a composite system in a mixed state $\hat{\rho}$ is defined as
\begin{equation}
    N(\hat{\rho}) = \frac{||\hat{\rho}^{T_B}||_1 - 1}{2}
\end{equation}
where $T_B$ represent the partial transpose on system $B$. The trace norm $||\cdot||_1$ is defined as the sum of the absolute values of the eigenvalues. The logarithmic negativity is instead:
\begin{equation}
    E(\hat{\rho}) = \log ||\hat{\rho}^{T_B}||_1
\end{equation}
\end{definition}

% https://arxiv.org/pdf/2402.09523.pdf

\section{Tensor network representation of quantum states}
Consider a many body wave function. Simply specifying its coefficients in a given local basis (for instance, the canonical computational basis in a quantum computing setup) does not provide any insight into the structure of entanglement among its constituents. In fact:
\begin{itemize}
\item[(i)]
We anticipate that this structure will vary depending on the dimensionality of the system, resulting in differences between 1D, 2D, or higher-dimensional scenarios.
\item[(ii)]
Indeed, the structure of entanglement is influenced by more subtle factors such as the criticality of the state and its correlation length.
\end{itemize}

To be more precise, let us consider a generic state describing $n$ qubits, namely $\ket{\psi}\in\mathcal{H}^{\otimes n}$,
where $\mathcal{H}$ is the local Hilbert space for a single qubit. When not differently specified, we always consider $\mathcal{H} = \rm{span}\{\ket{0},\ket{1}\}$, in terms of the eigenstates of the Pauli gate $\PauliZ$. In particular, in quantum computation it is customary to set $\ket{0}$ as the $+1$ eigenvector of $\PauliZ$, and $\ket{1}$ as the $-1$ eigenvector of $\PauliZ$.

The many body vector decomposes as\index{Many-body state}
\begin{equation}\label{chapt2_eq:psi_TN}
\includegraphics[width=0.5\textwidth,valign=c]{figs/chapt_2/psi_TN.pdf}
\end{equation}
where, unfortunately, the order-$n$ tensor $\psi_{x_1,\dots,x_n}$ lacks explicit information about the entanglement properties. Therefore, it is desirable to find a representation of quantum states where this information is explicit and readily accessible.

Tensor Networks provide a framework where this information is readily available; in a sense, Tensor Network states are inherently represented in a specific ``entanglement representation''.
The primary reason Tensor Networks serve as a key description of quantum many-body states is that the Hilbert space becomes exceedingly large, scaling exponentially with the number of local constituents (as $2^n$ in this case). Additionally, representing the state using coefficients in the computational basis is highly inefficient. The classical paradox example arises when we directly compare a generic random state with a random product state. In the case of a random product state, the associated tensor would simply be the product:\index{Many-body state!product state}
\begin{equation}
    \includegraphics[width=0.7\textwidth,valign=c]{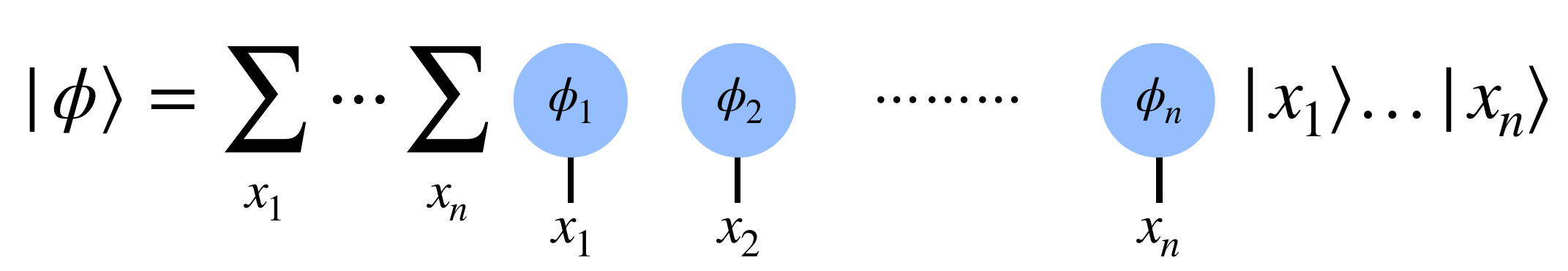}
\end{equation}
When fused together, the local tensors $\phi_j$ collectively generate an order-$n$ tensor that closely resembles the generic tensor $\psi_{x_1,\dots,x_n}$, being its entries basically indistinguishable from those of a completelly random order-$n$ tensor. Despite this, the complexity of a product state scales linearly with the number of qubits, with only $2n$ independent complex coefficients.

Actually, not all states within the Hilbert space of a many-body quantum system are equally important; some are more relevant than others. For instance, physical Hamiltonians often feature local interactions among constituents.
The locality of these interactions leads to significant consequences, such as the emergence of area laws for entanglement in low-energy states of local gapped Hamiltonians. In other words, these states exhibit short-range correlations.

\begin{definition}{Area law of bipartite entanglement entropy}{area_law}
 In quantum mechanics, a state adheres to an area law when the dominant contribution to its \textbf{entanglement entropy} increases at most proportionally to the boundary between two partitions.
 In one-dimensional systems, this implies, for instance, that the bipartite entanglement entropy would be strictly bounded.
\end{definition}

Moreover, locally manipulating (via unitary gates) shortly correlated one-dimensional states yields a class of highly valuable states for harnessing quantum resources and implementing quantum computing algorithms. In fact, the manifold generated after a series of local operations, whose number remains of the order of the system size, is exponentially small compared to the total Hilbert space.
Matrix Product States (MPS) precisely provide an efficient ansatz for this portion of the Hilbert space.
Numerous foundational articles and comprehensive reviews have been published on these topics over the past two decades. Here, we mention just a few of them,~\cite{PerezGarcia_2007,McCulloch_2007,Schollwock_2011,Orus_2014,Bridgeman_2017,Silvi_2019,Ran_2020, Evenbly_2022}, and we highly recommend that readers explore these works and seek out additional references cited in their respective bibliographies.

\subsection{Matrix Product States (MPS)}\label{subsec:MPS}
The ingredient which stays at the basis of a MPS\index{MPS} representation of a many-body quantum state is strictly related to the way the many-body coefficient $\psi_{x_1,\dots,x_n}$ entering in Eq.~(\ref{chapt2_eq:psi_TN}) can be manipulated as a tensor, basically as has been already shown in Chapter~\ref{chap1}.

Indeed, an efficient way of reshuffling the parameters of a quantum wave function, highlighting its entanglement content, is given by its \emph{Schmidt decomposition} (see Section~\ref{chapter2_sec:schmidt_decomposition}), which is a compact way of rewriting a quantum state of a system living in a bipartite universe.

Let us take the state in Eq.~(\ref{chapt2_eq:psi_TN}) and rewrite it as it is leaving in a the tensor product of two  Hilbert spaces $\mathcal{H}_{A}\otimes\mathcal{H}_{B}$,
such that $\mathcal{H}_{A} = {\rm span} \{\ket{a_j}\}$ and $\mathcal{H}_{B} = {\rm span} \{\ket{b_j}\}$. This can be done by arbitrarily fusing the indices of the tensor $\psi_{x_1,\dots,x_n}$ and reshape it as a matrix.
Then following what have been done in Section~\ref{chapter2_sec:schmidt_decomposition} the state can be rewritten as
\begin{equation}
\includegraphics[width=0.7\textwidth,valign=c]{figs/chapt_2/psi_schmidt.pdf}
\end{equation}
with \emph{Schmidt vectors} $\ket{\varphi^a_k}$ and $\ket{\varphi^b_k}$, and
real non-negative \emph{Schmidt values} $\Lambda_{kk}$.

Notice that, in the original representation of the state we were using ${\rm dim}(\mathcal{H}_{A})\times{\rm dim}(\mathcal{H}_{B})$ complex numbers to describe the full wave function.
However, if $\chi$ is the number of non-vanishing Schmidt (or singular) values, than, in the Schmidt basis, in addition to these real numbers, we are only using
$\chi\times [{\rm dim}(\mathcal{H}_{A})+{\rm dim}(\mathcal{H}_{B})]$ complex numbers to encode the wave function. Therefore, if $\chi$ is relatively small, or only few of the entries of $\Lambda_{kk}$ are relevant, we basically end up with a systematic way of reducing the complexity of the many-body state still retaining the leading correlations across the entire system.
In fact, from the relation with the reduced density matrices, we can easily compute the bipartite \textbf{entanglement entropy} between subsystem A and B
\begin{equation}
        S_{A/B} = -{\rm Tr}_A\hat\rho_A\log\hat\rho_A
    = -{\rm Tr}_B\hat\rho_B\log\hat\rho_B
    = -\sum_{k}\Lambda^{2}_{kk} \log \Lambda^{2}_{kk},
\end{equation}
which gives an indication of how much the original wave function is entangled. As already mentioned in section 2.3.2, a bound in the number $\chi$ of Schmidt values reflects in a maximum value of the entanglement entropy $S_{A/B} \leq \log\chi$, where we used the the fact that $\sum_k \Lambda^{2}_{kk} = 1$ due to the state normalisation.

However, if $\chi$ remains too large for the wave function to be efficiently handled by a classical computer, then we might question whether faithful compression is feasible. The Schmidt decomposition provides a solution to this fundamental query; specifically, we obtain an approximate state by retaining only the largest $\tilde{\chi} < \chi$ Schmidt vectors. As a matter of fact, the following state
\begin{equation}\label{chapter2_eq:psi_truncated}
\ket{\tilde\psi} = \sum_{k=1}^{\tilde\chi} \tilde\Lambda_{kk}
\ket{\varphi^{a}_{k}}\ket{\varphi^{b}_{k}}
\end{equation}
is exactly the one that minimise the norm-$2$ distance from the original state, namely $|| |\psi\rangle - |\tilde\psi\rangle ||^2$. Where in Eq.~\eqref{chapter2_eq:psi_truncated}, in order to keep the state normalized, we re-scaled the original singular values as
$\tilde\Lambda_{kk} \equiv \Lambda_{kk}/(\sum_{k=1}^{\tilde\chi}\Lambda_{kk}^2)^{1/2}$.

Now, since the bipartition we were considering was arbitrary, we can extend this reasoning to each qubit (or generic local Hilbert space) constituting the local degrees of freedom of the entire wave function. In practice, we can apply a process similar to what was described in Chapter~\ref{chap1} for decomposing a generic order-$n$ tensor, but this time to decompose a quantum state into a one-dimensional ansatz where locality is preserved and entanglement compression is naturally achieved. This ansatz is commonly referred to as the \textbf{Matrix Product State}.\index{Many-body state!matrix product state}

\begin{definition}{Matrix Product State}{MPS}
 We say that a quantum many-body state is a \textbf{Matrix Product State}
 when its wave function is written as a matrix product tensor network (cfr.\ Chapter~\ref{chap1}), thus having
\begin{equation}
    \includegraphics[width=0.75\textwidth,valign=c]{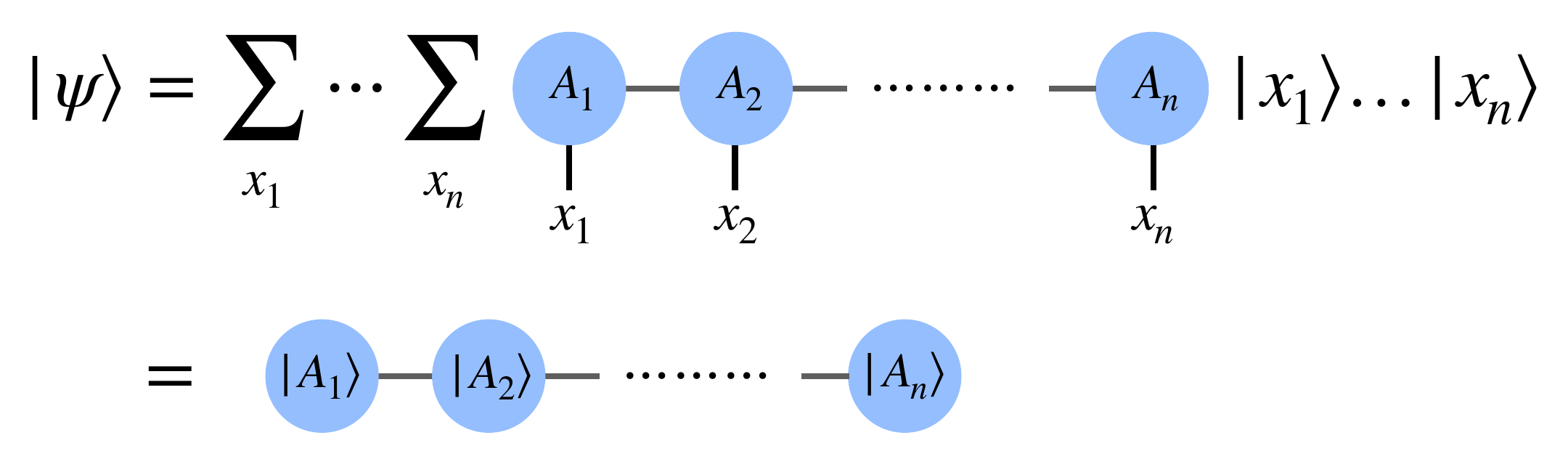}
\end{equation}
where in the last passage we have defined the following \textbf{vector-valued tensors}
$$
\includegraphics[width=0.5\textwidth]{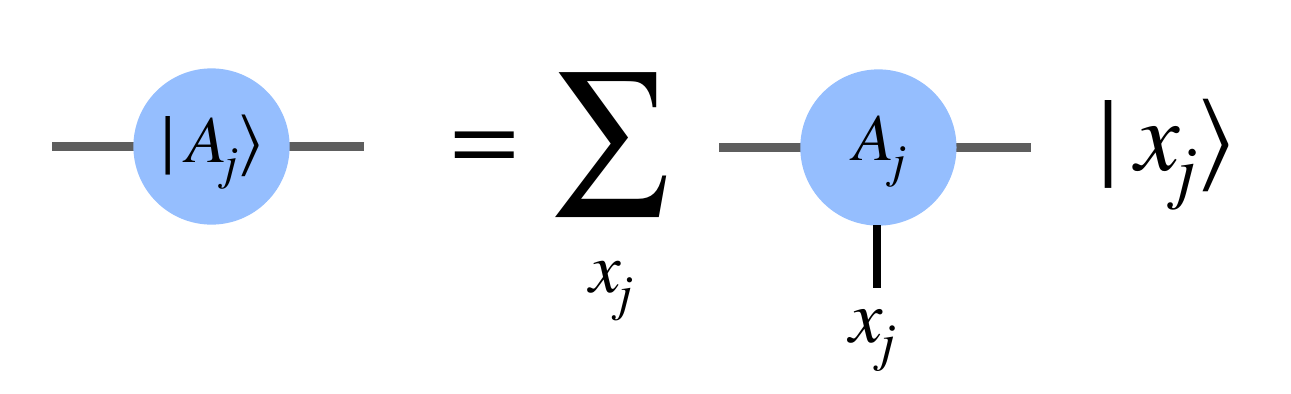}
$$
whose legs span auxiliary spaces whose dimension is typically termed the \textbf{bond or auxiliary dimension}, and denoted by $\chi$.
These legs guarantee the interconnection between the different local tensors, thereby enabling the encoding of correlations throughout the system. This capability extends beyond a non-correlated product state, which effectively corresponds to an MPS with bond dimension $\chi=1$.
\end{definition}

Given this representation, the norm of an MPS can be easily calculated as\index{MPS!norm}
\begin{equation}
    \includegraphics[width=0.55\textwidth,valign=c]{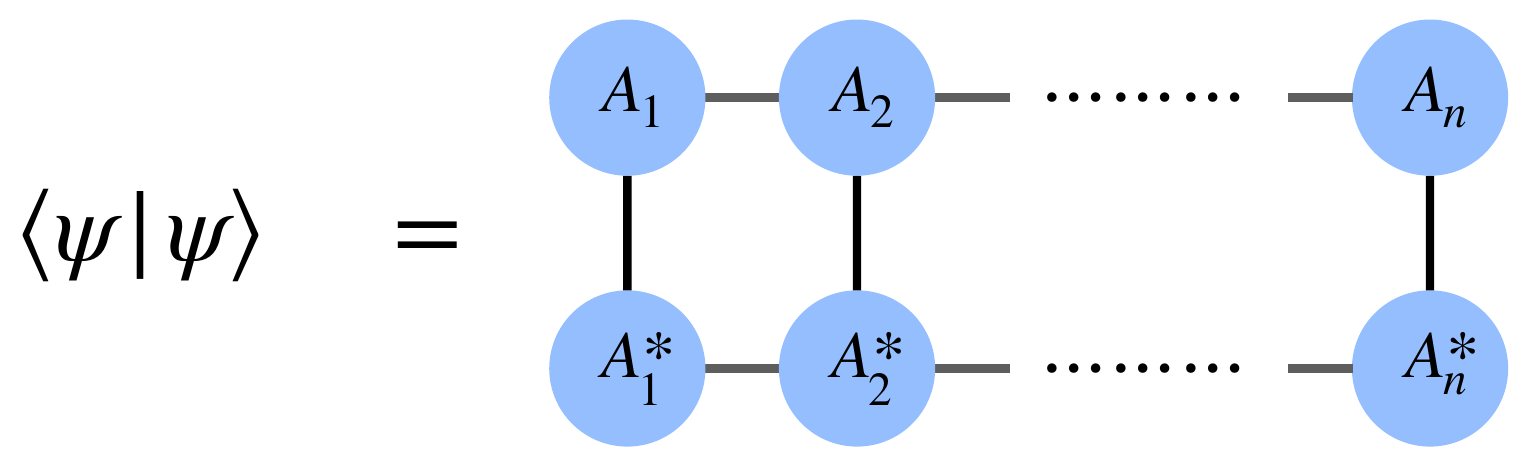}
\end{equation}
with a computational cost $O(n\chi^3)$. Here $ A_{j}^*$ is indicating the conjugate tensor.
Let us notice that since the bond dimension essentially denotes the dimension of the auxiliary space between two physical lattice sites, such as $j$ and $j+1$, it could potentially vary with position, i.e.\ $\chi\to\chi_j$.

\begin{definition}{MPS Transfer Matrix}{transfer_matrix}
 One of the central object in any MPS based description of a physical state is the so called \textbf{Transfer Matrix}\index{Transfer matrix}. It is defined as
 \begin{equation}
     \includegraphics[width=0.5\textwidth,valign=c]{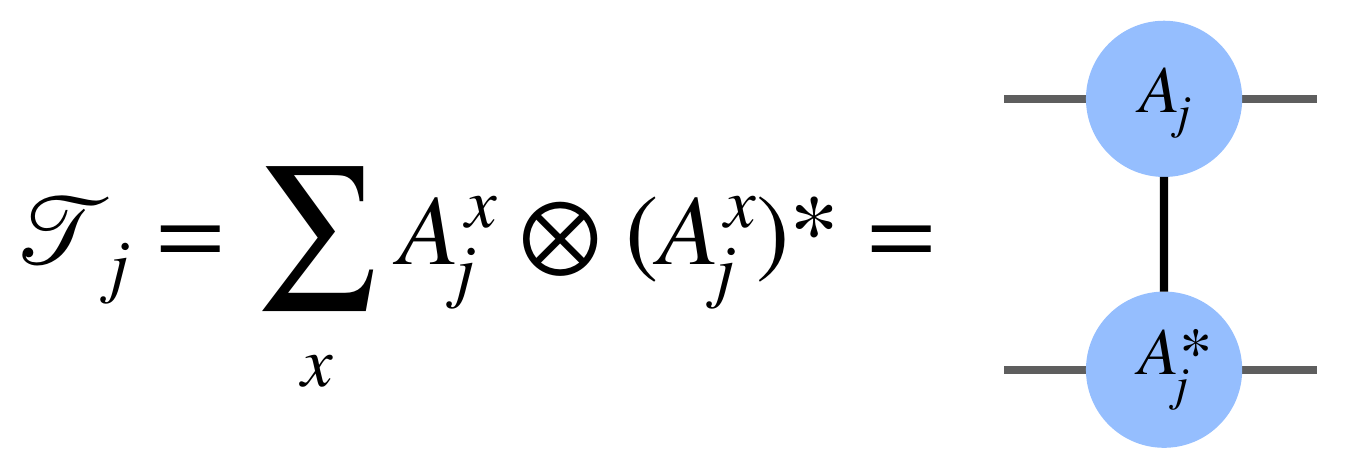}
 \end{equation}
as ``super'' operator which acts on the space of $\chi_j\times\chi_j$ matrices on its right or $\chi_{j-1}\times\chi_{j-1}$ matrices on its left, is defined as a completely positive map~\cite{Choi_1975}, with the MPS matrices serving as Kraus operators. Notably, the transfer matrix exhibits the property that its leading eigenvalue is a positive number $\eta$. In general, this correspond to a leading left (and right) eigenvector
such that
$$\includegraphics[width=0.7\textwidth,valign=c]{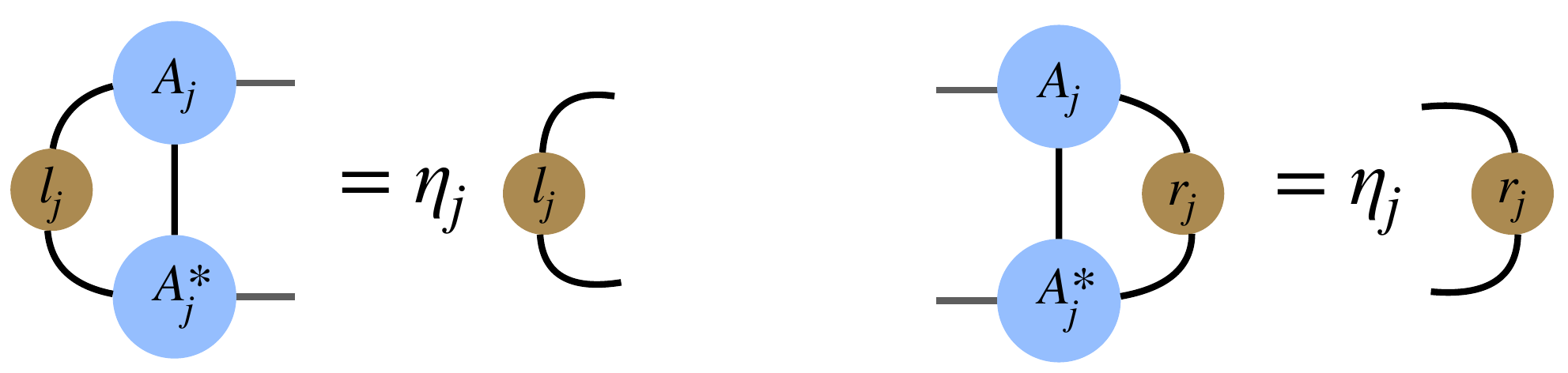}$$
where both $l$ and $r$ are hermitian matrices with non-negative eigenvalues and therefore which can be square decomposed as $l = L^{\dag}L$ and $r=RR^\dag$.

\end{definition}

\paragraph{MPS gauge freedom ---}
Whenever a Matrix Product Tensor ansatz is used to describe the many-body wave function of a quantum state, while the state is uniquely determined by the tensors $A_j$, the reverse is not necessarily true, as distinct tensors can yield the same physical state. This becomes apparent when considering the following gauge transformation\index{MPS!gauge transformation}:
\begin{equation}
     \includegraphics[width=0.5\textwidth,valign=c]{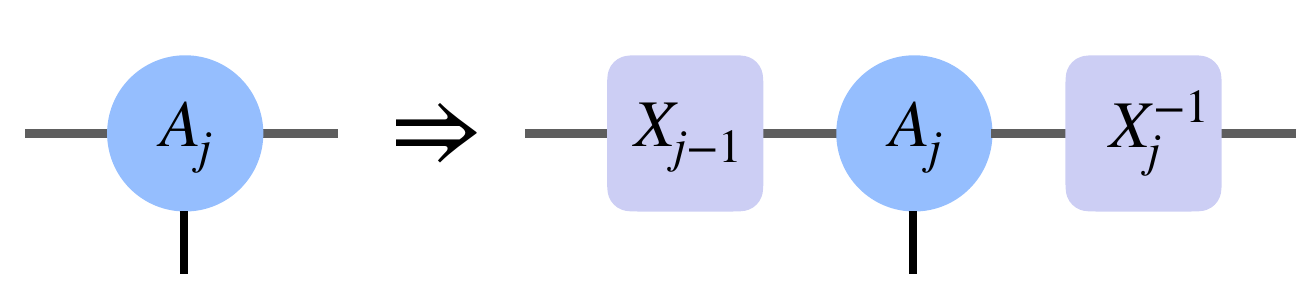}
 \end{equation}
 where $X_j$ is an invertible matrix.
Essentially, the previous transformation introduces a resolution of the identity $X^{-1}_{_j}X_{j}$ in each local auxiliary bond of the MPS wave function, effectively preserving the integrity of the entire state.
Indeed, it has been demonstrated~\cite{PerezGarcia_2007,PerezGarcia_2008} that this is the sole freedom in the parametrization, and it can be constrained (partially) by imposing canonical forms on the MPS tensors $A_j$.

As a matter of fact, for a finite system MPS with tensors $\{ A_{1}, \dots, A_n\}$, starting from the leftmost tensor, and proceeding iteratively on $j$, we can follow what have been done in Chapter~\ref{chap1}, but now employing the \textbf{QR decomposition}\index{QR decomposition} of the tensors in the following recursive way
\begin{equation}
     \includegraphics[width=0.5\textwidth,valign=c]{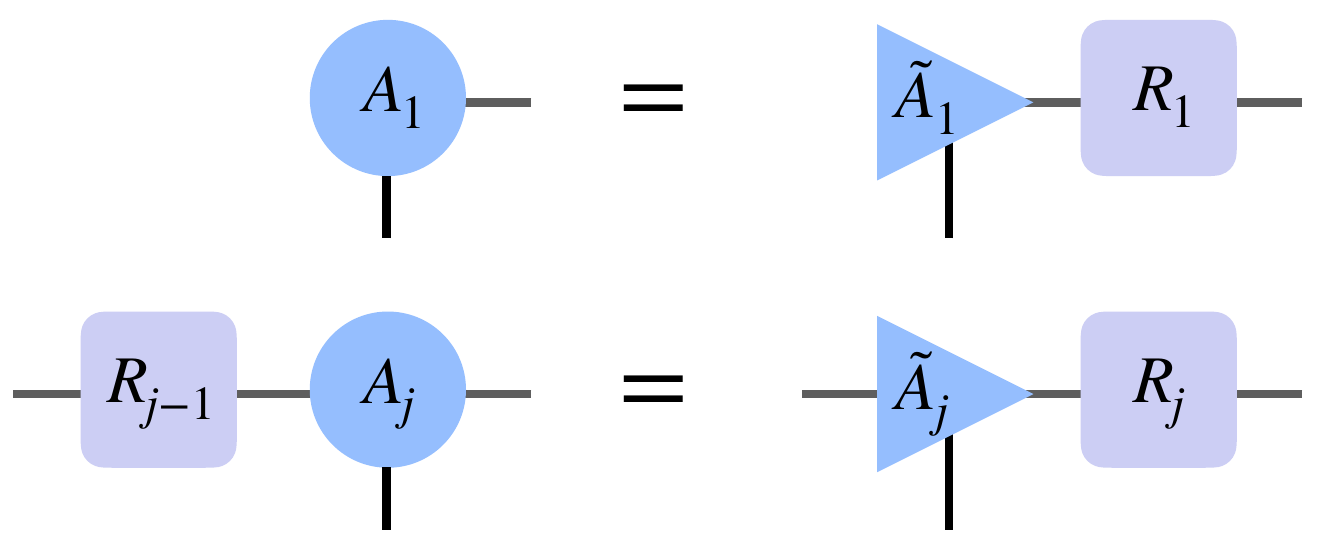}
 \end{equation}
 where we can identify the required Gauge transformation exactly with the upper triangular matrices $R_j$, such that the \textbf{Transfer Matrix} constructed with the new tensors $\tilde A_{j}$ has as leading left eigenvector the identity matrix, diagrammatically\index{MPS!left canonical}
$$
 \includegraphics[width=0.25\textwidth,valign=c]{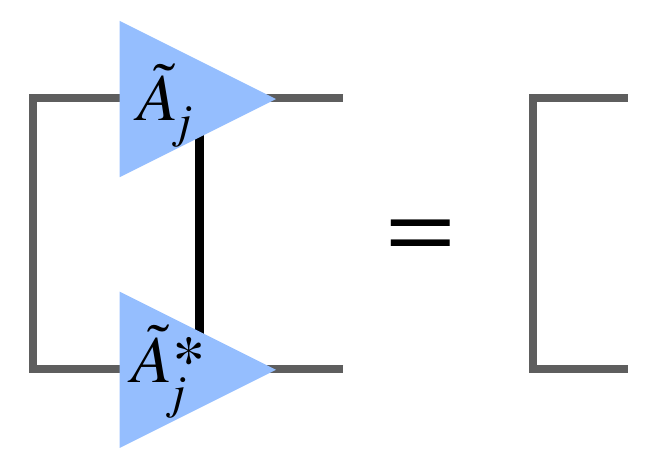}
$$
A similar iterative approach could lead to a completely right normalized decomposition where the transfer matrix now satisfies the following property\index{MPS!right canonical}
$$
 \includegraphics[width=0.25\textwidth,valign=c]{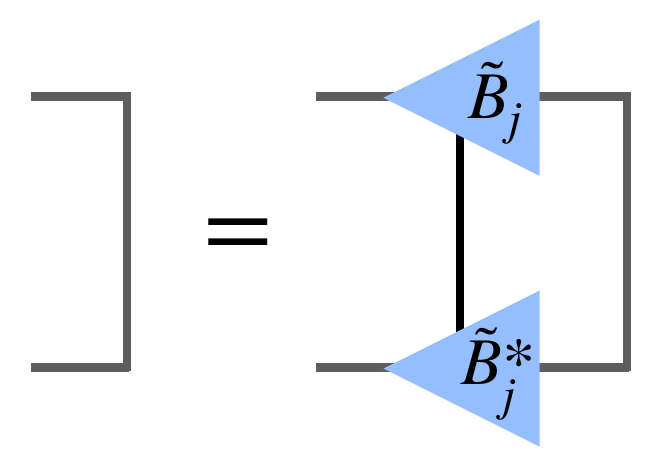}
$$
Mixing the two strategies, and thus orthogonalising in iterative way the left-subsystems tensors from the left boundary, and the right subsystems tensors for the right boundary, leads to the following, extremely useful
\textbf{Mixed-Canonical decomposition} of an MPS\index{MPS!mixed canonical}
\begin{equation}\label{eq:mixedcanon1}
 \includegraphics[width=\textwidth,valign=c]{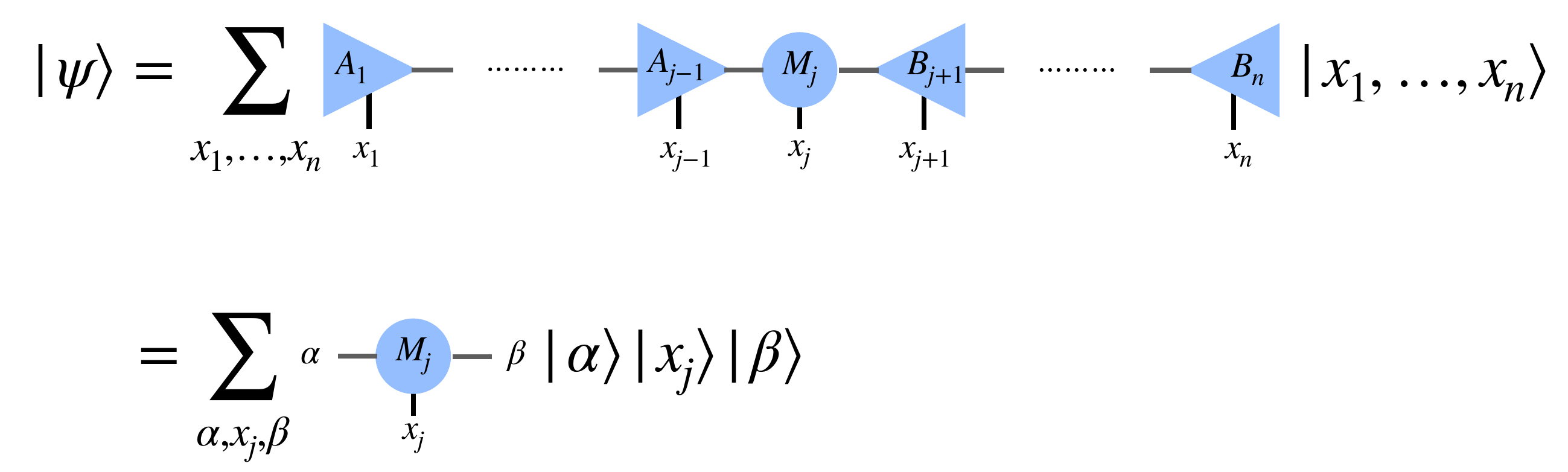}
\end{equation}
with left and right vectors
\begin{equation}\label{eq:mixedcanon2}
 \includegraphics[width=0.75\textwidth,valign=c]{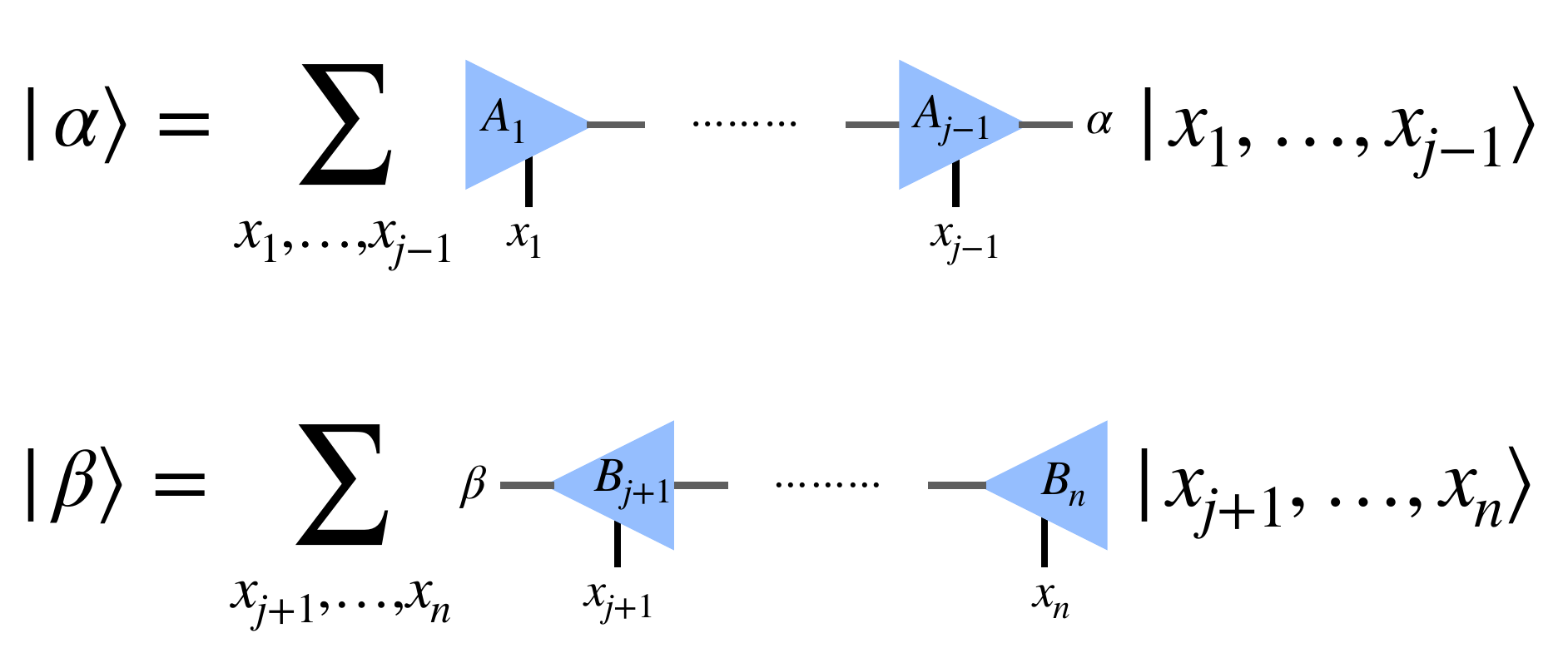}
\end{equation}
which satisfy $\braket{\alpha}{\alpha'} = \delta_{\alpha\alpha'}$
and $\braket{\beta}{\beta'} = \delta_{\beta\beta'}$ thanks to tensor properties.

\paragraph{MPS overlap \& expectation values ---}
Computing overlaps between quantum states or expectation values of a local observable is generally an exponentially hard problem as the system size increases. However, in the MPS formalism, this problem becomes polynomially complex because the complexity is essentially constrained by the bond dimension of the network representing the state. As we have seen with the norm of an MPS, the overlap between two MPS states\index{MPS!overlap}
$$
\includegraphics[width=0.75\textwidth,valign=c]{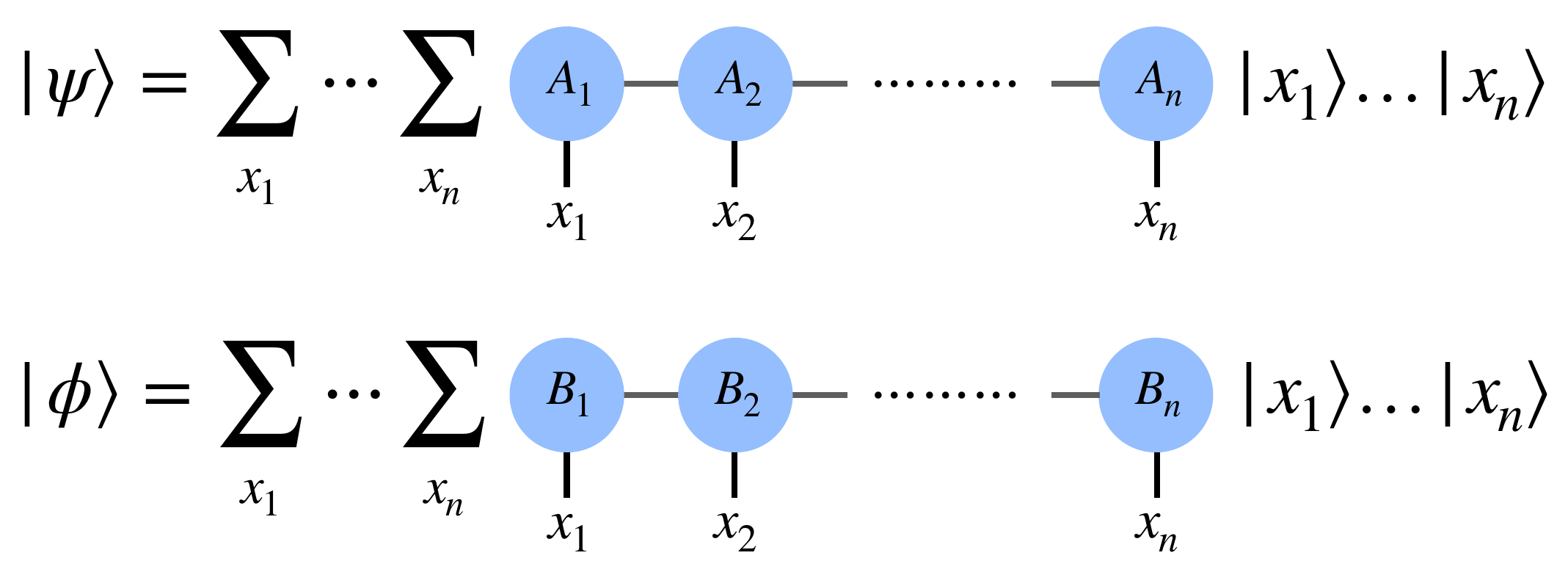}
$$
is easily computed as
$$
\includegraphics[width=0.5\textwidth,valign=c]{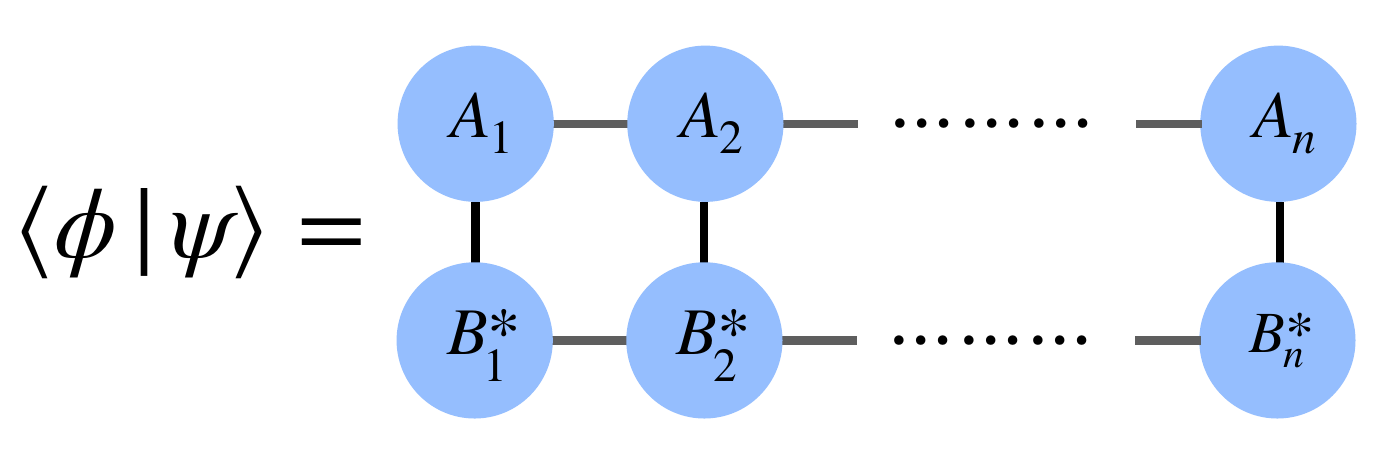}
$$
where we are not assuming any specific normalization of the MPS.
Let us stress again that the compression scheme of the Tensor Network is very important to achieve an optimal procedure. In fact, for any open boundary tensor network, the contraction should start from one of the two boundaries (for example, from the left), as outlined in the following steps:
$$
\includegraphics[width=\textwidth,valign=c]{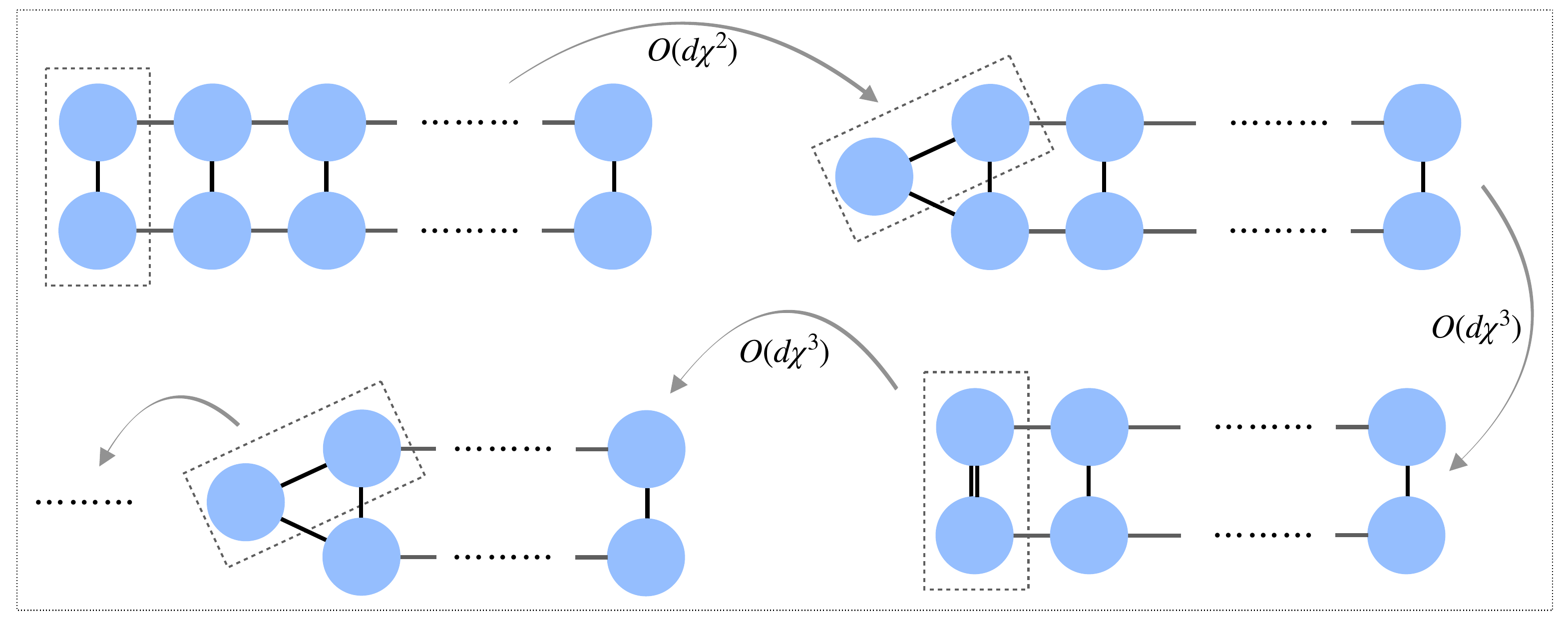}
$$
where we assume $d$ is the local physical dimension and $\chi$ is the auxiliary dimension of both the MPS wavefunctions.

Evaluating expectation values is as straightforward as computing overlaps. To keep the discussion concrete, let's suppose we want to compute the average of a Pauli string.\index{Pauli string}
$$
\includegraphics[width=0.9\textwidth,valign=c]{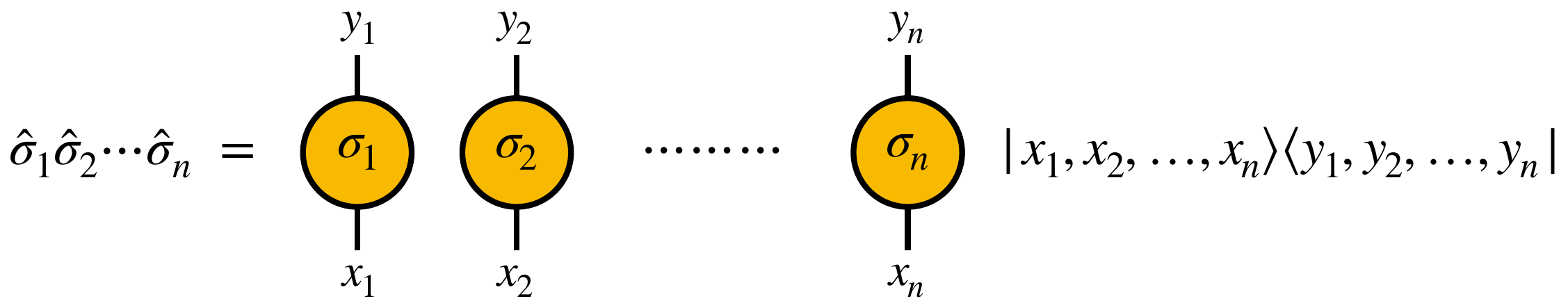}
$$
where $\hat{\sigma}_j \in \{\hat{I}, \hat{X}, \hat{Y}, \hat{Z}\}$. However, what we are going to discuss is valid for the tensor product of any local observable. As is typical with tensor networks, the expectation value of such an operator computed on an MPS state reduces to the following network contraction\index{MPS!Puali string average}:
$$
\includegraphics[width=0.6\textwidth,valign=c]{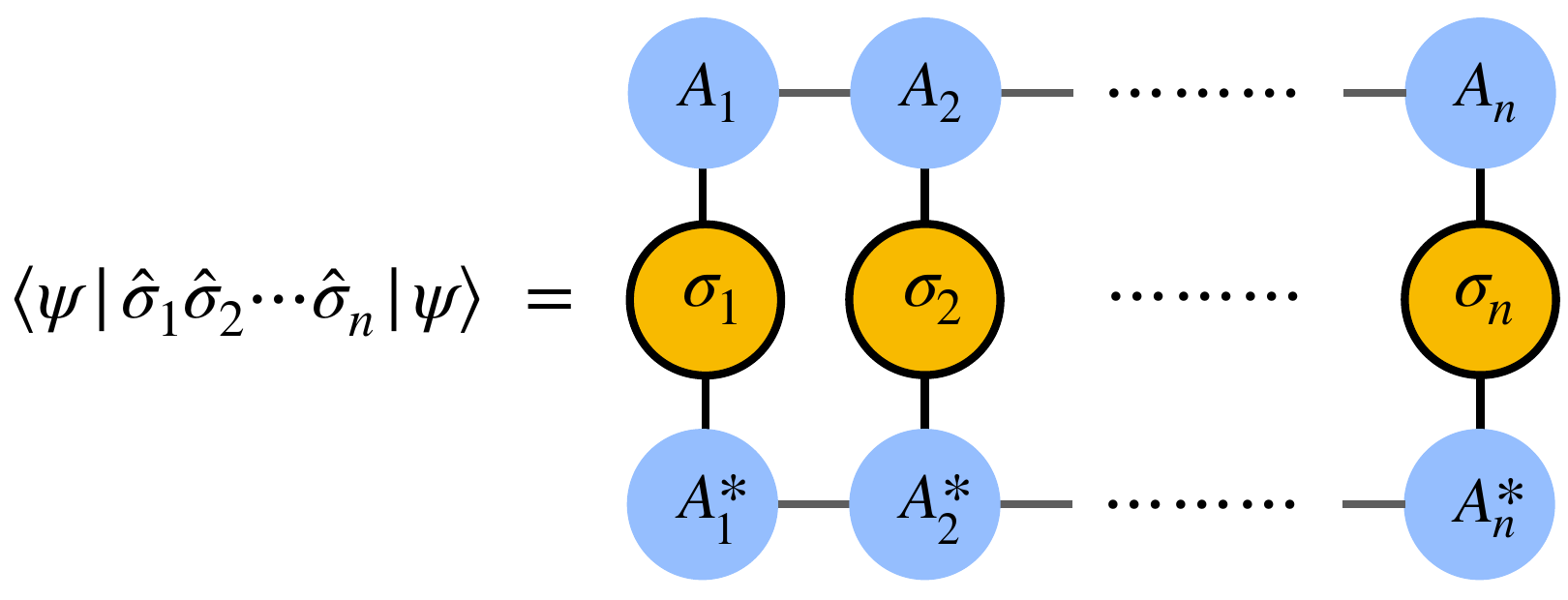}
$$
Notice that, in this case as well, the contraction scheme is important. We must follow the same procedure outlined for the computation of the overlap, which involves starting from one of the two boundaries and sequentially updating the boundary matrix. Therefore, when adding the generic lattice site $k$, we compute the updating step as follows:
$$
\includegraphics[width=0.45\textwidth,valign=c]{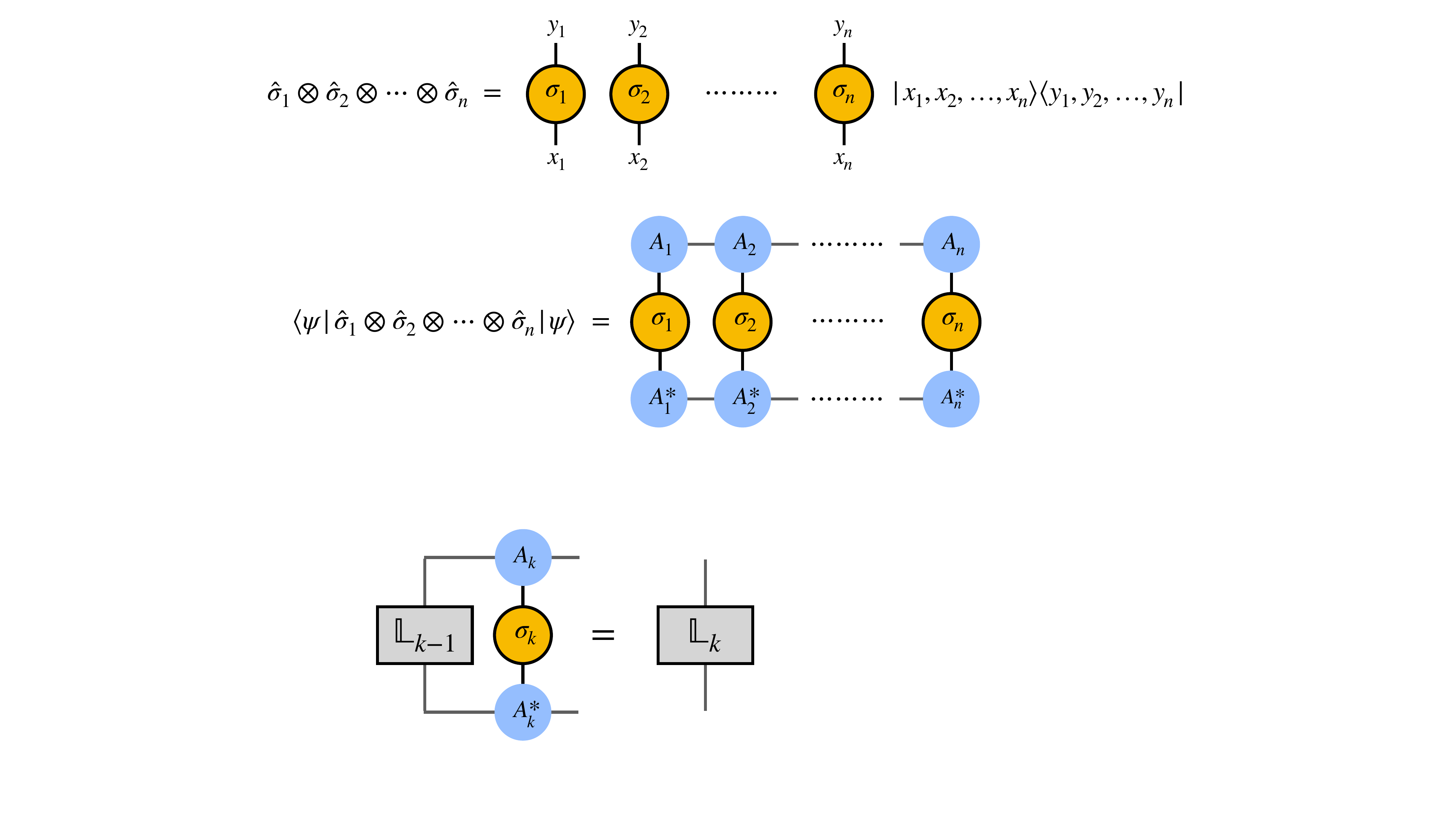}
$$
which basically correspond to compute $\mathbb{L}_{k} = \sum_{i,j} (\sigma_{k})_{ij} \, (A_{k}^{i})^{\dag}\cdot \mathbb{L}_{k-1}\cdot A_{k}^{j}$
thus keeping the computational cost of a single step $O(d^2\chi^3)$.
In such a recursive algorithm, $\mathbb{L}_{0} = (1)$ is a $1 \times 1$ identity matrix (or a scalar if you wish); and finally, $\mathbb{L}_{n} = \bra{\psi}\hat{\sigma}_1 \hat{\sigma}_2 \cdots \hat{\sigma}_n\ket{\psi}$ is the desired expectation value.

\subsection{Matrix Product Operator (MPO)}
Matrix Product Operators\index{MPO} were originally introduced by Juan Ignacio\linebreak Cirac~\cite{Verstraete_2004} and by Guifré Vidal~\cite{Zwolak_2004} as a Tensor Network Ansatz for density matrices, and they basically constitute the operator analogue of MPS.

Let us consider fro the moment $n$ lattice sites ordered in a one-dimensional chain. Each local Hilbert space $\mathcal{H}$ having dimension $d$ (with $d=2$ for qubits) so that quantum many-body state is living in $\mathcal{H}^{n}$ with total dimension $d^n$.

Any \textbf{linear operator} mapping a state $\ket{\psi}\in\mathcal{H}^{ n}$ to another state
$\ket{\psi'}\in\mathcal{H}^{ n}$ is an object leaving in the tensor product $\mathcal{H}^{ n} \otimes \mathcal{ H^*}^{ n}$,
so as in general we have\index{Observables}
\begin{equation}\label{eq:operator}
   \includegraphics[width=0.65\textwidth,valign=c]{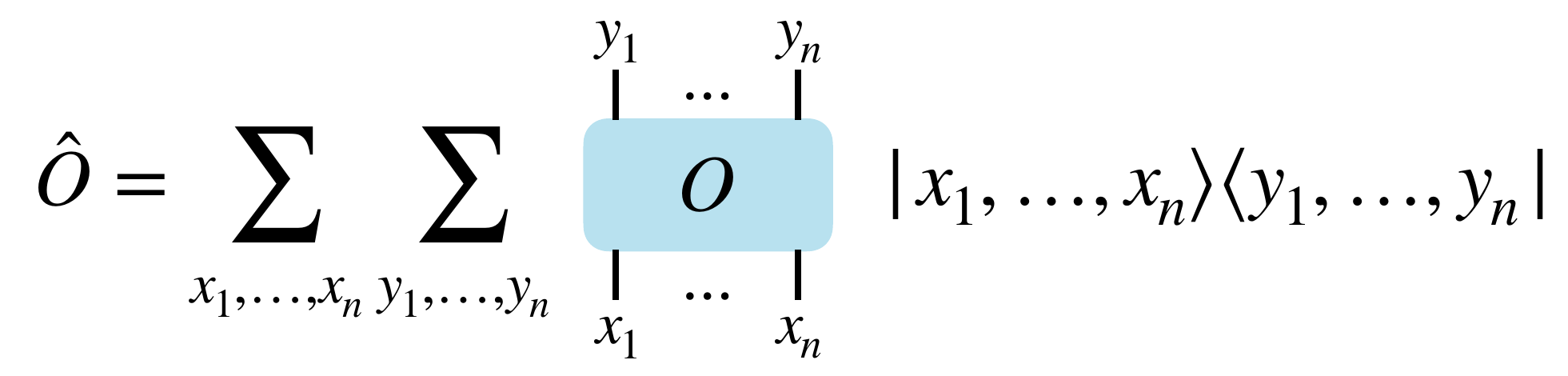}
\end{equation}
so it is natural to inquire whether the operator can be represented using the same tensor decomposition discussed in the previous section. Indeed, blindly applying the procedure outlined in Chapter~\ref{chap1} for the tensor in Eq.~\eqref{eq:operator} leads to a similar decomposition of the operator's coefficient. We can fuse the indices corresponding to each local Hilbert space $(x_j,y_j)$ and then perform the standard SVD procedure, resulting in:
\begin{equation}\label{eq:mpo}
   \includegraphics[width=0.85\textwidth,valign=c]{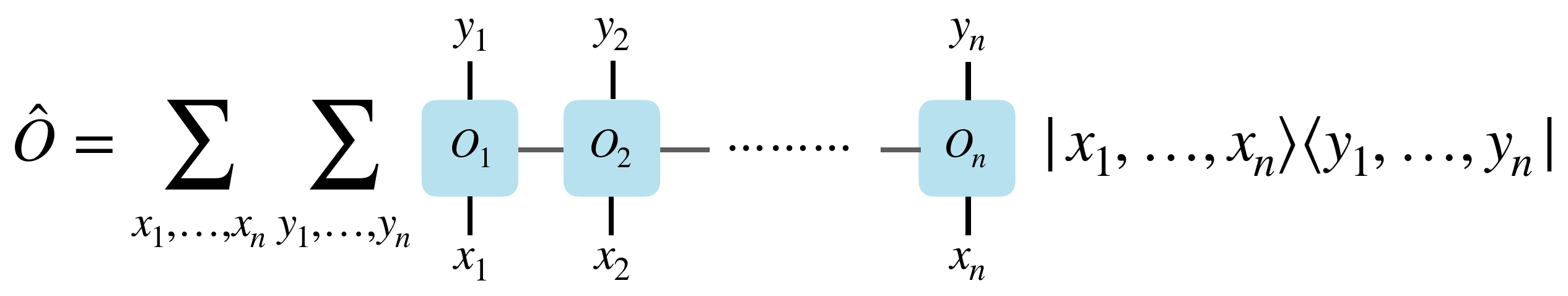}
\end{equation}
where we reshuffled back each fused index, so that $O_{j}^{x_jy_j}$ are matrices like the $A_{j}^{x_j}$ entering in the MPS, with the key distinction staying in the fact that, as representations of operators, they require both outgoing and incoming physical indices.
Notice that, in general, a systematic procedure based on iterative SVD from a many-body representation of the operator itself requires an exponentially large amount of resources and is unlikely to be feasible in practical scenarios.
Instead, there are cases where an MPO representation of an operator is easy to construct in a systematic way.

The analogous of product state in the realm of the operators are \emph{tensor product of local operators}, which have the straightforward representation
$$
\includegraphics[width=0.9\textwidth,valign=c]{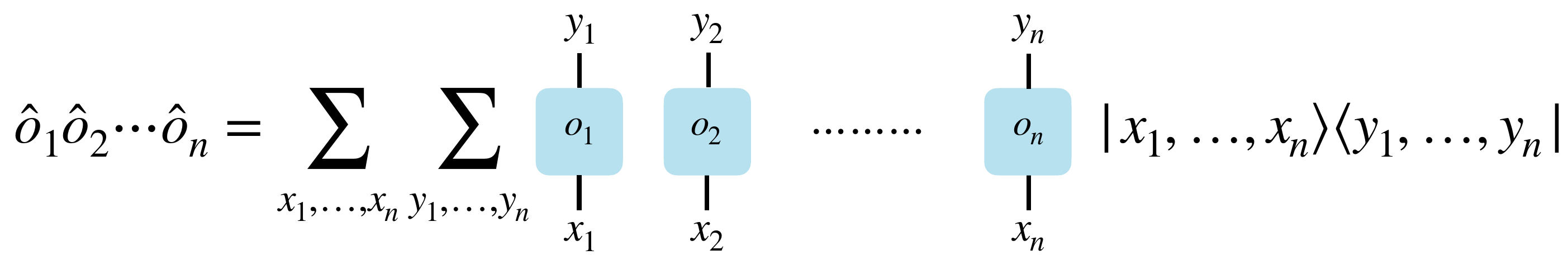}
$$
with local bond dimension equal to one.
As a further example, an highly non-local operator whose MPO
representation can be written straightforwardly is the
\textbf{projector to a MPS state}
$$
\includegraphics[width=0.9\textwidth,valign=c]{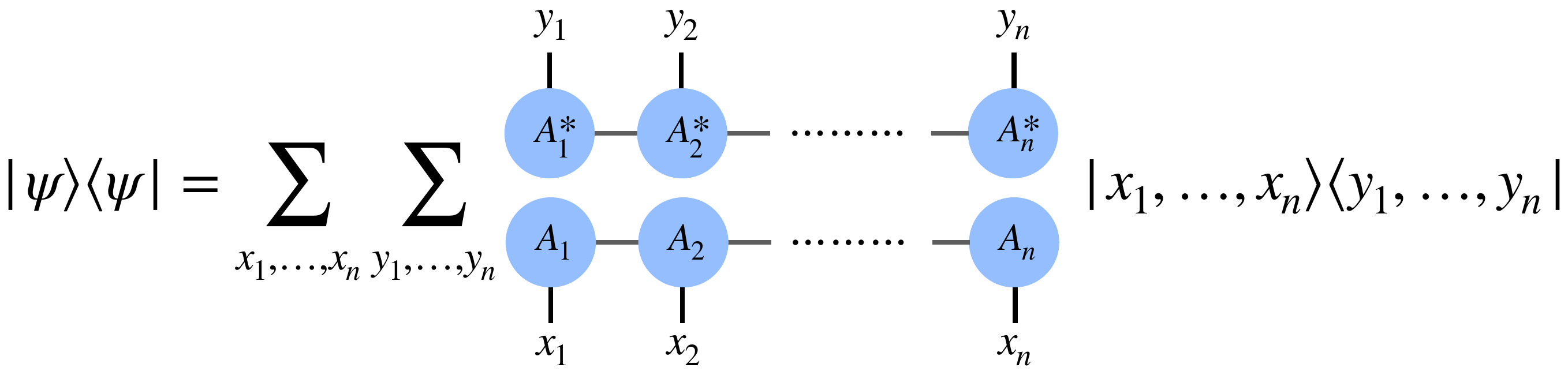}
$$
where now the by fusing the two auxiliary spaces, we can identify the local MPO matrices $W_{j}^{x_jy_y} = A^{x_j}_j\otimes (A^*_{j})^{y_j}$ with bond dimension given by the square of the MPS bond dimension.

\begin{example}{Multiply MPO to MPO or MPS}{mpo_mps_contractions}
Let us for a moment forget about normalisation and suppose we have the following MPS and MPO:
$$
\includegraphics[width=0.75\textwidth,valign=c]{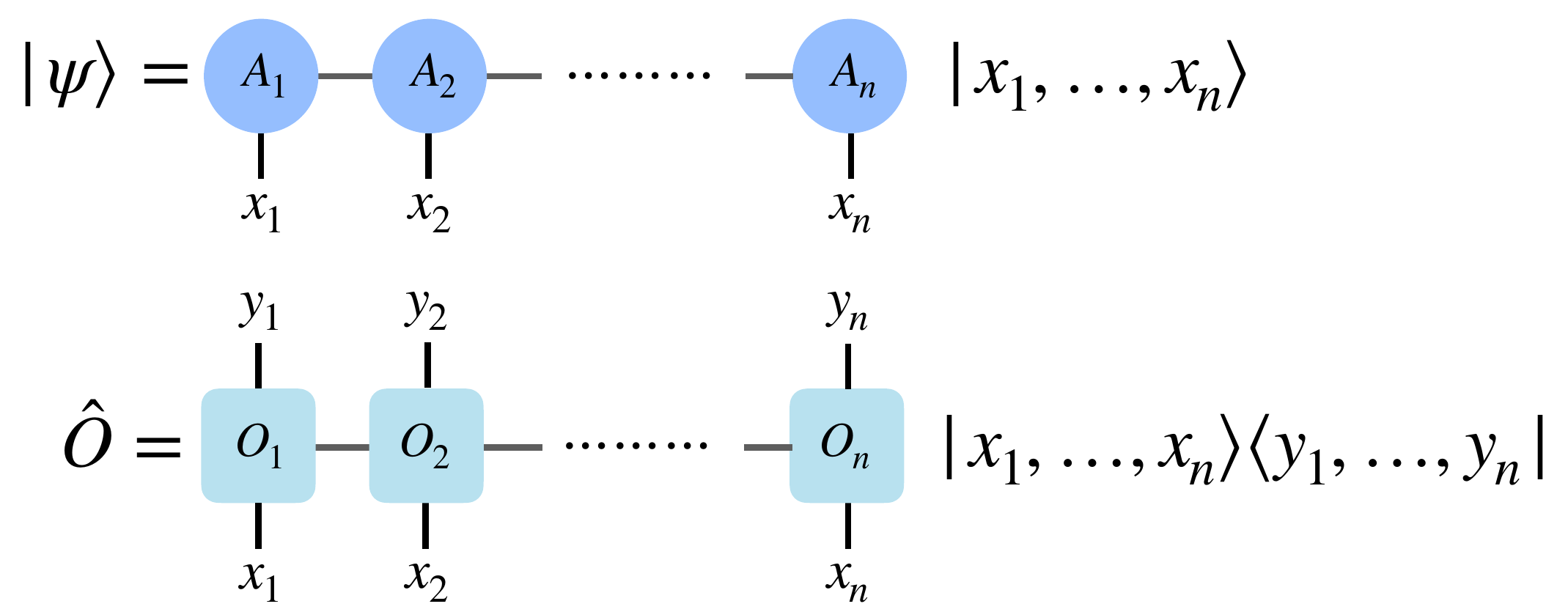}
$$
with ${\rm dim}(A_{j}^{x_j}) = \chi_{j-1}\times \chi_{j}$
and ${\rm dim}(W_{j}^{x_jy_j}) = D_{j-1}\times D_{j}$,
and where we use the implicit summation convention for repeated indices.
Applying the operator $\hat O$ to the state $\ket{\psi}$
actually results in the following contraction
$$
\includegraphics[width=\textwidth,valign=c]{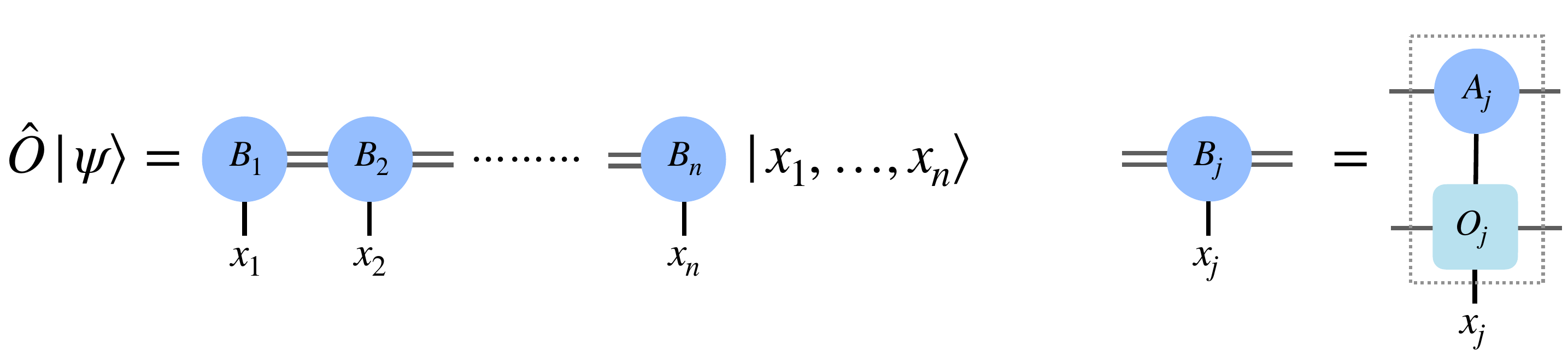}
$$
which is a new MPS state with matrices $B^{x_j}_{j}$
whose auxiliary dimensions are
${\rm dim}(A_{j}^{x_j}) = D_{j-1}\chi_{j-1}\times D_{j}\chi_{j}$. Notice that, a very similar procedure applies for the product of two operators $\hat O$ and $\hat V$ both having an MPO representation:\index{MPO!product}
$$
\includegraphics[width=0.8\textwidth,valign=c]{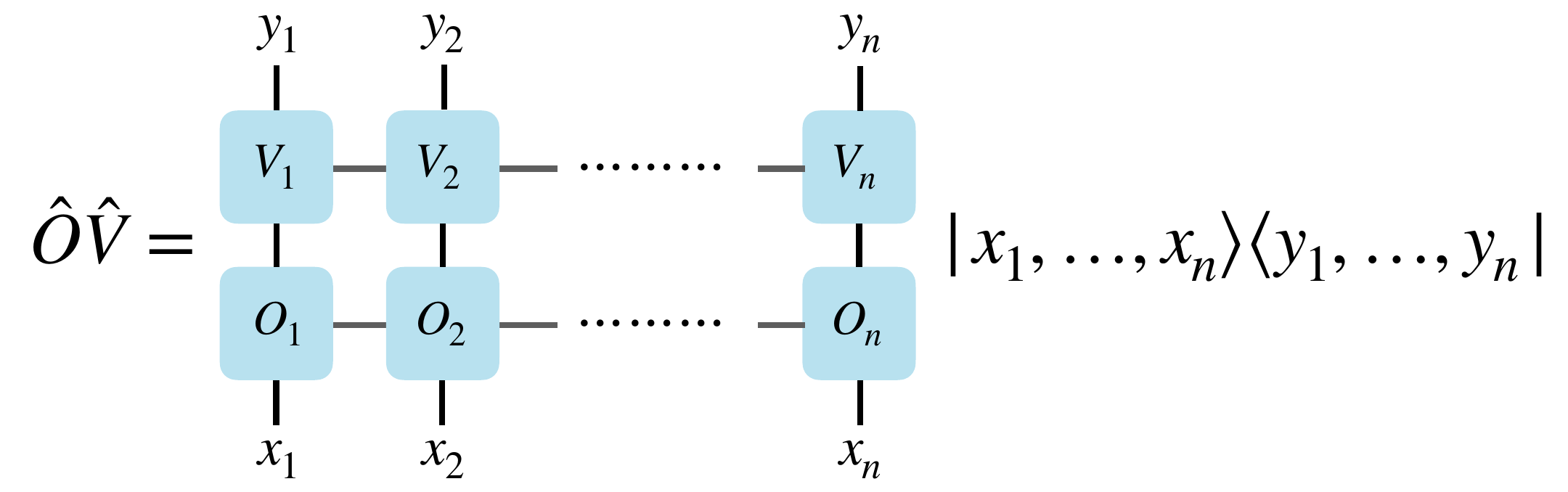}
$$
where the order of the physical indices contraction is relevant since in general $[\hat O,\hat V]\neq 0$.
\end{example}

\paragraph{Relevant finite-dimensional MPO ---}
Constructing an exact compact MPO representation for certain physically relevant operators might initially appear daunting. However, analogous to the case of low-entangled states that admit an exact MPS representation, whenever ``local'' operators are involved (such as Hamiltonians with short-range interactions), an exact finite-dimensional MPO representation exists.
Similarly to what we already did for MPS, we can rearrange a matrix product operator in terms of operator-value matrices,
as follow
\begin{equation}
    \includegraphics[width=0.9\textwidth,valign=c]{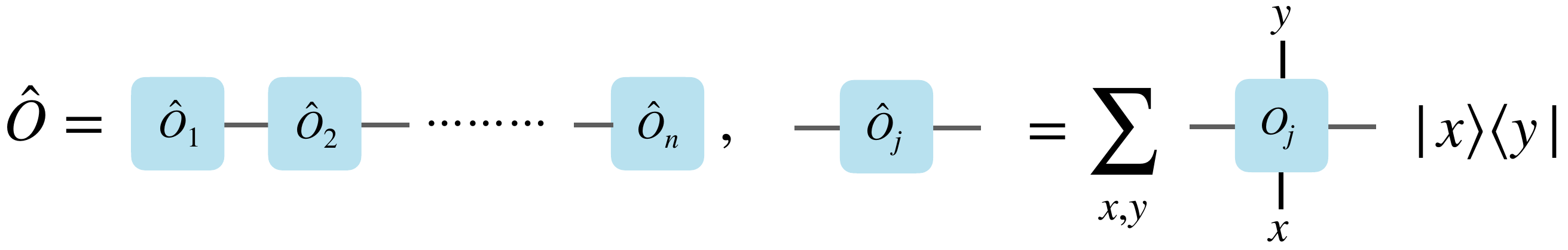}
\end{equation}
This representation is very useful to understand how to systematically construct exact MPO of relevant operators. In fact when considering the addition of two operators, $\hat W$ and $\hat V$, with both MPO representations, the resulting MPO is formed by the direct sum of the local operator-value matrices for all sites  $1 < j < n$, with the exception of the boundary sites where we add row and column vectors. Essentially, we get\index{MPO!direct sum}
$$
\includegraphics[width=\textwidth,valign=c]{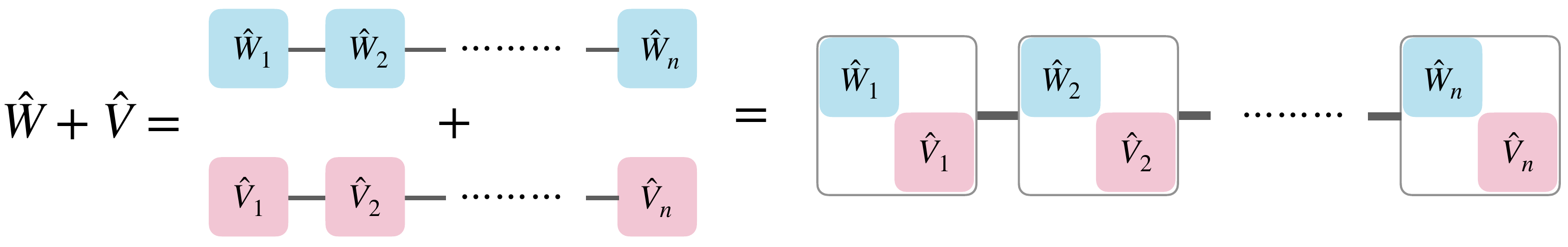}
$$
where basically we are using block diagonal matrices to take into account the independent effects of both operators.
However, for ``local'' operators, the previous representation is sub-optimal, and we can systematically construct much better representations.

In the following we present a list of example of operators which admit an exact MPO representation.
\begin{enumerate}

\item
Let us start from a simplest example of a local Hamiltonian acting on $n$ qubits
\begin{equation}\label{eq:H_local_1}
\hat H = \sum_{j=1}^{n} h_{j} \hat Z_{j}
\end{equation}

This compact form of Hamiltonian can be written in a more general tensor product notation,
\begin{equation}
\hat H = \sum_{j=1}^{n} \hat{I}_1 \otimes \ldots \otimes \hat{I}_{j-1} \otimes h_{j} \hat Z_{j} \otimes \hat{I}_{j+1} \otimes \ldots \otimes \hat{I}_n
\end{equation}

In this notation, it is evident that the size of the matrix corresponding to the operator $\hat{H}$ is $d^n \times d^n$, where each term in the sum is an exact product operator, i.e., an MPO with auxiliary dimension equal to one. Clearly, a suitable MPO representation of the Hamiltonian~(\ref{eq:H_local_1}) requires an auxiliary space with dimension $D > 1$. The question then becomes: what is the minimal extra resource needed to store all the information in a suitable operator-valued matrix product form $\hat{O}_1\hat{O}_2 \cdots \hat{O}_n$.

If we temporarily set aside the boundary vectors $\hat{O}_1$ and $\hat{O}_n$ and focus on the bulk matrices, we can ask ourselves what these matrices need to accomplish and remember:
\begin{itemize}
    \item Insert the identity operator regardless of the lattice site.
    \item Insert the operator $\hat{Z}$ and remember having done so.
    \item After the previous step, insert only the identity operator.
\end{itemize}

Actually, we can accomplish, and visualise the effect of the operator-value matrices, via the following \textbf{auxiliary-state diagram}\index{MPO!auxiliary-state diagram}
$$
\includegraphics[width=0.65\textwidth,valign=c]{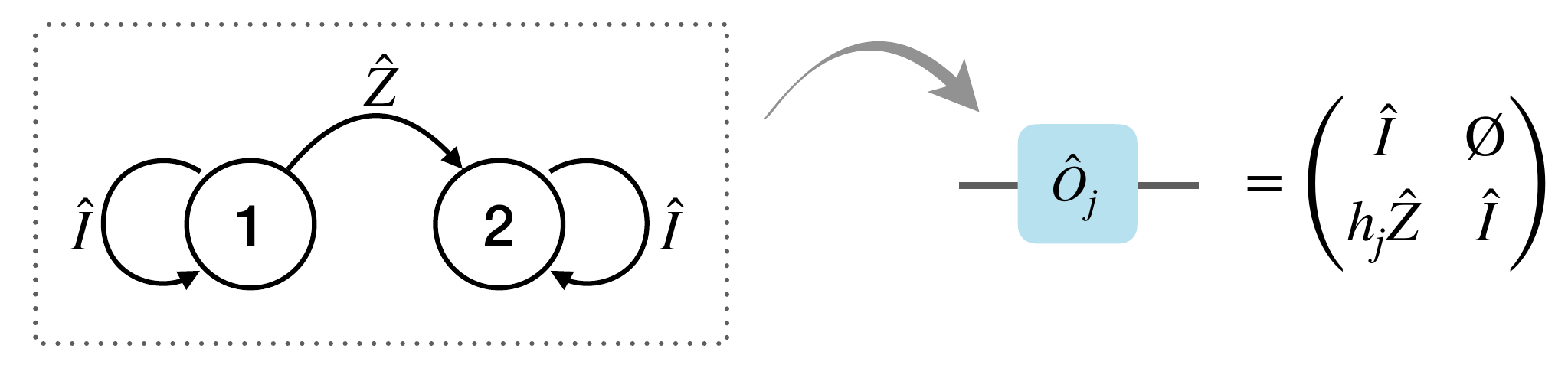}
$$
Finally, the boundary vectors can be easily obtained from the bulk matrix, by just piking up the last row and the first column, namely having
$\hat{O}_1 = \begin{pmatrix}
h_1\hat Z & \hat{I}
\end{pmatrix}$
and
$(\hat{O}_n)^{t} = \begin{pmatrix}
\hat{I} & h_n\hat Z
\end{pmatrix}$.

\item
Let us consider now an interacting Hamiltonian like the Heisenberg one with anisotropy (i.e.\ the XXZ Hamiltonian)
\index{MPO!XXZ Hamiltonian}
\begin{equation}\label{eq:H_XXZ}
\hat H = \sum_{j=1}^{n-1}
\hat X_{j}\hat X_{j+1}
+ \hat Y_{j}\hat Y_{j+1}
+ \Delta \hat Z_{j}\hat Z_{j+1},
\end{equation}
where for simplicity we are considering the anisotropy $\Delta$
uniform along the chain.
To find the MPO representation, let's focus on a generic neighboring interaction term $\sum_j \hat{S}_{j} \hat{S}_{j+1}$. Practically, as before, we need to insert the identity operator. Then, if the bulk matrix inserts the operator $\hat{S}$, immediately after we have to insert the same operator again. After this step, we complete the operator chain with the identity. This need to be done for each of the interaction terms. Using the auxiliary-state diagram we get
$$
\includegraphics[width=0.8\textwidth,valign=c]{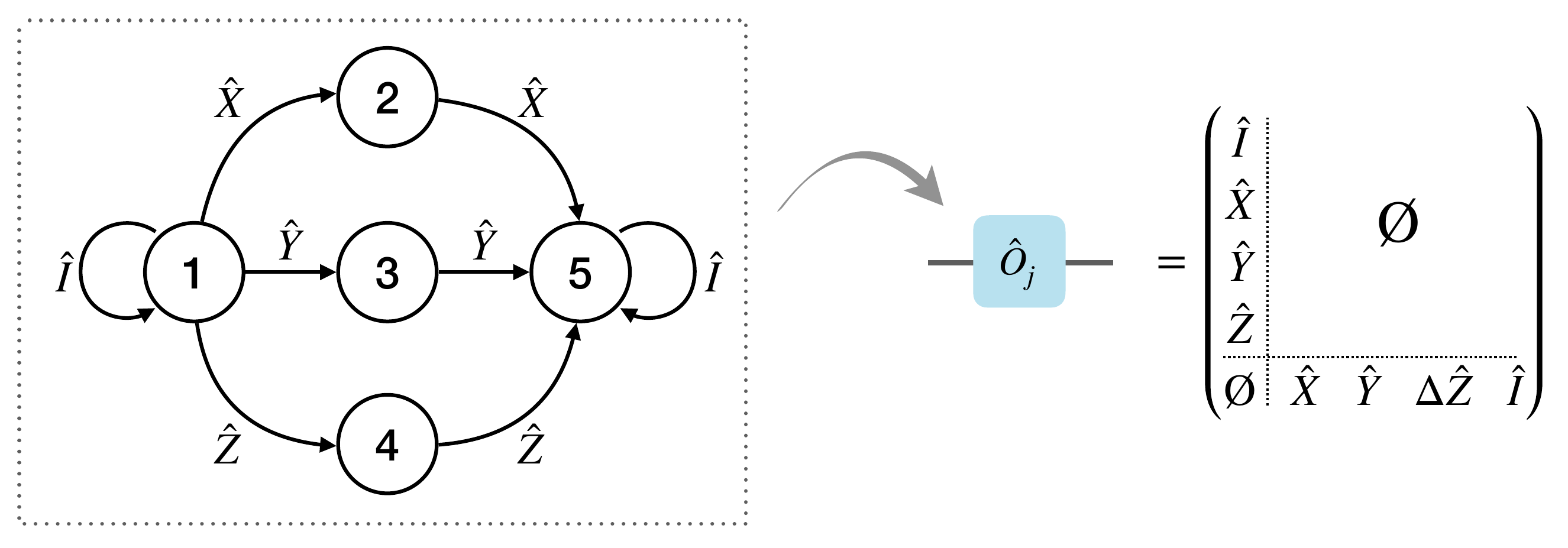}
$$
and similar consideration as before applies for the boundary vectors.

\item
Finally, we want to address arbitrary \textbf{long-range interacting Hamiltonian}.\index{MPO!Long-range Hamiltonian} Let' us start with the preliminary example of a fully connected spin system, such that
\begin{equation}
\hat H = \sum_{j>i}^{n} \hat S_i \hat S_j,
\end{equation}
and $\hat S$ here stays for a generic local operator.
A generic operator entering in that sum can be written explicitly as
 $
\cdots \hat I \otimes \hat I
\otimes \hat S \otimes \hat I^{|j-i|} \otimes\hat S
\otimes \hat I \otimes \hat I \cdots$
for any possible distance $|j-i| \in \{1,\dots,n-1\}$.
The only difference between a simple neighboring site interaction $\hat{S}_{j} \hat{S}_{j+1}$ is that, after inserting the operator $\hat{S}$ at any lattice site $j$, our MPO must allow for the insertion of an arbitrary number of identity operators until the second insertion of the operator $\hat{S}$ occurs. Thereafter, we complete our chain with the remaining identities. In terms of auxiliary-state diagram this reads
$$
\includegraphics[width=0.8\textwidth,valign=c]{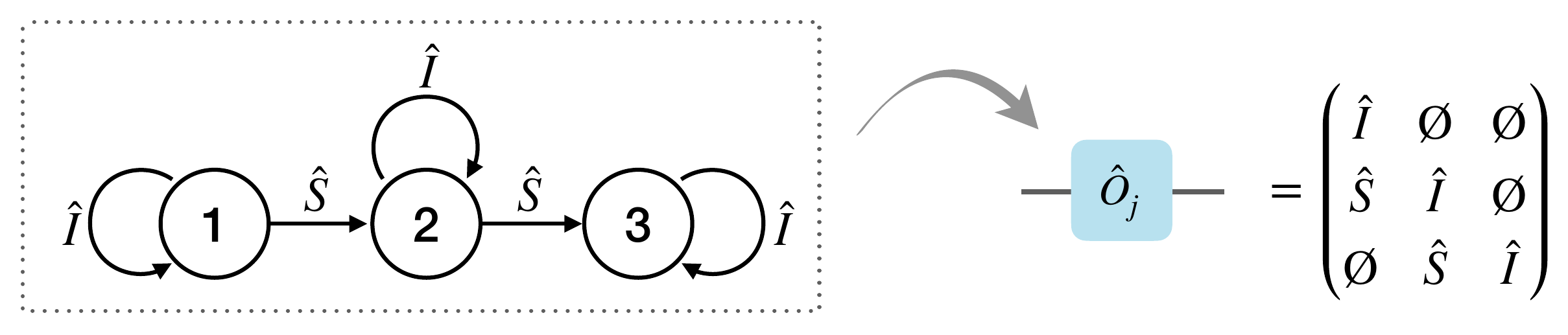}
$$
Now it is clear that, if we have exponentially decaying interactions such that
\begin{equation}
    \hat H = \sum_{j>i}^{n} e^{-|j-i|/\xi} \hat S_i \hat S_j
    = \sum_{j>i}^{n} \lambda^{|j-i|} \hat S_i \hat S_j
\end{equation}
with $\lambda \equiv \exp(-1/\xi)$, we can slightly modify the state diagram of the fully connected Hamiltonian by introducing an additional numerical coefficient $\lambda$ for each identity operator inserted by the auxiliary state $2$, and another $\lambda$ when transitioning from state $2$ to state $3$. This modification leads to the intriguing result that exponentially decaying interactions can be exactly encoded into an MPO with an auxiliary dimension of $D=3$:
\begin{equation}
   \hat O_j = \begin{pmatrix}
\hat I & \varnothing & \varnothing\\
\hat S & \lambda \hat I & \varnothing\\
\varnothing & \lambda \hat S & \hat I
\end{pmatrix}.
\end{equation}
This results is extremely useful, because it allows to efficiently represent generic interacting Hamiltonians on a finite chain
\begin{equation}
    \hat H = \sum_{j>i}^{n} J(j-i-1) \hat S_{i} \hat S_{j}.
\end{equation}
In fact, once we approximate the interactions as a sum of $K$ different exponential $J(r) \simeq \sum_{k=1}^{K} \alpha_{k} \lambda_{k}^{r}$, we can rewrite the generic Hamiltonian~as
\begin{equation}
    \hat H = \sum_{k=1}^{K}\sum_{j>i}^{n} \alpha_{k} \lambda_{k}^{j-i-1} \hat S_{i} \hat S_{j},
\end{equation}
and using the previous argument we can easily draw the following
auxiliary-state diagram
$$
\includegraphics[width=0.9\textwidth,valign=c]{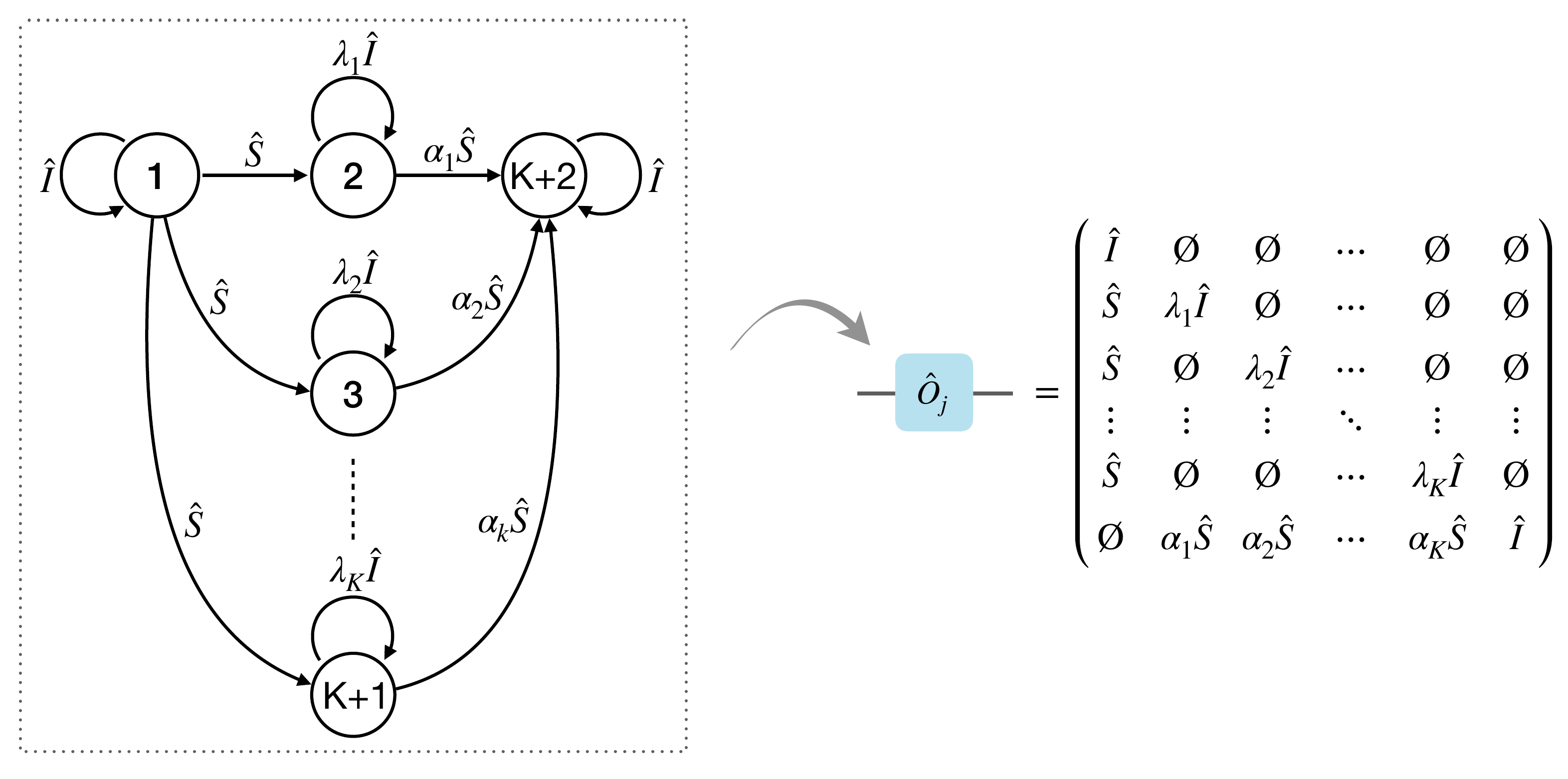}
$$
which results in a MPO with auxiliary dimension $D=2+K$.
\end{enumerate}

\biblio
\chapter{Quantum computing with Tensor Networks}\label{chap3}
\epigraph{I am made and remade continually. Different people draw different words from me.}{Virginia Woolf}%, The Waves, 1931}

Quantum computing protocols essentially involve preparing a system in a specific state $\ket{\psi}$, evolving it through a circuit composed of logical quantum gates, and finally performing measurements on the resulting states. In a more general setting (adaptive quantum circuits) it is also possible to perform mid-circuit measurements, conduct classical computations on these outcomes, and subsequently apply unitary gates conditionally, depending on the results of these classical computations.

In this chapter, we will see how each of these three steps can be efficiently replicated using the MPS formalism, thereby enabling the creation of classical simulators of quantum computers. Of course, since quantum computers are generally believed to perform tasks considered NP-hard for classical computation, these classical simulators necessarily involve a certain level of approximation.

\section{State preparation}

In this section, we address the challenge of preparing a generic state with quantum unitary gates and/or Tensor Network representation. State preparation is a fundamental task in quantum computation and quantum information. However, it is generally an inefficient process, often requiring an exponential number of gates relative to the number of qubits. To illustrate this, let us consider a quantum computer with $n$-qubit register initially in the computational basis state $\ket{00\dots 0}$, and examine the preparation of an arbitrary quantum state.
In general, we wish to prepare a generic many-body state as the following
\begin{equation}
\ket{\psi} = \sum_{\pmb{x}} \psi_{\pmb{x}} \ket{\pmb{x}}
\end{equation}
where $\psi_{\pmb{x}}$ are complex coefficients that define our target state. This preparation typically involves a sequence of rotations and entangling operations. For example, the state preparation circuit might employ a series of $\hat{R}_{\pmb{n}}(\alpha)$ gates to initialize the qubits, followed by controlled-NOT (CNOT) gates to create the necessary entanglement.

Despite the apparent simplicity in specifying such a state, the actual implementation can become highly complex. The number of required operations grows exponentially with the number of qubits, posing significant practical challenges. In the following subsections, we will explore various techniques and algorithms that have been developed to mitigate this complexity, enabling more efficient state preparation for specific classes of states.

We note that, similar to classical computation, preparing a quantum state with a quantum computer generally requires an exponential number of operations. However, quantum computing has a significant advantage in terms of memory requirements. Specifically, a quantum computer can efficiently store a wave vector with $n$ qubits, determined by $2^n$ complex numbers, which are the coefficients of its expansion over the computational basis. In contrast, a classical computer needs $O(2^n)$ bits to store these $2^n$ complex numbers.

This demonstrates the substantial memory efficiency of a quantum computer: it could perform the same task using only $n$ qubits. This memory efficiency directly impacts the complexity of quantum state preparation. While classical systems require exponentially increasing resources, quantum computers leverage their qubit-based architecture to manage state vectors efficiently.

\enlargethispage*{\baselineskip}
Furthermore, the connection between quantum state preparation and Matrix Product States (MPS) complexity offers additional insights. The auxiliary dimension in MPS reflects the entanglement present in the state, and states with lower entanglement can be represented with smaller auxiliary dimensions, reducing computational overhead. This connection suggests that quantum computers, which naturally handle entangled states, can efficiently prepare and manipulate states that would be complex to manage classically.

In essence, the exponential memory advantage of quantum computing, coupled with the understanding of MPS complexity, underscores the potential of quantum systems to outperform classical counterparts in handling complex, high-dimensional quantum states.

\subsection{Product state}
In certain scenarios, a specific wave function can be prepared efficiently. An operation on a quantum computer is considered efficient if it requires a number of elementary gates that is polynomial in the number of qubits. For example, creating an equal superposition of all states in the computational basis,\index{Many-body state!product state}
$$
\frac{1}{\sqrt{2^n}} \sum_{\pmb{x}}\ket{\pmb{x}},
$$
can be achieved by applying $n$ Hadamard gates, one to each qubit, starting from the initial state $\ket{00\dots 0}$.

The key insight here is that the previous state is simply a product state along a different direction, which can be efficiently prepared by applying local single-qubit unitaries. In general, any product state, or Matrix Product State (MPS) with bond dimension \( \chi = 1 \), can be readily prepared using the following procedure, assuming the quantum platform allows for arbitrary single-qubit rotations:
\begin{equation}
\includegraphics[width=0.8\textwidth,valign=c]{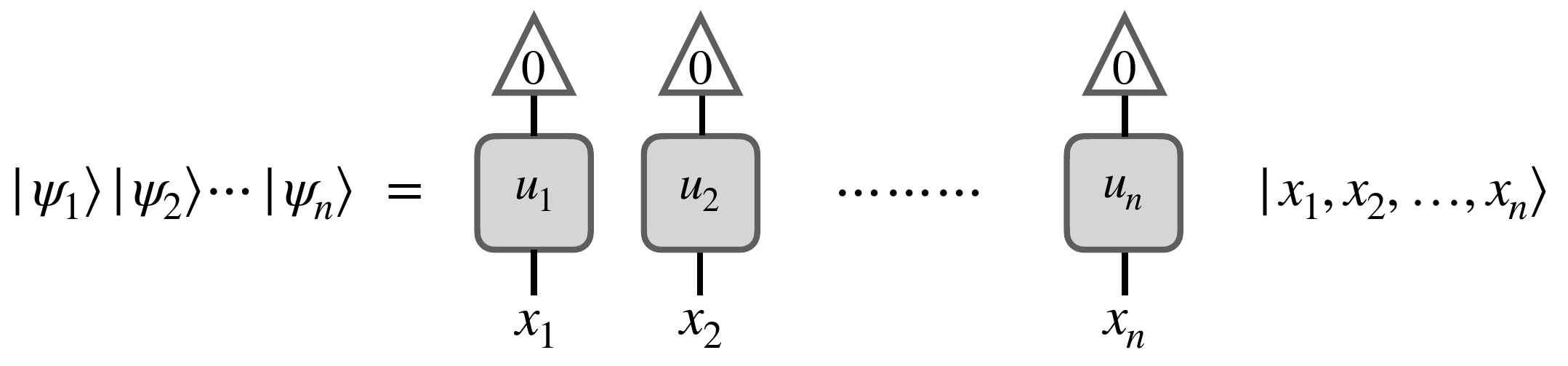}
\end{equation}
where $u_j$ denotes a single-qubit unitary operation applied to the $j$-th qubit. This approach leverages the ability to independently manipulate each qubit to create an entanglement-free state representation suitable for various quantum computing tasks.

\subsection{MPS to quantum circuit}

In the realm of universal quantum computation, it is well-established that any quantum state leaving in an $n$-qubit register (which can be extended to quantum qudits) can fundamentally be expressed as a global unitary transformation applied to an elementary computational basis state. This foundational principle asserts that the many-body wave function can be succinctly represented~as
$$
\includegraphics[width=0.6\textwidth,valign=c]{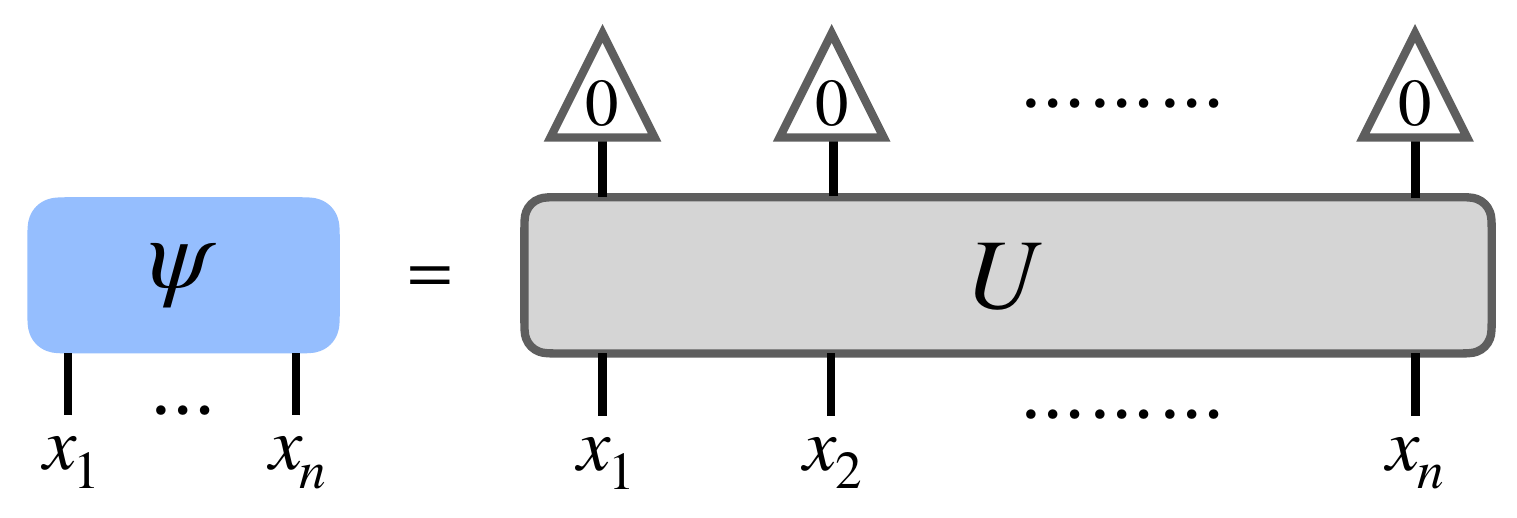}
$$
However, for an $n$-qubit state, finding the global unitary transformation $\hat{U}$ such that
$\ket{\Psi} = \hat{U}\ket{00\cdots 0}$ is generally quite challenging. Additionally, having this matrix as a global transformation is practically useless for quantum computing purposes. Instead, we desire the unitary operator $\hat{U}$ to be constructed from only one- and two-qubit gates, i.e., gates that can be feasibly implemented on a quantum device.

Here, we reduce the complexity of this problem by considering only \textbf{states that have an MPS representation with a local bond dimension equal to the dimension of the physical indices}~\cite{PhysRevLett.95.110503}. In other words, for local qubits, we want $\boldsymbol{\chi = 2}$ for all bonds except at the boundaries, of course. This approach can be easily generalized to qudits, where the physical dimension is now $d$~\cite{PhysRevA.101.032310}.

Therefore let us consider an MPS and assume to have it right-normalized (this condition can always been satisfied by means of a gauge transformation as we already have seen). The state wave function looks like
$$
\includegraphics[width=0.6\textwidth,valign=c]{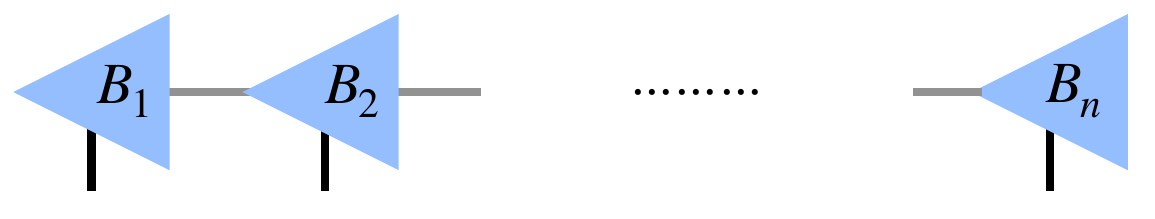}
$$

Now let's take  a specific site $j$ in the bulk $(1 < j < n)$ and reshape the tensor $B_j$ into a rectangular matrix of dimensions $2 \times 4$ by mapping
$(B_j)^{x}_{\alpha,\beta} \rightarrow (B_j)_{\alpha, (x,\beta)}$. Under this transformation, the right-normalization condition becomes $B_j B_j^{\dag} = I$.
However, $B_j^{\dag} B_j$ does not necessarily equal the identity matrix, indicating that $B_j$ is an \emph{isometric} matrix rather than a unitary one. To transform $B_j$ into a unitary matrix $u_j$, we introduce an additional artificial index $y$ (representing an auxiliary ), as follows:
$$
 (B_j)_{\alpha, (x,\beta)} \to (u_j)_{(\alpha,y)(x,\beta)}
$$
This can be achieved by augmenting the two $4$-component orthonormal row vectors with two additional orthonormal vectors to complete the basis. This ensures the condition that
$(u_j)_{(\alpha,0)(x,\beta)} = (B_j)_{\alpha, (x,\beta)}$.
This condition can be effectively represented using the following tensor diagram:
\begin{equation}
\includegraphics[width=\textwidth,valign=c]{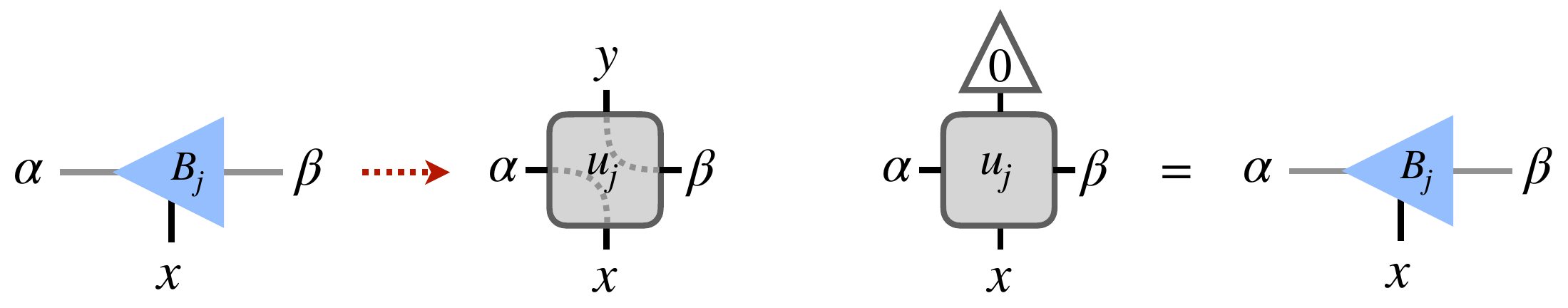}
\end{equation}

Notice that, in principle, we can combine the column and row indices to construct the unitary matrices in any order. However, we have chosen to maintain a specific order, where the auxiliary qubit is tensorized with the row auxiliary-dimension leg in reverse order compared to the column tensorization order. This choice, represented by the curved dotted wires in the graphical representation, allows us to clearly identify each virtual qubit Hilbert space line.

Finally a different situation arises at the boundaries. When $j=1$, the matrix $B_1$ is of dimension $1 \times 4$ (essentially, it is a single vector of length $4$). To embed this into a unitary matrix of size $4 \times 4$, we need to introduce two additional artificial qubits.
Conversely, at the opposite boundary where $j=n$, the tensor $B_n$ has dimensions $2 \times 2$, thus being inherently a square matrix and thus already a single-qubit unitary. The full graphical transformation looks like\index{MPS!circuit representation}
\begin{equation}
\includegraphics[width=\textwidth,valign=c]{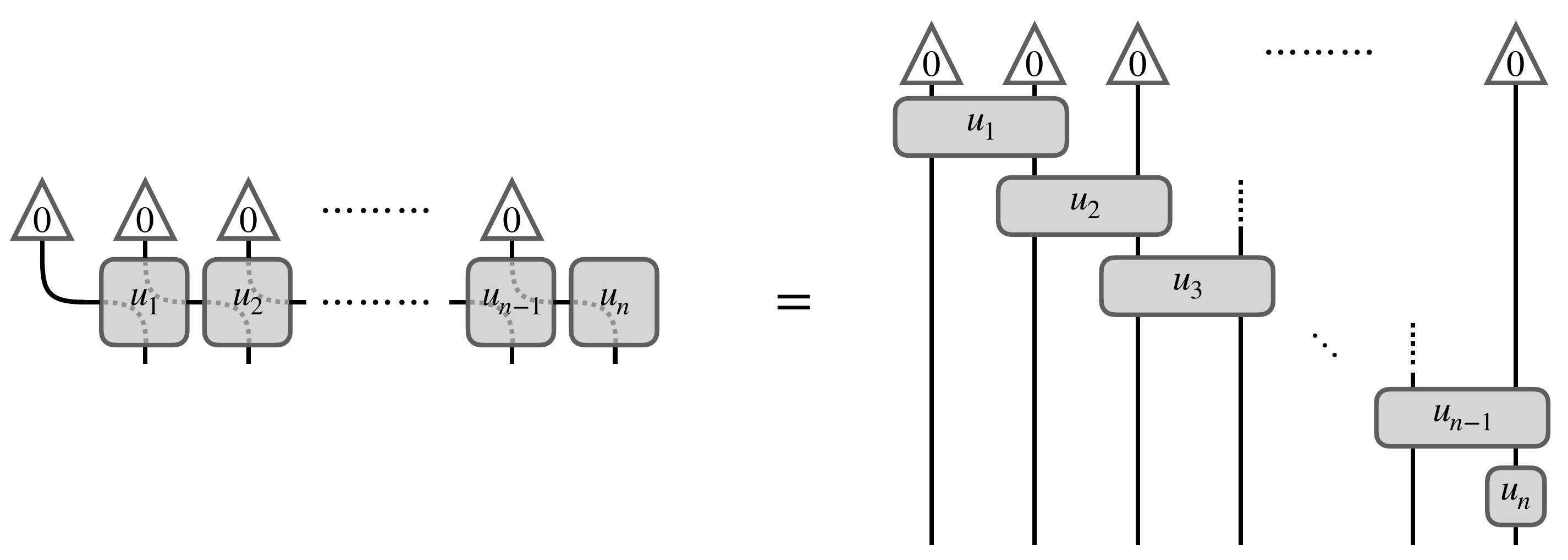}
\end{equation}
which forms a valid staircase quantum circuit, consisting solely of single-qubit and two-qubit gates acting on neighboring sites.

In the following we present two examples of non-trivial quantum states that admit a simple MPS representation.

\begin{example}{GHZ state as an MPS}{GHZ}
In the field of quantum information theory, a Greenberger--Horne--Zeilinger (GHZ)\index{GHZ state} state represents a unique type of entangled quantum state encompassing at least three qubits. The study of the four-particle GHZ state was first conducted by Daniel Greenberger, Michael Horne, and Anton Zeilinger in 1989~\cite{greenberger2007going}. These states may exhibit highly non-classical properties; however, it turns out that for an arbitrary number of qubits, they admit a very simple MPS (and thus, in a sense, ``classical'') representation.

Indeed, the standard MPS of the GHZ state for $n$ qubits is given as
\begin{equation}
    \ket{{\rm GHZ}} = \frac{1}{\sqrt{2}}
\begin{pmatrix}
    1 & 1
\end{pmatrix}
\begin{pmatrix}
    \ket{0} & 0 \\
    0 & \ket{1}
\end{pmatrix}^n
\begin{pmatrix}
    1 \\
    1
\end{pmatrix}
=\frac{1}{\sqrt{2}}
\left(\ket{00\dots 0} + \ket{11\dots 1}\right)
\end{equation}
which is an exact representation with bond dimension $\chi=2$. Graphically it corresponds to a sequence of copy tensors
\begin{equation}\label{chapt3_eq:GHZ}
\includegraphics[width=0.75\textwidth,valign=c]{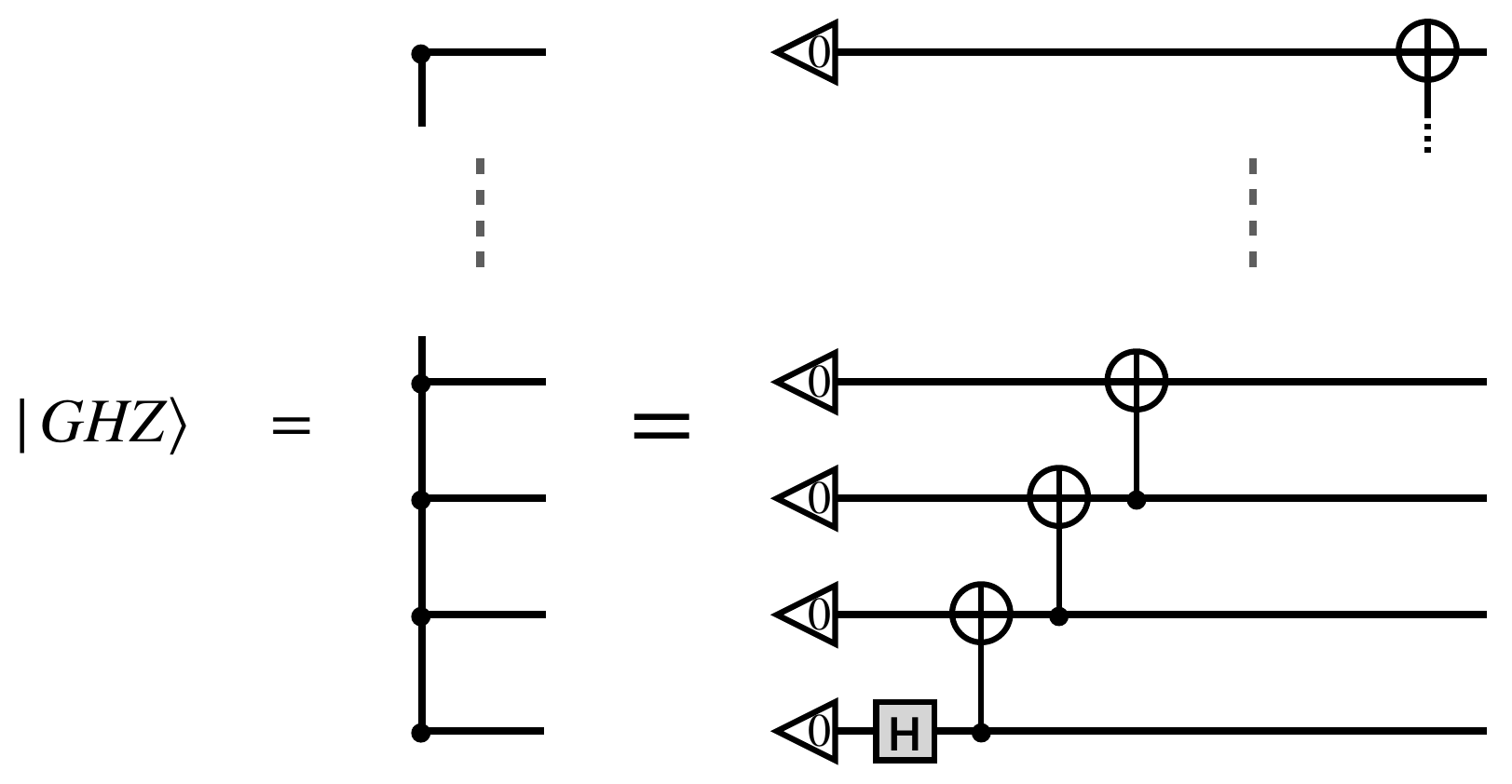}
\end{equation}
where in the last passage we transformed to sequential CNOT operation exploiting the following identity
$$
\includegraphics[width=0.35\textwidth,valign=c]{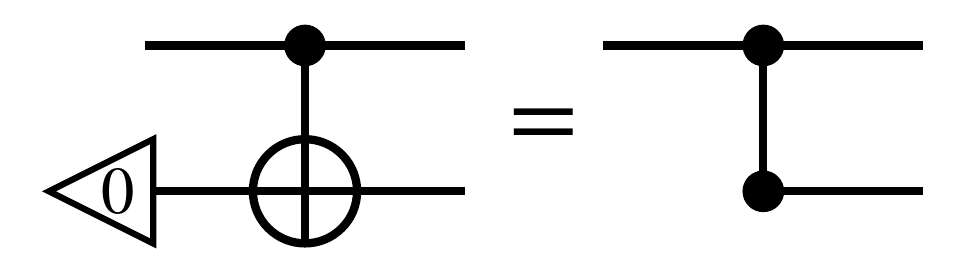}
$$

Let us conclude by noting that the ``classical'' nature of the GHZ state is further affirmed by the fact that, as we have seen, it has been constructed using only Clifford unitaries, making it a \emph{stabilizer state}. We refer the reader to Chapter \ldots  to learn more about \emph{stabilizer states} and \emph{magic}.

\end{example}

\begin{example}{W state as an MPS}{Wstate}
The $\rm{W}$ state\index{W state} is an entangled quantum state of three qubits, represented in bra-ket notation as follows:
$$
|\rm{W}\rangle = \frac{1}{\sqrt{3}} \left( |001\rangle + |010\rangle + |100\rangle \right)
$$
This state is remarkable for representing a specific type of multipartite entanglement and appears in several applications in quantum information theory. Particles prepared in this state exhibit the properties described by Bell's theorem, which asserts that no classical theory of local hidden variables can reproduce the predictions of quantum mechanics. The W state has been introduced for the first time by Wolfgang D\"ur, Guifré Vidal, and Ignacio Cirac in 2002~\cite{PhysRevA.62.062314}.

The W state exemplifies one of the two classes of three-qubit states that cannot be separated into independent subsystems. The other class is exemplified by the $3$-qubit GHZ state.
These two states, $|\rm{W}\rangle$ and $|{\rm GHZ}\rangle$, cannot be transformed into each other, even probabilistically, via LOCC (local operations and classical communication). Therefore, they represent fundamentally different types of tripartite entanglement.

The concept of the W state has been extended to  $n$ qubits, referring to a quantum superposition where each term has equal coefficients, and exactly one qubit is in the state $|1\rangle$ while the rest are in the $|0\rangle$ state; it can be easily written as an MPS
\begin{eqnarray}
    \ket{{\rm W}_n} & = & \frac{1}{\sqrt{n}}
\begin{pmatrix}
    0 & 1
\end{pmatrix}
\begin{pmatrix}
    \ket{0} & 0 \\
    \ket{1} & \ket{0}
\end{pmatrix}^n
\begin{pmatrix}
    1 \\
    0
\end{pmatrix} \\
& = & \frac{1}{\sqrt{n}}
\left(\ket{100\dots 0} + \ket{010\dots 0}+\ket{001\dots 0}+
\dots+\ket{000\dots 1}\right)\nonumber
\end{eqnarray}

\end{example}

\section{Quantum circuits}\label{subsec:circuits}

The state preparation is followed by the evolution of this initial state in time through a quantum circuit. In a traditional manybody setup the evolution is driven by the system Hamiltonian in continuous time. Quantum circuit on the other hand evolves the initial state in discrete time by application of sequence of unitary quantum operator, $|\psi\rangle_t = U_t |\psi\rangle_{t-1}$. The final evolution is given as $|\psi\rangle_t = U(0\rightarrow t) |\psi\rangle_{0}$ where $U(0\rightarrow t) = U_t U_{t-1} \ldots U_2 U_1$. Here, $U$ can be a generic operator however we will consider a special case of brickwork circuit where each $U$ is composed of alternating two site local unitary gates,
\begin{equation}
U_t = \begin{cases}
             \otimes_i \hspace{0.1cm}u_t^{2i+1,2i+2}  & i \in 0,1,2\ldots \\
             \otimes_i \hspace{0.1cm}u_t^{2i,2i+1}  & i \in 1,2,3\ldots
       \end{cases}
\end{equation}

Here each $u$ is a $4\times 4$ unitary matrix acting in the local Hilbert space of a pair of neighboring spins. The following figure illustrates the unitary evolution of a trivial initial state with a brickwork circuit where the discrete unitary evolution induces entanglement in the initial weakly entangled pure state over time.
$$
\includegraphics[width=0.60\textwidth,valign=c]{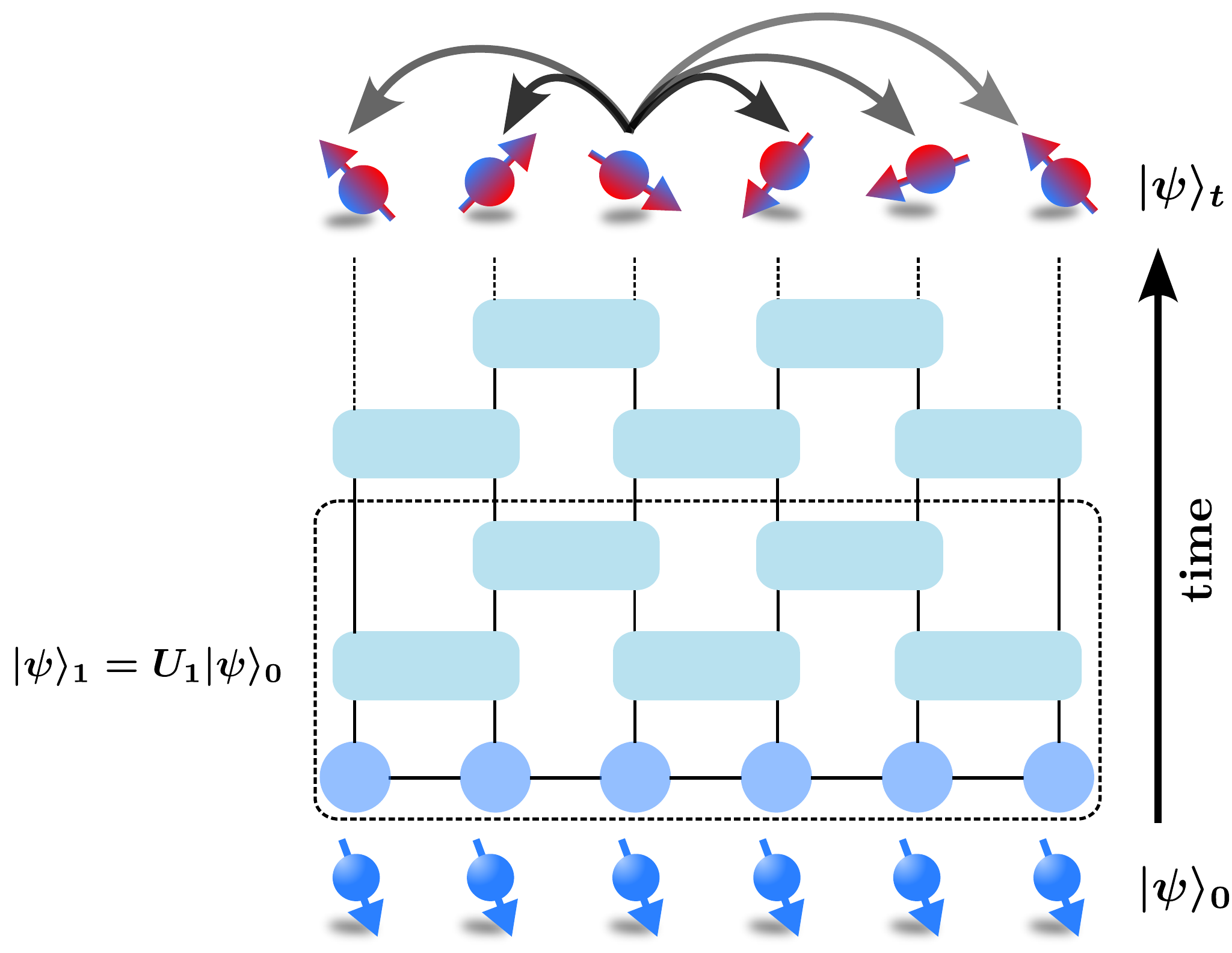}
$$

Here, the section inside the dotted rectangle represent the evolution of the initial state $|\psi\rangle_0$ over a unit time. Repeated evolution over similar units will generate the full unitary dynamics. With initial state as an MPS the brickwork Quantum circuit is a tensor network that can be evaluated by sequential contraction of the $u$ matrices with the MPS at specific sites. The optimal algorithm for evaluating such tensor network architecture is the time evolving block decimation (TEBD) algorithm~\cite{TEBD}\index{TEBD}. The TEBD algorithm was introduced to solve the continuous Hamiltonian dynamics and involved the Suzuki-Trotter decomposition of the evolution operator $e^{-i\Delta t H}$ into a product of two site unitary operators.\index{Trotter decomposition} Applying TEBD algorithm in a brickwork circuit evolution circumvents the Trotter error as the operators are already in the form of product of two site unitaries. TEBD constitutes an efficient sequential application of two site unitaries onto the MPS state. The evolution through a unit time can be written as,
\begin{equation}
    |\psi\rangle_t = U_t |\psi\rangle_{t-1} \rightarrow u^{1,2}_t \ldots \otimes u^{j,j+1}_t \otimes \ldots u^{N-1,N}_t |\psi\rangle_{t-1}.
\end{equation}

Conventionally, the TEBD algorithm is performed iteratively starting from the left edge of the MPS and moving towards the right edge and back while applying the two site unitary operators on the corresponding sites. Furthermore, the MPS is kept in a mixed canonical form (see subSection~\ref{subsec:MPS}) through out the evolution as the canonical form is preserved under the application of unitary operator. In the following example we will thoroughly outline the core of TEBD algorithm which is the application of a two site unitary gate onto the MPS.

\begin{example}{Local unitary updates of MPS with TEBD}{TEBD}
In this example we will thoroughly explain the local unitary update during the operation $|\psi\rangle_t = U_t |\psi\rangle_{t-1}$. Specifically, we will consider the operation $u_t^{i,i+1}|\psi\rangle_{t-1}$ as highlighted inside the dotted rectangle in figure below.
$$
\includegraphics[width=0.80\textwidth,valign=c]{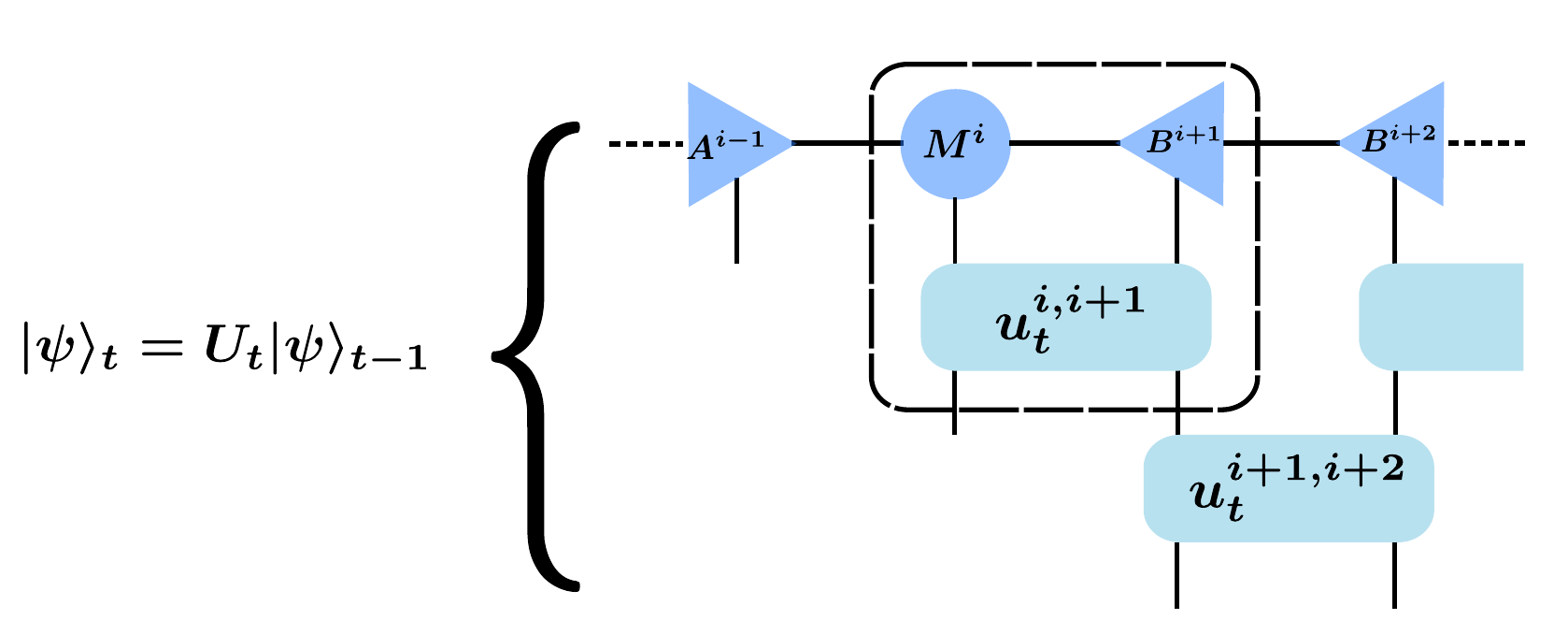}
$$

The state is in mixed canonical form (see equation~\eqref{eq:mixedcanon1}),
\begin{align*}
|\psi\rangle &= \sum_{\alpha,x_i,x_{i+1},\beta}  \big[M^{x_i}B^{x_{i+1}}\big]_{\alpha,\beta}|\alpha\rangle |x_i\rangle |x_{i+1}\rangle |\beta\rangle\\
&= \sum_{\alpha,x_i,x_{i+1},\beta}  T^{x_i,x_{i+1}}_{\alpha,\beta}|\alpha\rangle |x_i\rangle |x_{i+1}\rangle |\beta\rangle
\end{align*}
where $|\alpha\rangle$ and $|\beta\rangle$ are the left and right canonical vectors as defined in equation~(\ref{eq:mixedcanon2}). In this representation applying the two site unitary operator onto the state is now straightforward,
\begin{align*}
u^{\tilde{x}_i,\tilde{x}_{i+1}}_{x_i,x_{i+1}}|\psi\rangle &= \sum_{\alpha,x_i,x_{i+1},\beta}  u^{\tilde{x}_i,\tilde{x}_{i+1}}_{x_i,x_{i+1}} T^{x_i,x_{i+1}}_{\alpha,\beta}|\alpha\rangle |x_i\rangle |x_{i+1}\rangle |\beta\rangle\\
&= \sum_{\alpha,x_i,x_{i+1},\beta}\Tilde{T}^{x_i,x_{i+1}}_{\alpha,\beta}|\alpha\rangle |x_i\rangle |x_{i+1}\rangle |\beta\rangle .
\end{align*}

The operations outlined above are shown in the figure below,
$$
\includegraphics[width=1.00\textwidth,valign=c]{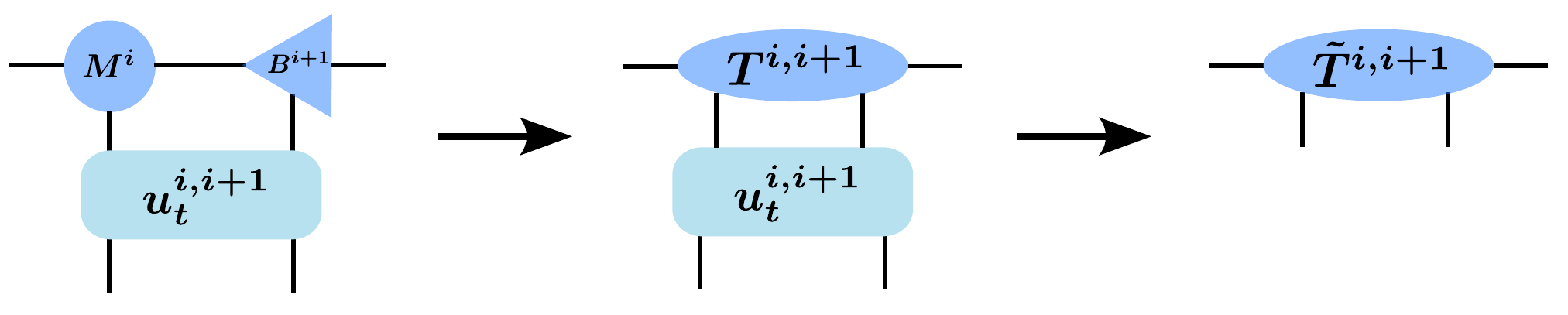}
$$

Applying two site unitary operator on site $i$ and $i+1$ increases the entanglement at those bonds, in other words the bond dimension between sites $i$ and $i+1$ increases from $\chi$ to $d \chi$, where $d$ is the local Hilbert space dimension ($d = 2$ for spin $1/2$ system). Successive application of local unitaries onto the bond during the evolution will exponentially increase the bond dimension rendering it computationally intractable. To this end the bond dimension at every bond is capped from above by $\chi_{\mathrm{max}}$. This truncation process is carried out by doing a Singular Value Decomposition (SVD)\index{Singular Value Decomposition} of the tensor $\tilde{T}$ as
$$
\Tilde{T}^{x_i,x_{i+1}}_{\alpha,\beta} \xrightarrow{reshape} \Tilde{T}_{[x_i\alpha],[x_{i+1}\beta]} \xrightarrow{SVD} \Tilde{A}^{x_i}_{\alpha,k}\tilde{\Lambda}^{i,i+1}_{k,k}\tilde{B}^{x_{i+1}}_{k,\beta},
$$

where, $\Tilde{A}^{x_i}$ is right normalized tensor,$\tilde{\Lambda}^{i,i+1}$ is a diagonal matrix, and $\tilde{B}^{x_{i+1}}$ is right normalized tensor. Therefor by construction the MPS remains mixed canonical after the application of local unitaries. The matrix $\tilde{\Lambda}^{i,i+1}$ has descending diagonal entries whose squares sums to unity $\sum_k \tilde{\lambda}_{k,k}^2$. The truncation process happens at this stage where we keep only the $\chi_{\mathrm{max}}$ largest values and discard the rest. This incurs a truncation error, $\sum_{k>\chi_{\mathrm{max}}} \tilde{\lambda}_{k,k}^2$. Following the truncation the matrix is normalized. The following figure illustrates the decomposition and truncation process,
$$
\includegraphics[width=1.00\textwidth,valign=c]{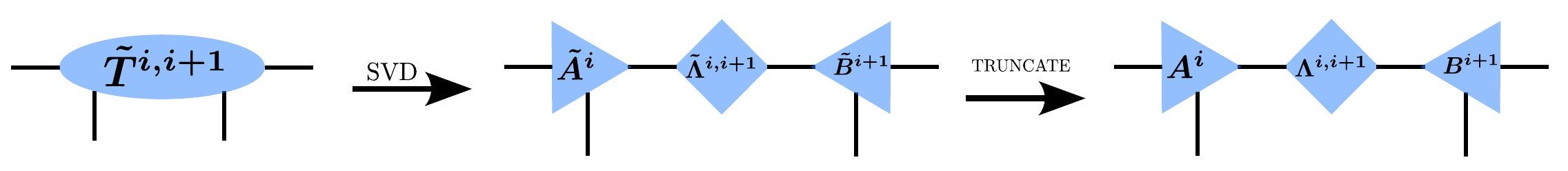}
$$

The state $|\psi\rangle$ remains in a mixed canonical form at the end of a local unitary update. The re-normalized diagonal matrix $\Lambda^{i,i+1}$ is now contracted with the tensor $B^{x_{i+1}}$ to build $M^{x_{i+1}}$. This brings us to the beginning of the outlined process just shifted by a site. Now following the same process we apply the unitary operator $u^{x_{i+1},x_{i+2}}$ on sites $i+1$ and $i+2$ respectively as shown in the figure below,
$$
\includegraphics[width=1.00\textwidth,valign=c]{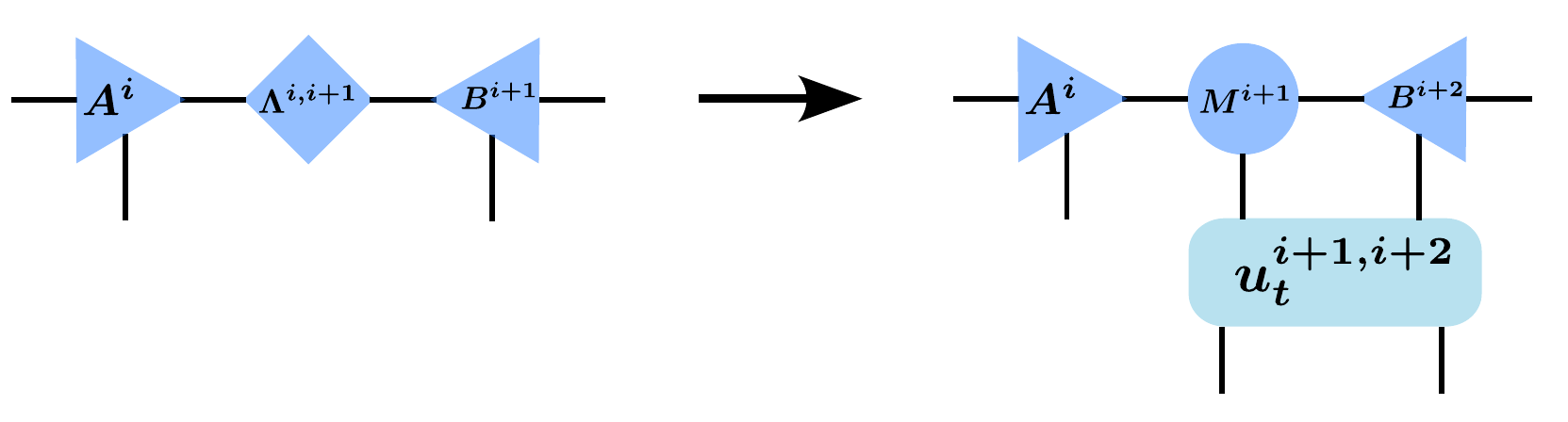}
$$

We proceed on a similar fashion by sequentially applying two site unitaries from left to right until we reach the right most site. This completes a unit left-to-right TEBD sweep and evolves the state through a unit time. This is followed by a similar sequential application of local unitaries starting from the rightmost site and moving toward left called right-to-left TEBD sweep. TEBD involves an alternative left-to-right and right-to-left sweeps.
\end{example}

\section{Measurements}\label{sec:Measurements}

Measurements are the essential readout of any quantum computation, allowing to extract (classical) information from a quantum state $\ket{\psi}$.\index{Measurement}

Suppose $\ket{\psi}$ is represented as an MPS, and we aim to perform a measurement. We will specifically focus on the simplest type of measurements: projective measurements of $\PauliZ_i$ operators on qubits indexed by $i \in I$, where $I$ is a subset of ${1, 2, \ldots, N}$ representing the qubits to be measured.

Since the operators $\{ \PauliZ_i \}_{i \in I}$ commute with each other, we can perform the measurements in any order we prefer. First, we will consider the qubit with the smallest index $i$ (i.e.\ the one closer to the left boundary of the MPS). Notice that
\begin{equation}
  \hat{P}_{0}^{(i)} = \frac{\Id + \PauliZ_i}{2} \qquad \hat{P}_{1}^{(i)} = \frac{\Id - \PauliZ_i}{2}
\end{equation}
are projectors on subspaces where qubit $i$ is respectively $\ket{0}$ or $\ket{1}$. We have therefore to compute outcomes probabilities
\begin{equation}
  p_0^{(i)} = \frac{1 + \expval{\PauliZ_i}{\psi}}{2} \qquad p_1^{(i)} = \frac{1 - \expval{\PauliZ_i}{\psi}}{2} \, .
\end{equation}
Evaluating the expectation value $\expval{\PauliZ_i}{\psi}$ becomes particularly simple if we have the orthogonality center of the MPS on site $i$. Indeed, if sites $j<i$ are left normalized and sites $j>i$ are right normalized, we get\index{Measurement!MPS scheme}
\begin{equation}\label{eq:mps_local_exp_value}
\includegraphics[width=\textwidth,valign=c]{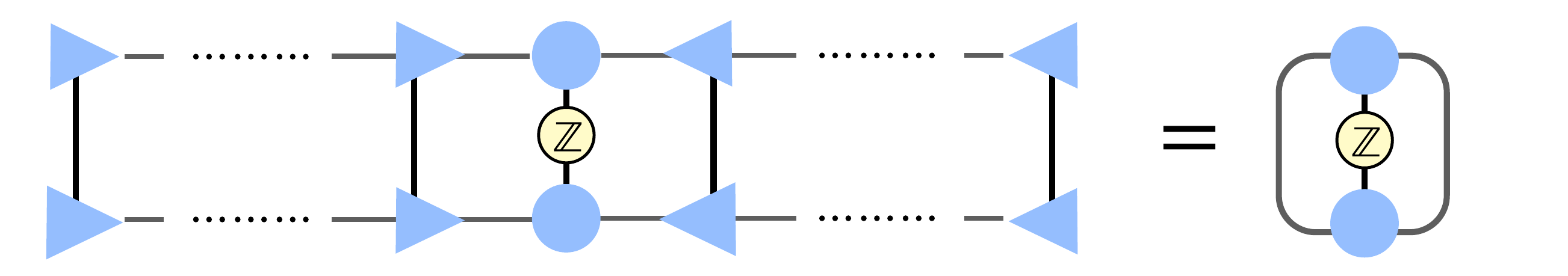}
\end{equation}
This contraction can be evaluated efficiently at a cost
$O(\chi^3)$. Having obtained $p_0, p_1$, the task is to simulate the wave function collapse with these probabilities. This can be done numerically by drawing a uniform (pseudo-)random variable $\xi \in [0,1]$ and setting the outcome $x$ to $0$ if $\xi < p_0$, and to $1$ if $\xi > p_0$. Once the outcome $x$ is determined, we must apply the measurement postulate (Eq.~\eqref{eq:jump_measurements}) to project $\ket{\psi}$ into the eigenspace corresponding to $x$ and then renormalize the state. This is achieved by setting the $\chi \times \chi$ matrix $A_i^{1-x}$ to $0$, effectively projecting away the component corresponding to $1-x$ in the MPS tensor $A_i$. For the normalization, $A_i$ is simply divided by $\sqrt{p_x^{(i)}}$ (which has already been evaluated). At this point, however, it is necessary to restore the correct MPS gauge by shifting the orthogonality center from $i$ to $i+1$. This can be done simply with a $QR$ decomposition of the tensor $A_i$ (after projecting it), and incorporating $R$ into the tensor $A_{i+1}$.
These last steps can be graphically represented as follows:\index{Measurement!MPS update}
\begin{equation*}
\includegraphics[width=\textwidth,valign=c]{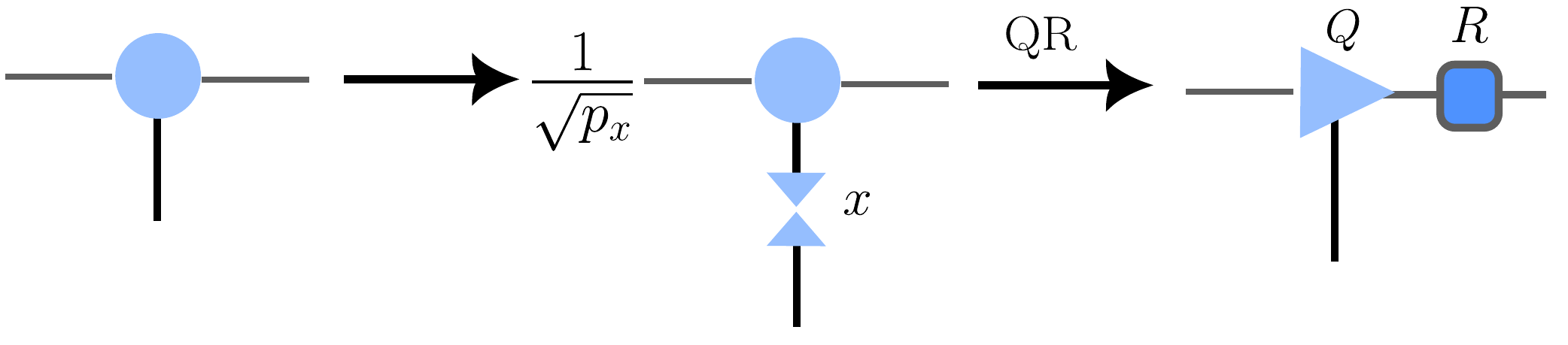}
\end{equation*}
Starting from a right-normalized MPS $\ket{\psi}$, we can then perform an entire sweep of QR decomposition, iteratively bringing the tensors into a left-norma\-lized form. Each time we reach a site $i \in I$, we need to measure $p_0$ and apply the outlined algorithm, extracting the outcome and projecting the MPS tensor,
before moving on to site $i+1$. The total cost of this algorithm is thus $O(N_{\text{meas}} N \chi^3)$, where $N_{\text{meas}}$ is the number of measurements we want to extract.

Simplifications occur, however, in the case where $I$ coincides with the set of all qubits $1, 2, \ldots  \, N$.
Referring to Eq.~\eqref{eq:mps_local_exp_value}, this is because if all sites $j<i$ have been measured, the bra and ket sides of the expectation value factorize. Consequently, the auxiliary line to the left of the right-hand side can be split, which simplifies the complexity of the evaluation. To be more precise, Eq.~(\ref{eq:mps_local_exp_value}) becomes
\begin{equation}\label{eq:mps_local_exp_value_2}
\includegraphics[width=\textwidth,valign=c]{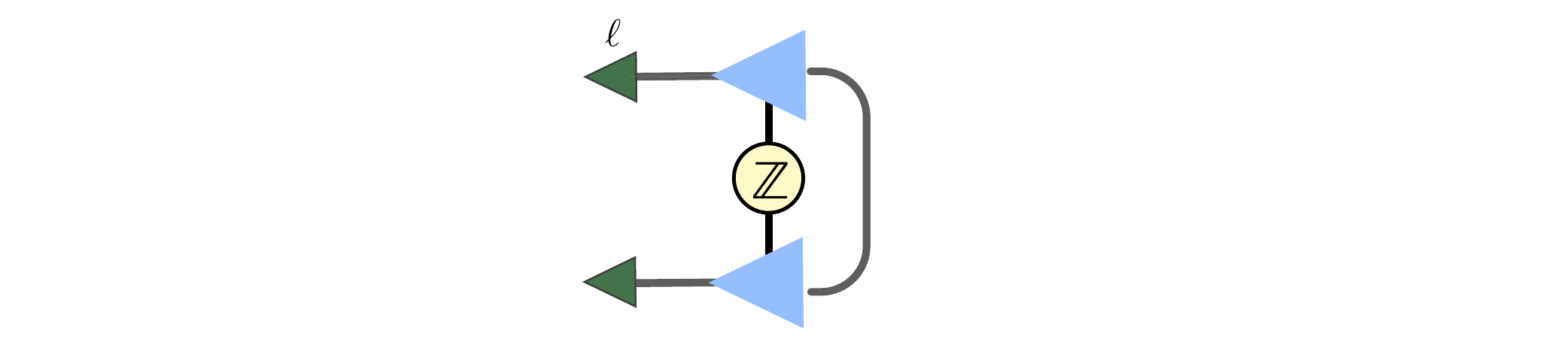}
\end{equation}
where we introduced a suitable left environment vector $\ell$. Notice that evaluating Eq.~\eqref{eq:mps_local_exp_value_2} have cost $O(\chi^2)$, instead of $O(\chi^3)$. Once we get $p_0, p_1$, we project $A_i$ as before, but without the need of a QR decomposition. Indeed, it is enough to update the environment vector as
\begin{equation}
\includegraphics[width=\textwidth,valign=c]{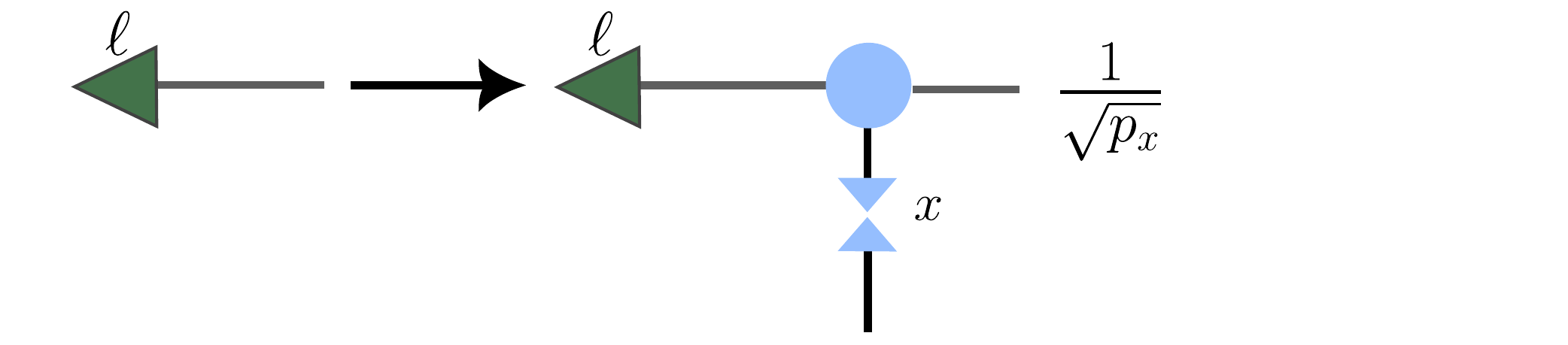}
\end{equation}
Again, we can start from a right-normalized MPS $\ket{\psi}$ and perform a sweep through all sites.
The total cost is therefore $O(N_{\text{meas}} N \chi^2)$. This improved sampling algorithm has been firstly proposed in ref.~\cite{Stoudenmire_2010}.\index{MPS!projective sampling}

Measuring all $\PauliZ_i$ operators is equivalent to sampling bit-strings $\pmb{x}$ from the computational basis, where each string is drawn with a probability given by $p(\pmb{x}) = |\langle \pmb{x}|\psi \rangle|^2$.\index{Measurement!bit-string sampling} This probability reflects the likelihood of obtaining the configuration $\pmb{x}$ in a measurement of the quantum state $\ket{\psi}$.\footnote{For a more detailed examination of the sampling algorithm and its connection to conditional probabilities, we recommend referring to Chapter~\ref{chap5}, where these concepts are thoroughly explored in the context of constructing the so-called minimal entangled typical thermal states (METS).}

Collecting these samples, or shots, is the typical method experimentalists use to measure (diagonal) observables. For instance, if an observable $\hat{O}$ is defined as
$\hat{O} = \sum_{\pmb{x}} O_{\pmb{x}} \ket{\pmb{x}} \bra{\pmb{x}}$, then $\expval{\hat{O}}{\Psi} = \sum_{\pmb{x}} O_{\pmb{x}}  p(\pmb{x})$, and therefore one can estimate $\expval{\hat{O}}{\Psi}$ as a sample average based on outcomes of $\pmb{x}$.
This situation is especially common when addressing classical problems embedded into quantum systems and proves useful numerically even when we lack an explicit MPO representation for $\hat{O}$.

Similar to what we have seen, every Pauli operator can be measured. For instance, to measure $\PauliX_i$, we utilize the identity $\PauliX = \Hgate \PauliZ \Hgate^{\dag}$. Thus, the measurement process can be expressed as:
\begin{equation}
\frac{\Id \pm \PauliX_i}{2} \ket{\psi} = \Hgate_i \hat{P}_{0,1}^{(i)} \ket{\tilde{\psi}} \quad \text{and} \quad p_{0,1} = \expval{\frac{\Id \pm \PauliX_i}{2}}{\psi} = \expval{\hat{P}_{0,1}^{(i)}}{\tilde{\psi}},
\end{equation}
where $\ket{\tilde{\psi}} = \Hgate^{\dag}_i \ket{\psi}$.
This method involves first applying the local gate $\Hgate^{\dag}_i$ to the MPS tensor $A_i$ and then calculating $p_0$ and $p_1$ as described. If the outcome is $x=0$, for example, we apply $\hat{P}_{0}^{(i)}$ to $\ket{\tilde{\psi}}$ and then transform it back with $\Hgate$.

\section{Quantum Platforms}
In the year 2000, theoretical physicist David DiVincenzo proposed a set of conditions, known as the DiVincenzo Criteria, which are necessary for building a proper fault-tolerant quantum computer \cite{DiVincenzo2000}. The requirements are the following\begin{enumerate}
    \item A scalable system with well-characterized qubits.
    \item The ability to initialize the state of the qubits to a simple fiducial state, such as $\ket{00 \ldots  0}$.
    \item Long relevant decoherence times, much longer than the gate operation time.
    \item An \emph{universal} set of quantum gates.
    \item A high-fidelity readout method.
\end{enumerate}
It is immediately clear that there is a certain level of contradiction in these requirements: on one hand, we need the quantum computer to be effectively protected from external influences to maintain its coherence, but on the other hand, we must interact with it intensively to set up the initial state, carry out the intended unitary evolution, and measure the final state. Nonetheless, in the last few years remarkable progress has been achieved thanks to the improvements in the control of quantum-mechanical systems, which allowed the transition of quantum hardware from laboratory curiosities to technical realities \cite{Cheng2023}.

We are going to focus on hardware platforms that show promise in realizing large-scale quantum computers, including superconducting qubits, neutral atoms and trapped-ions. In presenting these hardware implementations, our aim is not to favor any particular platform. Instead, we intend to demonstrate the potential application of tensor-networks methods in simulating the physics of these processors. It is important to notice that the field of quantum computing is still in a developmental phase, where multiple hardware platforms are being explored and tested for their efficacy in implementing quantum algorithms. Each of these platforms has its own set of advantages and challenges, making it unclear which one will ultimately prove to be the most effective for practical quantum computing. On the other hand, by effectively employing the physics and connectivity of quantum hardware, tensor-networks can serve as a crucial tool for advancing quantum technologies.

\subsection{Superconducting Qubits}
Superconducting qubits are advanced nonlinear circuits that utilize Josephson junctions\index{Josephson junction} to create quantized electromagnetic fields in the microwave frequency domain. These circuits are distinguished by their ability to be configured to exhibit customizable \emph{atom-like} energy spectra, thanks to the tunable parameters of their circuit elements. Often referred to as \emph{artificial atoms} due to their engineered energy levels, they are essential for quantum computing, offering high controllability and low noise in quantum logic operations. To function effectively, superconducting qubits require operation at cryogenic temperatures, typically around 10 millikelvin, to minimize thermal fluctuations. This is achieved using dilution refrigerators.

From a quantum circuit electrodynamics point of view, a Josephson junction consists of a nonlinear inductor $L_J = \phi_0/(2\pi) I_C \cos(\phi)$ in parallel with a capacitor $C_J$, where $I_c$ is the critical current, $\phi_0 = h/(2e)$ is the superconducting flux quantum, and $\phi$ is the phase difference across the junction. The two key parameters of the Josephson junction are the Josephson energy $E_J = \phi_0 I_c/(2\pi)$ and the charging energy $E_C = (2e)^2/(2C_J)$.

Among all possible superconducting qubits, transmons are a widely used variant. In their design, a shunt capacitor $C_S$ is added in parallel to the Josephson junction to decrease the charging energy, resulting in $E_C = (2e)^2/(C_J+C_S)$, and ensuring $E_J \gg E_C$. The single-qubit Hamiltonian for transmons is given~by:
\begin{equation}
    H = E_C n^2 - E_J \cos(\phi)
\end{equation}
where $n$ is the number of Cooper pairs traversing the junction, and $n$ and $\phi$ are conjugate variables with $[n, \phi] = i$. The shunt capacitor allows the low-energy eigenstates to be considered by expanding the potential term in a power series:
\begin{equation}
    -E_J \cos(\phi) \approx -\frac{1}{2} E_J \phi^2 + \frac{1}{24} E_J \phi^4
\end{equation}
The first term represents a quantum harmonic oscillator with equidistant energy levels, while the quartic term introduces anharmonicity, particularly affecting the gap between the first and second excited states. This non-linearity allows the definition of a qubit in the subspace of the ground state $|0\rangle$ and the first excited state $|1\rangle$. The Hamiltonian projected in the two-state basis is:
\begin{equation}
    H_Q = \frac{1}{2}(\epsilon \sigma_z - \Delta \sigma_x)
\end{equation}
with $\epsilon = -E_C$ and $\Delta = E_J$. The energy splitting of the eigenvalues is $\Omega = \sqrt{\epsilon^2 + \Delta^2}$. By varying the voltage on the Josephson junction with pulses, the parameters $\epsilon$ and $\Delta$ can be manipulated in time to generate single-qubit gates.

Microwave pulses are integral to the control and measurement of transmon qubits. These pulses, typically shaped with specific envelopes (like Gaussian or sine-squared), drive transitions between qubit states by resonating at particular frequencies that correspond to the qubit's energy levels. A typical control pulse might have a duration of around 10 nanoseconds and is carefully calibrated in terms of amplitude and phase to ensure high-fidelity operations. In superconducting quantum processors, the CNOT gate is typically implemented using a method called the cross-resonance gate. In this method, the control qubit is driven at the resonance frequency of the target qubit. This drive induces a coupling between the qubits, creating an interaction described by a Hamiltonian of the form:
\begin{equation}
    H = \frac{\omega}{2} \sigma_z \otimes \sigma_x
\end{equation}
where $\omega$ is the drive amplitude. By carefully adjusting the amplitude and duration of the microwave drive, an entangling interaction is created that, combined with single-qubit rotations, realizes the CNOT gate.

For measurement, microwave pulses are used in conjunction with resonators coupled to the qubits. A readout pulse, often several microseconds long, is applied to the resonator. The qubit's state influences the resonator's response, altering the reflected or transmitted microwave signal. This signal is then analyzed to infer the qubit's state.

\subsection{Neutral-Atoms Qubits}

Neutral atom quantum computers have emerged as a promising platform for both near-term applications and long-term fault-tolerant universal quantum computation. These systems utilize atoms trapped in optical tweezers\index{Optical tweezers} to create highly controllable qubits. The qubits are encoded in the electronic states or nuclear spin states of the atoms, which offer long coherence times and high-fidelity operations. Each atom is individually trapped by focused laser beams, forming a highly organized array that can be manipulated with precise laser pulses.

We start by considering the so-called \emph{digital addressing} of the neutral-atom arrays. In this setting the qubit states are encoded in the two hyperfine ground-states of the system $\{\ket{g}\equiv \ket{0}$, $\ket{h} \equiv \ket{1}\}$, corresponding to $F=1$ and $F=2$ which for atoms such as rubidium have a very long lifetime. An additional quantum state dubbed Rydberg state $\ket{r}$ mediate the interaction between the atoms using the phenomenon of the Rydberg blockade.\index{Rydberg blockade}

Single-qubit gates are implemented by using tightly focused laser beams to drive transitions between the internal states of the atoms, creating rotations around the Bloch sphere. Indeed, a transition, driven by an external pulse, between two energy levels $\{\ket{g},\ket{h}\}$ or $\{\ket{g},\ket{r}\}$ of an atom in the array can be mapped to a time-dependent spin $1/2$ Hamiltonian
\begin{equation}
    H^D(t) = \frac{1}{2} \mathbf{\Omega}(t) \cdot  \hat{\pmb{\sigma}}.
\end{equation}
The rotation vector is given by $\mathbf{\Omega}(t) = (\Omega(t)\cos\phi, -\Omega(t)\sin\phi, -\delta(t))^T$ where $\Omega$ is the Rabi frequency of the transition while $\delta$ represent the detuning. The driving hamiltonian therefore implements a rotation around the axis  $\mathbf{n} = \mathbf{\Omega}/\abs{\Omega}$  with angular velocity $\abs{\mathbf{\Omega}}=\sqrt{\Omega^2+\delta^2}$. In the case of a resonant pulse $\delta = 0$ of duration $\tau$ the Hamiltonian evolution generates the following unitary\index{Unitary operator!single qubit gate}
\begin{equation}
    U = \exp{-i\frac{\theta}{2}(\cos\phi \hat{X}-\sin\phi\hat{Y}) } = R_z(-\phi)R_x(\theta)R_z(\phi),
\end{equation}
where we defined
\begin{equation}
    \theta = \int_0^{\tau}dt \Omega(t)dt.
\end{equation}
Finally, applying a further rotation around the $z$-axis of an angle $\gamma+\phi$ we obtain an arbitrary single-qubit gate
\begin{equation}
    U(\gamma,\theta,\phi) = R_z(\gamma)R_x(\theta)R_z(\phi).
\end{equation}

Multi-qubit entangling gates are realized through the Rydberg blockade mechanism. In this method, atoms are excited to high-lying Rydberg states where their interactions are significantly enhanced, allowing for the creation of entangled states across multiple qubits. The Hamiltonian describing the ground-rydberg transitions is given by\begin{equation}
    \mathcal{H}^{gr}(t) = \sum_i \left(H_i^{D}(t) + \sum_{i<j} \frac{C_6}{R_{ij}^6}\hat{n}_i\hat{n}_j\right).
\end{equation}
where $\hat{n}_i$ denotes the projector $\dyad{r}_i$ on the $i$-th atom. Where the second term, which is the long-range dipole-dipole interactions via Rydberg states with strength $C_6$, implements the so-called Rydberg blockade effect which prevents the simultaneous excitation of the two nearby atoms in the state $\ket{rr}$.

\begin{example}{Implementation of the CZ and CNOT gates}{CZ Implementation}
 In neutral-atom devices the natural native two-qubits gate is given by the controlled-Z (CZ) gate. There are several implementations of the CZ gate\index{Controlled-Z gate}
 \begin{equation}
    \rm{CZ} = \begin{pmatrix}
        1 & 0 & 0 & 0\\
        0 & 1 & 0 & 0\\
        0 & 0 & 1 & 0\\
        0 & 0 & 0 & -1
    \end{pmatrix}.
\end{equation}
We illustrate one realized with a sequence of three pulses on the control-target two-qubit basis, namely $\{\ket{0_c0_t},\ket{0_c1_t},\ket{1_c0_t},\ket{1_c1_t}\}$, which reads\begin{equation}
    \rm{CZ} = e^{i\pi} R_{x}^{(c)}(\pi)\otimes\mathbb{I}^{(t)}\cdot \mathbb{I}^{(c)}\otimes R_{x}^{(t)}(2\pi) \cdot R_{x}^{(c)}(\pi)\otimes\mathbb{I}^{(t)}.
\end{equation}
We note the following\begin{itemize}
    \item Initial state $\ket{0_c0_t}$. The control qubit is excited to the Rydberg state $\ket{r}$ with a pulse $\pi$. The second pulse, because of the Rydberg blockade, leaves the target state unchanged. The final pulse brings the system back to the initial state with a phase $e^{i\pi}=-1$.
     \item Initial state $\ket{0_c1_t}$. The second pulse is off-resonant because of the energy difference between $\Delta E = E_h-E_g$, the target state is unchanged. The first and third pulses bring the system back to the intial state with a phase $e^{i\pi}=-1$.
    \item Initial state $\ket{1_c0_t}$. The first and third pulses are off-resonant, the control state is unchanged. The second pulses brings the system back to the intial state with a phase $e^{i\pi}=-1$.
    \item Initial state $\ket{1_c1_t}$. all the pulses are off-resonant, the system state remains unchanged with no phase factor.
\end{itemize}
We finally notice that the CZ gate, composed with two additional Hadamard gates, can be used to implement a CNOT by applying the following sequence of gates $ \mathbb{I}^{(c)} \otimes {\rm H}^{(t)} \cdot {\rm CZ} \cdot \mathbb{I}^{(c)} \otimes {\rm H}^{(t)}$\index{CNOT!implementation}
\end{example}

\subsection{Trapped-Ions Qubits}

We finally address the case of trapped-ion qubits.\index{Trapped ions} Individual ions, such as ytterbium or calcium, are confined in electromagnetic traps using a combination of static and radio-frequency electric fields. These ions have long-lived electronic states that serve as qubits, which are manipulated using laser or microwave fields. The ions are held in traps where electrodes create a potential well that confines them in a linear chain, with mutual Coulomb repulsion maintaining their separation and allowing for individual addressing.

Single-qubit gates, as for neutral atoms, are implemented by applying laser pulses that induce transitions between the qubit states, with the pulses typically shaped and timed precisely to drive rotations around the Bloch sphere. For instance, resonant laser fields at the qubit transition frequency can perform rotations $ R_x(\theta) $ and $ R_y(\theta) $, where $ \theta $ is the rotation angle determined by the pulse duration and amplitude.

Entangling gates, such as the CNOT gate, are often realized using the Mølmer-Sørensen interaction~\cite{PhysRevLett.82.1971} or the Cirac-Zoller gate~\cite{PhysRevLett.74.4091}. The Mølmer-Sørensen gate uses bichromatic laser fields to couple the internal states of two ions via their shared motional modes, creating an effective spin-spin interaction that entangles the qubit states. The Mølmer-Sørensen gate applies a state-dependent force to the ions, displacing their motional states conditioned on their internal states, and carefully timing these interactions to entangle the ions' internal states.

%\begin{example}{Implementation of the CNOT interaction}{Mølmer-Sørensen interaction}
%We can write something similar to the CZ gate for neutral atoms
%\end{example}

Measurement of trapped-ion qubits is performed using state-dependent fluorescence. A laser tuned to a transition that only one of the qubit states can absorb causes the ion to scatter photons if it is in that state. The presence or absence of fluorescence, detected by a photomultiplier or a CCD camera, indicates the qubit state with high fidelity.

% https://arxiv.org/pdf/quant-ph/0002077.pdf DiVincenzo Paper
% Review di Nori sulle quantum platforms https://link.springer.com/article/10.1007/s11467-022-1249-z

% Rydberg & Atomi in Cavita' https://quantum-journal.org/papers/q-2022-01-24-629/
% https://quantum-journal.org/papers/q-2020-09-21-327/pdf/

% Transmons & QED https://qiskit.org/textbook/ch-quantum-hardware/transmon-physics.html

\biblio

%%%%%%%%%%%%%%%%%%%%%%%%%%%%%%%%%%%%%%%%%%%%%%%%%%%%%%%%%%%%%

\part{Applications}
Tensor networks (TNs) have emerged as one of the most powerful tools for simulating complex quantum systems, particularly when dealing with the exponential growth of Hilbert space in many-body quantum mechanics. This Part of the book looks into the practical applications of tensor networks in modeling quantum dynamics from optimization problems to open system dynamics and resource theory and non-stabiliserness. We specifically focus into their application for addressing a subset of key challenges in quantum computation, particularly in the context of many-body systems. These problems, while not exhaustive due to the authors' preferences, expertise, and space constraints, are closely related by their shared reliance on tensor network techniques and the fundamental principles of quantum mechanics. Through these carefully selected topics, we aim to provide a representative exploration of how tensor networks can be employed to efficiently simulate quantum dynamics and optimize complex quantum systems.

Here, we will highlight the key frameworks, methods, and challenges tackled through the lens of tensor networks, setting the stage for a deeper exploration of these concepts in subsequent chapters.

\paragraph{Unitary Evolution of Hamiltonian Dynamics ---}
The accurate simulation of quantum systems evolving under unitary transformations governed by time-dependent Hamiltonians remains a core challenge in quantum physics. Tensor networks, particularly the Matrix Product States (MPS) and Matrix Product Operators (MPO) formalisms, offer a scalable approach to representing quantum states and operators, allowing for simulations that would otherwise be computationally intractable. One of the fundamental methods employed is the \emph{Time-Dependent Variational Principle (TDVP)}, which provides an efficient means to approximate the evolution of quantum states with a fixed bond dimension. TDVP has been extensively studied and applied in the context of one-dimensional systems, where its ability to control truncation errors is particularly valuable for long-time simulations.

In addition, tensor networks have found widespread application in \emph{Quadra\-tic Unconstrained Binary Optimization (QUBO)}  problems, which are integral to classical and quantum optimization algorithms. These problems can be mapped to Ising-like models, enabling the use of TN methods to solve real-world optimization challenges that are NP-hard.

\paragraph{Quantum Annealing and Digitized Approaches ---}
One of the most intriguing applications of TNs is in \emph{Quantum Annealing (QA)}, a method designed to find the ground state of a Hamiltonian by slowly varying a control parameter. QA can be effectively simulated on both quantum and classical devices. The \emph{Digitized Quantum Annealing (dQA)} approach, a discretized version of QA, allows for a gate-based implementation on quantum simulators. Tensor networks provide a natural framework for dQA simulations, where they allow for the compression of the exponentially growing Hilbert space into more manageable forms. Notably, TN methods have been employed to simulate dQA with a high degree of precision, as demonstrated in studies where MPS and MPO techniques are used to efficiently capture the state evolution under quantum annealing protocols.

Furthermore, \emph{Quantum Approximate Optimization Algorithm (QAOA)}, a variational quantum algorithm, leverages principles of both QA and gate-based quantum computation. Tensor networks can be used to analyze the performance of QAOA, optimizing the classical parameters that drive the quantum circuit. This provides an avenue for classical simulations of quantum algorithms aimed at solving combinatorial optimization problems.

\paragraph{Open Tensor Networks and Open System Dynamics ---}
In real-world quantum systems, interactions with the environment cannot be ignored, giving rise to \emph{open quantum systems}. TNs have been successfully adapted to simulate such systems by extending traditional MPS techniques to handle mixed quantum states and non-unitary evolutions. \emph{Tensor Network and Density Operators} techniques allow for an efficient representation of mixed states, where MPO formalism plays a crucial role. A prominent approach within this context is the \emph{Minimally Entangled Typical Thermal States (METTS)} algorithm, which uses tensor networks to sample typical thermal states, providing insight into quantum thermodynamics.

Incorporating environmental effects often requires the use of the \emph{Lindblad Master Equation}, a formalism that describes the evolution of open quantum systems. TN-based algorithms such as the \emph{Locally Purified Tensor Network} method offer a pathway to simulate open quantum dynamics in an efficient manner, enabling the study of decoherence, dissipation, and other non-unitary processes.

Finally, TNs are crucial for simulating quantum measurement protocols in open systems, particularly when dealing with \emph{monitored dynamics}, where the evolution of the system depends on continuous measurement outcomes. These \emph{unravelling schemes}, in combination with TN methods, provide an effective means to model complex quantum behavior in open systems.

\paragraph{Tensor Networks and Quantum Magic  ---}
The last part of this section addresses the interplay between tensor networks and quantum ``magic'' --- a term that refers to the non-stabilizer nature of quantum states, which plays a crucial role in the power of quantum computation beyond Clifford operations. Understanding how tensor networks can capture and quantify the non-stabilizerness of a quantum system is essential for leveraging their full potential in quantum information processing.

The chapter begins by introducing the stabilizer formalism, which underpins much of the classical simulability of certain quantum systems. Stabilizer states, those that can be expressed using Pauli and Clifford operators, can be simulated efficiently on classical computers, thanks to the Gottesman-Knill theorem. This efficiency, however, highlights the limitation of such states from the perspective of quantum computational advantage. Truly leveraging the power of quantum computation requires moving beyond stabilizer circuits --- into the domain of nonstabilizer, or ``magic'' states.

Quantum magic is identified as a critical resource for achieving quantum advantage. This concept refers to non-classicality beyond stabilizer operations and is essential for quantum speedup. Measuring the degree of nonstabilizerness is crucial for evaluating how far a quantum system departs from stabilizer form, thus revealing its true computational power.

Following this foundational introduction, the chapter explores measuring magic in tensor network states, where tensor network representations provide efficient descriptions of quantum many-body systems.  The sampling techniques and replica methods outlined in this chapter serve as foundamental tools for quantifying the magic present in tensor network states, allowing for a deeper understanding of their complexity and power.

The latter part of the chapter introduces advanced techniques for handling tensor networks in the presence of magic. In particular, the development of Clifford-enhanced Matrix Product States ($\mathcal{C}$MPS) offers a novel way to represent complex quantum states by combining the structure of stabilizer states with the flexibility of tensor networks. This hybrid approach is extended to incorporate stabilizer MPOs (Matrix Product Operators) and Clifford-dressed TDVP (Time-Dependent Variational Principle), providing a framework that balances computational efficiency with the need to capture nonstabilizer dynamics.

\chapter{Unitary Evolution of Hamiltonian Dynamics}\label{chap4}
\epigraph{Life is a dynamic rather than a static process, and when we don't change it kills us. It's not running away, it's moving on.}{Irvine Welsh}

Much of Quantum mechanics involves solving the Schr\"{o}dinger's equation,
\begin{equation}\label{eq:schrodinger}
    \frac{\partial}{\partial t}\ket{\psi} = -i\hat{H}\ket{\psi}.
\end{equation}

This fundamental equation dictates the unitary evolution of a pure quantum state, akin to how Newton's laws describe the evolution of classical systems. This seemingly harmless first-order time differential equation is analytically tractable only for a select few problems. Particularly for many-body systems, the curse of dimensionality strikes, and its unitary dynamics become an extremely challenging problem in quantum mechanics. Several popular numerical methods have been developed over time, each with certain benefits and drawbacks compared to others. Tensor network methods are a recent addition to this list. In the first two introductory chapters, we introduced tensor networks and their efficiency in representing one-dimensional quantum many-body systems, with states written as MPS and the Hamiltonian as MPO. Furthermore in the second chapter we also gave a comprehensive introduction to the TEBD algorithm (see Section~\ref{subsec:circuits}) which is the first tensor network based algorithm to calculate the unitary dynamics of pure quantum state. The TEBD algorithm decomposes the full Hamiltonian into a sequence of two-site local unitary gates. In this regard it is also a suitable algorithm for the discrete evolution of the quantum state via a sequence of random local unitaries. In this chapter we will focus in another approach for unitary evolution of a pure quantum state, specifically the one that preserves the structure of the Hamiltonian and is continuous in time, namely the Time-Dependent Variational Principle (TDVP) algorithm. The TDVP~\cite{TDVP1,TDVP2} is a variational approach based on the idea of MPS manifold and tangent space.

\section{Time-Dependent Variational Principle (TDVP)}

With the quantum state written as an MPS the Schr\"{o}dinger's equation can be written as,
\begin{equation}\label{eq:schrodinger}
    \frac{\partial}{\partial t}\ket{\psi(A(t))} = -i\hat{H}\ket{\psi(A(t))},
\end{equation}
where,
\begin{equation}
    \includegraphics[width=0.75\textwidth,valign=c]{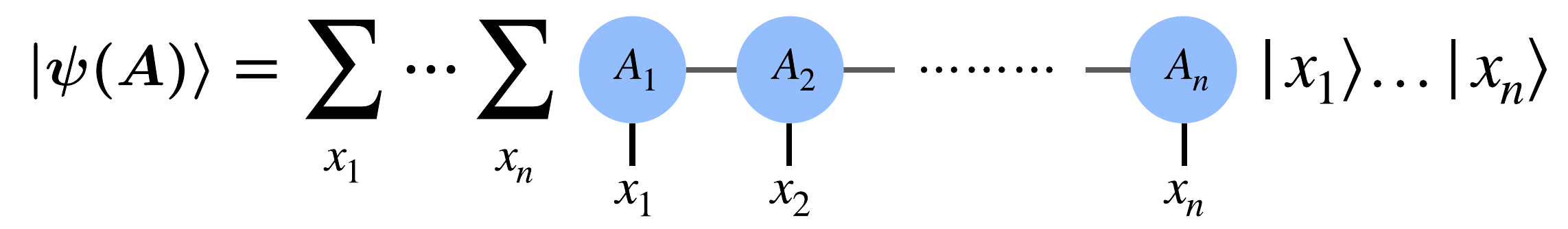}
\end{equation}

The MPS $\ket{\psi(A)}$ resides in a smooth manifold $\mathbb{M}_{\chi}$, which is a subset of the Hilbert space, $\mathbb{M}_{\chi} \subset \mathbb{H}$. This manifold $\mathbb{M}_{\chi}$ is characterized by the constraint that it exclusively contains MPS with a fixed bond dimension $\chi$. It should be noted that although $\mathbb{M}_{\chi}$ is a subset of the Hilbert space it is not a vector space because adding two MPS with bond dimension $\chi$ results in a MPS of a larger bond dimension and will therefore not lie within $\mathbb{M}_{\chi}$. Furthermore, applying a Hamiltonian onto the MPS also increases the bond dimension, therefore the right hand side of the Schr\"{o}dinger's equation~\eqref{eq:schrodinger} leads the MPS out of the manifold. The central idea of TDVP\index{TDVP} is to constrain the unitary evolution within the manifold $\mathbb{M}_{\chi}$ or in terms of MPS language we restrict the growth of the bond dimension to $\chi_{\text{max}} = \chi$. This is achieved by projecting the right hand side of the Schr\"{o}dinger's equation, $-i\hat{H}\ket{\psi(A(t))}$ back into the tangent space $\mathbb{T}_{A}\mathbb{M}$ defined at point $\ket{\psi(A(t))}$. The figure below demonstrates this process, the smooth curved space is the manifold $\mathbb{M}_{\chi}$ embedded inside the Hilbert space $\mathbb{H}$, the blue plane is the tangent space $\mathbb{T}_{A}\mathbb{M}$ defined at point $\ket{\psi(A)}$, the bold arrow pointing out of the manifold is the right hand side of the Schr\"{o}dinger's equation which is projected back to the tangent space as shown by the dashed arrow, finally, the green curve that lies entirely within the manifold represents the optimal evolution of the quantum state.

$$
\includegraphics[width=0.90\textwidth,valign=c]{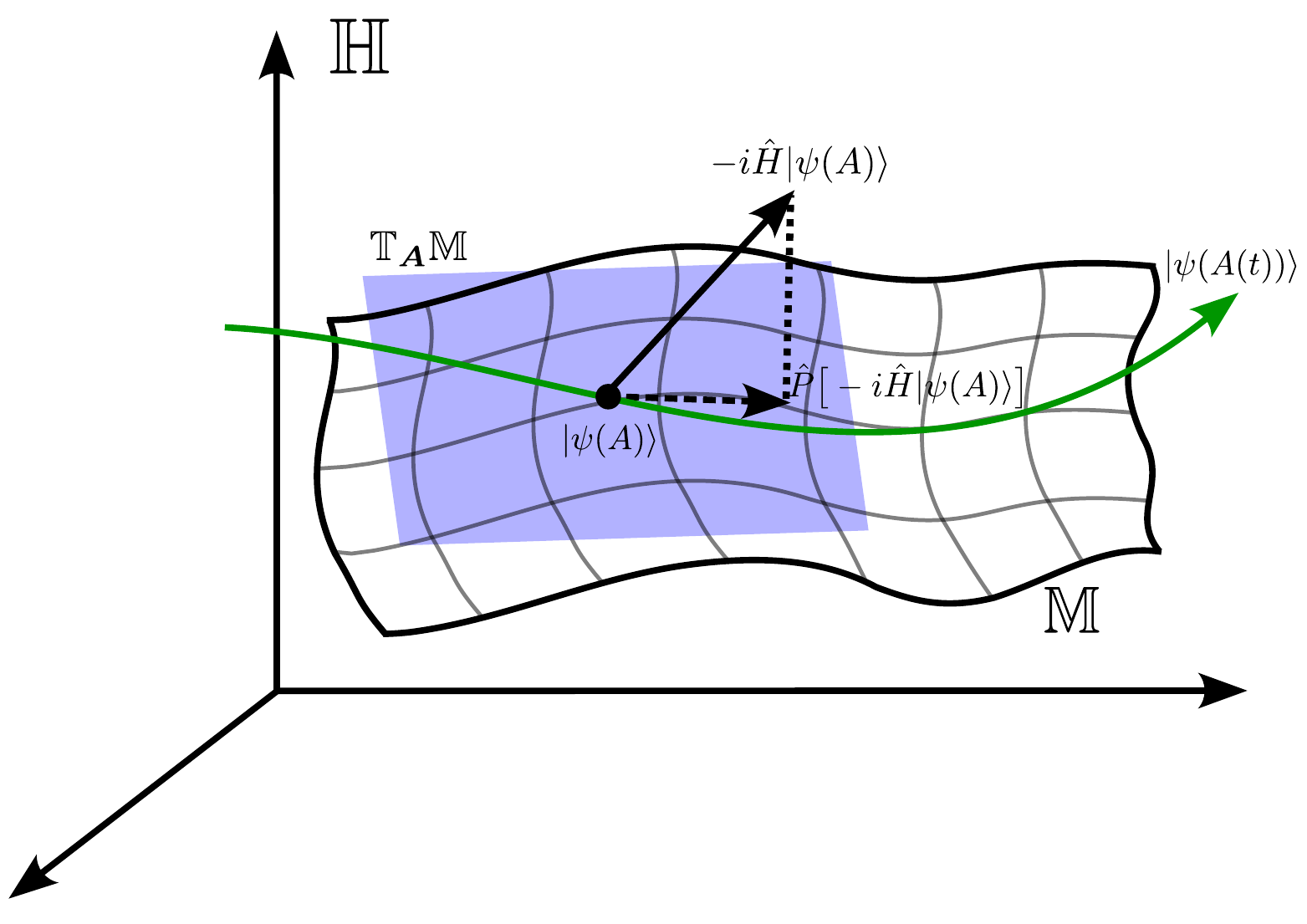}
$$

The effective Schr\"{o}dinger's equation such that the quantum state always remains in the manifold $\mathbb{M}_{\chi}$ is,
\begin{equation}\label{eq:TDVP_project_1}
    \frac{\partial}{\partial t}\ket{\psi(A(t))} = -i \hat{P}_{\mathbb{T}^{[2]}_{A}\mathbb{M}_{\chi}} \hat{H}\ket{\psi(A(t))},
\end{equation}

Here, the superscript $[2]$ signifies two site TDVP which is more accurate and adaptive compared to the one-site TDVP~\cite{TDVP2}. $\hat{P}_{\mathbb{T}^{[2]}_{A}\mathbb{M}_{\chi}}$ projects the MPS $-i\hat{H}\ket{\psi(A(t))}$ onto the tangent space $\mathbb{T}_A\mathbb{M}_{\chi}$ and is defined as,
\begin{equation}\label{eq:projector}
    \hat{P}_{\mathbb{T}^{[2]}_{A}\mathbb{M}_{\chi}} = \sum_{l=1}^{n-1} \hat{P}_L^{l-1} \otimes \mathbb{I}^l \otimes \mathbb{I}^{l+1} \otimes \hat{P}_R^{l+2} - \sum_{l=1}^{n-2} \hat{P}_L^l \otimes \mathbb{I}^{l+1} \otimes \hat{P}_R^{l+2}.
\end{equation}

The full derivation of the projector can be found in the following articles:~\cite{TDVP1,TDVP2}. The parts of the projector are defined as\index{TDVP!projectors}
\begin{equation}
    \includegraphics[width=0.8\textwidth,valign=c]{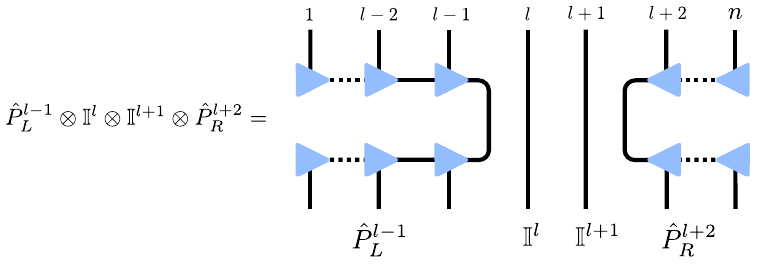}
\end{equation}
\begin{equation}
    \includegraphics[width=0.8\textwidth,valign=c]{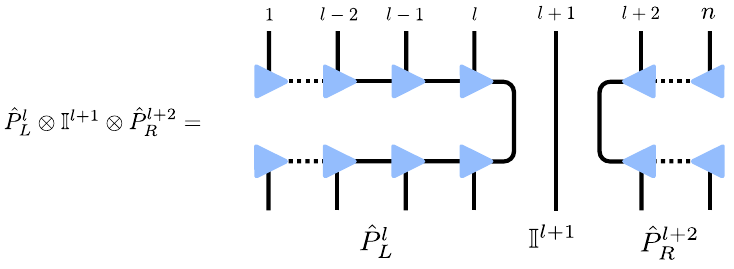}
\end{equation}

Replacing equation~\eqref{eq:projector} into equation~\eqref{eq:TDVP_project_1} we get,
\begin{equation}\label{eq:TDVP_project_2}
    \frac{\partial}{\partial t} \ket{\psi} = -i \sum_{l=1}^{N-1} \hat{P}_L^{l-1} \otimes \mathbb{I}^l \otimes \mathbb{I}^{l+1} \otimes \hat{P}_R^{l+2}\hat{H} \ket{\psi} + i \sum_{l=1}^{N-2} \hat{P}_L^l \otimes \mathbb{I}^{l+1} \otimes \hat{P}_R^{l+2} \hat{H} \ket{\psi}.
\end{equation}

The exact solution of equation~\eqref{eq:TDVP_project_2} is not tractable, however, the individual terms are integrable exactly. This kind of problem can be solved by Lie-Trotter splitting of the operators~\cite{splitting} and sequentially evolving a pair of differential equations for a time step of $\Delta t$,
\begin{align}
\label{eq:TDVP_project_forward}
    \frac{\partial}{\partial t} \ket{\psi} &= -i \hat{P}_L^{l-1} \otimes \mathbb{I}^l \otimes \mathbb{I}^{l+1} \otimes \hat{P}_R^{l+2}\hat{H} \ket{\psi},
\\
\label{eq:TDVP_project_backward}
    \frac{\partial}{\partial t} \ket{\psi} &= i \hat{P}_L^l \otimes \mathbb{I}^{l+1} \otimes \hat{P}_R^{l+2} \hat{H} \ket{\psi}.
\end{align}

This approach incurs a local time step error of $O(\Delta t^2)$. These two differential equations have opposite signs, this can be interpreted as a set of forward and backward evolution in time. Let us introduce a two-site map $\bar{\psi}^L_{l-1}\otimes\mathbb{I}^{l} \otimes \mathbb{I}^{l+1} \otimes\bar{\psi}^R_{l+2}$ and single-site map $\bar{\psi}^L_l\otimes \mathbb{I}^{l+1}\otimes\bar{\psi}^R_{l+2}$ with the following definitions,
\begin{equation}\label{eq:map2site}
\includegraphics[width=0.8\textwidth,valign=c]{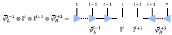},
\end{equation}
\begin{equation}\label{eq:map1site}
\includegraphics[width=0.8\textwidth,valign=c]{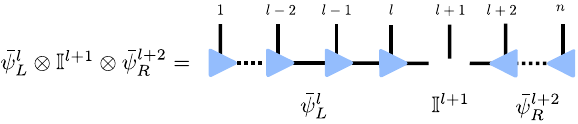}.
\end{equation}

We apply the two-site map~\eqref{eq:map2site} onto equation~\eqref{eq:TDVP_project_forward} and the single-site map~\eqref{eq:map1site} onto equation~\eqref{eq:TDVP_project_backward}. The resulting left hand side of these equations are two-site and single-site tensors given as,
\begin{equation}\label{eq:map2mps}
\includegraphics[width=0.95\textwidth,valign=c]{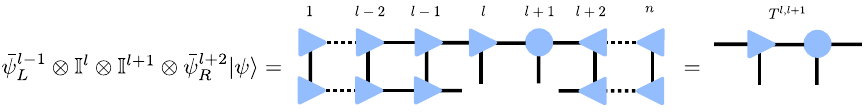},
\end{equation}
\begin{equation}\label{eq:map1mps}
\includegraphics[width=0.95\textwidth,valign=c]{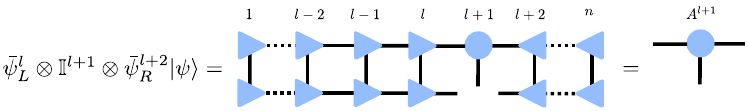}.
\end{equation}

Here we assumed that $\ket{\psi}$ is in a mixed canonical form with orthogonality center at site $l+1$ and exploited the left and right unitary properties of the canonical tensors. Similarly, the right hand side of these equations are also two-site and single-site tensors albeit more complex given by the following two diagrammatic equations,
\begin{equation}\label{eq:localTDVP2}
\includegraphics[width=0.95\textwidth,valign=c]{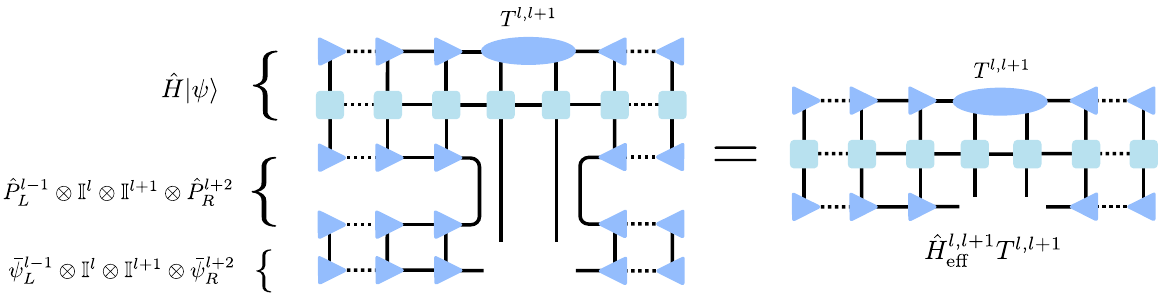},
\end{equation}
\begin{equation}\label{eq:localTDVP1}
    \includegraphics[width=0.9\textwidth,valign=c]{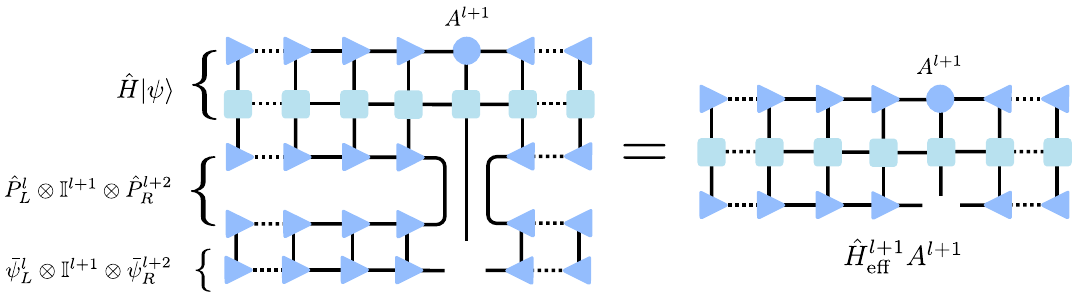}.
\end{equation}

Finally, the equations~\eqref{eq:TDVP_project_forward} and~\eqref{eq:TDVP_project_backward} becomes a pair of two-site and single-site effective equations given by,\index{TDVP!MPS equations}
\begin{align}
\label{eq:TDVP_project_forward_local}
    \frac{\partial}{\partial t} T^{l,l+1}(t) &= -i \hspace{0.1cm} \hat{H}_{\text{eff}}^{l,l+1} T^{l,l+1}(t) \rightarrow T^{l,l+1}(t) = e^{-i t \hspace{0.05cm} \hat{H}_{\text{eff}}^{l,l+1}} T^{l,l+1}(0),
\\
\label{eq:TDVP_project_backward_local}
    \frac{\partial}{\partial t} A^{l+1}(t) &= i \hspace{0.1cm} \hat{H}_{\text{eff}}^{l+1} A^{l+1}(t) \rightarrow A^{l+1}(t) = e^{it\hspace{0.05cm} \hat{H}_{\text{eff}}^{l+1}} A^{l+1}(t).
\end{align}

The two-site and single-site effective Hamiltonians $\hat{H}_{\text{eff}}^{l,l+1}$, $\hat{H}_{\text{eff}}^{l+1}$ are both rank-8 tensors as defined in the equations~\eqref{eq:localTDVP2} and~\eqref{eq:localTDVP1} respectively. The set of equations~\eqref{eq:TDVP_project_forward_local} and~\eqref{eq:TDVP_project_backward_local} can be integrated using iterative Lanczos exponential solver which has a computational cost of $O(\chi^3)$~\cite{Lancz1,Lancz2,Lancz5}. Furthermore, the error incurred during Lanczos\index{Lanczos} solver is well controlled and can be made arbitrarily small by sufficiently increasing the number of Krylov\index{Krylov} vectors. Solving these local equations forms the bottleneck of TDVP algorithm which comprises of solving these local equations iteratively over the system sites.

\begin{example}{Lanczos Exponential Solver}{Lanczos}
The Lanczos algorithm was initially developed for the purpose of tri-diagonalizing a Hermitian matrix. Essentially, the Lanczos process converts an $N \times N$ Hermitian matrix $M$ into an $N \times N$ tridiagonal matrix over the course of $n$ steps. The importance of this process lies in its ability to approximate the extreme eigenvalues and eigenvectors of the Hermitian matrix $M$ by using the extreme eigenvalues and eigenvectors of a $k \times k$ tridiagonal matrix after just $k \ll N$ iterations. The Lanczos algorithm starts with an $N \times N$ Hermitian matrix $M$ and an initial guess vector $v_{\text{input}}$. After $k$ iterations, it produces a $k \times k$ tridiagonal matrix $T_{k\times k}$ along with a set of orthogonal column vectors $V_{N \times k}$. The algorithm can be described concisely as follows:\\

\begin{algorithmic}[1]\label{alg:Lancz}
\State $v_1$ = $v_{\text{input}}/\abs{v_{\text{input}}}$ \Comment{input vector}
\For{$i \in [2,k+1]$}
\State $v_i$ = $M \times v_{i-1}$
\For{$j \in [i-2,i-1]$} \Comment{Building tridiagonal matrix}
\If{$j \geq 1$}
\State $T_{i-1,j} = \text{dot}(v_i,v_j)$
\State $T_{j,i-1} = T_{i-1,j}^*$

\EndIf
\EndFor
\For{$f \in [1,i-1]$} \Comment{Full reorthonormalization}
\State $v_i = v_f - \text{dot}(v_i,v_f)\times v_f$
\State  $v_i = v_i /\abs{v_i} $
\EndFor
\EndFor
\State \bf{return} $T$, $[v_1,v_2,\ldots,v_k]$
\end{algorithmic}
\vspace{0.5cm}
The set of orthonormal vectors $V_{n\times k} = [v_1,v_2,\ldots,v_k]$ are known as Krylov vectors that forms the basis of Krylov subspace~\cite{Lancz1,Lancz3}. The number $k$ is known as the Krylov dimension.\\

Finally, the matrix exponential applied to a vector can be approximated by the tridiagonal matrix as,
\begin{equation}\label{eq:lancz_exp}
    e^{M} \times v_{\text{input}} \approx V_{n\times k} \times e^{T_{k\times k}} \times \mathbb{I}_{k\times k}[:,1],
\end{equation}
where, $\mathbb{I}_{k\times k}[:,1]$ is the first column of the $k\times k$ identity matrix~\cite{Lancz5}.

\paragraph{Note:} in equations~\eqref{eq:TDVP_project_forward_local} and~\eqref{eq:TDVP_project_backward_local} $\hat{H}_{\text{eff}}^{l,l+1}$, $\hat{H}_{\text{eff}}^{l,l+1}$, $ T^{l,l+1}$, and $ A^{l+1}$ are not matrices and vectors but rather high rank tensors. In this case the Lanczos algorithm can be adapted for exponential of a higher rank tensor applied to another higher rank tensor.
\end{example}

\begin{example}{TDVP algorithm}{TDVP_algorithm}
The TDVP algorithm is fairly similar to the Density Matrix Renormalization Group (DMRG) algorithm~\cite{Schollwock_2011} in the sense that it involves sequential sweeping from left-to-right and right-to-left through the system sites. The right-to-left sweep is adjoint to the left-to-right sweep, giving a second-order symmetric integrator scheme so that the error time step error becomes of order $O(\Delta t^3)$. In the algorithm we then set $\Delta t \rightarrow \Delta t/2$ and consider the left-to-right and right-to-left sweep as a single time step~\cite{TDVP2}. The TDVP algorithm can be decomposed into three main parts;

\begin{enumerate}
    \item \textbf{Initialization}: This part involves setting up the system ready before the sweeps. The requirements are as follows:
    \begin{enumerate}
        \item The initial state $\ket{\psi(t=0)}$ in right canonical form. This can be achieved by a sequence of SVD/QR decompositions starting from the $n^{th}$ site to the first site.
        $$
        \includegraphics[width=0.5\textwidth,valign=c]{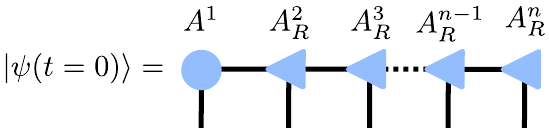}
        $$
        \item The Hamiltonian $\hat{H}$ as an MPO.
        $$
        \includegraphics[width=0.45\textwidth,valign=c]{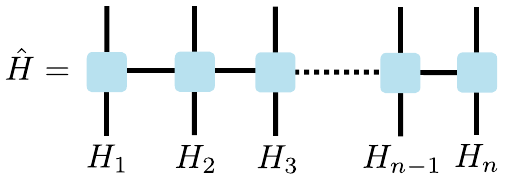}
        $$
        \item Right environment tensors $\{R_i\}_{i=3}^{i=n}$.
        $$
        \includegraphics[width=0.82\textwidth,valign=c]{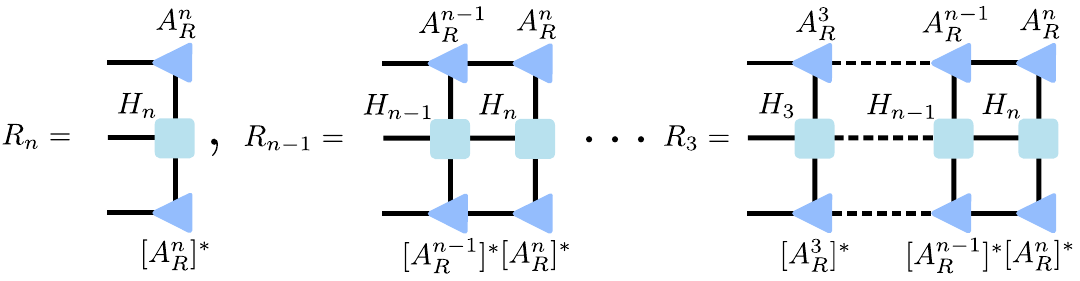}
        $$
    \end{enumerate}

    \item \textbf{Left-to-right sweep}

    \begin{algorithmic}[1]\label{alg:lefttoright}
    \For{$l \in [1,n-1]$}
    \State $T^{l,l+1} \gets A^lA^{l+1}_R$

    $$
    \includegraphics[width=0.6\textwidth,valign=c]{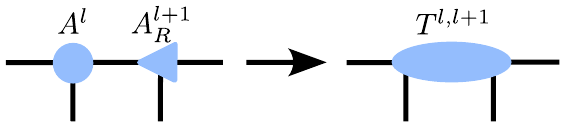}
    $$

    \State Solve the forward local evolution:  $T^{l,l+1}(t+\Delta t/2) \gets e^{-i\frac{\Delta t}{2} \hat{H}_{\text{eff}}^{l,l+1}} T^{l,l+1}(t)$, where $\hat{H}_{\text{eff}}^{l,l+1}$ is defined as,

    $$\includegraphics[width=0.5\textwidth,valign=c]{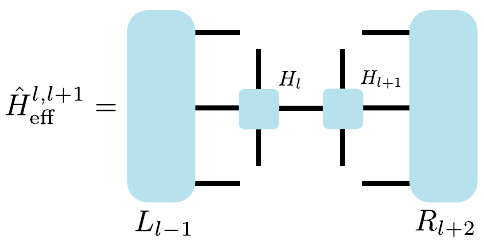}
    $$

    Here, $L_{l-1}$ is the $l-1^{th}$ left environment tensor akin to the right environment tensors. Note that $L_0$ and $R_{n+1}$ corresponds to a scalar 1.0. They are build successively as,

    $$\includegraphics[width=0.6\textwidth,valign=c]{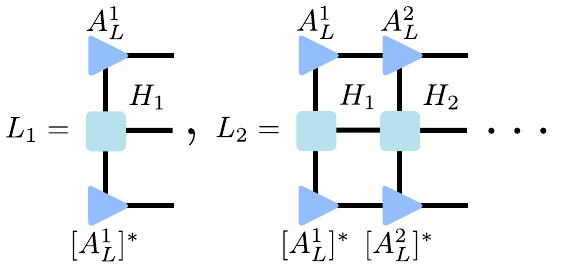}
    $$

    \State $A_{L}^{l}S^{l,l+1}\tilde{A}_R^{l+1} \gets T^{l,l+1}$ \Comment{SVD and truncation}

    $$\includegraphics[width=0.7\textwidth,valign=c]{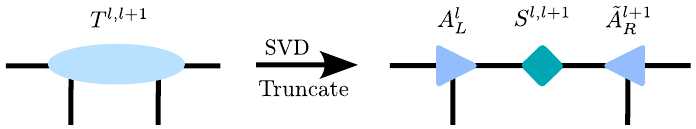}
    $$

    \State $S^{l,l+1}\tilde{A}_R^{l+1} \gets A^{l+1}$
    $$
    \includegraphics[width=0.6\textwidth,valign=c]{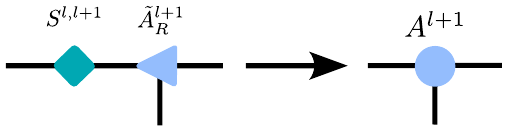}
    $$
    \If{$l \neq n-1$}
    \State $L_{l} \gets L_{l-1}[A^l_{L}]^*H_{l}A^l_L$
    $$
    \includegraphics[width=0.5\textwidth,valign=c]{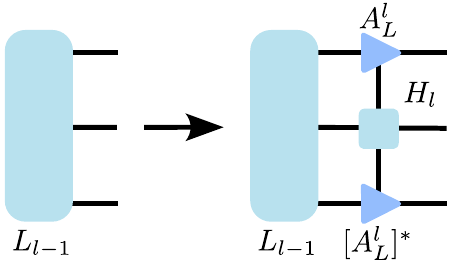}
    $$
    \State Solve the backward local evolution:  $A^{l+1}(t-\Delta t/2) \gets e^{i\frac{\Delta t}{2} \hat{H}_{\text{eff}}^{l+1}} A^{l+1}(t)$, where $\hat{H}_{\text{eff}}^{l+1}$ is defined as,
    $$
    \includegraphics[width=0.45\textwidth,valign=c]{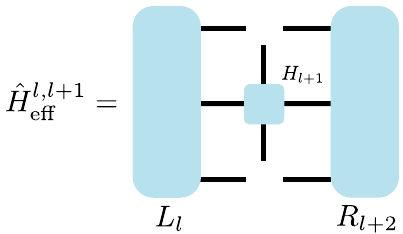}
    $$
    \State Delete $R_{l+2}$
    \EndIf
    \EndFor
    \end{algorithmic}

    \item \textbf{Right-to-left sweep}: At the end of left-to-right sweep the MPS is in left canonical form and we have a set of left environment matrices $\{L_i\}_{i=1}^{n-1}$. All the steps involved in the right-to-left sweep are equivalent to left-to-right sweeps except for the direction of iteration. So we will outline only the algorithm without the diagrams.

    \begin{algorithmic}[1]\label{alg:righttoleft}
    \For{$l \in [n,2]$}
    \State $T^{l-1,l} \gets A_L^{l-1}A^{l}$

    \State $T^{l-1,l}(t+\Delta t/2) \gets e^{-i\frac{\Delta t}{2} \hat{H}_{\text{eff}}^{l-1,l}} T^{l-1,l}(t)$

    \State $ \tilde{A}_{L}^{l-1}S^{l-1,l}A_R^{l} \gets T^{l-1,l}(t + \Delta t)$ \Comment{SVD and truncation}

    \State $\tilde{A}_L^{l-1}S^{l-1,l} \gets A^{l-1}$

    \If{$l \neq 2$}
    \State $R_{l} \gets R_{l+1}[A^{l}_{R}]^*H_{l}A^{l}_R$

    \State $A^{l-1}(t-\Delta t/2) \gets e^{i \frac{\Delta t}{2} \hat{H}_{\text{eff}}^{l-1}} A^{l-1}(t)$

    \State Delete $L_{l-2}$
    \EndIf
    \EndFor
    \end{algorithmic}

    \item Steps 2 and 3 evolves the state by a time step $\Delta t$: $\ket{\psi(t+\Delta t)} \gets \ket{\psi(t)}$. The full evolution is achieved by successive iteration of these steps.
\end{enumerate}
\end{example}

The TDVP algorithm encounters four main types of errors\index{TDVP!errors}. The first error emerges during the projection of the exact Schrödinger equation onto~\eqref{eq:TDVP_project_1}, stemming from the limited bond dimensions of the auxiliary indices in MPS tensors. To gauge this error, one can observe how TDVP data converges as the bond dimensions increase. The second error is related to the finite time step when approximating the differential equation~\eqref{eq:TDVP_project_2} with a series of solvable local differential equations~\eqref{eq:TDVP_project_forward} and~\eqref{eq:TDVP_project_backward}. For the second-order integrator described earlier, this error scales as $O(\delta t^3)$ per time-step. The third type of error is due to an insufficient number of Krylov vectors in solving the local differential equations~\eqref{eq:TDVP_project_forward_local} and~\eqref{eq:TDVP_project_backward_local}. This error can be made arbitrarily small by using a sufficiently large number of Krylov vectors in the Lanczos exponential solver. The final source of error arises from the truncation of the singular values in the local MPS tensors. Reducing the time step $\Delta t$ will decrease the time step error and the error incurred during solving local differential equations. The other two errors, projection error and truncation error are however independent of the discrete time step and therefore taking smaller $\Delta t$ will wind up more overall projection and truncation error since it require more discrete steps to reach the final state. In a real simulation the optimal $\Delta t$ is often determined by several round of hit and trials.

\section{Quadratic unconstrained binary optimization}

The Quadratic Unconstrained Binary Optimization (QUBO)\index{QUBO} problem is a form of optimization where the goal is to find the values of binary variables that minimize (or maximize) a quadratic objective function. This is pivotal in various scientific and engineering fields, such as finance, logistics, machine learning, and especially quantum computing, because of its compatibility with quantum annealers and certain quantum algorithms.

The QUBO problem involves the optimization of the following function
\begin{equation}
    f(\mathbf{x}) = \sum_{ij} x_i Q_{ij} x_j + \sum_i c_i x_i
\end{equation}
where $c_i$ represent the weight associated to each component and $Q_{ij}$ the weights associated with each pair of indices, defining the structure of the problem, and $x_j \in \{0,1\}$ are binary variables.

The optimal solution, $\mathbf{x}^*$, minimizes the objective function and satisfies the condition
\begin{equation}
  \forall \mathbf{x} \in \{0,1\}^n \quad f(\mathbf{x}^*) \leq f(\mathbf{x}).
\end{equation}

In its most general case, the QUBO problem is NP-hard, meaning it cannot be efficiently solved by any algorithm that runs in polynomial time. Before dwelling on how to map QUBO problems in a quantum mechanical problem, let us give some examples of such problems.

\paragraph{Unweighted graphs} --- A graph\index{QUBO!unweighted graphs} is a mathematical structure that is defined as a pair $G = (V, E)$, where $V$ represents the set of vertices (or nodes) and $E$ represents the set of edges (or links) between these vertices. The vertices can represent entities such as points, objects, or locations, while the edges represent the relationships or connections between pairs of vertices.

One common way to represent a graph is through its adjacency matrix $A$. The adjacency matrix $A$ is a square matrix of size $n \times n$, where $n$ is the number of vertices in the graph. The entries of $A$ are defined as follows
\begin{equation}
A_{ij} =
\begin{cases}
1 & \text{if there is an edge between vertex } i \text{ and vertex } j, \\
0 & \text{otherwise}.
\end{cases}
\end{equation}
For undirected graphs, the adjacency matrix $A$ is symmetric, meaning $A_{ij} = A_{ji}$. In contrast, for directed graphs, the adjacency matrix may not be symmetric, as the presence of an edge from vertex $i$ to vertex $j$ does not necessarily imply the presence of an edge from vertex $j$ to vertex $i$.
\begin{equation}\label{fig:examplegraph}
    \includegraphics[width=.33\linewidth,valign=c]{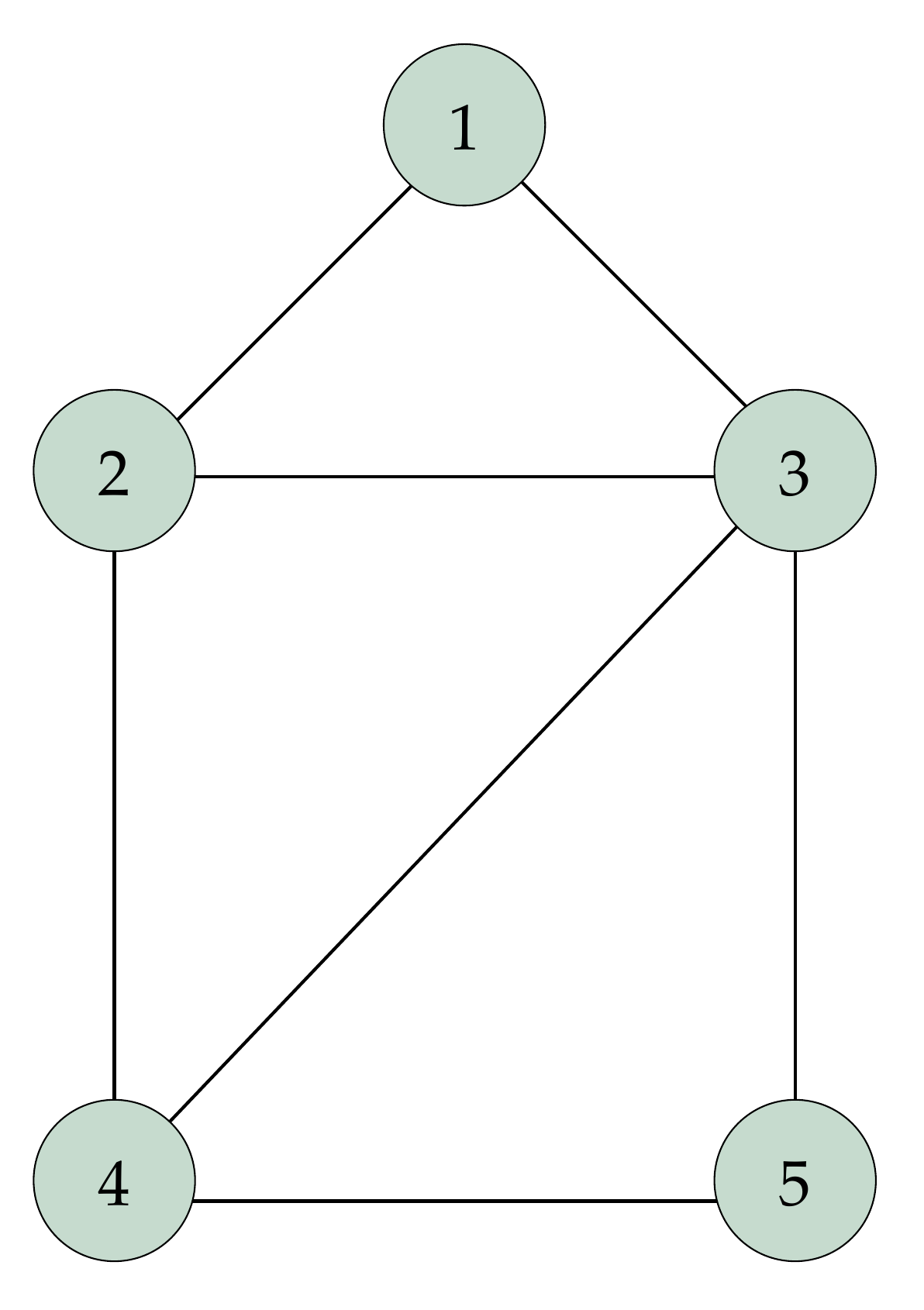} \qquad A = \begin{pmatrix} 0 && 1 && 1 && 0 && 0\\
1 && 0 && 1 && 1 && 0 \\
1 && 1 && 0 && 1 && 1 \\
0 && 1 && 1 && 0 && 1 \\
0 && 0 && 1 && 1 && 0
\end{pmatrix}
\end{equation}

\paragraph{Weighted graphs ---}
In addition to simple graphs, there are also weighted graphs, where each edge is assigned a numerical value or weight.\index{QUBO!weighted graphs} A weighted graph can be represented by a triplet $G = (V, E, w)$, where $V$ is the set of vertices, $E$ is the set of edges, and $w: E \to \mathbb{R}$ is a weight function that assigns a real number to each edge.

In the context of a weighted graph, the adjacency matrix $A$ is often extended to include these weights. Specifically, the weighted adjacency matrix of a weighted graph is defined as follows
\begin{equation}
W_{ij} =
\begin{cases}
w_{ij} & \text{if there is an edge between vertex } i \text{ and vertex } j, \\
0 & \text{otherwise}.
\end{cases}
\end{equation}

Here, $w_{ij}$ represents the weight of the edge between the vertex $i$ and the vertex $j$. For undirected weighted graphs, the adjacency matrix remains symmetric, that is, $W_{ij} = W_{ji}$, while for directed weighted graphs, the matrix may still be asymmetric depending on the direction and weights of the edges.

Weighted graphs are particularly useful in modeling real-world problems where relationships between entities have different strengths, costs, or capacities. Examples include road networks where edges represent distances or travel times, communication networks where edges represent bandwidth or latency, and financial networks where edges represent transaction amounts.

\paragraph{Max-Cut problem ---} The Max-Cut problem\index{QUBO!max-cut} is a fundamental problem in graph theory and combinatorial optimization. Given an undirected graph $G = (V, E)$, the objective of the Max-Cut problem is to partition the set of vertices into two disjoint subsets such that the number of edges between the two subsets is maximized. Equivalently, the problem seeks to maximize the weight of the cut, where the weight is defined as the sum of the weights of the edges that have endpoints in different subsets. For instance, the Max-Cut of~\eqref{fig:examplegraph} is the following
\begin{equation}
\includegraphics[width=.4\linewidth,valign=c]{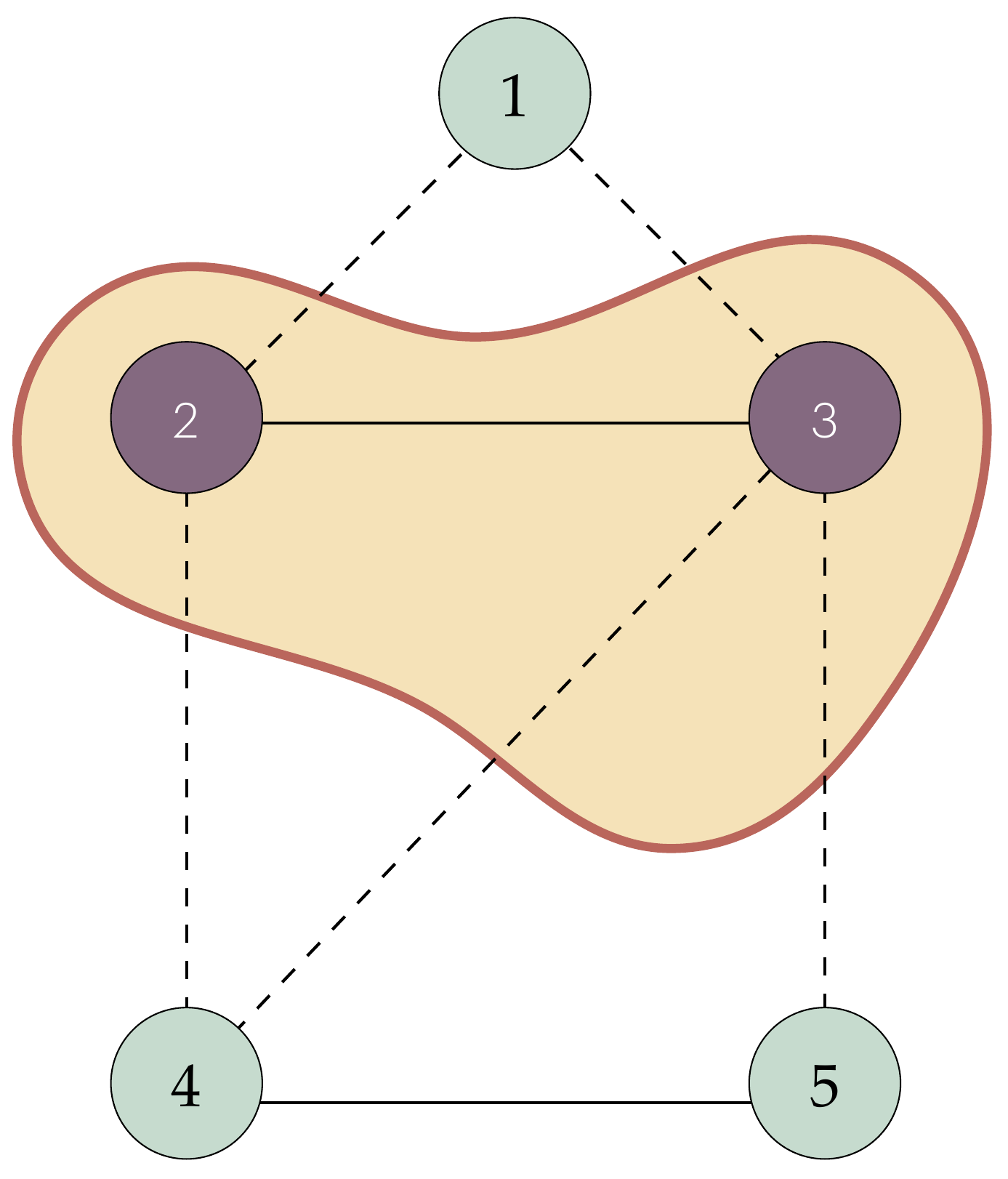}
\end{equation}

It is straightforward to express the Max-Cut problem as a QUBO optimization. We begin by assigning a binary variable $x_j$ to each vertex $j$, where $x_j = 0$ if the vertex is in the first subset and $x_j = 1$ if it is in the second subset. Then, it is easy to figure out that the matrix $Q$ in the MaxCut problem is related to the adjacency matrix $A$ of the undirected graph. Indeed, we have to minimize the function\begin{equation}
    f(\mathbf{x}) = - \sum_{i,j}A_{ij}(2x_i-1) (2x_j-1).
\end{equation}
which is such that for every connected $x_i \neq x_j$ contributes $-1$ to the value of $f(\mathbf{x})$, while every $x_i = x_j$ contribute with $+1$. Equivalently, discarding an unnecessary additive constant the problem is equivalent to find the $\mathbf{x}^*$ which minimizes\begin{equation}
    f(\mathbf{x}) = \sum_{ij}A_{ij}\left(2x_ix_j -x_i-x_j\right)
\end{equation}

\paragraph{Traveling salesman problem ---} The Traveling Salesman Problem (TSP) is a classic optimization problem in the field of combinatorial mathematics and computer science.\index{QUBO!traveling salesman problem} It involves determining the most efficient route for a seller who must visit a given set of cities, stopping at each city exactly once, and then returning to the starting point. The challenge lies in finding the shortest possible route that accomplishes this task, given the distances or costs between every pair of cities.

The problem is known for its computational complexity; as the number of cities increases, the number of possible routes grows factorially, making an exhaustive search impractical for large instances. Again, as for the Max-Cut problem, despite its simplicity in formulation, TSP is NP-hard, meaning that no efficient algorithm is known to solve all instances of the problem in polynomial time.

The TSP can be formulated as a QUBO optimization problem. For $N$ cities at distances $d_{ij}$, we define a set of $n = N^2$ binary variables $x_{i,\mu}$, where $i$ represents the index of a city, and $\mu$ denotes the position of that city in the sequence of the route. Specifically, if $x_{2,3} = 1$, it indicates that the second city is the third one visited in the route.

To ensure that the solution represents a valid tour, where each city is visited exactly once and every position in the sequence corresponds to exactly one city, we impose the following penalties or constraints
\begin{equation}
    \sum_{i=1}^N \left(1 - \sum_{\mu=1}^N x_{i,\mu}\right)^2 , \qquad \sum_{\mu=1}^N \left(1 - \sum_{i=1}^N x_{i,\mu}\right)^2 .
\end{equation}
These equations guarantee that each city is assigned a unique position in the route and that each position in the route is occupied by only one city. The first equation ensures that every city appears exactly once in the sequence, while the second equation ensures that each position in the route is filled by one and only one city. If some cities are not reachable directly from other cities, also include the equation
\begin{equation}
    \sum_{ij} \left(1-A_{ij}\right) x_{i,\mu} x_{j,\mu+1}
\end{equation}
where $A$ is the adjacency matrix of the unweighted graph. Finally, we have to include the cost of the route in the objective function, which is given by\begin{equation}
    h\sum_{ij} d_{ij} \sum_\mu^N x_{i,\mu}x_{j,\mu+1}.
\end{equation}

The scaling factor $h$ has to be small enough that it is never favorable to violate that the solution represents a valid tour, that is, $0< h\max{d}<1$. The QUBO problem that solves the TSP is therefore the following \begin{align}
    f(\mathbf{x}) &= \sum_{i=1}^N \left(1 - \sum_{\mu=1}^N x_{i,\mu}\right)^2 + \sum_{\mu=1}^N \left(1 - \sum_{i=1}^N x_{i,\mu}\right)^2 +\notag \\ &\quad + \sum_{ij} \left(1-A_{ij}\right) x_{i,\mu} x_{j,\mu+1} +   h\sum_{ij} d_{ij} \sum_\mu^N x_{i,\mu}x_{j,\mu+1}
\end{align}

% è scritto in https://www.frontiersin.org/journals/physics/articles/10.3389/fphy.2014.00005/full

\subsection{Mapping to Ising model}
The QUBO framework is relevant in the field of quantum computing since it is computationally equivalent to ground state search of a classical Ising model, which is described by the Hamiltonian function\index{Ising model}
\begin{equation}
H(\sigma) = -\sum_{\langle i,j\rangle} J_{ij} \sigma_i \sigma_j -  \sum_j h_j \sigma_j
\end{equation}
where the parameters $h_j, J_{ij}$ are real values applicable to all pairs $(i,j)$. As usual, in the Ising model, the spin variables $\sigma_j$ are binary, taking values from $\{-1,+1\}$.

These spin variables are usually arranged in a lattice, allowing only adjacent pairs to interact significantly. Applying the transformation $\sigma \mapsto 2x-1$, we can equivalently express the Ising model as a QUBO problem. The equivalent QUBO function is then given by
\begin{equation}
\begin{aligned}
f(x) &= \sum_{\langle i,j\rangle} -J_{ij}(2x_i-1)(2x_j-1) + \sum_{j} h_j(2x_j-1) \\
&= \sum_{\langle i,j\rangle} (-4J_{ij}x_ix_j + 2J_{ij}x_i + 2J_{ij}x_j - J_{ij}) + \sum_{j}(2 h_jx_j -  h_j) \\
&= \sum_{\langle i,j\rangle} (-4J_{ij}x_ix_j) + \sum_j \left(\sum_{\langle i,k=j\rangle}(2J_{ki} + 2J_{ik}) + 2 h_j \right)x_j - \sum_{\langle i,j\rangle}J_{ij} - \sum_{j} h_j,
\end{aligned}
\end{equation}
where we define the QUBO matrix $Q$ as
\begin{equation}
Q_{ij} = \begin{cases}
-4J_{ij} & \text{if } i \neq j \\
\sum_{\langle i,k=j\rangle}(2J_{ki} + 2J_{ik}) + 2 h_j & \text{if } i = j
\end{cases}
\end{equation}
and the constant term $C$ as
\begin{equation}
C = -\sum_{\langle i,j\rangle}J_{ij} - \sum_{j} h_j.
\end{equation}
Since the constant $C$ does not influence the optimization of $\mathbf{x}^*$, it can generally be ignored during the optimization process, although it is critical for calculating the original Hamiltonian function's value.

\section{Quantum Annealing}

Recent explorations into adiabatic quantum optimization have sparked significant interest due to its potential to tackle QUBO problems. This quantum computational approach takes advantage of a dynamical adiabatic change in Hamiltonian parameters from a simple initial Hamiltonian $H_0$, whose ground state is easily realizable, to a problem-specific Hamiltonian $H_P$, whose ground state encodes the solution to the desired problem.

The method relies on the principle of keeping the quantum system in its ground state while slowly changing from $\hat H_x$ to $\hat H_z$ (with $[\hat H_x, \hat H_z] \neq 0$)
\begin{equation}
\hat H(t) = \left(1 - s\right) \hat H_x + s\, \hat H_z,
\end{equation}
with an interpolation parameter $s=s(t)\in[0,1]$ such that $s(0) = 0$ and $s(\tau) = 1$, $\tau$ being the total annealing time. According to the adiabatic theorem, if $T$ is sufficiently large, compared to the inverse square of the minimal spectral gap of $\hat H(s)$, the system remains in its lowest energy state throughout the evolution, resulting in a solution at $t = \tau$.

The efficiency of adiabatic quantum optimization has been under scrutiny, particularly concerning its potential to surpass classical computational speeds. The relationship between the system size $N$ and the time $\tau$, given by $\tau = O(\exp(\alpha N^\beta))$, suggests that maintaining the ground state, and thus avoiding transitions to excited states due to minute energy gaps (Landau--Zener transitions), demands exponentially increasing time as $N$ approaches infinity. Although solving NP-complete problems in polynomial time using adiabatic optimization remains implausible, the actual coefficients $\alpha$ and $\beta$ might be competitively lower than those in classical approaches, suggesting a possible advantages for specific problem classes.

In the case of a QUBO problem $\hat H_z$ is the quantum counterpart of the classical Ising model, which, as already pointed out, typically involve a system of $N$ spins $\sigma_i = \pm 1$ described by the Hamiltonian\index{Ising model!quantum annealing}
\begin{equation}
H = -\sum_{i<j}^N J_{ij} \sigma_i \sigma_j - \sum_{i=1}^N h_i \sigma_i,
\end{equation}
so that its quantum counterpart reads
\begin{equation}
\hat H_z = \hat H(\hat Z_1, \ldots, \hat Z_N),
\end{equation}
where $\hat Z_i$ are Pauli matrices acting on the $i$-th qubit. To initialize the system, the initial state is the ground state of
\begin{equation}
\hat H_x =
-\sum_{i=1}^N \hat X_i,
\end{equation}
creating a superposition of all possible states in the computational basis. This setup creates an opportunity for employing tensor network methods. In particular, $\hat H(s)$ has a clear MPO description and the time evolution can be implemented for exampple with a TDVP algorithm. This allows to simulate and benchmark the performance of quantum annealing processes.  This approach not only aids in verifying the efficacy of quantum annealing solutions but also enhances our understanding of the scalability and limitations of current quantum optimization technologies.

\subsection{Digitized Quantum Annealing (dQA)}
In its digitized form, Quantum Annealing (QA) can be implemented on digital quantum simulators or classically simulated through a process known as digitized Quantum Annealing (dQA)~\cite{Martinis_Nat16, Mbeng_dQA_PRB2019, Mbeng_arXiv2019}.\index{Digitized quantum annealing} This procedure discretizes the continuous time evolution of QA into $P$ time steps, each of duration $\delta t = \tau/P$, followed by a Trotter decomposition to separate the non-commuting Hamiltonian terms. The lowest-order Trotter splitting is given by\index{Trotter decomposition}
\begin{equation}\label{eq:trottertrotter}
    e^{-i \hat H(s_p) \,\delta t }
    \simeq
    e^{-i  (1-s_p) \hat H_x \, \delta t}\; e^{-i  s_p \hat H_z \, \delta t } + \mathcal{O}\big(\delta t\big)^2 \;,
\end{equation}
where $p=1,2, \dots, P$ and $s_p = t_p/\tau = p/P$. The parameters $\beta_p = (1-s_p)\delta t$ and $\gamma_p = s_p\delta t$ allow the previous expression to be compactly rewritten as
\begin{equation}
\begin{split}
    e^{-i  \hat H(s_p) \, \delta t } & \simeq \hat U_x(\beta_p) \,  \hat U_z(\gamma_p) + \mathcal{O}\big(\delta t\big) ^2 \\
    \hat U_x(\beta_p) & = e^{-i\beta_p \hat H_x},\quad
    \hat U_z(\gamma_p)  = e^{-i\gamma_p \hat H_z},
    \end{split}
\end{equation}
yielding a quantum state at the end of the digitized annealing process:
\begin{equation}\label{eq:dQA_intro}
|\psi_{P}\rangle =
e^{-i\beta_{P} \hat H_x} e^{-i\gamma_{P} \hat H_z} \cdots
e^{-i\beta_{p} \hat H_x} e^{-i\gamma_{p} \hat H_z}
\cdots e^{-i\beta_{1} \hat H_x} e^{-i\gamma_{1} \hat H_z} |\psi_0\rangle \;,
\end{equation}
where $|\psi_0\rangle = |+\rangle^{\otimes N}$ is the ground state of the transverse field term $\hat H_x$, and $|+\rangle = \frac{1}{\sqrt{2}} (|0\rangle + |1\rangle)$.

This framework allows the digitized QA to closely approximate real-time Quantum Annealing dynamics as $P \to \infty$ and $\delta t \to 0$, while maintaining $\tau = P \delta t$ finite. However, for finite $P$, the trade-off between Trotter errors and the annealing time $\tau$ becomes crucial~\cite{Mbeng_dQA_PRB2019}. Small $\delta t$ results in non-adiabatic evolution, while large $\delta t$ makes the Trotter approximation too rough, introducing spurious quantum correlations.

Despite the rough approximation introduced by Trotterization, numerical studies have shown that even for values of $\delta t$ on the order of $\mathcal{O}(1)$, the discrete-time evolution often yields surprisingly accurate results without needing a higher-order split-up. This makes it a practical approach for both simulation and implementation of QA dynamics on near-term quantum devices.

\subsection{dQA with Tensor Network}
A novel tensor network appraoch was introduced to efficiently simulate digitized Quantum Annealing (dQA) for a broad class of classical Hamiltonians or cost functions~\cite{Guglielmo_dQA&TN_2023}. These Hamiltonians take the form:
\begin{equation}
H(\pmb{\sigma}) = \sum_{\mu=1}^{N_{\xi}} \mathfrak{h} \, ( \pmb{\xi}^{\mu} \cdot \pmb{\sigma} ) \;,
\end{equation}
where the variables $\xi^{\mu}_i \in \{-1, +1\}$ (with $\mu=1,2,\dots, N_{\xi}$ and $i=1,2,\dots, N$) represent spin or qubit configurations, often referred to as ``patterns'', and $\mathfrak{h}$ is any sufficiently regular function.

Notice that the QUBO Hamiltonian can be easily rewritten in this form; specifically the quadratic term only --- the so called Hopfield model~\cite{mehlig2021machine, hertz1991introduction, doi:10.1073/pnas.79.8.2554, PhysRevA.32.1007} --- looks like\index{Hopfield model}
\begin{equation}\label{chapt4_eq:hopfield}
   %H^{\text{Hopfield}}(\pmb{\sigma}) =
   \sum_{i,j} J_{ij} \sigma_{i} \sigma_{j} =
   \sum_{\mu=1}^{N_{\xi}} \left(\pmb{\xi}^{\mu} \cdot \pmb{\sigma}\right)^2,
   \qquad J_{ij} =  \sum_{\mu=1}^{N_{\xi}} \xi^{\mu}_i \xi^{\mu}_j \, ,
\end{equation}
where a sufficiently large number of patterns $N_{\xi}$  have been chose to reconstruct the random couplings $J_{ij}$.

The methods discussed in this work are notably versatile, applying to any Hamiltonian that can be reformulated in this general structure, reminiscent of the cost functions used in simple discrete neural networks. This framework opens up new possibilities for efficiently simulating complex quantum systems within the scope of dQA. In the following sections, we will explore this methodology in more detail.

Let us begin by noting that the initial state $\ket{\psi_0} =|+\rangle^N$ is a simple product state, i.e.\ a Matrix Product State (MPS) with a bond dimension of $\chi = 1$.
\begin{equation}
\includegraphics[width=0.6\textwidth,valign=c]{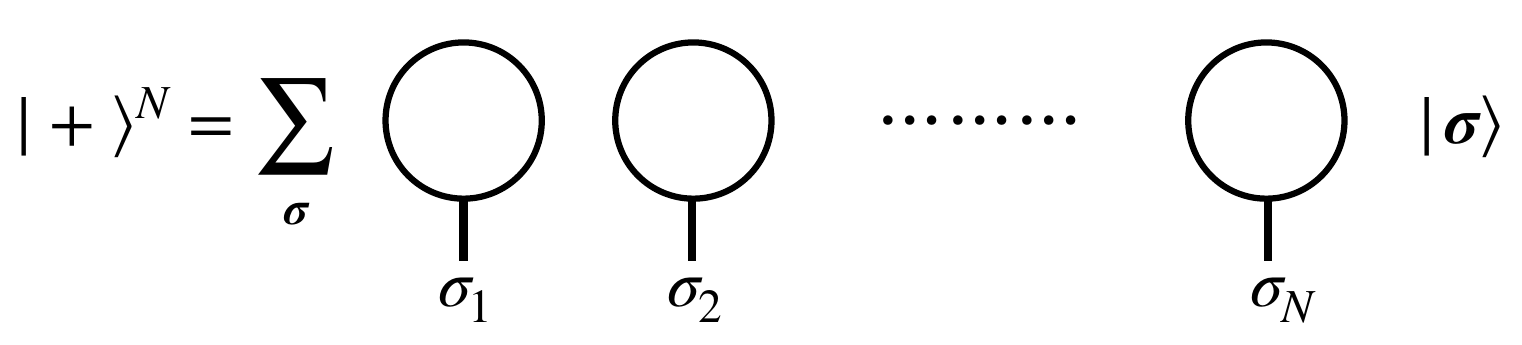}
\end{equation}

The next objective is to express the unitaries
$\hat U_z(\gamma_p)$ and $\hat U_x(\beta_p)$
as Matrix Product Operators (MPOs).

\paragraph{The $\hat U_x$ gate ---}
Let's begin with $\hat U_x$ which admits an elementary decomposition into an MPO of bond dimension $\chi=1$,
namely in the computational basis one has
\begin{equation}
\includegraphics[width=0.6\textwidth,valign=c]{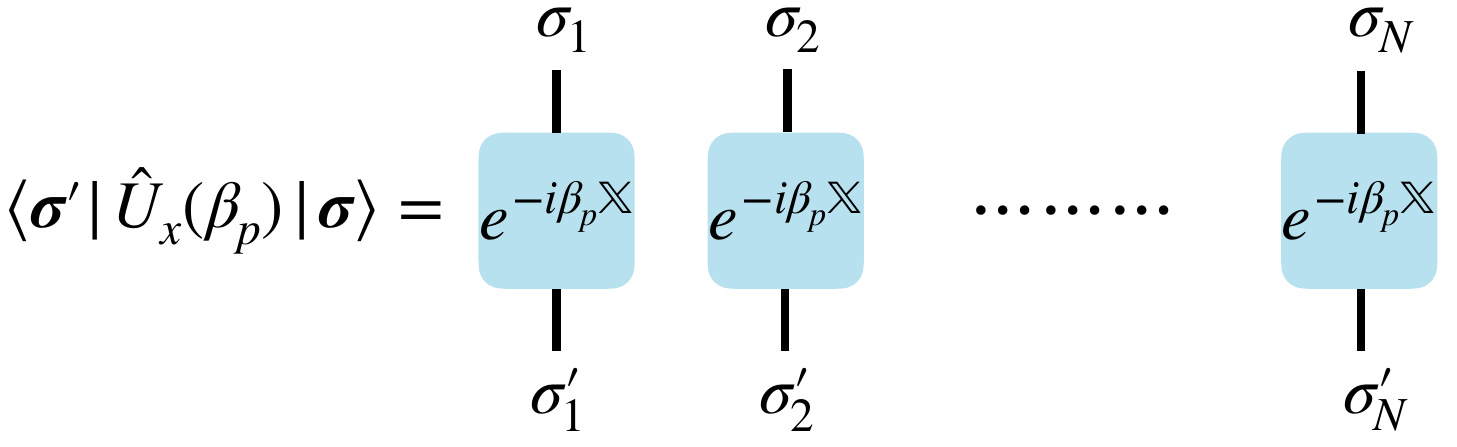}
\end{equation}
and has we know now very well, applying those operators to an MPS is trivial and it is not increasing the auxiliary dimension of its representation.

\paragraph{The $\hat U_z$ gate ---}
The real novelty comes in decomposing $\hat U_z$ as an MPO.
Since the target Hamiltonian encodes  different pattern, we can first decompose the total unitary in $N_{\xi}$ layers
\begin{equation}
\hat U_z(\gamma_p) =
\prod_{\mu=1}^{N_{\xi}} \hat U^{\mu}_z(\gamma_p) =
\prod_{\mu=1}^{N_{\xi}}
e^{-i\gamma_p\mathfrak{h}(\pmb{\xi}^{\mu} \cdot \hat{\pmb{Z}})},
\end{equation}
with $\hat{\pmb{Z}}=\{\hat Z_1,\dots, \hat Z_{N}\}$ a vector of Pauli matrices. These operators are diagonal in the computational basis with entries
\begin{equation}
\langle\pmb{\sigma}'|\hat{U}_z^{\mu}(\gamma_p) |\pmb{\sigma}\rangle=
\delta_{\pmb{\sigma'\sigma}}
e^{-i\gamma_p\mathfrak{h}(\pmb{\xi}^{\mu} \cdot \pmb{\sigma})}
\end{equation}

Let us now take advantage of the particular structure of the Hamiltonian $\hat H_z$. By its very nature, this Hamiltonian depends on the spin configuration $\pmb{\sigma}$ only through the following variables:
\begin{equation}
x^{\mu}(\pmb{\sigma})  = \frac{N - \pmb{\xi}^{\mu} \cdot \pmb{\sigma}}{2}
=
\sum_{j=1}^N \left( \frac{1 - \xi^{\mu}_j \sigma_j }{2} \right),
\end{equation}
which quantify the number of bits in $ \pmb{\sigma} $ that differ from those in $ \pmb{\xi}^{\mu} $. This expression represents the well-known Hamming distance between the configurations $ \pmb{\sigma} $ and $ \pmb{\xi}^{\mu} $, where
$x^{\mu} \in \{0, 1, \dots, N\}$.

This observation is critical, as it constitutes the only assumption underpinning the approach. Since $x$ (for any pattern $ \mu $) is a discrete integer variable taking values in $\{0, 1, \dots, N\}$, we can represent any function by applying the Discrete Fourier Transform (DFT). In particular for what we concern we get
\begin{eqnarray}
\langle\pmb{\sigma}'|\hat{U}_z^{\mu}(\gamma_p) |\pmb{\sigma} \rangle & = & \delta_{\pmb{\sigma'\sigma}}
\frac{1}{\sqrt{N+1}} \sum_{k=0}^N \widetilde{U}_{kp} \,
e^{i \frac{2\pi}{N+1} k x^{\mu}(\pmb{\sigma})} \\
\widetilde{U}_{kp} &=& \frac{1}{\sqrt{N+1}} \sum_{x=0}^N
e^{-i \frac{2\pi}{N+1} k x} \, e^{-i\gamma_p \mathfrak{h}(N-2x)},\label{chapt4_eq:Uz_fourier}
\end{eqnarray}
where Fourier components $\widetilde{U}_{kp}$ implicitly depend on the angle $\gamma_p$, and can be treated as a matrix with dimensions $(N+1) \times P$.

Through the use of Fourier decomposition, an ``efficient'' representation of $\hat U^{\mu}_z(\gamma_p)$ as a MPO becomes possible. In fact, the wave numbers $k = 0, 1, \dots, N$ naturally emerge as ``auxiliary indices'' within the MPO formalism. This enables the transformation of the operator into the form
\begin{equation}
\includegraphics[width=0.7\textwidth,valign=c]{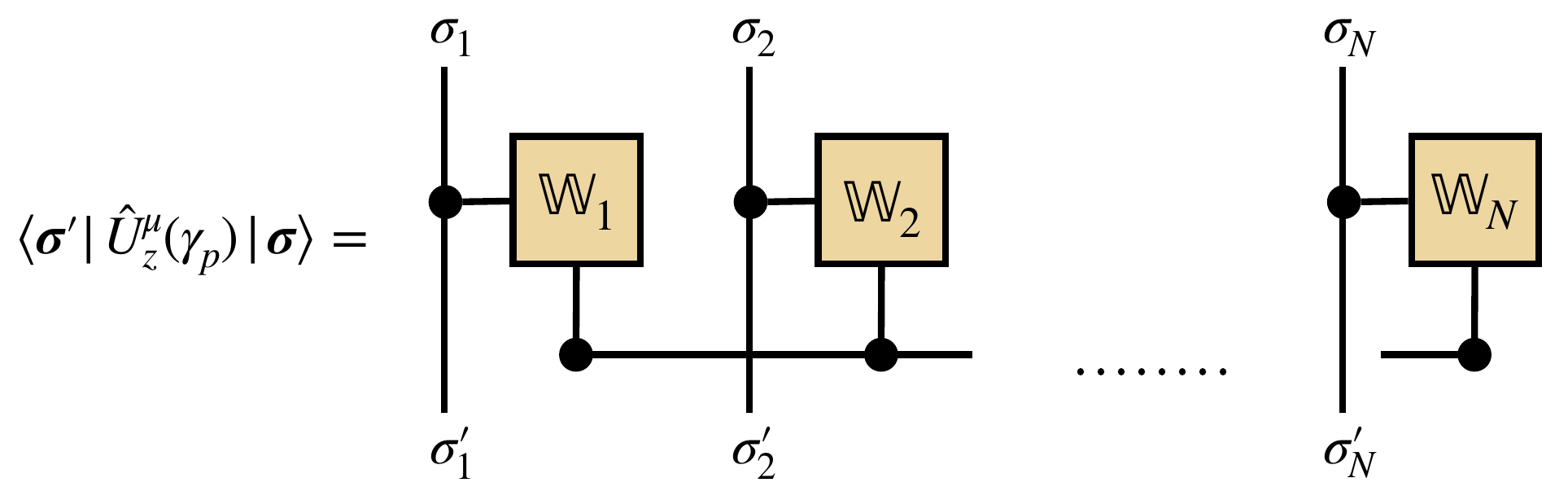},
\end{equation}
where the tensors $\mathbb{W}_j$,
have the following entries
\begin{equation}\label{chapt4_eq:Wj_mpo}
\includegraphics[width=0.45\textwidth,valign=c]{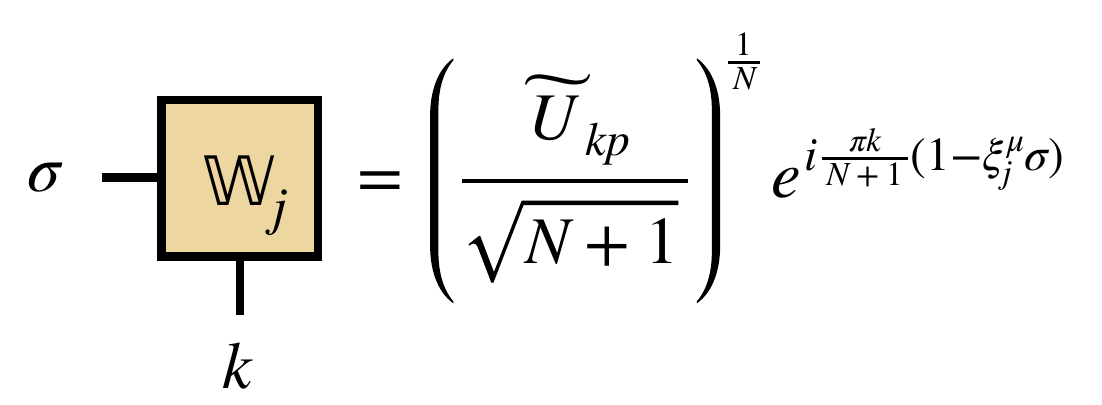},
\end{equation}

The tensors $\mathbb{W}_j$ are notably diagonal in both the auxiliary indices, meaning that they essentially depend on only a single auxiliary index $k$ and one physical spin $\sigma_j$,
since they only depends on the Pauli matrix $\hat Z_j$.
As usual, at the left (right) boundary, the tensor $\mathbb{W}_1$ ($\mathbb{W}_{N}$) reduces to a row (column) vector in the auxiliary space.
In addition, it's important to recognize that they also implicitly rely on the angle $\gamma_p$ and the pattern index $\mu$, even though these dependencies aren't explicitly shown in the notation. This formulation captures the critical factors needed for efficient tensor manipulations while abstracting the underlying dependencies.\\

Given these findings, the complete digitized quantum annealing (dQA) time evolution can be faithfully depicted as a 2D Tensor Network, as illustrated in the following figure:
$$
\includegraphics[width=0.8\textwidth,valign=c]{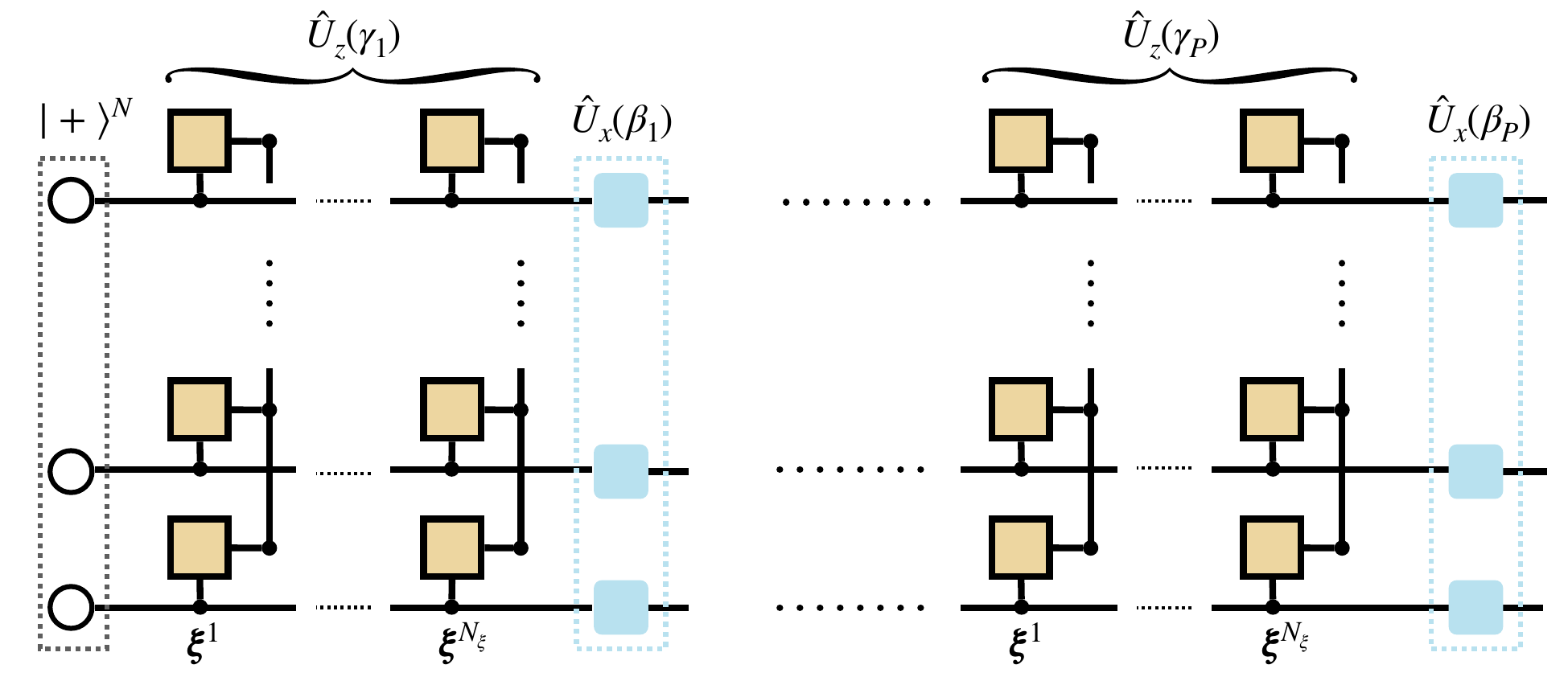}
$$

This network effectively models the application of successive Matrix Product Operators (MPOs) to the initial Matrix Product State (MPS) $\ket{\psi_0}$, which initially possesses a trivial structure.

This observation forms the foundation of an MPS-based algorithm for simulating dQA classically, outlined in the following pseudo-code:

\begin{algorithm}[H]
\caption{dQA with MPS.}\label{chapt4_alg:dQA}

\textbf{Input}: the dQA parameters $\tau, \, P, \,  \delta t = \tau / P$, the MPS maximum bond dimension $\chi$

\begin{algorithmic}[1]
\State Compute the Fourier matrix $\widetilde{U}_{k,p}$ (Eq.~\eqref{chapt4_eq:Uz_fourier}) of dimension $(N+1) \times P$
\State Define the initial state $\ket{\psi_0}$ as an MPS of bond dimension $\chi=1$ and set $\ket{\psi} = \ket{\psi_0}$
\For{($p=1$, $p\leq P$, $p\!+\!+$)}
  \For{($\mu=1$, $\mu\leq N_{\xi}$, $\mu\!+\!+$)}
     \State apply $\hat U_z^{\mu}(\gamma_p)$ to the current MPS $\ket{\psi}$ and if needed \underline{compress} the bond dimension to $\chi$
  \EndFor
  \State apply $\hat U_x(\beta_p)$ to the current MPS $\ket{\psi}$
\EndFor
\end{algorithmic}

\textbf{Output}: the final optimized MPS $\ket{\psi}$

\end{algorithm}

A significant challenge in this approach is the rapid growth of the bond dimension of the MPS as the number of Trotter steps $P$ increases. To mitigate this, a suitable \textbf{compression strategy} needs to be implemented (see Example~\ref{exmp:mps_compression}). This compression is crucial for maintaining computational feasibility, as a critical input for the algorithm is the predefined maximum bond dimension.

Let' finally mention that the MPO structure based on the DFT strategy offers an additional advantage: it enables the Hamiltonian itself to be expressed as an MPO with a bond dimension of $\chi = N+1$. This is a noteworthy feature, as it facilitates the \emph{precise computation} of the classical cost function --- specifically, the expectation value of $\hat H_z$ --- for \emph{any} MPS. Consequently, while running the algorithm, one can accurately monitor the evolution of the cost function over time.

\begin{example}{Variational compression of a MPS}{mps_compression}\index{MPS!variational compression}
As already explained in Chapter~\ref{chap1} and Chapter~\ref{chap2}, the SVD is a foundamental tool for performing a compression sweep over an MPS.
Here we outline another procedure which is more accurate and is based on a variational optimisation.
In practice, the compression technique restricts the evolving system to the manifold of MPS with a predefined maximum bond dimension $\chi$. This can be accomplished using an iterative approach, as outlined in works like Refs.~\cite{Schollwock_2011, Saberi_2009}, which efficiently solves the optimization problem step-by-step at each lattice site.

In practice, lets suppose we want to compress an MPS state with bond dimension $D$, such that
$|\psi\rangle =\sum_{\sigma_1,\dots,\sigma_N} A^{\sigma_1}\dots A^{\sigma_N}|\sigma_1,\dots,\sigma_N\rangle$. We basically want to find a new set of optimal tensors $\{\tilde A^{\sigma_j}\}$  with bond dimension $\chi \ll D$, such that the compressed state will be $|\tilde\psi\rangle =\sum_{\sigma_1,\dots,\sigma_N}\tilde A^{\sigma_1}\dots \tilde A^{\sigma_N}|\sigma_1,\dots,\sigma_N\rangle$. This is achieved by solving the following local variational equations
\begin{equation}\label{chapt4_eq:compress}
\frac{\partial}{\partial \tilde{A}^{\sigma_j}}
||  |\tilde\psi\rangle - |\psi\rangle ||^2
=
\frac{\partial}{\partial \tilde{A}^{\sigma_j}}
\left(
\langle\tilde\psi|\tilde\psi\rangle - \langle\tilde\psi|\psi\rangle
\right) = 0.
\end{equation}
In practice, this process involves carrying out a series of localized optimizations by sweeping across all the sites in the system multiple times, denoted by the number of sweeps $N_{sweeps}$. Each sweep progressively refines the approximation, ensuring convergence to a more accurate solution.

To optimize the tensor at position $j$, the best method involves having the MPS for $|\tilde\psi\rangle$ in its mixed canonical form relative to that specific site. This ensures that the optimal update is simply:
\begin{equation}
\includegraphics[width=0.9\textwidth,valign=c]{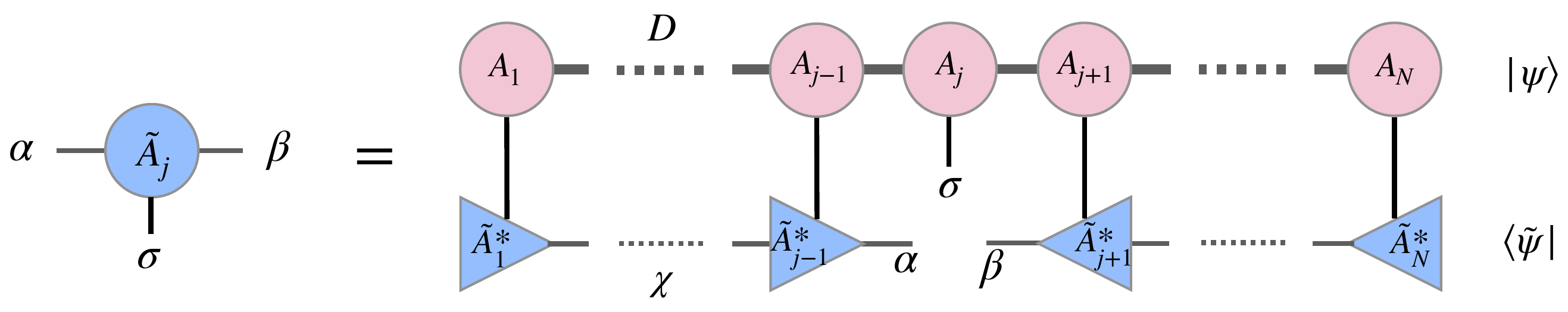}
\end{equation}

Here, the contraction on the right-hand side has a computational cost which scales as $N D \chi^2 + N D^2 \chi \sim N D^2 \chi$, when $D \gg \chi$. Notably, unlike the compressed state $|\tilde\psi\rangle$, the known state $ |\psi\rangle$ doesn't necessarily need to be in mixed canonical form. After computing $\tilde{A}^{\sigma_j}$, the next step depends on the direction of the sweep. If sweeping from right-to-left, one performs a QR decomposition, while for a left-to-right sweep, an LQ decomposition is used to move efficiently to the next lattice site at position $j+1$ or $j-1$ respectively. This ensures that the MPS remains in a canonical form appropriate for the chosen direction of the sweep.

As with most iterative procedures, this method benefits from a good initial representation of $|\tilde\psi\rangle$, which, in this case, is provided through the canonical SVD compression.

\paragraph{Optimised compression for dQA ---}\index{Digitized quantum annealing!optimised compression}
Whilst this strategy is generic, when applied within the dQA algorithm the state $\ket{\psi}$ is obtained by applying a unitary evolution operator to the MPS from the previous step, expressed as $|\psi\rangle = \hat U_z^{\mu}(\gamma_p) |\phi\rangle$. We learned that, when an MPO is applied to an MPS, the bond dimensions increase as their product, meaning $D = (N+1) \chi $ at regime. As a result, applying blindly the previous scheme, the cost of computing the local optimised tensor becomes $\sim N^3 \chi^3$, where $\chi$ is the maximum bond dimension fixed in the dQA algorithm.
However, the compression efficiency can be greatly improved by exploiting the particular diagonal structure of the MPO.
By explicitly writing the action of the MPOs $\hat W_j$ in Eq.~(\ref{chapt4_eq:Wj_mpo}) we easily get
\begin{equation}
\includegraphics[width=0.9\textwidth,valign=c]{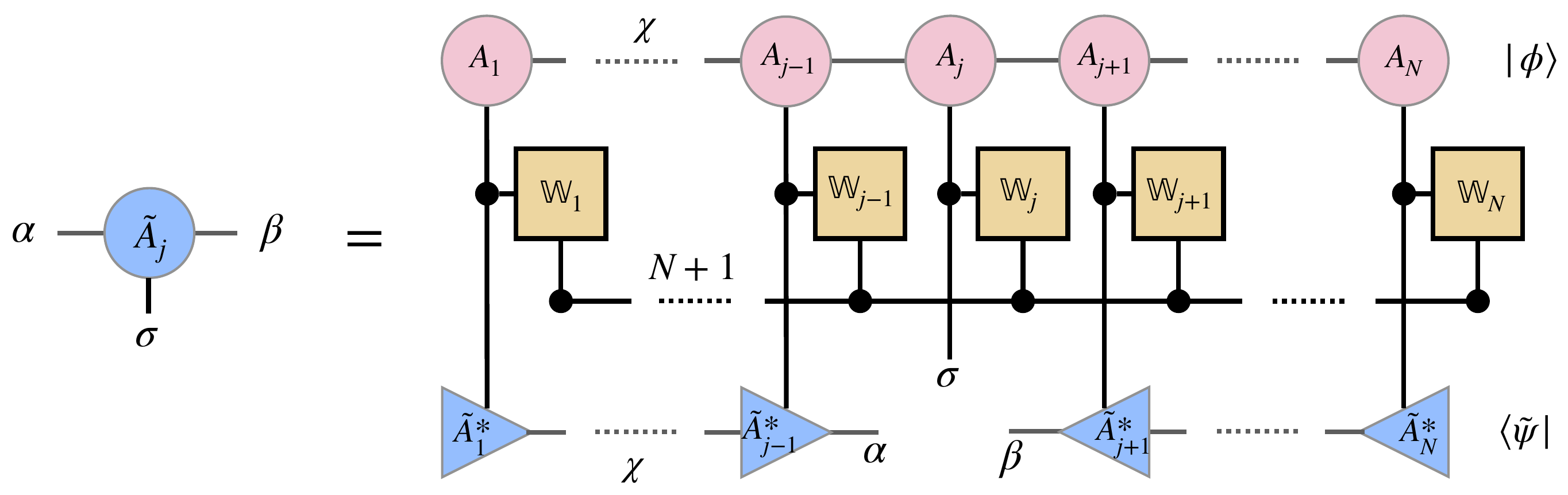}
\end{equation}
reducing the computational cost to $\sim N^2 \chi^3$, since we have basically to compress $N+1$ times a tensor network which has a computational cost $\sim N\chi^3$.

\end{example}

\subsection{Quantum approximate optimization algorithm}

The Quantum Approximate Optimization Algorithm (QAOA) is a variational quantum algorithm that builds upon the digitized quantum annealing (dQA) process by treating the parameters $\boldsymbol{\beta} = {\beta_{1},\dots,\beta_{P}}$ and $\boldsymbol{\gamma} = {\gamma_{1},\dots,\gamma_{P}}$ as variational variables. This approach transforms dQA into a Variational Quantum Algorithm (VQA), with QAOA playing a crucial role in modern quantum computing research~\cite{Farhi_arXiv2014, VQA_review}.

In the context of QAOA, as in the previous sections, the goal is to minimize a problem-specific cost function, typically encoded in a Hamiltonian $\hat{H}_P$, by evolving an initial state $\ket{+}^{\otimes N}$ through a sequence of alternating unitary operators. These unitaries correspond to the time-evolved versions of the mixing Hamiltonian $\hat{H}_x$ and the problem Hamiltonian $\hat{H}_z$, with the evolution parameters being the variational angles $\boldsymbol{\beta}$ and $\boldsymbol{\gamma}$. The alternating application of these Hamiltonians is mathematically described as:
\begin{equation}
\ket{\psi(\boldsymbol{\beta}, \boldsymbol{\gamma})} = \prod_{i=1}^{P} e^{-i \beta_i \hat{H}_x} e^{-i \gamma_i \hat{H}_P} \ket{+}^{\otimes N}.
\end{equation}

The variational nature of QAOA comes from tuning the parameters $\boldsymbol{\beta}$ and $\boldsymbol{\gamma}$ so that the expectation value of the problem Hamiltonian $\hat{H}_P$ in the evolved state $\ket{\psi(\boldsymbol{\beta}, \boldsymbol{\gamma})}$ is minimized
\begin{equation}
\expval{ \hat{H}_P }{\psi(\boldsymbol{\beta}, \boldsymbol{\gamma})}.
\end{equation}
This expectation value serves as a cost function to be minimized using classical optimization techniques. The success of the algorithm depends on finding optimal parameters and determining their quantity,  balancing exploration and exploitation. As the number of layers $P$ increases, QAOA approaches the performance of adiabatic quantum computing, but even with small $P$, it provides useful approximations, making it suitable for near-term quantum devices.

In QAOA the entanglement between qubits grows with circuit depth and problem complexity, often requiring large bond dimensions in MPS to accurately capture this entanglement~\cite{Dupont2022PRA}. This challenge highlights the classical hardness of simulating QAOA for deeper circuits, as the entanglement barrier increases. As a result, simulating QAOA on classical devices often involves a trade-off between accuracy and computational cost.

Nevertheless, TN methods serve as an excellent classical benchmark to evaluate the performance of quantum algorithms such as QAOA. By adjusting the bond dimension, TN simulations can provide upper bounds on the entanglement that classical methods can handle. Comparing the performance of quantum hardware with TN benchmarks allows us to quantify the quantum advantage. Thus, TN methods offer a practical way to calibrate the difficulty of QAOA problems and assess quantum hardware performance in realistic scenarios.

\biblio
\chapter[Open tensor network]{Open tensor network: from thermal states to open system dynamics}\label{chap5}
\epigraph{No man ever steps in the same river twice, for it is not the same river and he is not the same man.}{Heraclitus}

Until now, aside from a brief introduction to density matrices, we have primarily discussed scenarios in which the state of a quantum system is fully described by a single unit vector $|\psi\rangle$ in its Hilbert space. However, in many real-world situations, the exact state vector $|\psi\rangle$ is not known. Consider, for instance, a beam of thermal neutrons emitted from a radioactive source. In this scenario, we do not have specific information about the energy of each neutron. Instead, we only know the overall distribution of their energies, which might follow a canonical distribution at a given temperature. This exemplifies a situation where our knowledge of the system is incomplete.

In this chapter, we will consider scenarios where a single unit vector in Hilbert space is insufficient to fully describe the system under investigation. This situation often arises when the system is in thermal equilibrium at a non-zero temperature or interacts with an environment about which we lack detailed knowledge. Additionally, continuous monitoring of the system by external probes can influence its dynamics in a non-unitary manner, necessitating a more comprehensive description. There are numerous situations where a pure state description falls short; here, we will focus on a select few that are particularly pertinent to tensor network methods and quantum computing. These include thermal equilibrium states, open quantum systems dynamics, and measurement-induced dynamics, highlighting their relevance and application in contemporary quantum research.

\section{Open system recap}
Continuing from the introduction of this chapter, our lack of information about the state of a quantum system is mathematically represented by the fact that we cannot assign a unique state in the Hilbert space to our system. Instead, we must describe it as an ensemble of states, each associated with a certain probability:\index{Statistical mixture}
\begin{equation}
    \{\ket{\psi_1}, p_1\}, \; \{\ket{\psi_2},p_2\}, \; \dots, \;
    \{\ket{\psi_D},p_D\}
\end{equation}
such that $p_k \geq 0 \, \forall k \in \{1, \dots, D\}$, and $\sum_{k=1}^{D} p_k = 1$ to ensure a proper probability distribution. It is important to note that the states $|\psi_k\rangle$ are not necessarily orthogonal to each other.

It is fundamentally impossible to describe a statistical mixture using a single ``typical'' vector; instead, it can be described using either a distribution of ``typical'' vectors or just a ``typical'' operator, commonly known as the \textbf{density operator} or \textbf{density matrix}.\index{Density matrix}

\begin{definition}{Statistical Mixture}{mixture}
We describe a statistical mixture, also known as a mixed state, when instead of having a single pure state representing the system, we have an ensemble of states with corresponding probabilities $\{\ket{\psi_k}, p_k\}$, such that $\sum_k p_k = 1$. Practically, this ensemble is represented in quantum mechanics through the \textbf{density operator}:
\begin{equation}\label{chapt5_eq:mixture}
    \hat{\rho} = \sum_k p_k \ketbra{\psi_k},
\end{equation}
where $\Tr(\hat{\rho}) = 1$.

\end{definition}

According to the rules of quantum mechanics, the density operator provides the probabilities of the associated statistical mixture. Specifically, for a pure state $\ket{\psi_k}$ with probability $p_k$, this probability can be computed using the density operator as $p_k = \Tr(\hat{\rho} \ketbra{\psi_k})$.

In addition, consider any observable $\hat{O}$, which can be decomposed in terms of its eigenvectors $\ket{o_j}$ and eigenvalues $o_j$:
\begin{equation}
    \hat{O} = \sum_{j} o_j \ketbra{o_j}.
\end{equation}

The probability of obtaining the outcome $o_j$ when measuring the observable $\hat{O}$ on the state described by the density operator $\hat{\rho}$ is given by:
\begin{equation}
    p(o_j) = \Tr(\hat{\rho} \ketbra{o_j}) = \sum_{k} p_k |\braket{o_j}{\psi_k}|^2.
\end{equation}

This formula encapsulates how the density operator effectively captures the statistical distribution of measurement outcomes over the ensemble of states. The expectation value of the observable $\hat{O}$, which represents the average measurement outcome, is then given by:
\begin{equation}
    \langle \hat{O} \rangle = \sum_{j} o_j p(o_j) = \Tr(\hat{O} \hat{\rho}).
\end{equation}

To summarize, the density operator $\hat{\rho}$ not only encodes the probabilities of different pure states in the mixture but also allows us to calculate the probabilities of measurement outcomes and the expectation values of observables in a straightforward manner. This is crucial for understanding the behavior of quantum systems where the state is not precisely known and must be described as a mixture of several possible states~\cite{Benenti2019}.

\paragraph{Mixed system unitary evolution ---}
When a mixed system evolve in time according to the quantum mechanics postulates (see Chapter~\ref{chap2}), assuming the evolution is induced then by a unitary transformation, aka the Schrodinger equation,
the time dependent statistical mixture reads
\begin{equation}
    \hat{\rho}(t) = \sum_k p_k |\psi_k(t)\rangle \langle \psi_k(t)|,
\end{equation}
where $\{|\psi_k(t)\rangle\}$ are the pure states evolving according to the Schrodinger equation and $\{p_k\}$ are the corresponding time-independent probabilities.

To derive the \emph{von Neumann equation}, which govern the time evolution of the density operator, we start from taking the time derivative of $\hat{\rho}(t)$ as written above:
\begin{equation}
    \partial_t \hat{\rho}(t) =
    \sum_k p_k \left[  \left(\partial_t|\psi_k(t)\rangle \right) \langle \psi_k(t)|
    + |\psi_k(t)\rangle\left(\partial_t\langle\psi_k(t)| \right) \right],
\end{equation}
then using the Schr\"odinger equation for the pure states we easily obtain:\index{von Neumann equation}
\begin{equation}\label{eq:von_neumann}
   \partial_t\hat{\rho}(t) = -i \left[ \hat{H}, \hat{\rho}(t) \right],
\end{equation}
where in the entire formulation we fixed $\hbar = 1$.
This equation shows that the time evolution of the density matrix $\hat{\rho}(t)$ is governed by the commutator of the Hamiltonian $\hat{H}$ with $\hat{\rho}(t)$. This formalism is crucial for describing the dynamics of mixed states in quantum mechanics, allowing for a complete representation of the statistical properties of quantum systems.

\paragraph{Thermal (or Canonical) ensemble ---}
What does a typical wavefunction of a quantum system look like at a finite temperature? In statistical mechanics, a fundamental concept is indeed the density matrix which describes the state of a system in thermal equilibrium. For a system with Hamiltonian $\hat{H}$ at an inverse temperature $\beta = 1/k T$, the density matrix is given by:\index{Thermal state}
\begin{equation}
    \hat{\rho} = \frac{1}{Z} \exp(-\beta \hat{H}),
\end{equation}
where $Z =\Tr{\exp(-\beta \hat{H})}$ is the \emph{partition function}.\index{Partition function}
This density matrix can be understood in several ways. For example, it can arise from averaging over an ensemble of pure states, from taking the long-time average of one system, or from the quantum mechanical entanglement with a heat bath, which produces mixed states. Regardless of the interpretation, the predictions of statistical mechanics depend only on $\hat{\rho}$.

However, in general, even though, as we have seen in Chapter~\ref{chap2}, local Hamiltonians often admit simple MPO representations, exponentiating an MPO is typically a challenging task. This difficulty is analogous to exponentiating a matrix, which, in general, if it lacks special properties, is nontrivial.

A very trivial example of thermal state which is Hamiltonian independent, is the \textbf{infinite temperature state}, i.e.\ $\beta = 0$, which in fact a part from a suitable normalisation, reduces to the \textbf{identity operator}
$$
\includegraphics[width=0.9\textwidth,valign=c]{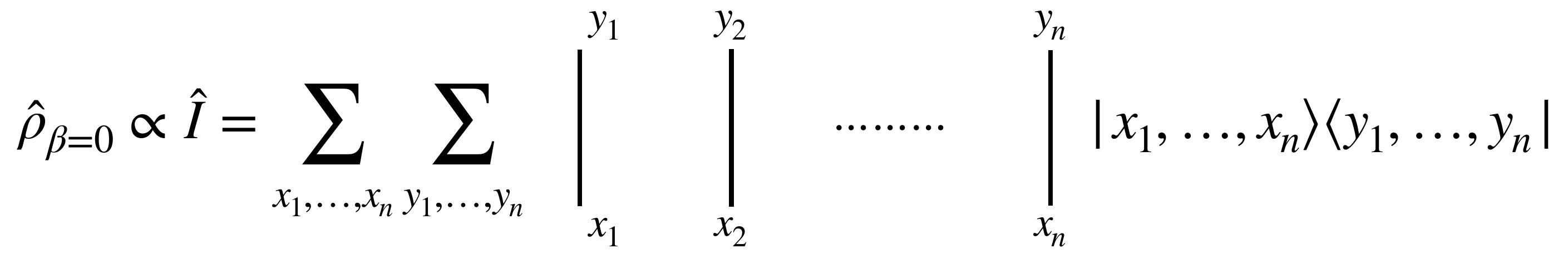}
$$
where the norm of the identity operator, for $n$ qubits state, is $Z = \Tr\hat I = 2^n$.

\begin{example}{Exact MPO for classical thermal states}{classical_thermal}
Let us consider the \textbf{classical ferromagnetic Ising model in one dimension}, in the presence of an external magnetic field $h$. Here, we assume homogeneous couplings between neighboring qubits. The Hamiltonian reads\index{Ising model}
\begin{equation}
\hat{H} = -J \sum_{j} \hat{Z}_{j}\hat{Z}_{j+1} - h \sum_{j} \hat{Z}_{j} \equiv \hat{H}_J + \hat{H}_h,
\end{equation}
where we arbitrarily choose to work with Pauli operators in the $\hat{z}$ direction, and we split the Hamiltonian into two commuting operators.
Here we consider a system with $n$ Spin-$1/2$ (or qubits) with open boundary conditions. We have already seen in Chapter~\ref{chap2} that this Hamiltonian admits an exact MPO representation. However, here we want to compute its \textbf{thermal density operator}
\begin{equation}
\exp(-\beta\hat{H}) = \exp(-\beta\hat{H}_J)\exp(-\beta\hat{H}_h),
\end{equation}
where we used the fact that $[\hat{H}_J, \hat{H}_h] = 0$. The magnetic contribution to the density operator is trivial due to the fact that each term acts locally; in practice, we have a product operator, namely an MPO with bond dimension $\chi_h = 1$.
$$
\includegraphics[width=0.35\textwidth,valign=c]{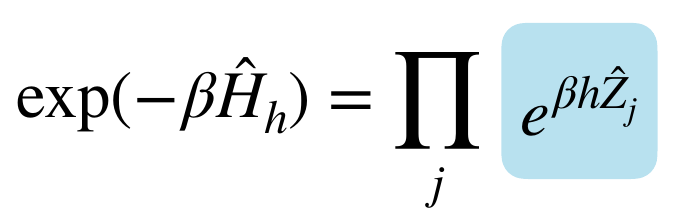}
$$
More interesting is the contribution from the neighboring coupling term, which reads (using the properties of Pauli matrices)
\begin{equation}\label{chapt5_eq:exp_beta_H1}
\exp (-\beta\hat H_J) = \prod_{j} [\cosh(\beta J) + \sinh(\beta J)\hat Z_{j}\hat Z_{j+1}].
\end{equation}
In fact, we can recast this as an MPO by explicitly rewriting each term in the product as a scalar product between two local neighboring vectors, i.e.,
\begin{equation}
\cosh(\beta J) + \sinh(\beta J)\,\hat Z_{j}\hat Z_{j+1} =
\begin{pmatrix}
\hat I_j & \hat Z_{j}
\end{pmatrix}
\cdot
\begin{pmatrix}
\cosh(\beta J)\, \hat I_j \\
\sinh(\beta J)\, \hat Z_{j+1}
\end{pmatrix},
\end{equation}
where we formally reintroduce the local identity operator.
Using this last result into the Eq.~\eqref{chapt5_eq:exp_beta_H1}, and collecting the terms acting on the same qubit, we can rewrite the exponential factor as an MPO with bond dimension $\chi_J = 2$
$$
\includegraphics[width=0.7\textwidth,valign=c]{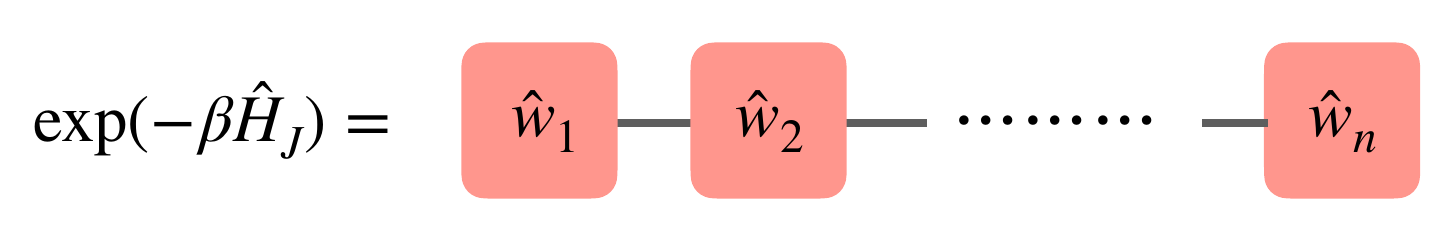}
$$
with uniform bulk tensors
\begin{equation}
\hat w_j =
\begin{pmatrix}
\cosh(\beta J)\,\hat I & \cosh(\beta J)\,\hat Z\\
\sinh(\beta J)\,\hat Z & \sinh(\beta J)\,\hat I
\end{pmatrix}
\end{equation}
and boundary tensors
\begin{equation}
\hat w_1 =
\begin{pmatrix}
\hat I & \hat Z
\end{pmatrix}, \quad
\hat w_n =
\begin{pmatrix}
\cosh(\beta J)\,\hat I \\
\sinh(\beta J)\,\hat Z
\end{pmatrix}.
\end{equation}

At this point, it is straightforward to combine the two MPO representations of the exponential factors into a single MPO representing $\exp(-\beta\hat{H})$ with bond dimension $\chi = \chi_J \chi_h = 2$, by following the rules of the tensor network contractions.

Let us finally conclude this example by considering the fact that, with this representation in hand, we can easily compute the \textbf{Partition Function} $Z = \Tr[\exp(-\beta\hat{H})]$ of this model.\index{Partition function} This computation corresponds to the normalization of the density operator. The connection with thermodynamics is established by identifying the \textbf{Free Energy} with the logarithm of the partition function, namely $\mathcal{F} = - \frac{\log \mathcal{Z}}{\beta}$. Specifically, taking the trace of the full MPO representation of the thermal density operator involves wrapping the physical wires locally for each qubit, thus obtaining
$$
\includegraphics[width=0.5\textwidth,valign=c]{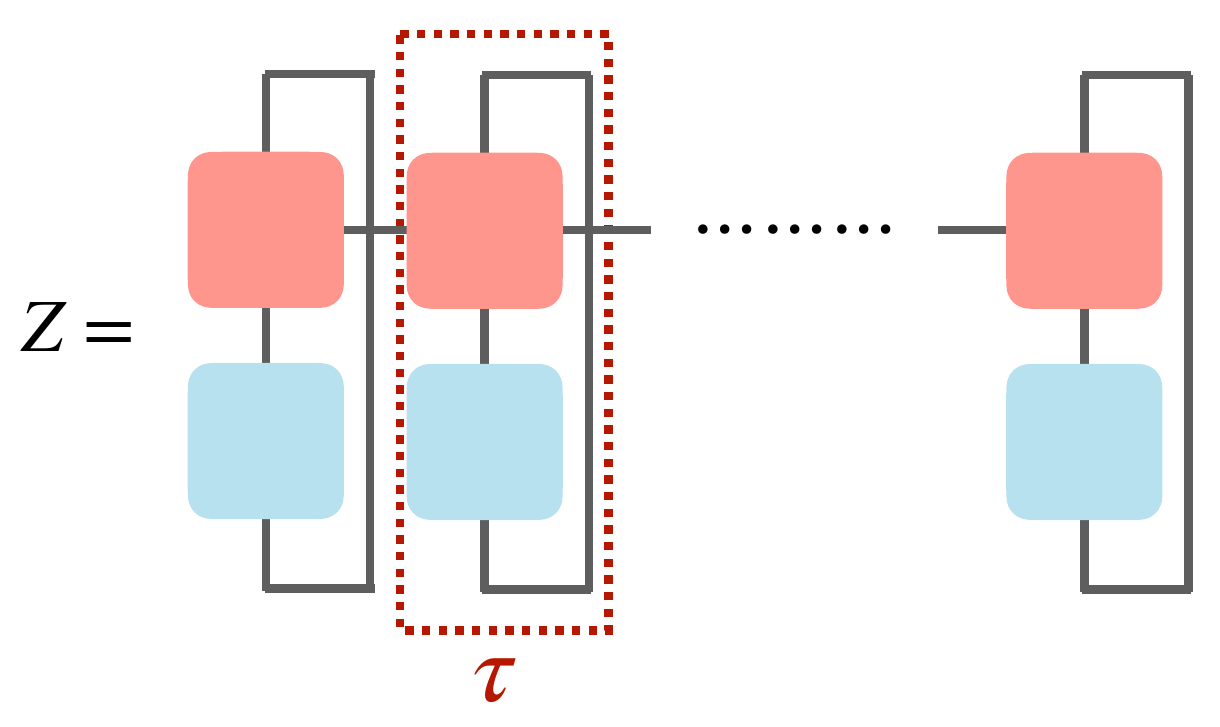}
$$
in terms of the $2\times 2$ classical transfer matrix (for the bulk)\index{Transfer matrix}
\begin{equation}
\tau = 2
\begin{pmatrix}
\cosh(\beta J)\cosh(\beta h) & \cosh(\beta J)\sinh(\beta h)\\
\sinh(\beta J)\sinh(\beta h) & \sinh(\beta J)\cosh(\beta h)
\end{pmatrix}.
\end{equation}
For instance, in the simplest scenario where the magnetic field vanishes ($h = 0$), $\tau$ simplifies to a diagonal matrix. By considering the boundary vectors, we obtain the well-known result from classical statistical mechanics:
\begin{equation}
Z_{h=0} = 2^n \cosh(\beta J)^{n-1}.
\end{equation}
The general case with a finite magnetic field can be solved by diagonalizing the classical transfer matrix, which is left as an exercise for the reader.
\end{example}

\paragraph{Finite temperature states ---}
Finite temperature states for systems on a lattice, such as registers of qubits, can be efficiently prepared by extending the canonical Time-Evolving Block Decimation (TEBD) scheme to handle mixed states and performing imaginary time evolution~\cite{Vidal_2004}. This extension allows for the simulation of thermal states by evolving the system in imaginary time, effectively capturing the thermal properties at any desired temperature.

In fact, we can map operators, such as the density matrix $\hat{\rho}$, into ``superkets'' $\dket{\rho}$, and superoperators, acting on linear operators, into linear mappings acting on the superkets\index{Density matrix!superket formalism}. For instance, the commutator in the Liouville equation of motion transforms as $[\hat{H}, \cdot] \to \hat{H} \otimes \hat{I} - \hat{I} \otimes \hat{H}^T$. Notice that, all those transformation are simply based on the \textbf{vectorization of matrices}.

\begin{example}{Vectorization of matrices}{superket}
\emph{Vectorization} of a matrix is a specific reshaping operation where an $m \times n$ matrix is transformed into a vector of length $mn$.
In tensor network terminology, this corresponds to \textbf{bending the physical wires} of an MPO to get an MPS, resulting in:\index{Vectorization}
$$
\includegraphics[width=0.9\textwidth,valign=c]{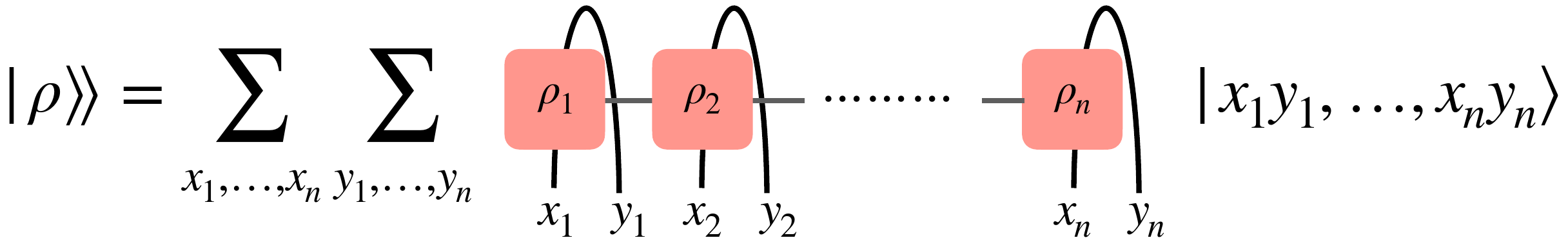}
$$
Similarly, the infinite temperature state, represented by the identity matrix, can be decomposed into a tensor product of non-normalized Bell states as follow:
$$
\includegraphics[width=0.9\textwidth,valign=c]{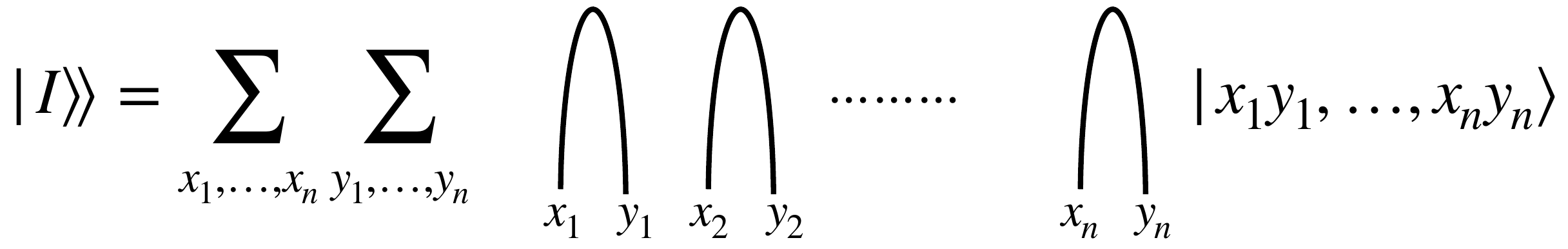}
$$

Vectorization basically is going to costruct a Hilbert space specifically for density matrices by defining an appropriate scalar product. This procedure effectively transforms matrices into vectors in this Hilbert space, where the density matrix $\hat \rho$ is represented as $\dket{\rho}$. This Hilbert space is known as the Fock-Liouville Space (FLS).

In this framework, the scalar product between two matrices, $\hat \phi$ and $\hat \rho$, is given by:
$
\dbraket{\phi}{\rho}
\equiv \text{Tr}[\hat\phi^\dagger \hat\rho].
$
With this setup, the Liouville superoperator becomes an operator acting within the FLS. The primary advantage of the Fock-Liouville Space is that it enables a convenient matrix representation for the evolution operator, facilitating the analysis and simulation of dynamic processes in quantum systems.
\end{example}

Before we delve into the procedure for using imaginary-time TEBD on superkets to generate finite temperature states, it is important to mention a key aspect of MPS-based algorithms. These algorithms typically preserve the standard ket-normalization, $\dbraket{\rho}{\rho}$, which results in a normalization condition at the operator level of $\Tr(\hat{\rho}^2)$. However, this may not be the desired norm to maintain throughout the simulation.

Now let's consider a nearest neighbor Hamiltonian $\hat{H}$ and an inverse temperature $\beta$. The thermal state of interest at this temperature can be constructed by using the superket formalism and simulating the imaginary time evolution from the completely mixed state,
\begin{equation}
e^{-\beta \hat H} = e^{-\beta \hat H/2} \, \hat I \, e^{-\beta \hat H/2}
\; \to \;
|e^{-\beta \hat H} \rangle\!\rangle = e^{-\beta\mathbb{\hat H}} | \hat I \rangle\!\rangle
\end{equation}
with $\mathbb{\hat H} = (\hat H \otimes \hat I + \hat I \otimes \hat H^t)/2$.

To efficiently compute the thermal state $\dket{\rho}$, we can utilize the Trotter decomposition to approximate the exponential operator $e^{-\beta \mathbb{\hat H}}$ as a sequence of operations involving only two adjacent sites. This method allows us to update the MPS using the Time-Evolving Block Decimation (TEBD) algorithm (see section 2.5.2).

The Trotter decomposition breaks down the exponential of a sum of operators into a product of exponentials of the individual operators. Specifically, for a Hamiltonian $\mathbb{\hat H}$ that can be written as a sum of terms $\mathbb{\hat H}_1$ and $\mathbb{\hat H}_2$ (where each term acts on different pairs of adjacent sites), we can approximate the exponential as:\index{Trotter decomposition}
\begin{equation}
e^{-\beta \mathbb{\hat H}} \approx \left(e^{-\Delta \beta \mathbb{\hat H}_1} e^{-\Delta \beta \mathbb{\hat H}_2}\right)^{\beta/\Delta \beta},
\end{equation}
where $\Delta \beta$ is a small imaginary time step. By iteratively applying this process, the thermal state $\dket{\rho}$ is constructed through a series of local updates involving only pairs of adjacent sites.

An important advantage of this approach is that a single run of the simulation generates the thermal state for any intermediate value of $\beta' \in [0, \beta]$. This is because the TEBD algorithm incrementally builds up the thermal state from the infinite temperature state (completely mixed state) to the desired temperature.
To implement the Trotter decomposition in practice, follow these steps (for a qubits system):
\begin{enumerate}
\item
\emph{Initialize the System ---}
Start with the infinite temperature state, which is represented as an MPS with bond dimension $\chi=1$,
each tensor describing the normalized Bell state $(\ket{00} + \ket{11})/\sqrt{2}$, so that to have a proper normalized superket MPS.
\item \emph{Apply Local Updates ---}
Sequentially apply the strings of local two-site operators $e^{-\Delta \beta \mathbb{\hat H}_1}$ and $e^{-\Delta \beta \mathbb{\hat H}_2}$ to the MPS. This can be done using the TEBD algorithm, which efficiently updates the MPS by considering one pair of adjacent sites at a time.
\item \emph{Repeat ---}
Iterate the above step $\beta/\Delta \beta$ times.
\end{enumerate}

Let us finally remember that, due to the anomalous normalisation of a thermal state in the superket MPS representation, namely
$\dbraket{\rho}{\rho} = \Tr{\hat\rho^2} = 1$,
we do have to compute the partition function\index{Partition function!superket}
\begin{equation}
%Z = \dbraket{I}{\rho} ,
\includegraphics[width=0.6\textwidth,valign=c]{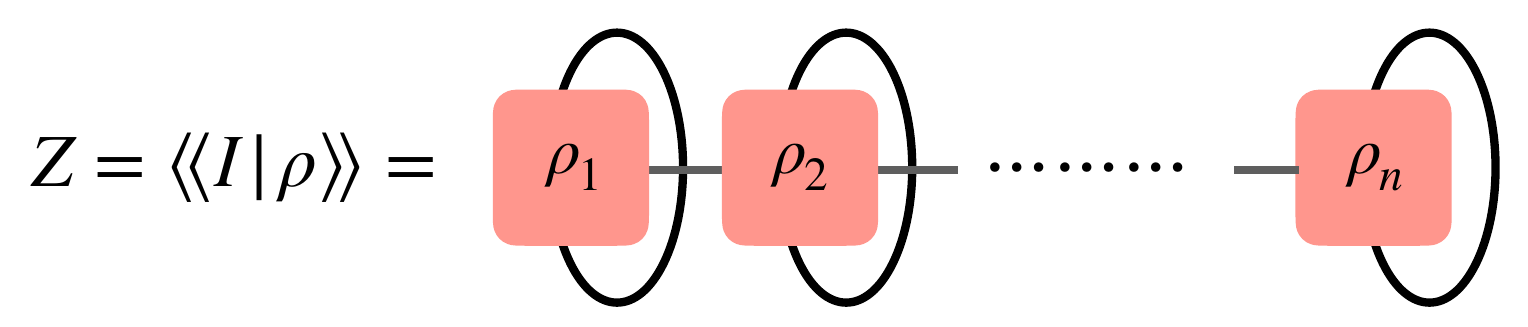}
\end{equation}
in order to properly measure observables
\begin{equation}
\langle \hat O \rangle = \frac{1}{Z}\Tr{\hat O \hat \rho} = \frac{\dbra{I} \hat O \otimes \hat I \dket{\rho}}{Z},
\end{equation}
where in the MPO formalism, the expectation value reads
\begin{equation}
\includegraphics[width=0.6\textwidth,valign=c]{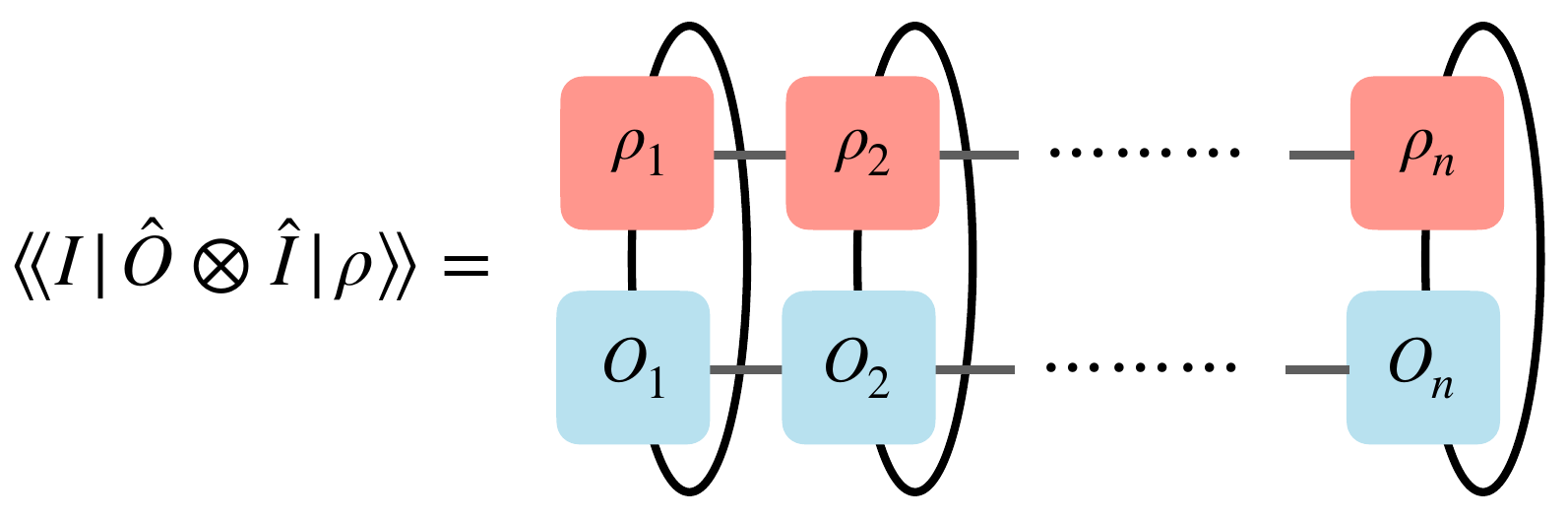}
\end{equation}

\subsection{Tensor Network and Density Operators}

In the language of Tensor Networks, any operator can, in principle, be represented as a Matrix Product Operator (MPO), provided we have sufficient computational resources. As discussed in Chapter~\ref{chap2}, this applies to any operator, including the density matrix. Therefore, the density matrix can be expressed in the computational basis as an MPO:
\begin{equation}
\includegraphics[width=0.9\textwidth,valign=c]{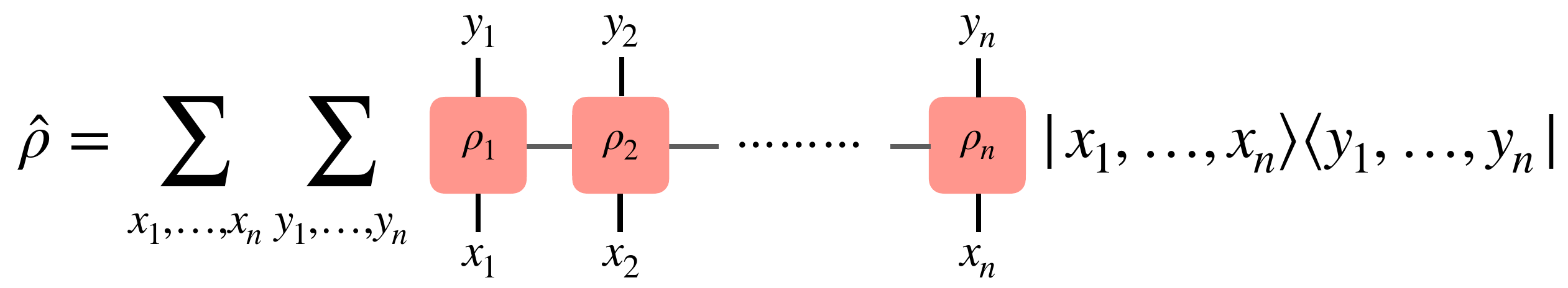}
\end{equation}
which in particular, by using the normalized Pauli tensors (see Definition~\ref{def:paulitensor}),
can be even rewritten in the normalized Pauli basis as
\index{MPO!Pauli basis}
\begin{equation}\label{chapt5_eq:rho_pauli}
\includegraphics[width=0.8\textwidth,valign=c]{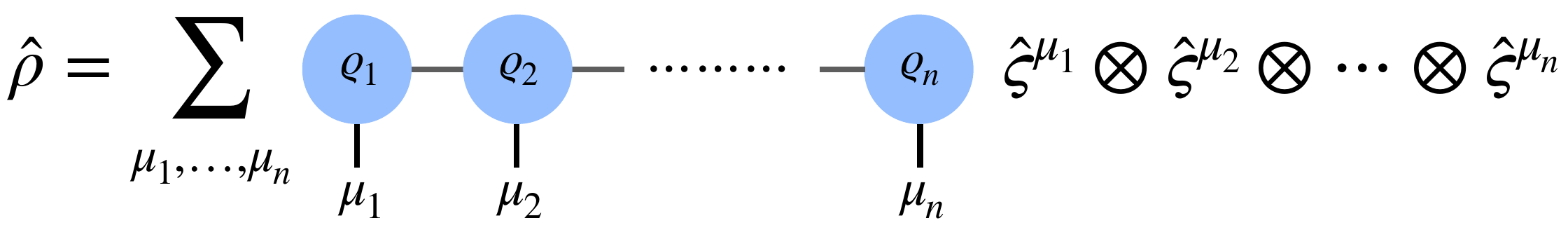}
\end{equation}
where using the properties of the Pauli tensor we can pass from one representation to the other via the following identity:
$$
\includegraphics[width=0.45\textwidth,valign=c]{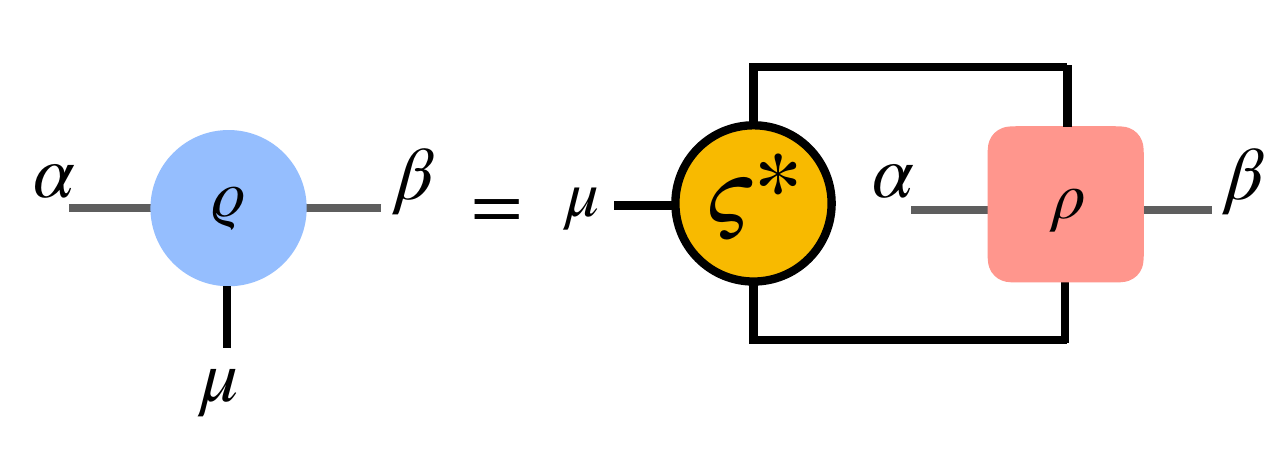}
$$

Notice that, this change of basis induced a trivial tensor transformation for the infinite temperature state, where indeed the local coefficient in the
Pauli basis reduce to $\varrho_{\mu}(\beta=0) = \Tr\hat\varsigma^{\mu} = \sqrt{2}\delta_{\mu0}$.

In addition, both representations have their advantages and disadvantages. For example, the density matrix is a Hermitian operator, which implies that we must enforce the condition $\hat \rho^{\dag} = \hat\rho$; this condition results in nontrivial constraints in the computational basis representation of the MPO. However, when using the Pauli basis representation, due to the Hermiticity of the basis itself, the condition simplifies to requiring only real numbers for the entries of the tensors representing the operator.

On the contrary, when the density matrix represents a global pure state, it reduces to a projector. Its representation in the computational basis splits into a very convenient tensor product of two MPS (see section 2.4.2), each representing the same state (apart from conjugation). However, when transitioning to the Pauli basis representation, this tensor structure is lost due to the local mixing of each tensor and its conjugate induced by the change of basis described above.

Moreover, from the very definition of a statistical mixture in Eq.~\eqref{chapt5_eq:mixture}, it is clear that a density operator is \emph{non-negative}, which implies that it can be rewritten as $\hat{\rho} = \hat{\Phi}\hat{\Phi}^{\dagger}$. It turns out that the MPO representation in the computational basis is more suitable for enforcing this condition, as we will see in the positive tensor network approach in Section~\ref{chapt5_ss:LPTN}.

\begin{definition}{normalized Pauli Tensor}{paulitensor}
For TN computations is very useful to introduce the so called
\textbf{normalized Pauli tensor} $\varsigma^{\mu}_{ij}$
which basically incorporate in one single
order-$3$ tensor all $3$ Pauli matrices plus the identity matrix, which form a complete basis for  single quibit  operators. Basically we have
in graphical notation (for $\mu\in\{0,1,2,3\}$)
$$
\includegraphics[width=0.5\textwidth,valign=c]{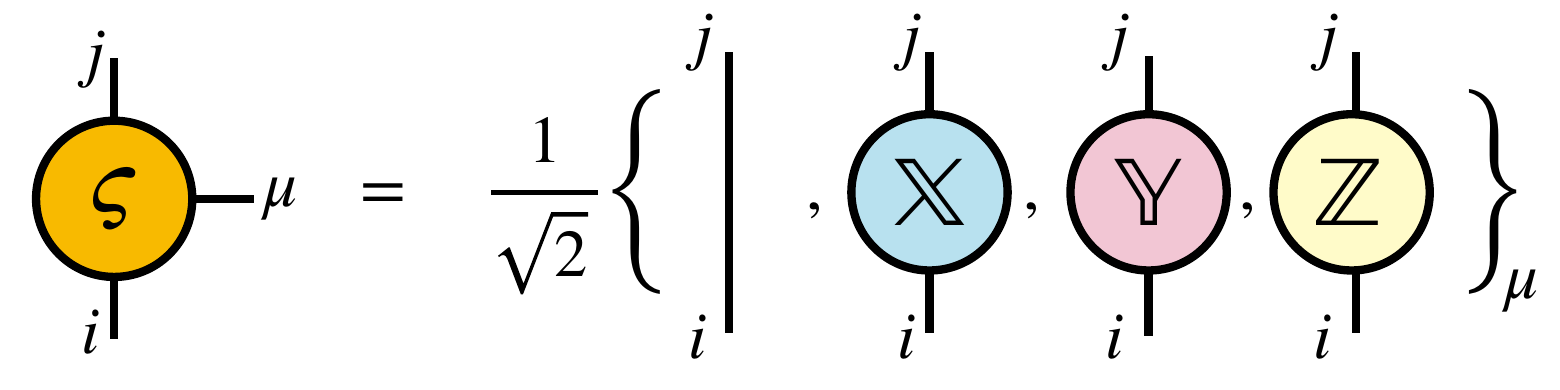}
$$
which satisfies the following orthogonality and completeness properties\index{Pauli matrices!orthogonality}\index{Pauli matrices!completeness}
$$
\includegraphics[width=\textwidth,valign=c]{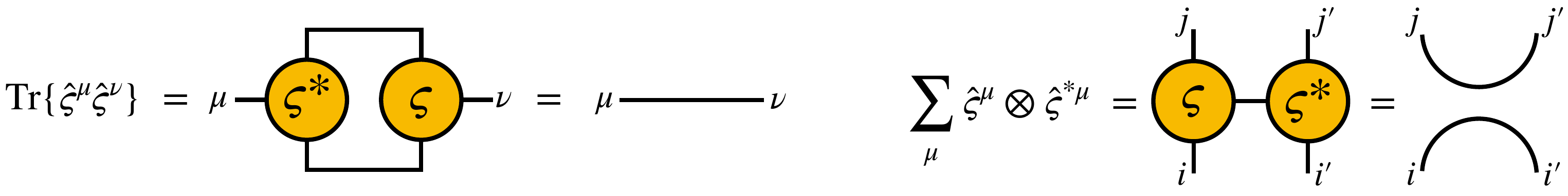}
$$
Notice that any operator can be decomposed in terms of the normalized Pauli tensor; for example in the single qubit case one gets
$$
\includegraphics[width=0.5\textwidth,valign=c]{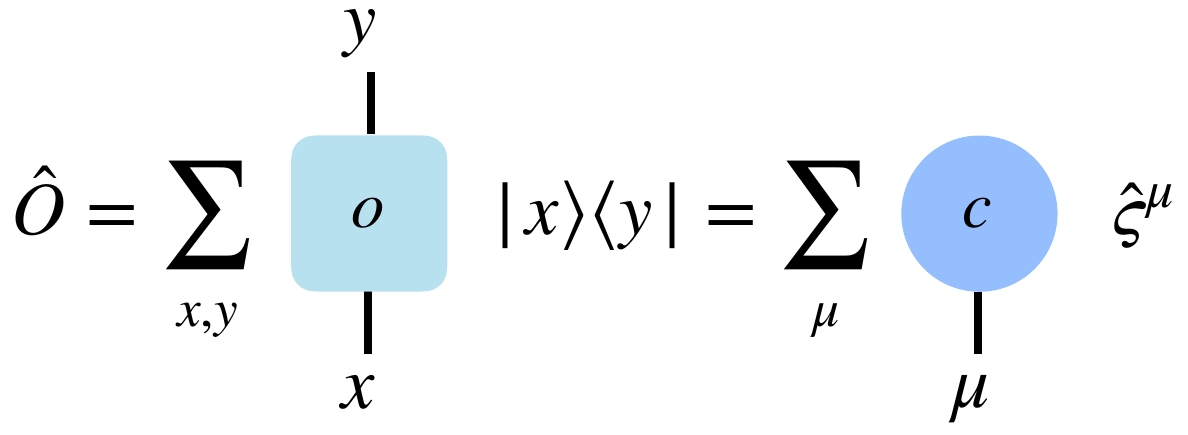}
$$
with the local tensors which are related by the following local transformation
$$
\includegraphics[width=0.3\textwidth,valign=c]{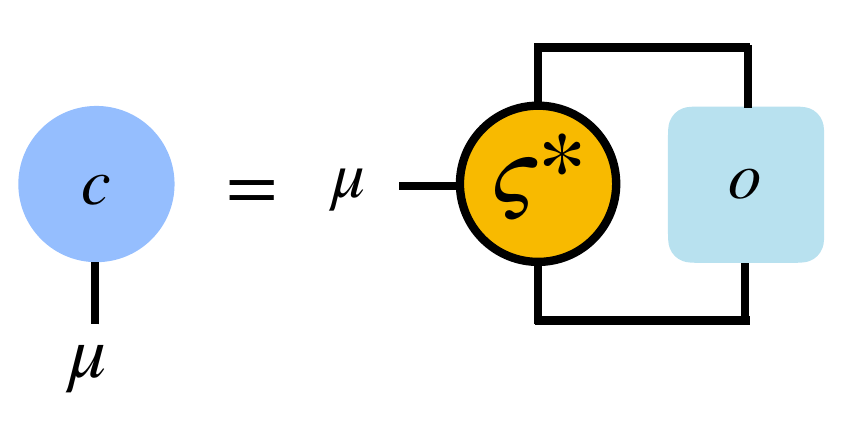}
$$
\end{definition}

\section{Minimally Entangled Typical Thermal State}\label{chapt5_s:METS}

Here we present a class of states originally introduced by Steven R. White in ref.~\cite{White_2009}, along with the related algorithm proposed by Steven R. White and Erwin M. Stoudenmire in ref.~\cite{Stoudenmire_2010}.

The approach is exactly been deviced to overcome the difficulty we have riced at the end of the previous section, and again it is inspired by the generic and very important question to describe a system at thermal equilibrium.

To make this concept more tangible, imagine we have a collection of pure states that represent different possible configurations of our system, as is typically the case for a mixture. Each pure state is like a snapshot of the system at a specific moment. When we average over many such snapshots, we get a picture of the system's behavior at thermal equilibrium.

For instance, consider a quantum system weakly coupled to a heat bath. As the system equilibrates with the heat bath, it reaches a state described by the density matrix $\hat{\rho}$. If we then remove the heat bath, the system retains this mixed state. From this perspective, we can think of the system's state as an ensemble of pure states that collectively generate $\hat{\rho} = \exp(-\beta \hat{H})/Z$.

This ensemble-of-pure-states viewpoint is particularly useful because it aligns with how we often simulate quantum systems. For example, when using techniques like diagonalization, the Density Matrix Renormalization Group (DMRG), or tensor network approaches, we can generate these pure states and efficiently compute finite temperature properties.

By proposing a set of idealized states, we argue that these states can be considered ``typical'' for a system in thermal equilibrium. The algorithm we introduce to generate these states not only helps us understand the system better but also provides a more efficient way to determine finite temperature properties of lattice models.

\subsection{Classical vs Quantum Typical Sample}

In quantum statistical mechanics, ``typical states'' refer to a set of states $\{|\phi(x)\rangle\}$ with associated unnormalized probabilities $P(x)$, from which we can sample to represent the thermal properties of a quantum system. The requirement for these states is that their ensemble should replicate the thermal density matrix of the system:
\begin{equation}
\sum_x P(x) |\phi(x)\rangle \langle \phi(x)| = e^{-\beta \hat H}
\end{equation}
This ensures that the expectation value of any observable $\hat A$ can be approximated by averaging $\langle \phi(x) | \hat A | \phi(x) \rangle$ over these typical states, with each state $|\phi(x)\rangle$ selected according to the probability $P(x)/Z$, with $Z=\sum_x P(x)$. This framework bridges classical and quantum statistical mechanics, ensuring that the sampled states are representative of the system's thermal properties.
Notice that the states $\ket{\phi(x)}$ need neither be orthogonal to each other nor constitute a complete basis set for the Hilbert space.

For classical lattice systems, typical states are \textbf{classical product states (CPS)}, which are configurations like $|x\rangle = \prod_j |x_j\rangle$, where $x_j$ denotes the state of the $j$-th site. For instance, in an Ising model, a CPS might be $\ket{\uparrow\downarrow\uparrow\downarrow\cdots}$, while in a register of qubits the analogous would be $\ket{0101\cdots}$. These states are intuitive and straightforward to generate numerically and experimentally, providing insight into the system's properties that might be obscured by the full density matrix.
However, in quantum lattice models (spin systems, qubits registers, etc.), CPS do not typically represent the system's thermal state, especially at zero temperature, where the ground state is generally not a CPS. This is because CPS lack the entanglement necessary to capture the correlations present in a quantum system's ground state or thermal state.

\paragraph{Energy Eigenstates ---}

An alternative approach might consider the energy eigenstates of the system. The energy eigenstates $|k\rangle$ and eigenvalues $E_k$ satisfy:
\begin{equation}
\hat \rho = \frac{1}{Z}\sum_k e^{-\beta E_k} \ketbra{k}
\end{equation}

Thus, the thermal density matrix can be expressed in terms of these eigenstates. However, these eigenstates should not be considered typical states for several reasons:

1. \emph{Equilibration and Dynamics}: Schrödinger highlighted that for large systems, especially at non-zero temperatures, the system does not equilibrate to any single energy eigenstate. The time required for such equilibration would be exponentially long in the number of particles $N$, making it impractical and unrealistic~\cite{schrodinger1989statistical}.

2. \emph{Sensitivity to Hamiltonian Uncertainties}: The energy eigenstates are exponentially sensitive to small changes in the Hamiltonian. Any slight uncertainty or perturbation in the system's parameters can drastically alter these states, making them unreliable for practical use in representing thermal properties.

3. \emph{Superposition States}: In phases with broken symmetry, eigenstates often form highly non-classical superpositions of states with different values of the order parameter. Such superpositions are not representative of typical states that one would expect to find in nature or experiments.

4. \emph{Entanglement and Decoherence}: Energy eigenstates tend to be highly entangled, which is contrary to the effects of decoherence in realistic physical systems. Decoherence tends to suppress such high entanglement, leading to states that are more classically correlated.

5. \emph{Computational Intractability}: From a computational perspective, obtaining energy eigenstates, especially for large systems, is highly challenging. The small energy level spacings require a full diagonalization of the Hamiltonian, which is computationally prohibitive for all but the smallest systems.

\paragraph{Constructing Typical States ---}

To address these issues, we can construct typical states that satisfy the thermal property requirement using a complete orthonormal basis $\{|x\rangle\}$. These states are defined as:
\begin{equation}
|\phi(x)\rangle = P(x)^{-1/2} e^{-\beta \hat H / 2} |x\rangle
\end{equation}
where $ P(x) = \bra{x} e^{-\beta \hat H}  \ket{x} $ is the unnormalized probability distribution. This  construction ensures that:
\begin{equation}
\langle \hat A \rangle = \frac{1}{Z} \Tr(\hat \rho \hat A) = \frac{1}{Z}\sum_x P(x) \bra{\phi(x)} \hat A \ket{\phi(x)}
\end{equation}
This approach avoids the need for computing the full spectrum of energy eigenstates, making it computationally feasible for large systems.

\emph{Minimally Entangled Typical Thermal States} (METTS) are a specific type of typical state designed to have minimal entanglement.\index{METTS} METTS are generated by choosing the initial set $\{|x\rangle\}$ as CPS. This minimization of entanglement makes METTS particularly suitable for numerical simulations using tensor network methods such as the Density Matrix Renormalization Group (DMRG). In fact, they provide a series of advantages:

1. \emph{Low Entanglement}: By design, METTS have low entanglement, simplifying their numerical representation and manipulation.

2. \emph{Physical Realism}: METTS incorporate decoherence effects naturally, making them more realistic for physical systems where such effects are significant.

3. \emph{Symmetry Breaking}: METTS can spontaneously break symmetries, which is important for studying systems with long-range order.

4. \emph{Revealing Short-Range Order}: Even in systems without broken symmetries, METTS can reveal underlying short-range order.

By leveraging the properties of METTS, we can effectively study the thermal properties of quantum systems, gaining insights that are computationally accessible and physically meaningful.

\subsection{METTS with Tensor Network}

The pure state method in ref.~\cite{Stoudenmire_2010} involves a straightforward algorithm for generating a series of Matrix Product States that are distributed correctly according to the probability $P(x)/Z$. The procedure is as follows: (1) at first select a random CPS
$\ket{x}$; (2) compute the METTS
$\ket{\phi(x)} = e^{-\beta \hat H / 2} \ket{x} P(x)^{-1/2}$,
store it, if not too much resource consuming, otherwise evaluate all observable of interest and store the results;
(3) Collapse to a new classical pure state
$\ket{y}$ from $\ket{\phi(x)}$ with probability
$p(x\to y) = |\braket{y}{\phi(x)}|^2$,
and return to step (2) for the next imaginary evolution.

\begin{example}{How METTS guarantees the correct distribution}{MEETS_ditribution}
Let us explore why the method presented in this section for producing METTS guarantees the correct distribution.

Consider an initial ensemble of classical product states (CPS) $|x\rangle$ distributed according to the probability $P(x)/Z$. If we select a CPS $|x\rangle$ at random from this ensemble and then proceed with the algorithm described previously, the probability of observing a specific CPS $|y\rangle$ at the end of step (3) is given by
\begin{equation}
\sum_{x}\frac{P(x)}{Z} p(x \to y) =
\sum_{x}\frac{P(x)}{Z} \left| \langle y | \phi(x) \rangle \right|^2.
\end{equation}

Substituting the definition
$|\phi(x)\rangle = P(x)^{-1/2} e^{-\beta \hat H / 2} \ket{x} $, we get
\begin{equation}
\sum_{x} \frac{\bra{y} e^{-\beta \hat H/2} \ket{x}\bra{x} e^{-\beta \hat H/2} \ket{y}}{Z}
= \frac{P(y)}{Z}.
\end{equation}
Thus, the probability of observing $|y\rangle$ aligns with the desired distribution $P(y)/Z$. The consistency of the resulting ensemble of $|y\rangle$ with the original distribution $P(x)/Z$ is ensured indeed by the detailed balance condition:
$P(x) p(x \to y) = P(y) p(y \to x)$.

Thus, the ensemble of CPS distributed according to $P(x)/Z$ remains invariant under this process (is basically a fixed point of the algorithm). Consequently, since each METTS $|\phi(x)\rangle$ is deterministically generated from a CPS $|x\rangle$ in step (2), the METTS themselves are also distribute according to $P(x)/Z$.
\end{example}

Notice that the aforementioned procedure generates a Markov chain of states. After a typical relaxation time, which may depend on the specific choice of the CPS basis states, the states $\{\ket{\phi(x)}\}$ are distributed according to the target distribution $P(x)/Z$. Consequently, any expectation value can be estimated by averaging over those states. In other words,
\begin{equation}
\langle \hat{A} \rangle =
\mathbb{E}_{x \sim P(x)/Z}
\left[ \bra{\phi(x)} \hat{A} \ket{\phi(x)}\right],
\end{equation}
where $\mathbb{E}_{x \sim P(x)/Z}$ is indicating the ensemble average.

We now discuss the practical implementation of these steps assuming we are working with $n$-qubits register. In this scenario, a CPS state is represented by a classical configuration of the register, namely $\ket{x_1,x_2,\dots,x_n}$ with $x_{j}\in\{0,1\}$.
Extension to any local base dimension is straightforward.

\paragraph{METTS generation ---}
The second step of the algorithm involves generating the typical thermal state $\ket{\phi(x)}$ from a classical product state $\ket{x}$. In the tensor network framework, this can be efficiently achieved using either the Time-Evolving Block Decimation (TEBD) or the Time-Dependent Variational Principle (TDVP) for Matrix Product States (MPS) in imaginary time. The choice of algorithm depends on the specific form of the Hamiltonian's interactions.
We refer the reader to Chapter~\ref{chap2} and Chapter~\ref{chap3} for a detailed explanation of the methods. Here, we will briefly review the TEBD approach for nearest-neighbor Hamiltonians.

For instance, for short-range interacting Hamiltonians with couplings only between neighboring sites, represented as $\hat{H} = \sum_i \hat{h}_{i,i+1}$, the imaginary time evolution for a small step $\Delta\beta$ can be approximated by
\begin{equation}
e^{-\Delta\beta \hat{H}} \simeq
e^{-\Delta\beta \hat{h}_{1,2}/2}
%e^{-\Delta\beta \hat{h}_{2,3}/2}
\cdots
e^{-\Delta\beta \hat{h}_{n-1,n}/2}
e^{-\Delta\beta \hat{h}_{n-1,n}/2}
%e^{-\Delta\beta \\hat{h}_{n-2,n-1}/2}
\cdots
e^{-\Delta\beta \hat{h}_{1,2}/2},
\end{equation}
with a Trotter error of order $\sim (\Delta\beta)^3$. This operator can be applied to an MPS by performing a left-to-right sweep followed by a right-to-left sweep, sequentially updating the local tensors and performing local Singular Value Decomposition (SVD). This process normalizes the state locally, effectively moving the central site of the mixed canonical form of the MPS. These sweeps are repeated until the desired inverse temperature $\beta/2$ is reached.

\paragraph{CPS sampling of an MPS ---}\index{MPS!projective sampling}
Here we assume $\ket{\phi}$ is in the right canonical form, i.e.\ with the orthogonality center at the first site. We have omitted the variable indicating the CPS state from which the METTS was generated, as it is irrelevant for the sampling procedure.

The goal is to sample a CPS state $\ket{x_1,\dots,x_n}$ with a probability
$P_{\phi}(x_1,\dots,x_n)$ $\equiv |\braket{x_1,\dots,x_n}{\phi}|^2$ which in general is an exponentially hard problem due to the Hilbert space size.

To overcome this difficulty, we rewrite the full probability in terms of conditional and marginal probabilities as
\begin{equation}\label{eq:chain_prob}
P_{\phi}(x_1,\dots,x_n)
= p_{\phi}(x_1)
p_{\phi}(x_2|x_1)
\cdots
p_{\phi}(x_n|x_1,\cdots, x_{n-1}),
\end{equation}
where
$
p_{\phi}(x_j|x_1,\cdots,x_{j-1}) =
\frac{p_{\phi}(x_1,\cdots,x_j)}
{p_{\phi}(x_1,\cdots, x_{j-1})}
$
is the probability that the state $\ket{x_j}$ occurs at position $j$ given that the classical configuration $\{x_1,\cdots,x_{j-1}\}$ has already been found at positions $\{1, \dots, j-1\}$, regardless of the outcomes in the rest of the system (i.e., marginalizing over all possible outcomes for the remaining qubits $\{j+1, \dots, n\}$).
We defined the marginal probability as
$
p_{\phi}(x_1,\dots,x_j)
=
\sum_{x_{j+1},\dots,x_n} |\braket{x_1,\dots,x_n}{\phi}|^2.
$.

The key advantage here is that the chain of conditional probabilities can be efficiently evaluated in an MPS state. To illustrate, let's start with the orthogonality center at the first site. The probability of the first outcome is given by:
\begin{equation}
\includegraphics[width=0.2\textwidth,valign=c]{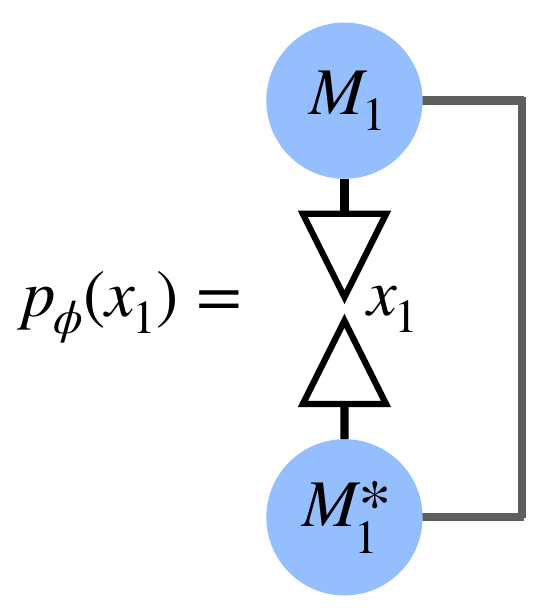}
\end{equation}
and we can sample $x_1$ based on this probability. Suppose the specific outcome $x_1$ is selected.\footnote{For a qubit system where $x_1 \in \{0,1\}$, we can uniformly draw $r \in [0,1]$. If $r \leq p_{\phi}(0)$, then $x_1 = 0$; otherwise, $x_1 = 1$.} The state is then updated by projecting the first qubit into this classical configuration, resulting in a new MPS state for the remaining qubits $\{2,\dots,n\}$:
\begin{equation}
\includegraphics[width=0.6\textwidth,valign=c]{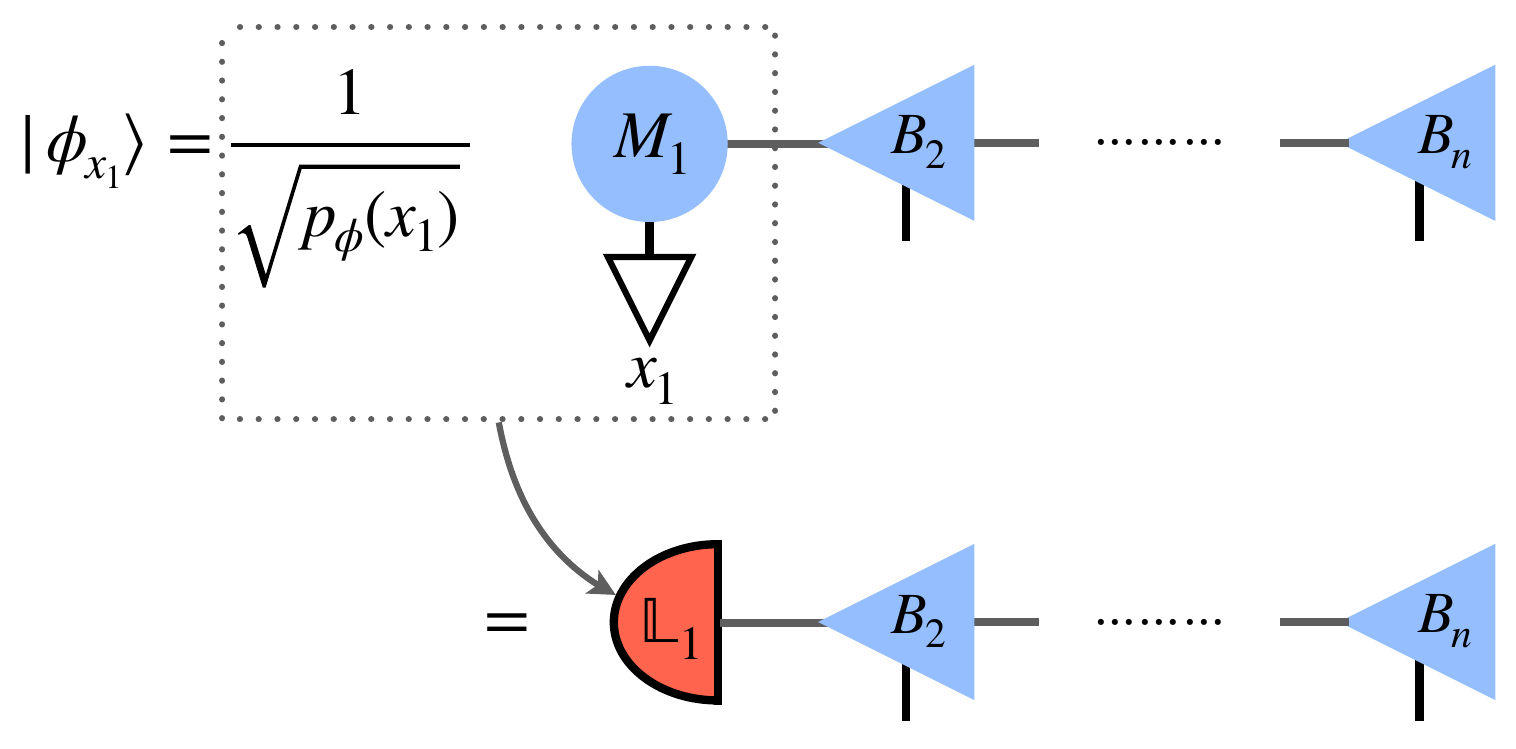}
\end{equation}
with an updated left-boundary vector $\mathbb{L}_j$ (where $j=1$ at the end of the first step).
Next, we compute the conditional probability for the second qubit given the first:
\begin{equation}
\includegraphics[width=0.45\textwidth,valign=c]{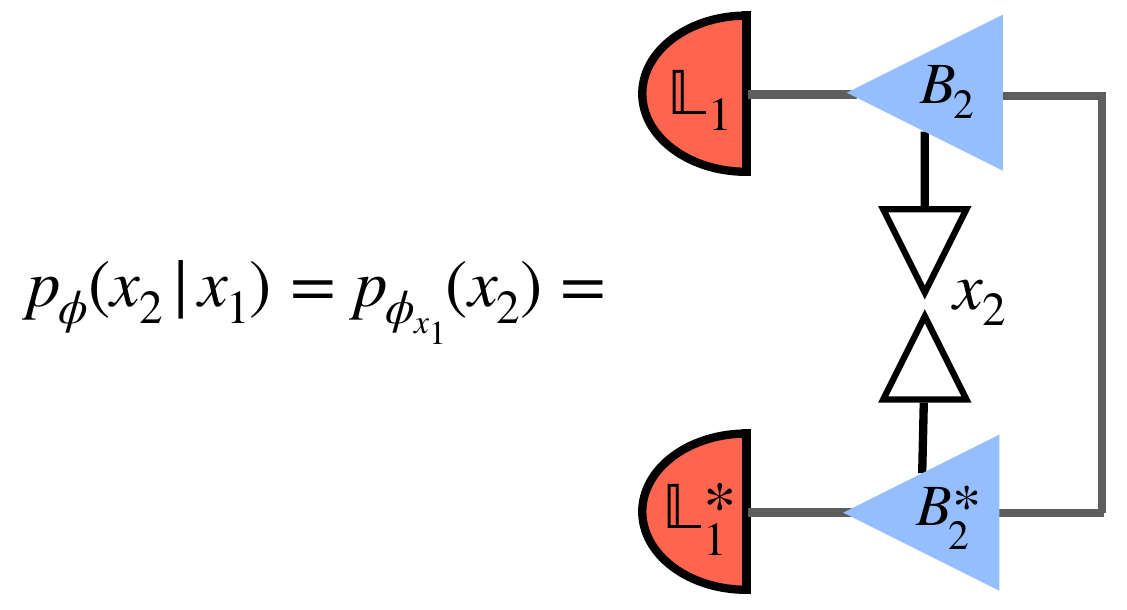}
\end{equation}
Using this probability, we sample $x_2$, following the same procedure as in the first step. Upon determining a classical configuration for the second qubit, we update the state by projecting the second qubit, yielding a new MPS for the remaining qubits $\{3,\dots,n\}$:
\begin{equation}
\includegraphics[width=0.7\textwidth,valign=c]{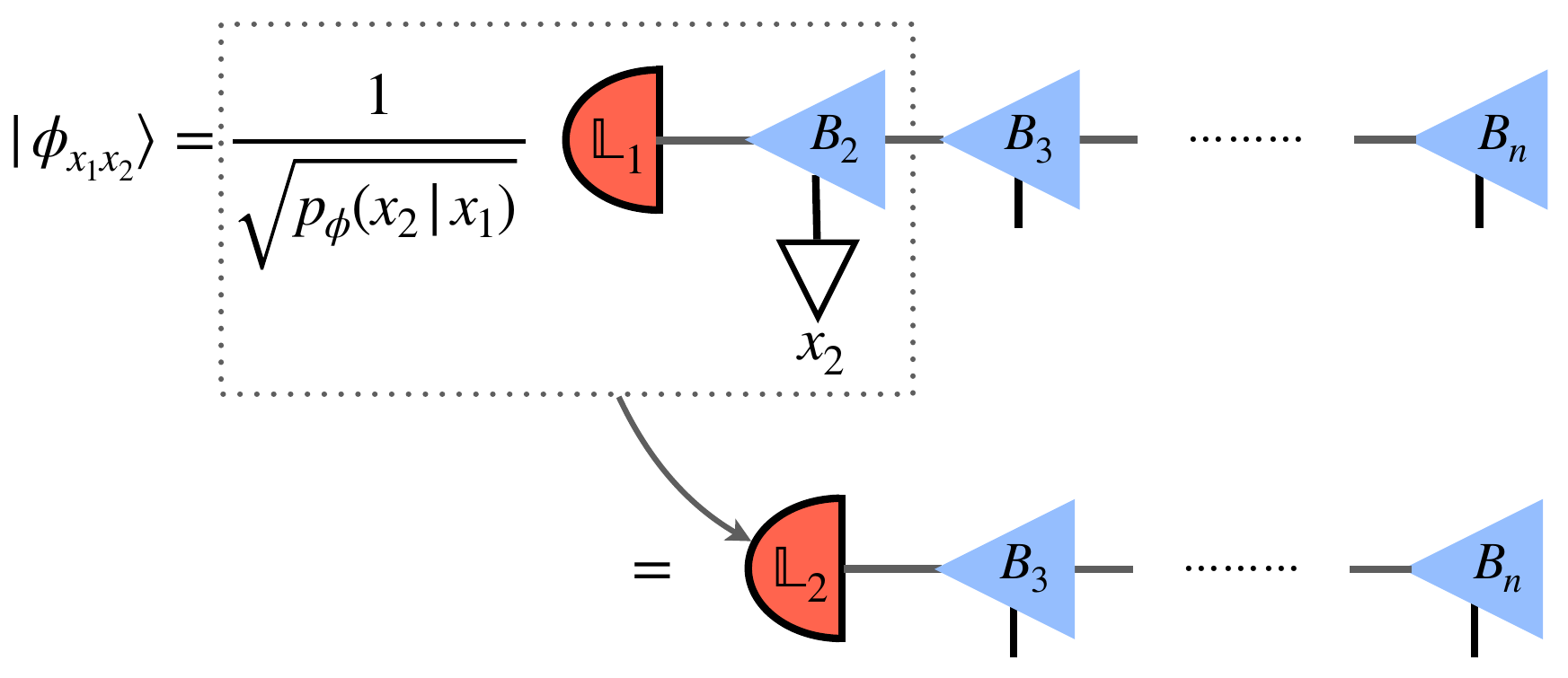}
\end{equation}

The outlined procedure can be reformulated as an \textbf{iterative algorithm} from the leftmost to the rightmost qubit of the MPS.
After initializing the boundary vector $\mathbb{L}_0 = (1)$, at each $j$-th step:
\begin{enumerate}
    \item Compute the conditional probability $p_{\phi}(x_j|x_1,\dots,x_{j-1}) = p_{\phi_{x_1\dots x_{j-1}}}(x_j)$.
    \item Generate a classical local configuration accordingly.
    \item Update the boundary vector: $\mathbb{L}_{j} = p_{\phi}(x_j|x_1,\dots,x_{j-1})^{-1/2}\mathbb{L}_{j-1}\cdot B^{x_j}$.
    \item Repeat steps (1) to (3) until the rightmost qubit is sampled.
\end{enumerate}

At the end of the algorithm, we obtain a CPS state $\ket{x_1,x_2,\dots,x_n}$ with probability $|\braket{x_1,\dots,x_n}{\phi}|^2$.

Note that the outlined algorithm requires only matrix-to-vector multiplications along the auxiliary space of the MPS, resulting in an overall computational complexity scaling as $\sim \chi^2$, where $\chi$ is the maximum bond dimension of the MPS.

\section{Lindblad dynamics}
In this section, we aim to explore the most general class of Markovian transformations that can be applied to density matrices. Up to this point, we have considered two primary modes of quantum evolution: coherent evolution and state collapse following measurement, as described by postulates of quantum mechanics (see Chapter~\ref{chap2}). Significant effort has been devoted to reconciling these two forms of quantum evolution, but a definitive unification remains elusive~\cite{Manzano2020}.

The central question we address is: What is the most general transformation that can be performed on a quantum system, and how can we describe this transformation using a dynamical equation?

To answer this, we focus on maps that transform density matrices into density matrices. We denote by $\varrho_\mathcal{H}$ the space of all density matrices associated to a Hilbert space $\mathcal H$.
We seek a map $\mathcal V: \varrho_\mathcal{H} \to \varrho_\mathcal{H}$ that acts on this space and maps it onto itself. For the map $\mathcal V$ to ensure that its output is still a valid density matrix, it must satisfy the following criteria:

\begin{itemize}
    \item \textbf{Trace Preservation:} The map must preserve the trace, meaning for any operator $\hat A$ in the space of linear operators acting on the Hilbert space $\mathcal{H}$, the trace of the transformed operator should equal the trace of the original operator, i.e.\ $\Tr[\mathcal{V}(\hat A)] = \Tr[\hat A]$.

    \item \textbf{Complete Positivity:} The map must be completely positive. This condition is more nuanced and requires further discussion, as it ensures that the map remains positive even when extended to a larger Hilbert space.
\end{itemize}

Maps that satisfy both of these properties are referred to as \emph{completely positive and trace-preserving maps} (CPT maps).\index{CPT maps} While trace preservation is straightforward and self-evident, complete positivity involves a more detailed definition.

\begin{definition}{Positive Map}{Pmap}
A \emph{positive map} $\mathcal{V}$ between two spaces of operators is a linear map that preserves the positivity of operators. Formally, let $\mathcal{B}(\mathcal{H})$ denote the space of bounded linear operators on a Hilbert space $\mathcal{H}$. A map $\mathcal{V}: \mathcal{B}(\mathcal{H}) \to \mathcal{B}(\mathcal{K})$, where $\mathcal{H}$ and $\mathcal{K}$ are Hilbert spaces, is called positive if:
\begin{equation}
\mathcal{V}(\hat A) \geq 0 \text{ whenever } \hat A \geq 0.
\end{equation}

In other words, $\mathcal{V}$ is positive if it maps positive semidefinite operators to positive semidefinite operators. Here, $\hat A \geq 0$ denotes that $\hat A$ is a positive semidefinite operator, meaning that $\langle \psi | \hat A | \psi \rangle \geq 0$ for all $|\psi\rangle \in \mathcal{H}$.
\end{definition}

\begin{definition}{Completely Positive Map}{CPmap}
A map $\mathcal{V}$ between spaces of operators is called \emph{completely positive} if it is positive and, additionally, remains positive when extended to larger spaces. Formally, let $\mathcal{B}(\mathcal{H})$ and $\mathcal{B}(\mathcal{K})$ denote the spaces of bounded linear operators on Hilbert spaces $\mathcal{H}$ and $\mathcal{K}$, respectively. A map $\mathcal{V}: \mathcal{B}(\mathcal{H}) \to \mathcal{B}(\mathcal{K})$ is completely positive if for every integer $n$ and for every matrix $\{\hat A_{ij}\}_{i,j=1}^n$ of operators $\hat A_{ij} \in \mathcal{B}(\mathcal{H})$, the following condition holds:
\begin{equation}
\mathcal{V} \otimes \mathbb{I}_{\mathcal{B}(\mathbb{C}^n)} \left( \sum_{i,j=1}^n \hat A_{ij} \otimes |i\rangle\langle j| \right) \geq 0,
\end{equation}
where $\mathbb{I}_{\mathcal{B}(\mathbb{C}^n)}$ is the identity map on $\mathcal{B}(\mathbb{C}^n)$, and $|i\rangle\langle j|$ are the standard matrix units in $\mathbb{C}^n$.
In simpler terms, $\mathcal{V}$ is completely positive if it remains positive when extended to act on a larger Hilbert space by tensoring with the identity on a finite-dimensional space.
\end{definition}

\paragraph{Lindblad equation ---}
A widely used approach for deriving the Lindblad master equation stems from the study of open quantum systems. In this context, the Lindblad equation describes the dynamics of a subsystem that is part of a larger, more complex system. For a detailed exposition of this derivation, one can refer to standard texts such as Breuer and Petruccione's \emph{The Theory of Open Quantum Systems}~\cite{Breuer2002} and Gardiner and Zoller's \emph{Quantum Noise}~\cite{Gardiner2000}.

The scope of this book is not to give a detailed derivation of such equation whilst explain how to integrate that equation via TN methods.

Let's just mention that a Lindblad master equation provides a powerful framework for describing the dynamics of the system's density matrix $\hat\rho$. This equation accounts for both the unitary evolution governed by the system's Hamiltonian and the effects of dissipation and decoherence due to interaction with the environment.
The Lindblad master equation can be expressed as:\index{Lindblad}
\begin{equation}\label{chapt5_eq:lindblad}
\partial_t \hat\rho(t) = -i [\hat H, \hat\rho(t)] + \sum_i \gamma_i \left( \hat L_i \hat\rho(t) \hat L_i^\dagger - \frac{1}{2} \{\hat L_i^\dagger \hat L_i, \hat \rho(t) \} \right),
\end{equation}
where:
\begin{itemize}
  \item $\hat\rho(t)$ is the density matrix of the system at time $t$.
  \item $\hat H$ is the Hamiltonian, representing the coherent evolution of the system.
  \item $\hat L_i$ are the Lindblad operators (known also as \textbf{jump operators}),\index{Jump operator} which model the interaction of the system with its environment and the resulting dissipative effects.
  \item $\gamma_i \geq 0$ are the damping rates, which quantify the strength of the dissipative processes.
\end{itemize}
If all the damping rates $\gamma_i$ are zero, the Lindblad equation reduces to the von Neumann equation~(\ref{eq:von_neumann}). This approach simplifies the description of the dynamics, making it easier to analyze and simulate the behavior of open quantum systems.

\subsection{MPO formalism}
When simulating the Lindblad dynamics of a system of qubits using Matrix Product Operator (MPO) techniques, two distinct approaches can be employed, each leveraging the unique structure of the quantum system's density matrix.

The first approach parallels the method used for imaginary time evolution in preparing thermal states. In this method, the density matrix $\hat\rho$ is treated as a ``superket'' $\dket{\rho}$, effectively doubling the Hilbert space by associating an ancillary qubit with each physical qubit. The resulting state is then approximated as a Matrix Product State (MPS), where each local tensor encompasses the physical and ancillary qubits in a tensor product space. This approach is intuitive as it extends the well-established MPS formalism to mixed states by simply reinterpreting the density matrix as a state vector in an enlarged Hilbert space.

The second approach retains the operatorial nature of the density matrix. Instead of converting $\hat\rho$ into a state vector, it is expressed directly in the Pauli matrix basis
as represented in Eq.~(\ref{chapt5_eq:rho_pauli}). This technique leverages the natural decomposition of quantum operators in terms of the Pauli matrices, allowing the density matrix to be represented as a linear combination of Pauli operators. The advantage of this approach lies in its direct connection to the operator algebra and the potential for a more compact and efficient representation, especially when dealing with qubits.

However, similarly to what happens in the computational basis representation, in this case as well, when treating the wave function associated with the density matrix as an MPS, the algorithm inherently preserves the norm square of the wave function. This preservation corresponds to maintaining $\Tr(\hat{\rho}^2)$. To properly compute expectation values, we need to compute first the partition function\index{Partition function!Pauli basis}
\begin{equation}
\includegraphics[width=0.6\textwidth,valign=c]{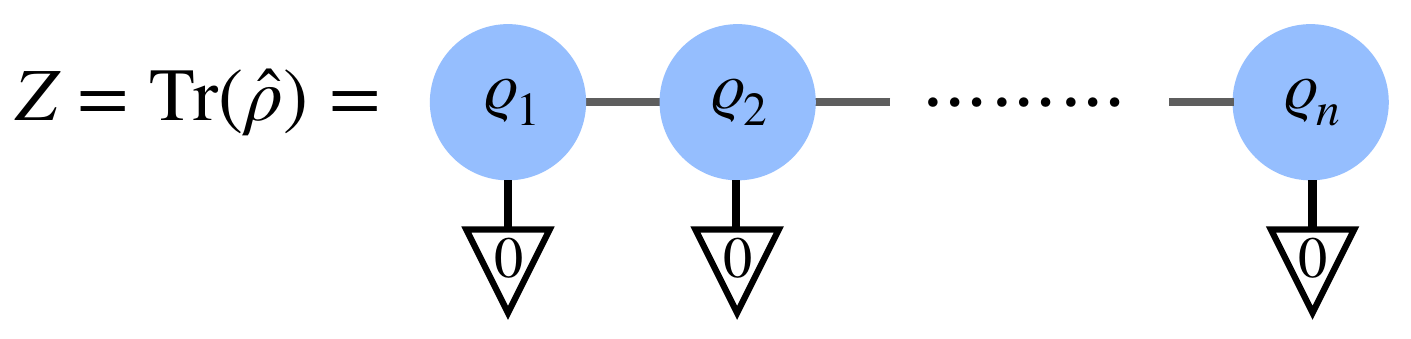}
\end{equation}
which allows us to restore the correct normalization of the density matrix. In this framework, due to the anomalous normalization of the density operator, the purity of a state, which quantifies how mixed the state is, will be given by
$\text{Purity} = Z^{-2}$.
This expression shows that the purity is inversely proportional to the square of the partition function, and reflects the impact of the normalization on assessing how close the state is to being pure.

In the following, we will provide a detailed exposition of this second approach, elucidating how the Pauli basis expansion facilitates the integration of Lindblad dynamics in a computationally efficient manner using the MPO formalism. We will also compare and contrast this with the superket method, highlighting the contexts in which each approach is most effective.

First and foremost, the integration of the Lindblad equation involves exponentiating the right-hand side of Eq.~(\ref{chapt5_eq:lindblad}). To accomplish this, and assuming that we are evolving a density matrix over a small time step $dt$, we apply a Trotter decomposition to the evolution super-operator. This decomposition separates the evolution into two distinct parts: the unitary part and the dissipative part, as follows:\index{Trotter decomposition}
$
\mathcal{U}[\cdot] = \mathcal{U}_{u}[\cdot]\mathcal{U}_{d}[\cdot]
$
where the unitary part acts as usual
\begin{equation}
\mathcal{U}_{u}[\hat\rho] =  e^{-i\hat H dt}\hat\rho e^{i\hat H dt}
\end{equation}
while the dissipative part can be formally written
as
\begin{equation}
\mathcal{U}_{d}[\hat\rho] = e^{\mathcal{D} [\hat\rho] dt}
\end{equation}
with
\begin{equation}\label{chapt5_eq:dissipator}
\mathcal{D} [\hat\rho]
= \sum_i \mathcal{D}_i [\hat\rho]
\equiv
\sum_i \gamma_i \left( \hat L_i \hat\rho(t) \hat L_i^\dagger - \frac{1}{2} \{\hat L_i^\dagger \hat L_i, \hat \rho(t) \} \right),
\end{equation}
where the sum runs over the local qubits.

\paragraph{Unitary part ---}
The unitary part of the evolution, for short-range interacting Hamiltonians, can be integrated using the TEBD scheme, but now in the Pauli basis representation. Specifically, given a Hamiltonian of the form $\hat{H} = \sum_{i} \hat{h}_{i,i+1}$, the elementary brick-wall evolution operator is
$\hat{u} \equiv e^{-i \hat{h}_{i,i+1} \, dt},$
which acts on the density matrix as $\hat{u} \hat{\rho} \hat{u}^{\dag}$. This induces a transformation of the local tensors in the Pauli-basis MPO formalism\index{Lindblad!MPO formalism} of the density matrix as follows:
\begin{equation}
\includegraphics[width=0.8\textwidth,valign=c]{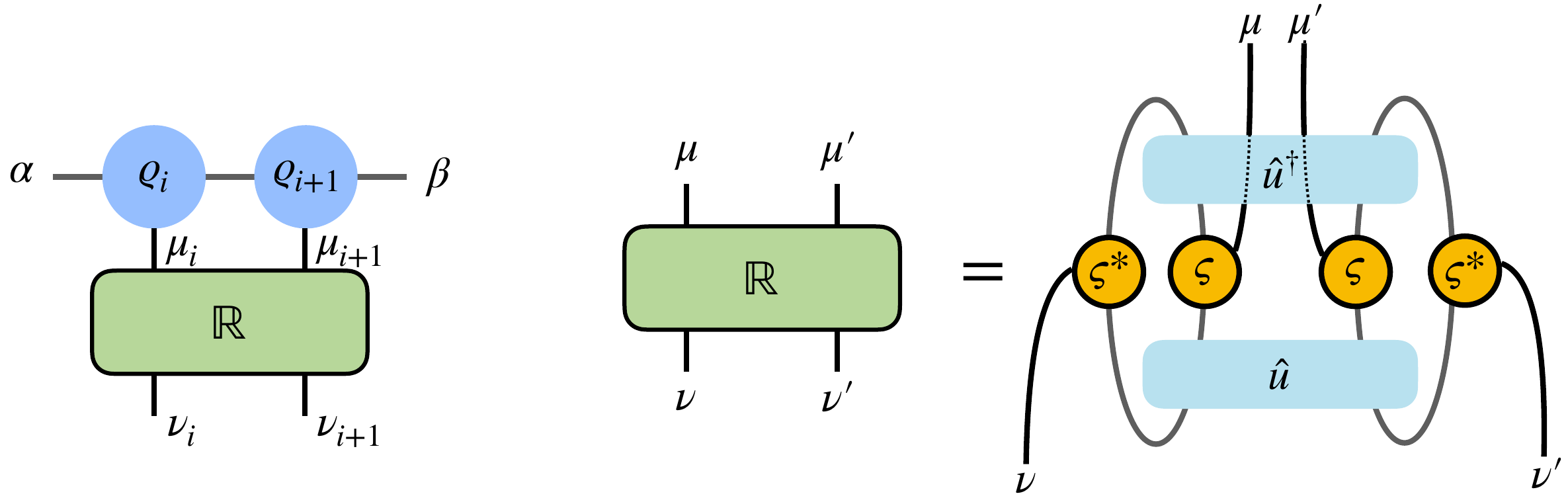}
\end{equation}
where we basically defined the Pauli-basis unitary tensor
\begin{equation}
\mathbb{R}^{\mu\mu'}_{\nu\nu'} =
\Tr[\hat\zeta^{\nu}\otimes\hat\zeta^{\nu'}
\hat u
\hat\zeta^{\mu}\otimes\hat\zeta^{\mu'}
\hat u^{\dag}
].
\end{equation}
As is customary in the TEBD algorithm, it is advantageous to apply the local tensor $\mathbb{R}$ when the MPO representation of the density matrix is in mixed canonical form with respect to one of the two sites involved in the local update. Once in this form, the standard TEBD procedure is followed by performing a truncated singular value decomposition (SVD). This SVD step results in a compressed pair of new local tensors, thereby maintaining the efficiency and accuracy of the MPO representation.

\paragraph{Dissipative part ---}
Frequently, the dissipative component of the Lindblad equation is comprised of a series of non-unitary operations applied to individual lattice sites. When this is the case, as in Eq.~(\ref{chapt5_eq:dissipator}), we can apply each local exponential superoperator $e^{\mathcal{D}_i [\cdot] dt}$ sequentially on the local tensor $\varrho_i$ of the Pauli-basis MPS representation of the density matrix.\index{Lindblad!MPO dissipator} This process can be effectively implemented by considering the superket reshaping of the Pauli basis, allowing for the computation of the resulting tensor.
\begin{equation}
\includegraphics[width=0.55\textwidth,valign=c]{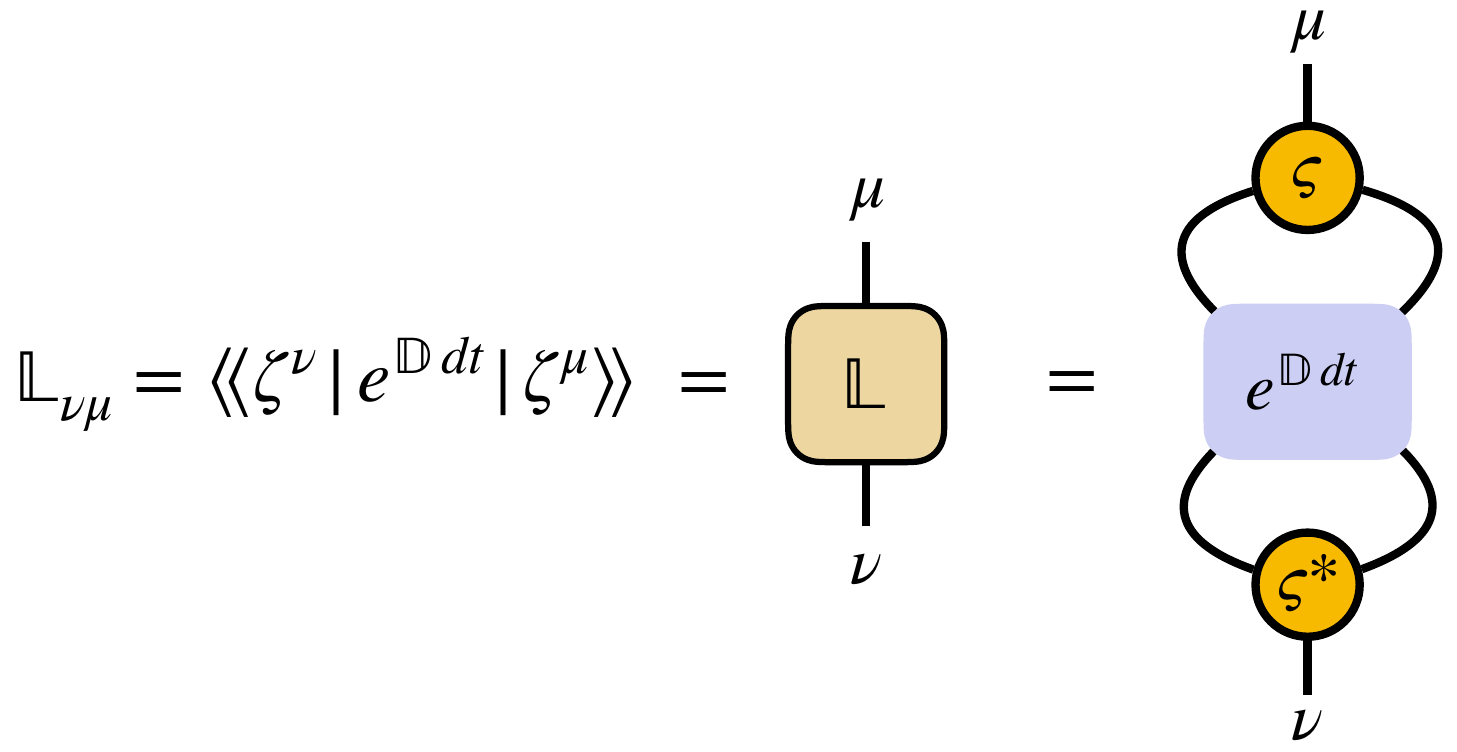}
\end{equation}
where $\mathbb{D} \equiv \gamma_i \left[ \hat L_i \otimes \hat L_i^* - \frac{1}{2} (\hat L_i^\dagger \hat L_i \otimes \hat I + \hat I \otimes \hat L_i^t \hat L_i^* ) \right]$, and we have omitted the subscript $i$ indicating the lattice site. Subsequently, we can update each local MPS tensor by contracting its physical index with this new tensor via the operation $\sum_{\mu}\mathbb{L}_{\nu\mu}\varrho^{\mu}_{\alpha\beta}$, without altering the auxiliary bond dimension.

\begin{example}{One-Body particle loss and gain in hard-core bosons}{HCboson_loss_gain}
As an example here we consider hard-core Boson particle jumping in a 1D lattice. These are particles that basically behave as bosons, a part from the fact that each lattice site cannot be occupied by more than one single bosons. As a matter of fact, the unitary dynamics is governed by an Hamiltonian which can be described in terms of a fictitious spin-$1/2$
\begin{equation}\label{chapt5_eq:H_hcboson}
\hat H = -\sum_{j} \left( \hat\sigma_{j}^{+}  \hat\sigma_{j+1}^{-}
+\hat\sigma_{j}^{-}  \hat\sigma_{j+1}^{+} \right)\,,
\end{equation}
where the local boson density is related to a fictitious $\hat z$-magnetisation via $\hat n_{j} = (1-\hat\sigma_{j}^{z})/2$.

The simple unitary part can be mapped to a non-interacting fermionic theory and the dynamics can be analytically solved. However, in a real scenario, each lattice sites may be imperfect, and incoherent fluctuations in the number of particles can occur due to a one-body loss and gain,
resulting in a Markovian evolution governed by the following Lindblad equation for the density matrix
\begin{equation}
\begin{split}
    \partial_t\Hat\rho=-i[\Hat H,\Hat\rho]&+\gamma_L\sum_{j}\left(\Hat L_{j}\Hat\rho \Hat L^\dagger_{j}-\dfrac{1}{2}\{\Hat L_{j}^\dagger \Hat L_{j},\Hat\rho\}\right)\\
    &+\gamma_G\sum_{j}\left(\Hat L^\dagger_{j}\Hat\rho \Hat L_{j}-\dfrac{1}{2}\{\Hat L_{j} \Hat L_{j}^\dagger,\Hat\rho\}\right),
\end{split}
\end{equation}
where $\gamma_L,\gamma_G > 0$ set the loss and gain rates,
and $\hat L_j = \hat\sigma^+_j$ removes a particles
from the site $j$.

Notice that in the previous equation one has two incoherent terms acting independently in each lattice site. Nonetheless, we can collect both single-site loss and gain and compute the total local exponential $\exp{\mathbb{D}\, dt}$ as in the main text
\begin{equation}
\mathbb{L}_{\nu\mu} =
\begin{pmatrix}
    1 & 0 & 0 & 0 \\
    0 & e^{-\gamma dt/2} & 0 & 0 \\
    0 & 0 & e^{-\gamma dt/2} & 0 \\
    (1-2n_s)(1-e^{-\gamma dt}) & 0 & 0 & e^{-\gamma dt}
\end{pmatrix},
\end{equation}
where we defined the total decay rate $\gamma = \gamma_L + \gamma_G$ and the stationary density in the steady-state
$n_s = \gamma_G/(\gamma_L+\gamma_G)$.
In fact, since the Lindbladian is uniform, one may expect $\hat\rho$ to relax to a homogeneous state. Indeed, it is easy to show that the stationary state is an infinite temperature state in the sector with a fixed number of particles
\begin{equation*}
    \hat\rho_{\infty} = \frac{1}{Z} e^{\mu \sum_j \hat n_j} = \prod_{j} \frac{e^{\mu \hat n_j}}{1+e^{\mu}}\,\,,\,\,\,\mu=\log\frac{\gamma_G}{\gamma_L}\,,
\end{equation*}
where the chemical potential is determined by the condition that the dissipator annihilates $\hat\rho_\infty$, as $\hat{H}$ trivially commutes with it because of the particle number conservation.

For more details on such setup, and a thorough comparison of the numerical TN results with a theoretical time-dependent Gaussian approximation of the density matrix we refer the reader to ref.~[CITE].

\end{example}

Collecting both the dissipative and the unitary part, ad using a time step of $dt/2$, we can easily implement the following second-order integration scheme whose graphycal representation looks like (evolution is from left to right)
$$
\includegraphics[width=\textwidth,valign=c]{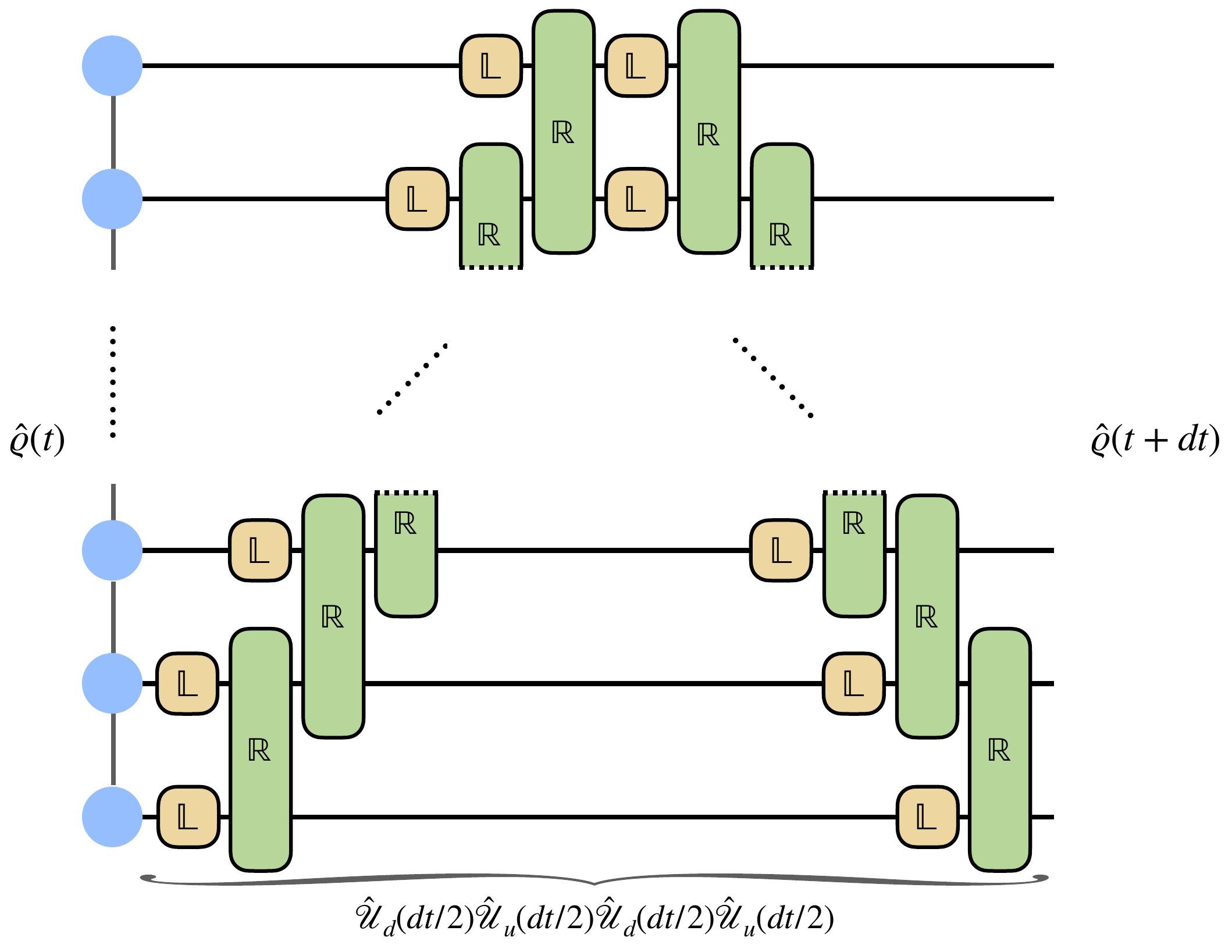}
$$

\subsection{Locally Purified Tensor Network}\label{chapt5_ss:LPTN}
As already mentioned at the beginning of this chapter, one of the condition which is sacrificed by both the ``superket'' and the Pauli-MPS approach outlined in the previous section, is the requirement that a well defined density operator needs to be \emph{non-negative}. And in fact, a more restrictive ansatz can be introduced which is non-negative by construction: the purification ansatz, know also as \textbf{Matrix Product Density Operator (MPDO)} ansatz~\cite{positiveTN2004,positiveTN2013, positiveTN2016}.\index{Density matrix!MPDO}

As a matter of fact, in quantum mechanics, it is well-established that for any mixed state $\hat\rho$, there exists a corresponding purification. This means that one can always find a pure state $\ket{\Phi_{SA}}$ in a larger Hilbert space, involving an additional \emph{ancillary} system, such that tracing out the ancillary degrees of freedom yields the original mixed state of the system. Mathematically, if the enlarged system is described by the pure state $\ket{\Phi_{SA}}$ in the combined Hilbert space $\mathcal{H}_S \otimes \mathcal{H}_A$, then
$
\hat\rho = \Tr_A \left(\ket{\Phi_{SA}}\bra{\Phi_{SA}}\right)
$
returns the original mixed state $\hat\rho$ on the system's Hilbert space $\mathcal{H}_S$.

The concept of purification is particularly powerful in the context of tensor networks, where it underpins the construction of Matrix Product Density Operators.\index{Purification} The purification ansatz leverages the idea of representing the purified state as a Matrix Product State on an extended system. Specifically, for every physical site in the system, an ancillary site of the same dimension is introduced, effectively doubling the physical dimension of each site.

This means that the Hilbert space of the combined system, originally of dimension $d$ per site (two for qubits), is extended to $d \times d$ per site (four for qubits) in the purified MPS. The enlarged site thus accommodates both the physical and ancillary degrees of freedom, leading to an MPS representation that efficiently encodes the purification
\begin{equation}
%\ket{\Phi_{SA}} =
%\sum_{x_1, \dots, x_n} \sum_{y_1,  \dots, y_n}
%A^{x_1 y_1}\cdots A^{x_n y_n}
%\ket{\boldsymbol{x}} \otimes \ket{\boldsymbol{y}}.
\includegraphics[width=0.75\textwidth,valign=c]{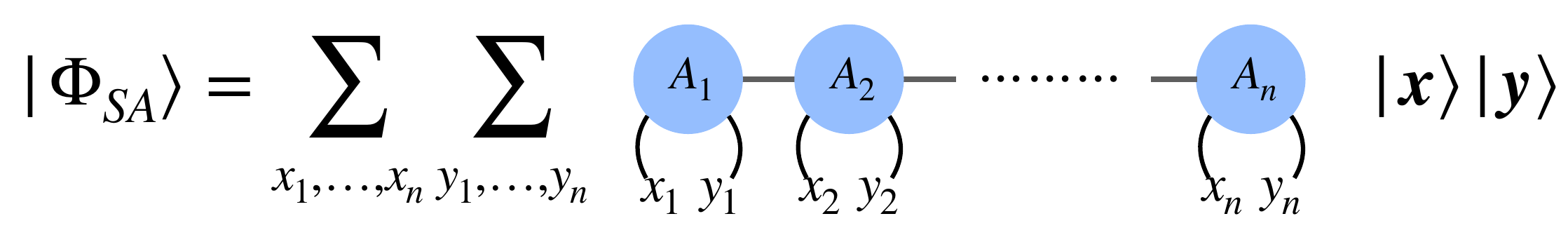}
\end{equation}

Here, the MPS describes the local purified state with indices $\boldsymbol{x}$ corresponding to the system's physical sites and indices $\boldsymbol{y}$ corresponding to the ancillary sites. The crucial point is that by tracing out the ancillary indices, one recovers the original MPDO representation of the mixed state, i.e.\ \begin{equation}
\includegraphics[width=0.9\textwidth,valign=c]{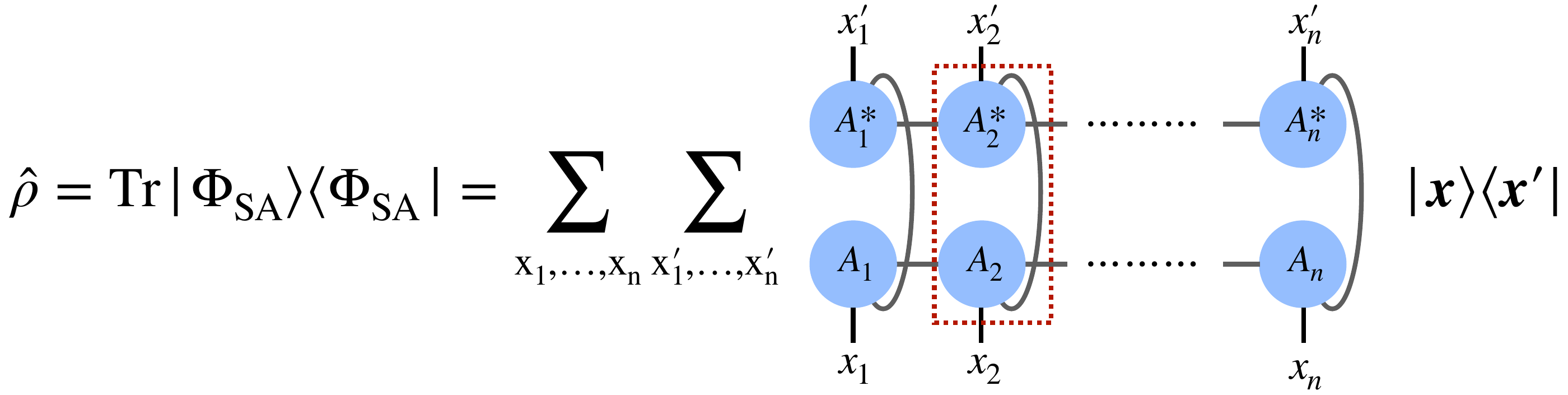}
\end{equation}

Since it is constructed from the partial trace of a normalized state, the resulting MPDO is a valid physical state. But it is not the most general MPO, because the tensors describing the density matrix MPO now have a very
peculiar structure,
$\rho_{\tilde\alpha\tilde\beta}^{xx'} = \sum_{y} A_{\alpha\beta}^{xy} (A_{\alpha'\beta'}^{x'y})^*$,
with the auxiliary indices $\tilde\alpha = (\alpha,\alpha')$ and $\tilde\beta = (\beta,\beta')$
originated by the fusion of the two auxiliary spaces. Notice that, in this representation, the normalisation of the purified state imply the usual normalisation of the density matrix $\langle\Phi_{SA}|\Phi_{SA}\rangle = \Tr(\hat\rho) = 1$.

In practice, this positive ansatz can be utilized effectively in numerical simulations, as will be demonstrated in the following scenarios. Before proceeding, it is important to observe that, much like the ``superket'' formulation of a density operator, the purification ansatz simplifies the computation of expectation values of local operators (or MPOs in general). Specifically, given that the density matrix can be expressed as
$
\hat{\rho} = \mathrm{Tr}_A \left(\ket{\Phi_{SA}}\bra{\Phi_{SA}}\right),
$
the expectation value of an operator $\hat{O}$ with respect to the density matrix reduces to the expectation value in the purified state:
$
\mathrm{Tr}\left(\hat{O} \hat{\rho}\right)
=
\bra{\Phi_{SA}} \hat{O}_S \otimes \hat{I}_A \ket{\Phi_{SA}},
$
where the operator $\hat{O}_S$ acts non-trivially only on the system indices, while $\hat{I}_A$ acts as the identity on the ancillary indices. Graphically one has:
$$
\includegraphics[width=0.5\textwidth,valign=c]{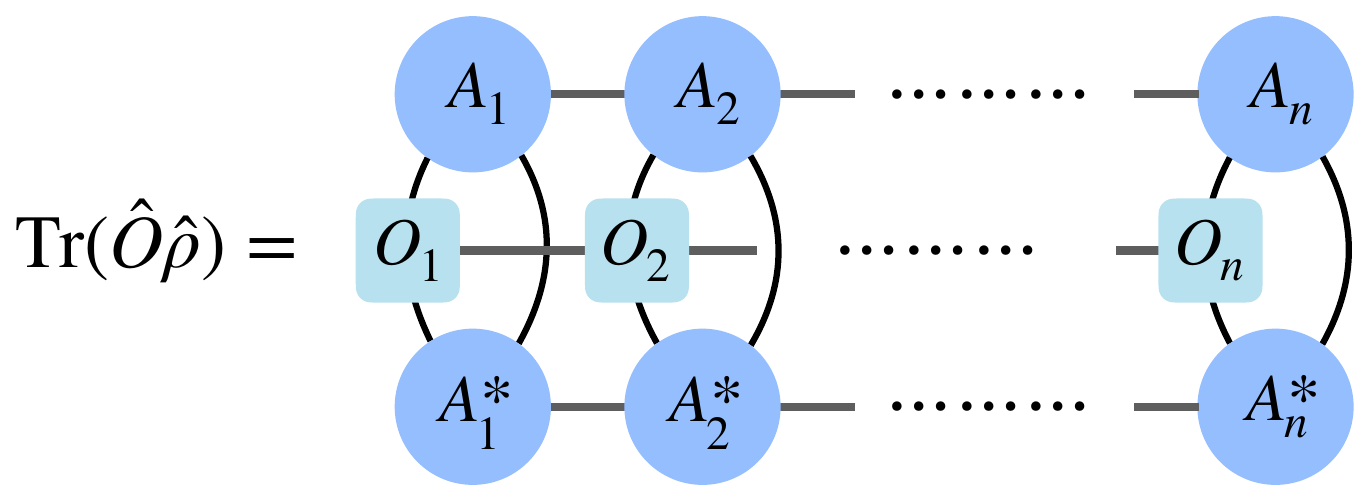}
$$

\subsubsection{Thermal state with positive TN}
The purification ansatz is very natural for thermal states. Indeed, a purification of a Gibbs ensemble is intimately related to the METS we have seen in Section~\ref{chapt5_s:METS}. In this scenario, the purified state reads
\begin{equation}
|\Phi(\beta)\rangle = \frac{1}{\Tr(e^{-\beta\hat H})}
\sum_{k} e^{-\beta E_k /2} \ket{E_k}_S\ket{E_k}_A
\end{equation}
where the sum run over all the energy eigenstates,
$\hat H \ket{E_k} = E_k \ket{E_k}$. Notice that such state can be rewritten as (a part from the normalisation)
\begin{equation}
|\Phi(\beta)\rangle \propto
(e^{-\beta \hat H/2}\otimes \hat I) | I\rangle_{SA}
\end{equation}
where basically the vector $\ket{ I}_{SA}$ is a reshaping of the identity operator (as we already seen in  the superket notation) and represents a maximally entangled state between system and ancilla.
Notice that, we have decided to operate with the Hamiltonian on the system only, however, due to the equivalence of system and ancilla, we may have acted on the ancilla as well by splitting the imaginary evolution
as $e^{-\beta \hat H/4}\otimes e^{-\beta \hat H^{t}/4}$.
The procedure we have outlined above can be graphically
represented as:\index{Thermal state!positive TN}
$$
\includegraphics[width=0.7\textwidth,valign=c]{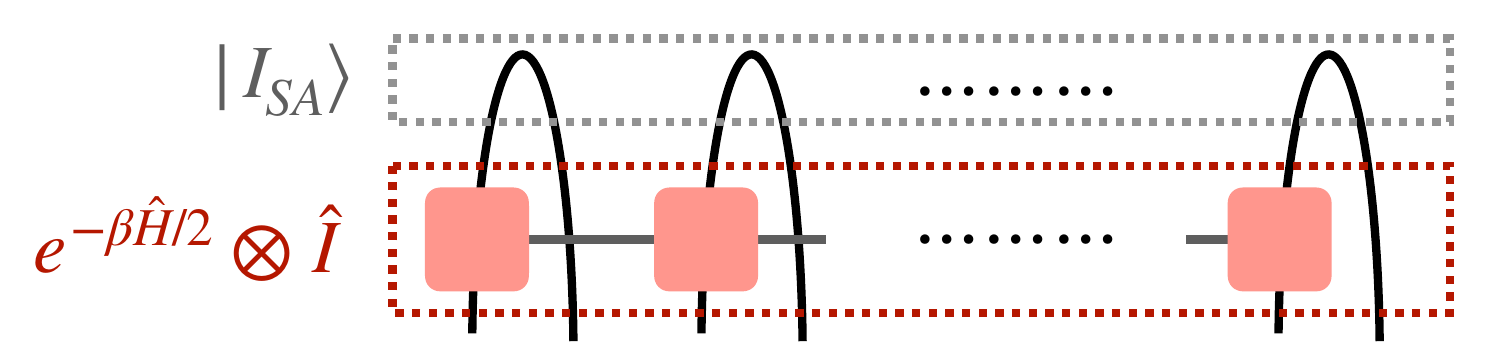}
$$
It is important to note that, within this Tensor Network (TN) representation, we assume the ability to express the operator $e^{-\beta \hat{H}/2}$ in Matrix Product Operator (MPO) form. However, as we are well aware, for both local and non-local Hamiltonians, it is often more practical to obtain the action of the exponentiated Hamiltonian $e^{-\beta \hat{H}/2}$ through imaginary-time evolution using techniques such as the Time-Evolving Block Decimation (TEBD) or the Time-Dependent Variational Principle (TDVP) schemes.

\subsubsection{Open system dynamics with positive TN}
When dealing with Lindblad dynamics\index{Lindblad!positive TN}, the positive TN approach can have both advantages and disadvantages.
We refer the reader to Eq.~(\ref{chapt5_eq:lindblad}),
where the dissipative part of the lindblad generator is assumed to act locally on each lattice site (or qubit).

When integrating the \textbf{unitary part of the dynamics}, the purified state can be evolved from time $t$ to $t+dt$ using the TEBD scheme, particularly for local Hamiltonians where $\hat{H} = \sum_{j} \hat{h}_{j,j+1}$. The update of two neighboring tensors is depicted as follows:

\begin{center}
\includegraphics[width=0.3\textwidth,valign=c]{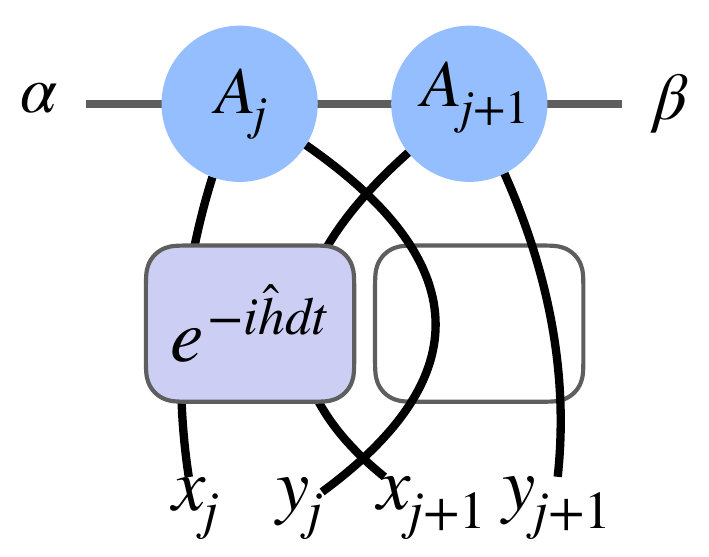}
\end{center}

In this representation, the identity operator acts trivially on the ancillary states. As is typical in the TEBD algorithm, this operation is followed by a singular value decomposition (SVD) to reshape the updated tensors into their original MPS form.

It's important to note that the unitary part of Lindblad dynamics is often the primary contributor to entanglement spreading across the system, leading to an unbounded growth in the auxiliary bond dimension. To mitigate this issue, one can utilize the fact that the ancillary system can be evolved unitarily according to any operator. For instance, Karrasch et al.\ in ref.~\cite{Karrasch_2012} demonstrated that the entanglement growth inherent to any time-dependent calculation can be significantly reduced if the ancillary degrees of freedom, which purify the statistical operator, are time-evolved with the physical Hamiltonian but with time reversed. This approach is locally translated into the following TN contraction:

\begin{center}
\includegraphics[width=0.3\textwidth,valign=c]{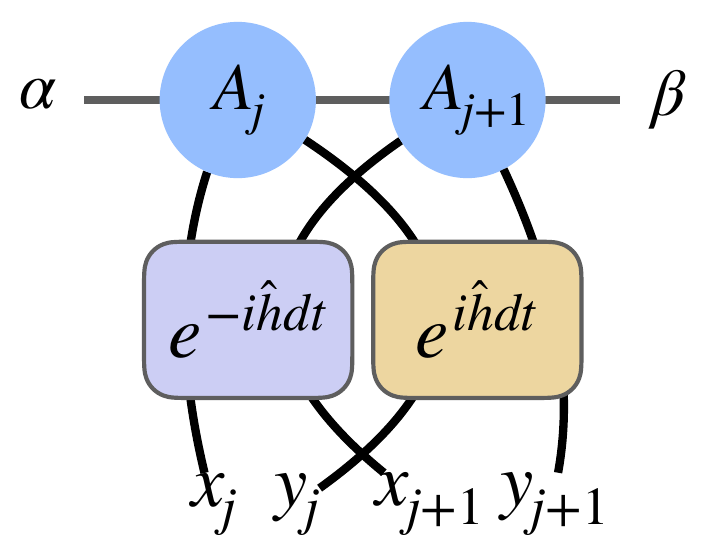}
\end{center}

However, this method requires the ancillary system to be a copy of the physical one (thus having the same local physical dimensions), which may not be applicable once a local dissipator comes into play.

An even more effective approach, independent of the dimensions of the ancillary system, involves exploiting the fact that the physical density matrix is unaffected by \emph{any} unitary operation performed on the ancillary system. In graphical notation, we can apply the local time-step operator $e^{-i \hat{h} dt} \otimes \hat{u}_{A}$ as follows:

\begin{center}
\includegraphics[width=0.3\textwidth,valign=c]{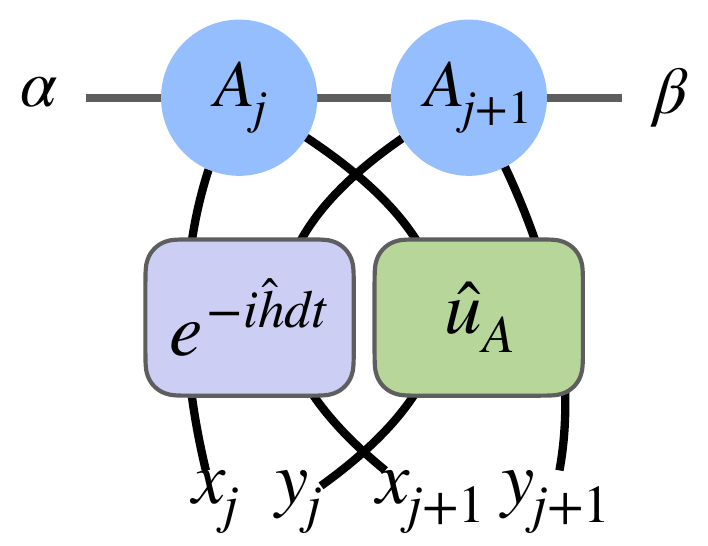}
\end{center}

Here, we seek the optimal unitary $\hat{u}_{A}$ that minimizes the bipartite entanglement entropy, effectively acting as the best local disentangler for the purified state $|\Phi_{SA}\rangle$.

After implementing the coherent part of the evolution, the next step is to apply the \textbf{dissipative part}, which is typically assumed to be local in space. This means that each dissipative operator acts independently on each lattice site. The dissipative operator, denoted as $e^{\mathbb{D}\,dt}$, is \textbf{completely positive}, allowing it to be decomposed into a set of Kraus operators $\{\hat K_q\}$ such that
$
e^{\mathbb{D}\,dt} = \sum_{q=1}^{k} \hat K_q \otimes  \hat K_q^*.
$
The action of this dissipative operator on the density matrix is realized through the contraction of the Kraus operators with the MPS tensors that describe the purified state. This process is represented graphically as follows:
\begin{center}
\includegraphics[width=0.8\textwidth,valign=c]{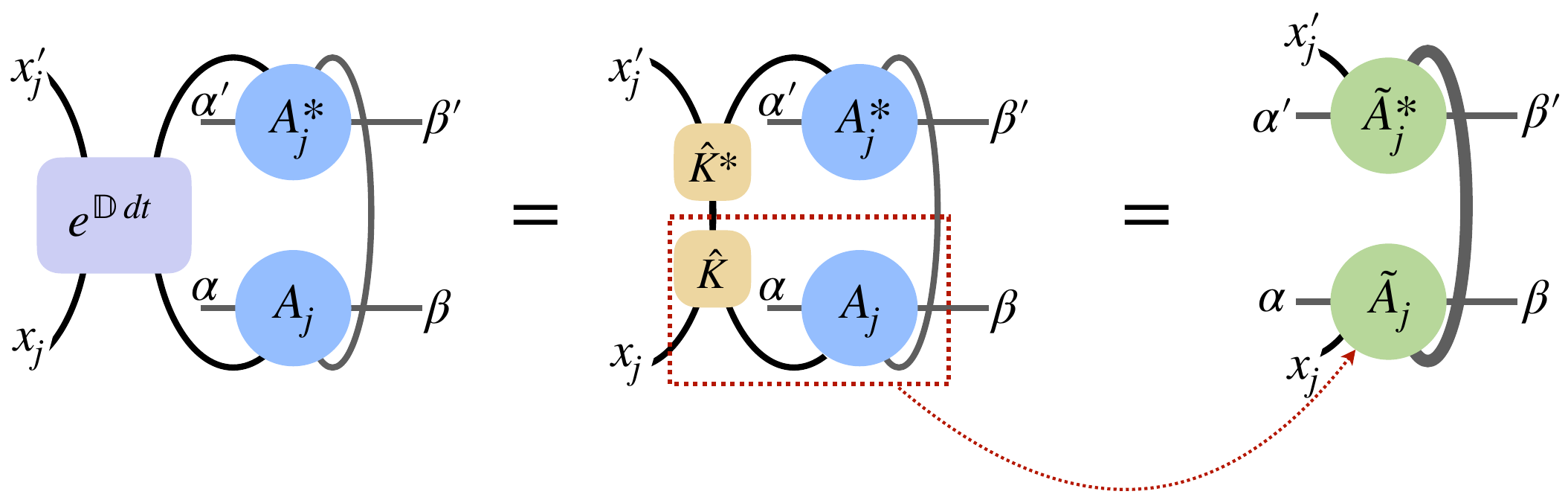}
\end{center}
Each time a dissipator layer is locally applied, the ancillary dimension  --- also known as the Kraus dimension $\kappa$ --- of the MPS increases by a factor of $k$.

The dissipative action tends to mix the state progressively. For example, the purified tensor network representation of a pure state (where the density matrix is a projector) has a Krauss dimension $\kappa = 1$, indicating no entanglement with the ancilla. However, as the dissipative part continues to act, this mixing increases the system-ancilla entanglement, leading to a rapid growth in the Kraus dimension $\kappa$.

To manage this growth, a dedicated compression procedure has been proposed, as outlined in ref.~\cite{positiveTN2016}. This method involves performing an SVD on the appropriately reshaped local tensor:
\begin{center}
\includegraphics[width=0.7\textwidth,valign=c]{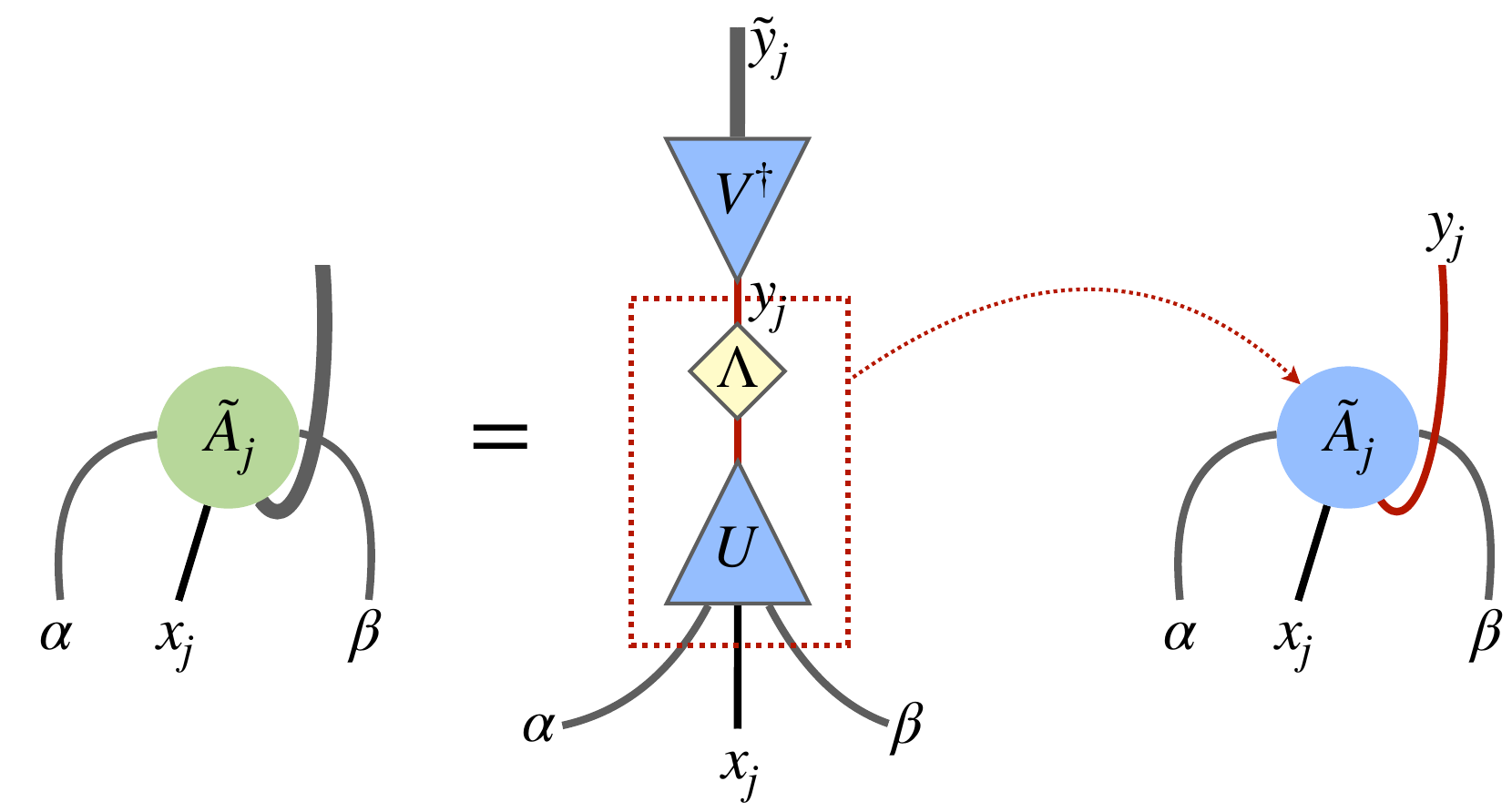}
\end{center}
In this step, the largest singular values up to a given threshold are retained, while the isometry $V^{\dag}$ is discarded, as it is irrelevant to the definition of the density operator after tracing out the ancillary degrees of freedom.

\section{Unravelling and monitored dynamics}
One way of simulating Lindblad dynamics~\eqref{chapt5_eq:lindblad} is through a technique known as \textbf{unraveling} the Lindblad equation\index{Lindblad!unraveling}. In this approach, the evolution of the quantum system is described as an ensemble of individual quantum trajectories. Each trajectory represents a possible realization of the system's dynamics. This method is computationally efficient for many systems since it avoids directly solving the full Lindblad master equation for the density matrix, which can be highly complex and resource-intensive, especially for large systems.

In the following we present two unravelings of the Lindblad equation for a monitored quantum dynamics.

\subsection{Measurement protocols}
Measurement in quantum mechanics is a fundamental aspect that significantly influences the state of a quantum system. The nature of quantum measurements can be broadly categorized into two types: projective (or strong) measurements, which are typically associated with quantum jumps, and weak measurements.

\paragraph{Quantum Jumps ---}\index{Quantum jumps} Consider a quantum many-body system in one dimension, with the total Hilbert space $ \mathcal{H} = \bigotimes_j \mathcal{H}_j $ being the tensor product of single-particle Hilbert spaces $ \mathcal{H}_j $. If the system is isolated from the environment, it evolves according to the Schrödinger equation
\begin{equation}
\ket{\psi(t)} = e^{-i\hat{H}t}\ket{\psi(0)},
\end{equation}
where $ \hat{H} $ is the full interacting Hamiltonian of the system and $ \ket{\psi(0)} $ is the initial state. Assume that the unitary dynamics are sporadically disrupted by local measurements. Each local Hilbert space $ \mathcal{H}_j $ is briefly randomly coupled to a measurement apparatus that measures a local observable $ \hat{O}_j = \sum_{k=1}^K o_k {\Pi}_j^{(k)} $, where $ o_k $ are the possible outcomes and $ {\Pi}_j^{(k)} $ are the corresponding projection operators. These measurements occur at discrete time intervals $ dt $ with a characteristic rate $ \gamma $. When a measurement occurs, the state $ \ket{\psi} $ is projected according to Born's rule
\begin{equation}
\ket{\psi} \rightarrow \frac{{\Pi}_j^{(k)}\ket{\psi}} {{\sqrt{\expval{{\Pi}_j^{(k)}}{\psi}}}},
\end{equation}
According to this protocol, the evolution of the many-body state $ \ket{\psi(t)} $ depends on the series of measurement events and their results, featuring occasional quantum jumps that are sudden transitions in the quantum system. Notice that the quantum state is pure during the entire evolution.

This monitored evolution can be depicted through digitized quantum dynamics. The idea involves designing a circuit that unitarily evolves the wave function over a time interval $dt$. Afterwards, a measurement is carried out as illustrated in the circuit below
\begin{equation}
    \includegraphics[width=.7\linewidth]{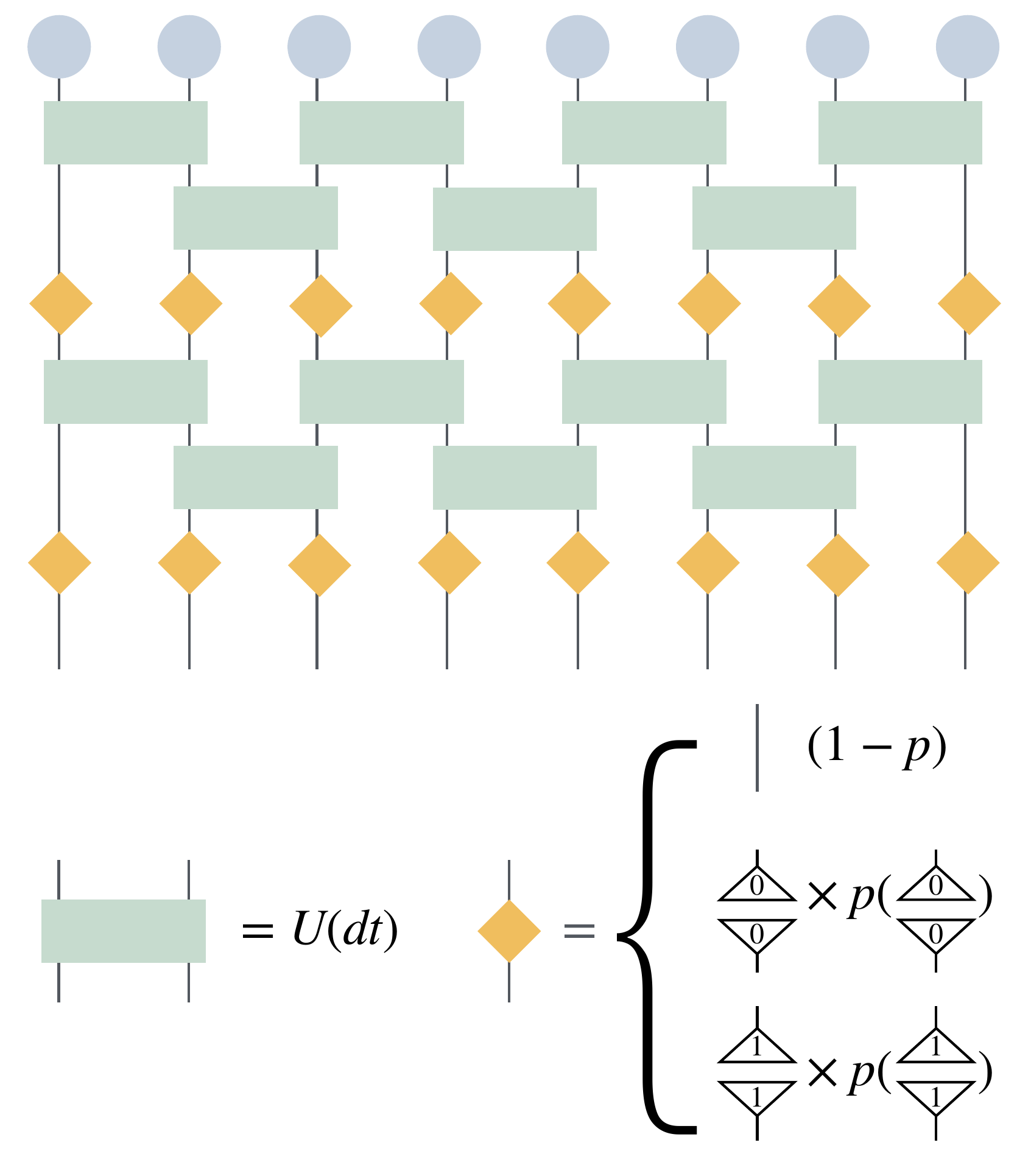}
\end{equation}
following this protocol\begin{equation}
    \ket{\psi_n} = \prod_{k=1}^n\left[ \mathcal{M} \prod_{i=1}^{N-1}U_{i,i+1}(dt)\right] \ket{\psi_0}
\end{equation}
where $\mathcal{M}$ implements the projective measurement on each site.

\paragraph{Weak-measurements ---}\index{Weak measurements} A weak measurement is a measurement that extracts partial information from a quantum system. The traditional way to describe a quantum measurement, that is a von Neumann projective measurements, is to write the state of the system into the eigenstates of a given observable $O$, namely
\begin{align}
\ket{\psi} = \sum _a c_a \ket{a}, \qquad O = \sum_n o_a \dyad{a}.
\end{align}
This measurement completely projects $\rho$ into an eigenstate $\ket{a}$ of the observable with probability $\abs{c_a}^2$, thus extracting maximal information.

Conversely, generalized measurements, which extract partial information, are described in terms of POVMs, i.e.\ positive operator-valued measures. Consider a set of operators $L_a$ such that $\sum_a L_a^\dagger L_a = \mathbb{I}$. The measurement process is described by transforming\begin{equation}
    \rho \longrightarrow \rho_f = \frac{L_a \rho L_a^\dagger}{\Tr{L_a \rho L_a^\dagger}}
\end{equation}
with probability $P(a) = \Tr{L_a \rho L_a^\dagger}$. Let us present the following simple model to provide a clear illustration of a weak projective measurement. Consider a two-level ancilla, represented by the eigenstates $\{\ket{+},\ket{-}\}$ of the Pauli matrix $\sigma_z$, initially prepared in the state\begin{equation}
    \ket{a} = \frac{\ket{+}+\ket{-}}{\sqrt{2}}\,.
\end{equation}
The ancilla is coupled to the system of interest, represented by the state $\ket{\psi_t}$. Let both the ancilla and the system evolve over a time $\Delta t$ under the unitary evolution operator,  $\hat{U}_{S+A}(\Delta t)$ \begin{equation}
    \hat{U}_{S+A}(\Delta t) \ket{\psi_{t}}\ket{\alpha} = \left(L_{+}\ket{\psi_t}\right)\ket{+}+\left(L_{-}\ket{\psi_t}\right)\ket{-}\end{equation}
where $L_{\pm}= \mel{\pm}{\hat{U}_{S+A}(\Delta t)}{a}$ act exclusively on the system's Hilbert space. Following this evolution, a projective measurement acts on the ancilla along the $z$-axis, resulting in the outcome $a = \pm 1$. As a result, the back-action of the measurement places the system in the state
\begin{equation}
    \ket{\psi_{t+\Delta t}} = \frac{L_{a}\ket{\psi_t}}{\sqrt{\mel{\psi_t}{L_{a}^\dagger L_{a}}{\psi_t} }}\, .
\end{equation}
Let us now consider the continuous limit $\Delta t \to 0$. We have to derive the explicit form of the operators $L_{\pm}$. If we want to measure the observable $X$, let us consider the coupling between $\mathcal{S} + \mathcal{A}$ of the form
\begin{equation}
    H_{S + A} =  H + \lambda X \sigma_y \;.
\end{equation}
We now take the limit $\Delta t \to 0$, scaling lambda in such a way that $\gamma = \lambda^2 \Delta t$ is kept constant.  Expanding the propagator $U$ we have
\begin{equation}
\begin{split}
    U_{S+A} &= e^{- i (\Delta t H + \sqrt{\gamma \Delta t} X \sigma_y)} \\ &= 1 - i \Delta t H - i \sqrt{\gamma \Delta t} X \sigma_y - \frac{1}{2} \gamma \Delta t O^2  + O (\Delta t^{3/2})
\end{split}
\end{equation}
We thus obtain for the $L_{\pm}$
\begin{equation}
    L_{\pm} = \frac{1}{\sqrt{2}} \left(1 - \imath \Delta t H \mp \sqrt{\gamma \Delta t} X - \frac1 2 \gamma \Delta t X^2  \right) \,.
\end{equation}
In order to compute the norm, we expand
\begin{equation}
    L_{a}^\dagger L_{a} = \frac 12 - a \sqrt{\gamma \Delta t} X \, ,
\end{equation}
while the corresponding probabilities become
\begin{equation}
    P(a) = \frac 12 - a \sqrt{\gamma \Delta t} \overline{X}. %\mean{X} \,.
\end{equation}
It follows that
\begin{equation}
    \begin{split}
    \ket{\psi_{t+ \Delta t}} &= \ket{\psi_t} - i H \Delta t \ket{\psi_t} - a \sqrt{\gamma \Delta t} ( X - \langle X \rangle) \ket{\psi_t} \\ &\quad + \frac 3 2  \gamma \Delta t \langle X \rangle^2 \ket{\psi_t}  - \gamma \Delta t \langle X  \rangle X \ket{\psi_t}- \frac 1 2 \gamma \Delta t X^2 \ket{\psi_t}
    \end{split}
\end{equation}
Finally, we have to put this in the form of a stochastic equation. To this aim we observe that the measurement outcome $a$ is a random variable that satisfies
\begin{equation}
    \overline{a} = - 2 \sqrt{\gamma \Delta t} \langle X \rangle  \;, \quad \overline{a^2} = 1 \;.
\end{equation}
Let us thus introduce $Y_t =  \sqrt{\Delta t} \sum_{t'\leq t} a$: in the limit $\Delta t \to 0$ this converges to a continous stochastic variable such that
\begin{equation}
\label{eq:dY}
    d Y = - 2 \sqrt{\gamma} \langle X \rangle d t + d \xi \,,
\end{equation}
where $\xi_t$ is a real Wiener process (i.e.\ $\overline{d \xi} = 0$ and $\overline{d \xi^2} = d t$).
Replacing $a \to d Y/\sqrt{\Delta t}$ and using~\eqref{eq:dY} in the limit $\Delta t \to 0$, we recover
\begin{equation}
\begin{split}
    d\ket{\psi} = - i H d t \ket{\psi_t} + \left(\sqrt{\gamma} ( X - \langle X \rangle ) d \xi - \frac \gamma 2 ( X - \langle \hat X \rangle)^2 d t \right) \ket{\psi_t}
\end{split}
\end{equation}

\biblio
\chapter{Tensor networks and quantum magic}\label{chap6}
\epigraph{\emph{``But I don't want to go among mad people'' Alice remarked. \\
``Oh, you can't help that'' said the Cat: ``We're all mad here. I'm mad. You're mad.''}}{Lewis Carroll}

Due to their unique characteristics and wide-ranging applications, the Clifford group and stabilizer states play a crucial role in quantum information theory. In quantum error correction, for instance, they form a robust framework that enables efficient error detection and correction, enhancing the stability and reliability of quantum computations. Recent advancements have further expanded the importance of Clifford unitaries, leading to a variety of applications in quantum computation and the learning of unknown quantum states.

One of the most striking features of stabilizer states is that, despite their potential extensive entanglement, they can be described and manipulated using classical computational resources. This classical tractability is attributed to the fact that operations within the Clifford group, when applied to stabilizer states, do not increase the computational complexity beyond classical limits.

In this chapter, we present a concise theoretical introduction to the stabilizer formalism and the Clifford group, focusing on the concept of magic, which is the quantum resource associated with the use of non-Clifford gates (or non-stabilizer states). We will then explore how a certain measure of magic, known as Stabilizer Rényi Entropies, can be computed for MPS states. Finally, we will discuss some recent proposals to incorporate the stabilizer formalism into Tensor Networks.

\section{Introduction to Stabilizer Formalism and Magic}
In this section, we introduce key concepts from the stabilizer formalism, a framework essential for understanding quantum error correction, efficient classical simulation of quantum systems, and the characterization of quantum resources beyond classical reach. These concepts underpin modern approaches to quantum computation and are foundamental for exploring the structure of quantum states.

\subsection{Pauli group and stabilizer groups}
\begin{definition}{Pauli group}{pauli group}
The Pauli group $\mathcal{P}_1$ is the $16-$elements group consisting of the following $2 \times 2$ matrices
\begin{equation*}
   \mathcal{P}_1 = \{ \pm \mathbb{1},  \pm i \mathbb{1}, \pm \PauliX, \pm i \PauliX, \pm \PauliY, \pm i \PauliY, \pm \PauliZ, \pm i \PauliZ \} \, .
\end{equation*}
The Pauli group $\mathcal{P}_N$ is the group obtained by taking the tensor product of any $N$ elements of $\mathcal{P}_1$, i.e.\
\begin{equation*}
   \mathcal{P}_N = \{ \pm \mathbb{1},  \pm i \mathbb{1}, \pm \PauliX, \pm i \PauliX, \pm \PauliY, \pm i \PauliY, \pm \PauliZ, \pm i \PauliZ \}^{\otimes N} \, .
\end{equation*}
\end{definition}

Below is a list of key properties of Pauli operators. Here and in the following subsections, we use the symbols $g$ (or $h$) to represent a generic Pauli string and highlight that these are elements of a group.
\begin{enumerate}[a)]
    \item The elements of $\mathcal{P}_N$ are Hermitian ($g^{\dag} = g$) or anti-Hermitian ($g^{\dag} = - g$)
    \item The elements of $\mathcal{P}_N$ have eigenvalues $\pm 1$ (if Hermitian), or $\pm i$ (if anti-Hermitian)
    \item If $g \in \mathcal{P}_N$, then $g^2 = \mathbb{1}$ (if $g$ is Hermitian) or $g^2 = - \mathbb{1}$ (if $g$ is anti-Hermitian)
    \item The elements of $\mathcal{P}_N$ are unitary
    \item If $g, h \in \mathcal{P}_N$, then either they commute $[g,h] = 0$ or they anti-commute $\{g,h\} = 0$
    \item With the exception of $\pm \mathbb{1}$, $\pm i\mathbb{1}$, the elements of $\mathcal{P}_N$ are traceless
\end{enumerate}

\begin{definition}{Stabilizers}{stabilizers}
We define \emph{stabilizer group} any subgroup of $\mathcal{P}_N$ that does not contain the element $- \mathbb{1}$. Given a stabilizer group $\mathcal{S}$, we define $V_{\mathcal{S}}$ as the set of $N-$qubits states $\ket{\psi}$ such that $g \ket{\psi}= \ket{\psi}, \forall g \in \mathcal{S}$ and $\forall \ket{\psi} \in V_{\mathcal{S}}$. $V_{\mathcal{S}}$ is by construction a subspace. It is called \emph{stabilizer subspace}. If it is one-dimensional, it defines a single state called \emph{stabilizer state}.
\end{definition}

Notice that in order to have a non trivial subspace $V_{\mathcal{S}}$ it is necessary to have $ - \mathbb{1} \notin \mathcal{S}$. Indeed otherwise we would have: $ (- \mathbb{1}) \ket{\psi} = \ket{\psi}, \forall \ket{\psi} \in V_{\mathcal{S}}$, which implies $V_{\mathcal{S}} = \{ 0 \}$. Moreover, it is important to notice that any stabilizer group is an \emph{abelian group}. Indeed, let us suppose that $ \exists g, h \in \mathcal{S}$ such that $[g,h] \neq 0$. Because of the observation $e)$, we have $\{g,h \} = 0 \, \Rightarrow \, gh = - hg$. Now, since $gh \in \mathcal{P}_N$, it can be either $(gh)^2 = \mathbb{1}$ or $(gh)^2 = -\mathbb{1}$. In the second case, we can conclude since we break the initial hyphotesis. In the first case, we have $(gh)(hg) = - (gh)(gh) = - (gh)^2 = - \mathbb{1}$, and therefore we find again the absurd.

Thus, we conclude stabilizers are abelian groups of the Pauli group, where all elements are Hermitian and square to the identity. Therefore, any stabilizer group $\mathcal{S}$ is generated by an independent set of $k$ (commuting) elements of the Pauli group $g_1, \ldots  ,g_k \in \mathcal{P}_N$, meaning that
\begin{equation*}
    \mathcal{S} = \Bigl\{ \prod_{j=1}^k \big( g_j \big)^{n_j} \, , \, n_j \in \{0,1\} \Bigl\} = \expval{g_1, \,  \ldots  ,g_k}
\end{equation*}
The size of $\mathcal{S}$ is therefore $|\mathcal{S}| = 2^k$.

\begin{definition}{Stabilizer projector}{Stabilizer projector}
Given a stabilizer group $\mathcal{S}$, we define the operator
\begin{equation*}
   P_{\mathcal{S}} = \frac{1}{|\mathcal{S}|} \sum_{g \in \mathcal{S}} g \, \, .
\end{equation*}
It can be easily proven that $P_{\mathcal{S}}$ is nothing but the orthogonal projection on $V_{\mathcal{S}}$. In fact, given $g \in \mathcal{S}$, $P_{g} = (\mathbb{1} + g)/2$ is the orthogonal projector on its eigenspace with eigenvalue $+1$. The projector on $\mathcal{S}$ is therefore
\begin{equation*}
   P_{\mathcal{S}} = \prod_{g \in \{ g_1, \ldots  ,g_k \} } P_g = \frac{1}{2^k} \sum_{\{ n \} } \prod_{j=1}^k \big( g_j \big)^{n_j} = \frac{1}{|\mathcal{S}|} \sum_{g \in \mathcal{S}} g  \, \, .
\end{equation*}
\end{definition}

Let $\ket{\psi}$ be a stabilizer state. Its density matrix operator $\rho = \ket{\psi} \bra{\psi}$ is the orthogonal projection on $\ket{\psi}$. Thus
\begin{equation}
   \hat{\rho} = P_{\mathcal{S}} = \frac{1}{|\mathcal{S}|} \sum_{g \in \mathcal{S}} g \, \, .
\end{equation}

\begin{lemma}{Dimension of the stabilizer group}{Dimension of the stabilizer group}
The dimension of the stabilizer subspace $V_{\mathcal{S}}$ of a stabilizer group generated by $k$ elements of $\mathcal{P}_N$ is $2^{N-k}$. Indeed, since $P_{\mathcal{S}}$ is an orthogonal projection its eigenvalues are $0$ or $1$ and its trace is equal to the number of $1$ eigenvalues, i.e.\ to the dimension of $V_{\mathcal{S}}$. We have $\Tr[P_{\mathcal{S}}]= 2^{-k} \sum_{g \in \mathcal{S}} \Tr[ g ]$ and, since $\Tr[ g ]$ is non zero only if $g = \mathbb{1}$, $\Tr[P_{\mathcal{S}}] =  2^{-k} 2^N$.
\end{lemma}

From this lemma, we conclude that the stabilizer group of an $N$-qubit stabilizer state is generated by exactly $N$ mutually commuting Pauli operators. The next question is whether stabilizer states span the entire $N$-qubit Hilbert space. The answer is no. In fact, since $\mathcal{P}_N$ has finite cardinality, the number of $N$ mutually commuting Pauli operators is also finite. For instance, the stabilizer states for $N=1$
are just six, namely
\begin{equation*}
    \ket{0} \, , \ket{1} \, , \frac{\ket{0} + \ket{1}}{\sqrt{2}} \, , \frac{\ket{0} - \ket{1}}{\sqrt{2}} \, , \frac{\ket{0} + i \ket{1}}{\sqrt{2}} \, , \frac{\ket{0} - i \ket{1}}{\sqrt{2}} \, .
\end{equation*}
In general, it can be shown that the cardinality of the set of pure stabilizer states is $|\St| = 2^N \prod_{k=0}^{N-1} \big( 2^{N-k} + 1)$.

\subsection{The Clifford group}

\begin{definition}{Clifford group}{Clifford group}
The Clifford group $\mathcal{C}_N$ is the group of $N-$qubits unitaries that map elements of the Pauli group into elements of the Pauli group
\begin{equation*}
      \mathcal{C}_N = \{ U \in \mathcal{U}_{2^{N} \times 2^{N}} \,  \text{ s.t. } \, U \mathcal{P}_N U^{\dag}=\mathcal{P}_N \} \, .
\end{equation*}
\end{definition}
It can be shown that the Hadamard gate $H$, the phase gate $S$ and the CNOT,\footnote{CNOT can be replaced with controlled $\PauliZ$.} i.e.\index{CNOT}\index{Clifford gates}
\begin{equation}
\includegraphics[width=.5\linewidth,valign=c]{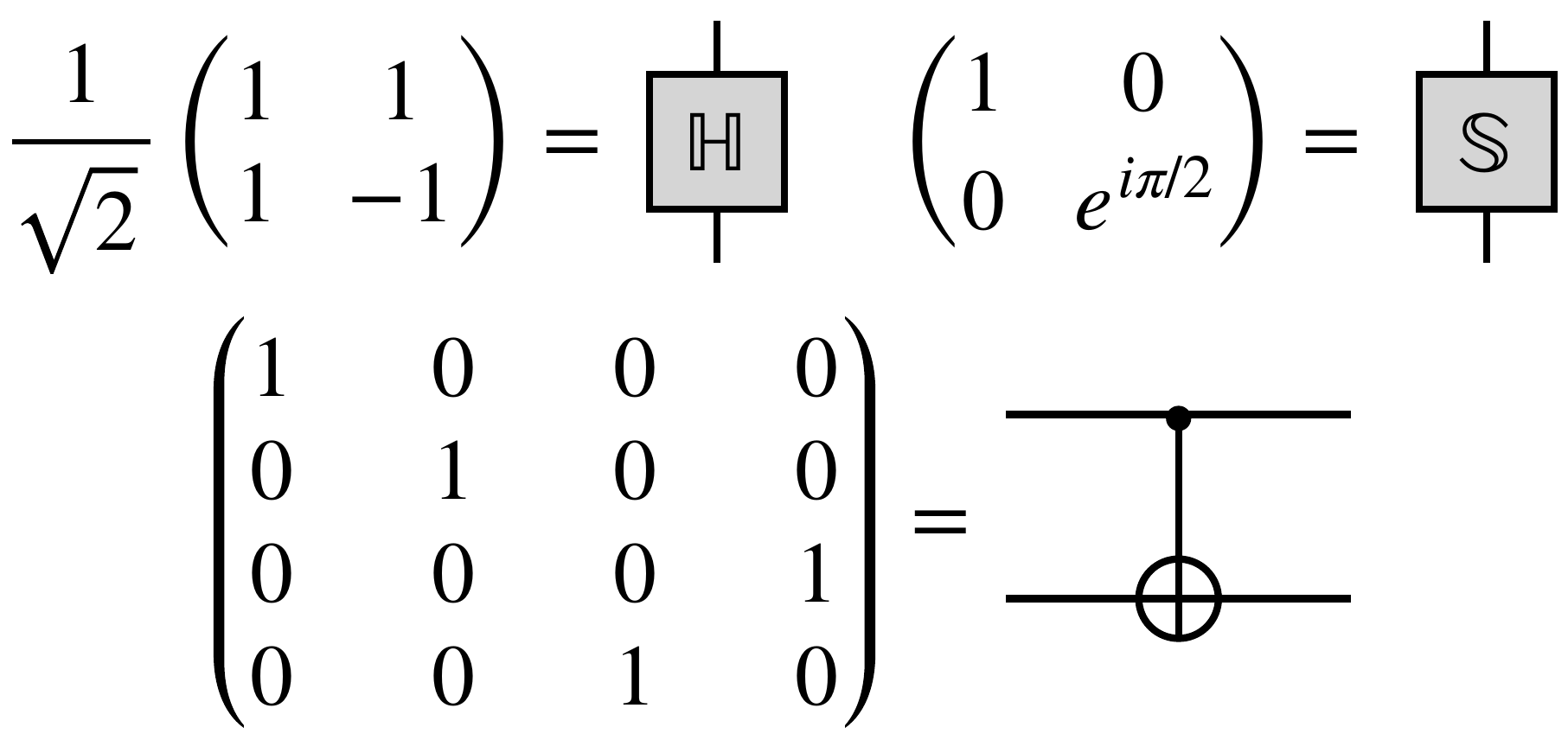}
\end{equation}
generate the entire Clifford group $\mathcal{C}_N$. Nevertheless, another gate is necessary to construct an universal set of gates, i.e.\ to span all the $N-$qubits unitaries. This is usually chosen to be
\begin{equation}
\includegraphics[width=.22\linewidth,valign=c]{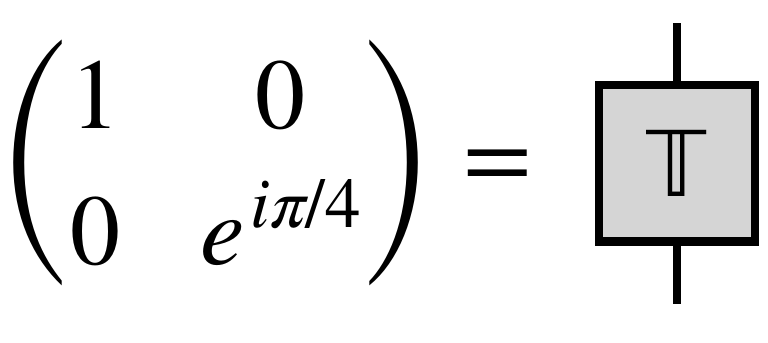}
\end{equation}
which is non Clifford, since for instance $T \PauliX T^{\dag} = \frac{1}{\sqrt{2}} \big(\PauliX + \PauliY)$. \\

Importantly, the uniform distribution on the Clifford group is a $3-$design, meaning that the average of any third order polynomial over $\mathcal{C}_N$ equals the average over the entire unitary group $U_N$. Consequently, Clifford group can be used for the simulation of random unitary circuits, allowing the extraction of information up to the third order moments.

The following result gives a connection between the Clifford group and the set of stabilizer states.

\begin{lemma}{Preparation of Stabilizer states}{Preparation of Stabilizer states}
A state $\ket{\psi}$ is a stabilizer if and only if it is equal to $C \ket{0}^{\otimes N}$, for some $C \in \mathcal{C}_N$. The second implication is very easy to show. Indeed,  $\ket{0}^{\otimes N}$ is clearly a stabilizer state having as generators
\begin{align*}
    \begin{split}
        g_1 & = \PauliZ \otimes \mathbb{1}  \otimes \ldots   \otimes \mathbb{1} \\
        g_2 & = \mathbb{1} \otimes \PauliZ \otimes \ldots   \otimes \mathbb{1} \\
        & \quad \qquad \ldots  \\
        g_N & = \mathbb{1} \otimes \mathbb{1} \otimes \ldots   \otimes \PauliZ \, . \\
    \end{split}
\end{align*}
Since the group $C \expval{g_1,g_2 \ldots  g_N} C^{\dag}$ stabilizes $\ket{\psi} = C \ket{0}^{\otimes N}$, we conclude that $\ket{\psi}$ is a stabilizer. The converse can also be shown easily.
\end{lemma}

Consider now any stabilizer subspace $V_{\mathcal{S}}$ and an unitary operator $U \in \mathcal{C}_N$. The action of $U$ maps $V_{\mathcal{S}}$ to $U V_{\mathcal{S}}$. It is not difficult to realize that $U V_{\mathcal{S}}$ is the stabilizer subspace for the group $U \mathcal{S} U^{\dag}$. Indeed, $\forall g' = U g U^{\dag}, \, \forall \ket{\psi'} = U \ket{\psi} \in U V_{\mathcal{S}}$ we have $g' \ket{\psi'} = g'  U \ket{\psi} = U g U^{\dag} U \ket{\psi} = U g \ket{\psi} = U \ket{\psi} = \ket{\psi'}$. This means that we can keep track of the action of any Clifford unitary simply by applying it to the generator set of $\mathcal{S}$.

\subsection{The tableau representation}
A simple but useful mapping exists between elements of $\mathcal{P}_1$  and the binary vector space $(\mathbb {Z} _{2})^2$. The mapping is given by the following tableau
\begin{equation*}
    \begin{cases}
        r(\mathbb{1}) \, \, = \big( 0 \, | \, 0 \big) \\
        r(\PauliX) = \big( 1 \, | \, 0 \big) \\
        r(\PauliY) = \big( 1 \, | \, 1 \big) \\
        r(\PauliZ) = \big( 0 \, | \, 1 \big) \\
    \end{cases}
\end{equation*}
and neglects the overall phases $\pm 1$ or $\pm i$ (thus, $r(-i \PauliX) = r(\PauliX)$). It is not difficult to show that $r(g h) = r(g) \oplus r(h)$, $\oplus$ denoting the sum modulo $2$.
The binary representation can be easily generalized to $\mathcal{P}_N$. Indeed, if we split $\pmb{r}(g)$ as $(\pmb{r}^{(1)}(g) \, | \, \pmb{r}^{(2)}(g))$, where $\pmb{r}^{(1)}$ and $\pmb{r}^{(2)}$ are vectors of length $N$, we can assign the value $1$ to the $j$th component of
$\pmb{r}^{(1)}(g)$ ($j=1,2 \ldots  \,  N$) iff the $j$th Pauli matrix in $g$ is $\PauliX$ or $\PauliY$, and the value $1$ to the $j$th component of
$\pmb{r}^{(2)}(g)$ iff the $j$th Pauli matrix in $g$ is $\PauliY$ or $\PauliZ$. Furthermore, if we define the matrix
\begin{equation*}
    \Lambda_N = \begin{pmatrix}
0_{N \times N} & \mathbb{1}_{N \times N} \\
\mathbb{1}_{N \times N} & 0_{N \times N} \\
\end{pmatrix}
\end{equation*}
of dimension $2 N \times 2 N$, we can realize that the symplectic inner product of two binary vectors evaluates whether the matrices commute or anticommute
\begin{equation*}
\big( \pmb{r}(g) \big)^T \Lambda_N \pmb{r}(h) = \begin{cases}
    0 \quad \text{if $g,h$ commute} \\
    1 \quad \text{if $g,h$ anticommute} \\
\end{cases} \, .
\end{equation*}

\begin{definition}{Generator matrix}{Generator matrix}
Given a stabilizer state, fully specified by $\mathcal{S}=\expval{g_1, g_2 \ldots  g_N}$, its \emph{generator matrix} is the $N \times 2N$ matrix
\begin{equation*}
    G = \begin{pmatrix}
\pmb{r}(g_1) \\
\pmb{r}(g_2) \\
\ldots  \\
\pmb{r}(g_N) \\
\end{pmatrix} \, .
\end{equation*}
This matrix only specifies the generators up to an overall phase.
\end{definition}

Notice that the stabilizer generators must be independent, meaning that no product of them can produce
$\mathbb{1}$. Since $\pmb{r}(g_i g_j) = \pmb{r}(g_i) \oplus \pmb{r}(g_j)$, no rows of $G$ can sum to zero and therefore $G$ must have full rank over the finite field $\mathbb{Z}_{2}$. Furthermore, since generators must commute they must satisfy $G^T \Lambda_N G = 0 $.

Let us notice that the Hadamard gate $H$ and the phase gate $S$ act as follows on the Pauli matrices
\begin{equation*}
\begin{cases}
    H \PauliX H^{\dag} = \PauliZ \\
    H \PauliY H^{\dag} = - \PauliY \\
    H \PauliZ H^{\dag} = \PauliX \\
\end{cases} \qquad
\begin{cases}
    S \PauliX S^{\dag} = \PauliY \\
    S \PauliY S^{\dag} = - \PauliX \\
    S \PauliZ S^{\dag} = \PauliZ\\
\end{cases}
\end{equation*}

In other words, $H$ interchanges $\PauliX$
and $\PauliZ$, giving $\PauliY$ a phase, and $S$ interchanges $\PauliX$ and $\PauliY$, possibly adding a phase, and leaves $\PauliZ$ unchanged.
In the tableau representation, this means:
\begin{align*}
\bigg(  \, \ldots  \, \, r_j^{(1)} \, \ldots   \, \bigg\rvert \,  \, \ldots  \, r_j^{(2)} \, \ldots  \,  \bigg) & \xleftrightarrow{H_j}  \bigg(  \, \ldots  \, \, r_j^{(2)} \, \ldots \, \, \bigg\rvert \,  \, \ldots  \, r_j^{(1)} \, \ldots  \,  \bigg) \\
\bigg( \, \ldots  \, r_j^{(1)} \, \ldots  \,  \, \bigg\rvert \, \, \ldots  \, r_j^{(2)} \, \ldots  \,  \bigg) & \xleftrightarrow{S_j}  \bigg( \, \ldots  \, r_j^{(1)} \, \ldots \,  \, \bigg\rvert \,  \, \ldots  \, \big( r_j^{(1)} \oplus r_j^{(2)} \big) \, \ldots  \,  \bigg) \, .
\end{align*}

Regarding the CNOT gate, we have ($\text{ CNOT } = \text{ CNOT }^{\dag}$)
{\small
\begin{equation*}
\begin{cases}
    \text{ CNOT }  ( \mathbb{1} \otimes \mathbb{1} ) \text{ CNOT } = \mathbb{1} \otimes \mathbb{1}  \\
    \text{ CNOT }  ( \mathbb{1} \otimes \PauliX ) \text{ CNOT } = \mathbb{1} \otimes \PauliX \\
    \text{ CNOT }  ( \mathbb{1} \otimes \PauliY ) \text{ CNOT } = \PauliZ \otimes \PauliY \\  \text{ CNOT }  ( \mathbb{1} \otimes \PauliZ )  \text{ CNOT } = \PauliZ \otimes \PauliZ \\
    \text{ CNOT }  ( \PauliX \otimes \mathbb{1} )  \text{ CNOT } = \PauliX \otimes \PauliX \\
    \text{ CNOT }  ( \PauliX \otimes \PauliX )  \text{ CNOT } = \PauliX \otimes \mathbb{1} \\
    \text{ CNOT }  ( \PauliX \otimes \PauliY )  \text{ CNOT } = \PauliY \otimes \PauliZ \\  \text{ CNOT }  ( \PauliX \otimes \PauliZ ) \text{ CNOT } = - \PauliY \otimes \PauliY \\
\end{cases}
\qquad
\begin{cases}
    \text{ CNOT }  ( \PauliY  \otimes \mathbb{1} ) \text{ CNOT } = \PauliY \otimes \PauliX \\
    \text{ CNOT }  ( \PauliY \otimes \PauliX ) \text{ CNOT } = \PauliY \otimes \mathbb{1} \\
    \text{ CNOT }  ( \PauliY  \otimes \PauliY ) \text{ CNOT } = - \PauliX \otimes \PauliZ \\  \text{ CNOT }  ( \PauliY  \otimes \PauliZ )  \text{ CNOT } = \PauliX \otimes \PauliY \\
    \text{ CNOT }  ( \PauliZ \otimes \mathbb{1} )  \text{ CNOT } = \PauliZ \otimes \mathbb{1} \\
    \text{ CNOT }  ( \PauliZ \otimes \PauliX )  \text{ CNOT } = \PauliZ \otimes \PauliX \\
    \text{ CNOT }  ( \PauliZ \otimes \PauliY )  \text{ CNOT } = \mathbb{1} \otimes \PauliY \\  \text{ CNOT }  ( \PauliZ \otimes \PauliZ ) \text{ CNOT } = \mathbb{1} \otimes \PauliZ \\
\end{cases}
\end{equation*}}\relax

This table can be summarized in following update rule for a CNOT where the qubit $k$ control the target qubit $j$ ($j,k \in \{1,2,\ldots  N \}$)
{\thinmuskip=2mu\medmuskip=1mu\thickmuskip=3mu
\footnotesize
\begin{equation*}
\bigg(  \, \ldots  \, \, r_j^{(1)} \oplus r_k^{(1)} \, \ldots  \, r_k^{(1)} \, \ldots  \,  \, \bigg\rvert \,  \, \ldots  \, r_j^{(2)} \, \ldots  r_k^{(2)} \oplus r_j^{(2)} \, \ldots \, \bigg) \, \xleftrightarrow{\text{CNOT}_{kj}} \, \bigg(  \, \ldots  \, \, r_j^{(1)} \, \ldots  \, r_k^{(1)} \, \ldots  \,  \, \bigg\rvert \, \, \ldots  \, r_j^{(2)} \, \ldots  r_k^{(2)} \, \ldots \,  \bigg) \, .
\end{equation*}}\relax
As a simple example, let us consider a system with $N=2$ qubits. The initial state is $\ket{00}$, which has $\{\mathbb{1}_1\PauliZ_2, \PauliZ_1 \mathbb{1}_2 \}$ as stabilizer generators. The stabilizer matrix is therefore
\begin{equation*}
    G = \begin{pmatrix}
    \begin{array}{cc|cc}
    0 & 0 & 1 & 0 \\
    0 & 0 & 0 & 1 \\
    \end{array}
    \end{pmatrix} \, .
\end{equation*}
If we apply the gate $H$ on the first qubit we can update the matrix $G$ as follows
\begin{equation*}
    G' = \begin{pmatrix}
    \begin{array}{cc|cc}
1 & 0 & 0 & 0 \\
0 & 0 & 0 & 1 \\
\end{array}
    \end{pmatrix} \, .
\end{equation*}
Then, we apply a CNOT, that gives
\begin{equation*}
    G'' = \begin{pmatrix}
    \begin{array}{cc|cc}
    1 & 1 & 0 & 0 \\
    0 & 0 & 1 & 1 \\
    \end{array}
    \end{pmatrix} \, .
\end{equation*}
Indeed, we have
\begin{equation*}
\text{CNOT}_{12} \big( H_1 \ket{00} \big) = \text{CNOT}_{12} \frac{\ket{00} + \ket{10}}{\sqrt{2}} =  \frac{\ket{00} + \ket{11}}{\sqrt{2}}
\end{equation*}
and the final state has $\{ \PauliX_1 \PauliX_2, \PauliZ_1 \PauliZ_2 \}$ as stabilizer generators, which are exactly the operators encoded in the tableau $G''$.

Other simple rules can be found to keep track of the overall sign $\pm 1$ and to update the generator matrix after a Pauli measurements. These observations lead to the following important Theorem~\cite{nielsen_chuang_2010}.\index{Gottesman-Knill theorem}

\begin{theorem}{Gottesman-Knill}{Gottesman-Knill}
Suppose we perform a quantum computation which involves only state preparations in the computational basis, Hadamard gates $H$, phase gates $S$, CNOT gates, Pauli gates and measurements of observables in the Pauli group, together with the possibility of classical control conditioned on the outcome of such measurements. Such a computation may be \emph{efficiently} simulated on a classical computer, meaning that there exists an Algorithm to do this classically
in a polynomial time, specifically at cost $\mathcal{O}(N^2 M)$ operations, where $N$ is the number of qubits and $M$ the number of operations.
\end{theorem}

\subsection{Entanglement}
Let us suppose to have a pure stabilizer state $\rho = \ket{\psi} \bra{\psi} = \frac{1}{2^N} \sum_{g \in \mathcal{S}} g$. We aim to assess the entanglement entropy $S_{A/B}$ associated to the partitioning of the system into a subsystem $A$ consisting of $N_A$ qubits and a complementary subsystem $B$ containing $N_B = N - N_A$ qubits. We start by computing the reduced density matrix $\rho_A=\Tr_B[\rho]$:
\begin{equation*}
   \hat{\rho}_A = \frac{1}{2^N} \sum_{g \in \mathcal{S}} \Tr_B[g]  \, \, .
\end{equation*}
Only some elements $g$ of the group $\mathcal{S}$ give a non vanishing contribution, namely the Pauli strings that can be written as $g = g_A \otimes \mathbb{1}_B$. Notice that this subset $\mathcal{S}_A$ is also a stabilizer group, since $(g_A \otimes \mathbb{1}_B) (g_A' \otimes \mathbb{1}_B) = (g_A g_A') \otimes \mathbb{1}_B$. Let us denote $k_A$ the number of independent generators of $\mathcal{S}_A$. Now we have
\begin{equation*}
   \hat{\rho}_A = \frac{1}{2^N} 2^{N_B}\sum_{g \in \mathcal{S}_A} g = \frac{1}{2^{N_A}} 2^{k_A} \frac{1}{2^{k_A}} \sum_{g \in \mathcal{S}_A} g \, .
\end{equation*}
The last part of the expression is the projector on the stabilizer subspace of $\mathcal{S}_A$, thus
\begin{equation}\label{eq:projectorstabA}
   \hat{\rho}_A = \frac{1}{2^{N_A - k_A}} P_{\mathcal{S}_A} \, \, .
\end{equation}
Now, since eigenvalues of $P_{\mathcal{S}_A}$ are only $1$ (with degeneracy $2^{N_A - k_A}$) and $0$ (with degeneracy $2^{N_A} - 2^{N_A - k_A}$), we get\index{Entanglement!stabilizer state}
\begin{equation*}
   S_{A/B} = - 2^{N_A - k_A} \frac{1}{2^{N_A - k_A}} \log \frac{1}{2^{N_A - k_A}} = (N_A - k_A) \log 2  \, \, .
\end{equation*}
How to compute $k_A$ in practice given the generator matrix $G$ of $\ket{\psi}$? It is not difficult to show that $k_A$ is simply the rank of the matrix obtained by setting to $0$ all the entries of $G$ corresponding to the subsystem $B$. Indeed, this operation corresponds to projecting the original generators on the space of operators having the form $g_A \otimes \mathbb{1}_B$.

\subsection{Nonstabilizerness as a quantum resource}

As we saw, a quantum computation involving only stabilizer states (or equivalently, Clifford circuits and measurements in the computational basis) can be simulated classically.
Therefore, one has to consider nonstabilizer protocols to reach a true quantum advantage. In this sense, the ``nonstabilizerness'' can be considered as a key ``resource'' of the quantum world, making it more powerful of the classical one. This intuitive picture can be made mathematically rigorous by exploiting the meta-theory of quantum resources. In this framework, one usually introduce a set of ``free-operations'' that by construction do not generate resource. In the case of entanglement these are the Local Operations and classical Communication (see Definition~\ref{def:locc}). In the case of the nonstabilizerness, also dubbed \emph{quantum magic}\index{Magic}, they are the following stabilizer operations:\footnote{Sometimes an additional condition, known as \emph{strong monotonicity}, is also imposed. It can be expressed as  $\sum_k p_k \mathcal{M}(\hat{P}_k \hat{\rho} \hat{P}_k) \leq \mathcal{M}(\hat{\rho})$ and it ensures that $\mathcal{M}$ does not increase, on average, when the experimenter has the ability to post-select multiple outcomes of a quantum measurement.}
\begin{itemize}
    \item Clifford unitaries, i.e.\ $\hat{\rho} \rightarrow \hat{U} \hat{\rho} \hat{U}^{\dag}$, with $\hat{U} \in \mathcal{C}_N$
    \item Composition with stabilizer states, i.e.\ $\hat{\rho} \rightarrow \hat{\rho} \otimes \hat{\rho}_S$, where $\hat{\rho}_S$ is a stabilizer state
    \item Measurements in the computational basis, i.e.\ $\hat{\rho} \rightarrow \sum_k \hat{P}_k \hat{\rho} \hat{P}_k$, where $\hat{P}_k$ are the projectors on the computational basis ($\sum_k \hat{P}_k = \mathbb{1}$)
    \item Discarding some qubits, i.e.\ $\hat{\rho} \rightarrow \Tr_A[\hat{\rho}]$, where $A$ is any subsystem
    \item The above operations conditioned on the outcomes of a measurement
\end{itemize}
The goal is typically to quantify the resources stored in a state $\hat{\rho}$ using a measure $\mathcal{M}(\hat{\rho})$, often referred to as a \emph{monotone}. This measure must not increase under arbitrary free operations. Specifically, $\mathcal{M}$ is considered a \emph{magic monotone} if and only if it satisfies the condition $\mathcal{M}\big( \mathcal{E}(\hat{\rho}) \big) \leq \mathcal{M}\big( \hat{\rho} \big)$ for any quantum channel $\mathcal{E}$ composed of the stabilizer operations listed above. If a monotone $\mathcal{M}$ can be identified and calculated, it can be used to distinguish between states that can or cannot be prepared from an initial state using free operations. The resource needed to prepare a state can be injected in the initial state, for instance having access to many copies of the \emph{magic state}\index{Magic state}
\begin{equation*}
  \ket{T} = \frac{1}{\sqrt{2}} \big( \ket{0} + e^{i \frac{\pi}{4}} \ket{1} \big) \, .
\end{equation*}
Examples of known genuine magic monotones are the following\index{Magic!Monotones}.
\begin{enumerate}
    \item\index{Robustness of magic} The robustness of magic~\cite{PhysRevLett.118.090501,Heinrich2019robustnessofmagic}, defined as
\begin{equation}
    R(\rho) = \min \big\{ \sum_{\alpha=1}^{|\St|} |x_k| \text{      s.t.  } \hat{\rho} = \sum_{\alpha} x_{\alpha} \hat{\rho}_{\mathcal{S}}^{(\alpha)} \big\} \, ,
\end{equation}
where $\rho_{\mathcal{S}}^{(\alpha)} = \ket{\psi^{(\alpha)}} \bra{\psi^{(\alpha)}}$, $\alpha=1,2 \ldots  \, |\St|$, are all the pure stabilizer states over $N$ qubits.
    \item\index{Min-relative entropy of magic} The min-relative entropy of magic~\cite{Veitch_2014,Bravyi_2019,Liu_2022}, which for pure states reads
    \begin{equation}
    D(\ket{\psi}) = - \log \bigg( \text{max}_{\alpha} |\expval{\psi^{(\alpha)}|\psi}|^2 \bigg) \, ,
    \end{equation}
    with the same notation as above.
\end{enumerate}

\noindent
However, both these quantity are very difficult to handle in practice, both in experiments and in numerical simulations, since their evaluation requires an optimization over an exponentially large space.

\subsection{Stabilizer Rényi Entropies}
In this Section, we will introduce the Stabilizer Rényi Entropies\index{Stabilizer Rényi Entropies} ~\cite{PhysRevLett.128.050402}, a simpler way of measuring the quantum magic. First, we introduce the projective Pauli group.

\begin{definition}{Projective Pauli group}{{Projective Pauli group}}
The projective Pauli group is the standard Pauli group $\mathcal{P}_N$ modulo global phases, i.e.\ $\tilde{\mathcal{P}}_N = \{ \mathbb{1}, \PauliX, \PauliY, \PauliZ \}^{\otimes N}$.
\end{definition}

From now on, we will indicate as $\hat{\pmb{\sigma}}$ a generic Pauli string belonging to $\tilde{\mathcal{P}}_N$.

\begin{definition}{Stabilizer Rényi Entropies}{Stabilizer Rényi Entropies}
Given a $N-$qubits state $\ket{\psi}$ and a Rényi index $n > 0$, we define the \emph{Stabilizer Rényi Entropy} (SRE) as
\begin{equation}
    M_{n} ( \ket{\psi} ) = \frac{1}{1 - n} \log \bigg( \sum_{\hat{\pmb{\sigma}} \in \tilde{\mathcal{P}}_N} \frac{1}{2^N} \expval{\hat{\pmb{\sigma}}}{\psi}^{2 n} \bigg) \, .
\end{equation}
To understand the relation with usual Rényi entropies, it is useful to introduce the density matrix $\hat{\rho} = \ket{\psi}\bra{\psi}$ and the function $\Pi_{\rho}(\hat{\pmb{\sigma}}) = \frac{1}{2^N} (\Tr[\hat{\rho} \hat{\pmb{\sigma}}])^{2}$. Let us observe that
\begin{align*}
\begin{split}
\sum_{\hat{\pmb{\sigma}}} \Pi_{\rho}(\hat{\pmb{\sigma}}) &= \frac{1}{2^N} \sum_{\hat{\pmb{\sigma}}}  \Tr[\hat{\rho} \hat{\pmb{\sigma}}] \cdot \Tr[\hat{\rho} \hat{\pmb{\sigma}}] = \\ &=\Tr \big[ \hat{\rho} \cdot \frac{1}{2^N} \sum_{\hat{\pmb{\sigma}}}  \Tr[\rho \hat{\pmb{\sigma}}] \hat{\pmb{\sigma}} \big] = \Tr[\hat{\rho}^2] = 1 \, .
\end{split}
\end{align*}
We used the decomposition of the matrix $\rho$ in terms of $\{ \frac{\hat{\pmb{\sigma}}}{\sqrt{2^N}} \}_{\hat{\pmb{\sigma}} \in \tilde{\mathcal{P}}_N} $, which is a complete base set for the Hermitian matrices of size $2^{N}$. We also use the fact that for pure states $\Tr[\hat{\rho}^2] = 1$. Thus, we can consider $\Pi_{\rho}(\hat{\pmb{\sigma}})$ as a probability distribution. We have
\begin{equation}
    M_{n} ( \ket{\psi} ) = \frac{1}{1 - n} \log \bigg( \sum_{\hat{\pmb{\sigma}}} \big(\Pi_{\rho}(\hat{\pmb{\sigma}}) \big)^{n} \bigg) - N \log 2 \, ,
\end{equation}
that shows that $M_{n} ( \ket{\psi} )$ is the $n$-Rényi entropy of $\Pi_{\rho}(\hat{\pmb{\sigma}})$, apart from a constant. Notice that, to extend the definition of $M_{n}$ to arbitrary density matrices $\rho$, we have to normalize with the purity, i.e.\
\begin{equation}
\Pi_{\rho}(g) = \frac{1}{2^N} \frac{(\Tr[\rho \hat{\pmb{\sigma}}])^{2}}{\Tr[\rho^2]} \, \, .
\end{equation}

\end{definition}

\hspace{5 cm}
\begin{enumerate}[a)]
\item  $M_{n}(\ket{\psi})=0$ for stabilizer states. Indeed, it is easy to show that $\expval{\hat{\pmb{\sigma}}}{\psi}=+1$ if $\hat{\pmb{\sigma}}$ belongs to the stabilizer group $\mathcal{S}$ of $\ket{\psi}$, $\expval{\hat{\pmb{\sigma}}}{\psi}=-1$ if $- \hat{\pmb{\sigma}} \in \mathcal{S}$, and $\expval{\hat{\pmb{\sigma}}}{\psi}=0$ in all other cases. Consequently, if $\ket{\psi}$ is a stabilizer state, $\expval{\hat{\pmb{\sigma}}}{\psi}^{2 n} = 1$ for $2^N$ strings, while it is $0$ for the others, and $\sum_{\hat{\pmb{\sigma}}} \frac{1}{2^N} \expval{\hat{\pmb{\sigma}}}{\psi}^{2 n} = 1$.

\item In general, $M_{n}(\ket{\psi})=0$ iff $\ket{\psi}$ is a stabilizer and $M_{n}(\ket{\psi}) > 0$ otherwise.

\item The SRE are invariant under Clifford unitaries. Indeed, even if the Pauli strings are reshuffled by a Clifford unitary, the probability values $\Pi_{\rho}(\hat{\pmb{\sigma}})$ remain unchanged and consequently also the values of the Rényi entropies.

\item The SRE are additive, i.e.\ $M_{n}(\ket{\psi} \otimes \ket{\phi})=M_{n}(\ket{\psi}) + M_{n}(\ket{\phi})$. This is a trivial consequence of the fact that the sum over Pauli strings on a system can be factorized into the sum of the Pauli strings of two subsystems.
\end{enumerate}

These properties have recently attracted a lot of attention to SREs, considered as a potential simpler monotone for magic.
However, it has been shown in ref.~\cite{Haug2023stabilizerentropies} that, for Rényi indices in the range $0 \leq n < 2$, Stabilizer Rényi Entropies (SREs) fail to behave as monotones under stabilizer protocols involving measurements in the computational basis, even when considering pure states. Additionally, for any Rényi index, SREs do not adhere to the strict monotonicity criteria with respect to computational-basis measurements.
Only recently, it has been definitively established in ref.~\cite{leone2024} that stabilizer entropies are monotonic for $n\geq 2$ within the framework of magic-state resource theory, specifically for pure states. Furthermore, it has been demonstrated that linear stabilizer entropies act as strong monotones in this context. Additionally, these stabilizer entropies have been extended to mixed states as valid magic monotones using convex roof constructions, ensuring their applicability beyond pure-state settings as well.

\section{Measuring the magic of Tensor Network states}
In this section, we explore various approaches to quantify the magic in tensor network states, with a particular emphasis on Matrix Product States (MPS). Among these, Stabilizer Rényi Entropies (SREs) stand out as the most viable option for tensor network representations. This is because, unlike other proposed measures of magic, SREs can be efficiently computed using tensor network techniques, making them especially suitable for large-scale many-body systems. While alternative methods have been explored, they tend to encounter significant computational challenges in the tensor network framework, making SREs the preferred tool for assessing nonstabilizerness in this setting --- at least based on the current understanding and available techniques.

\subsection{Replica methods}\label{chapt6_sec:replica_methods}
Replica methods are based on the simple fact that the expectation value of an operator to the power $k$, with $k$ an integer number, can be seen as the expectation value of a replicated operator over a replicated state. For instance;
\begin{equation}
    \expval{\hat{\pmb{\sigma}}}{\psi}^{k} = \big( \langle \psi | \big)^{\otimes k} \hat{\pmb{\sigma}}^{\otimes k} \big( | \psi \rangle \big)^{\otimes k} \, .
\end{equation}
Following ref.~\cite{Haug_2023}, we will explore how these methods can be applied to evaluate the SRE (we will restrict to integer $n > 1$). Notice that the number of replicas $k$ equals $2n$. Now, if $\ket{\psi}$ is an MPS with bond dimension $\chi$, the replicated state $\ket{\psi}^{\otimes k}$ is also an MPS with larger bond dimension $\chi' = \chi^{k}$ and physical dimension $d'=d^{k}$ ($d=2$ for qubits). Indeed, one can simply merge together the $k$ replicated MPS, as shown here for $k=2$:
\begin{equation*}
\includegraphics[width=.7\linewidth]{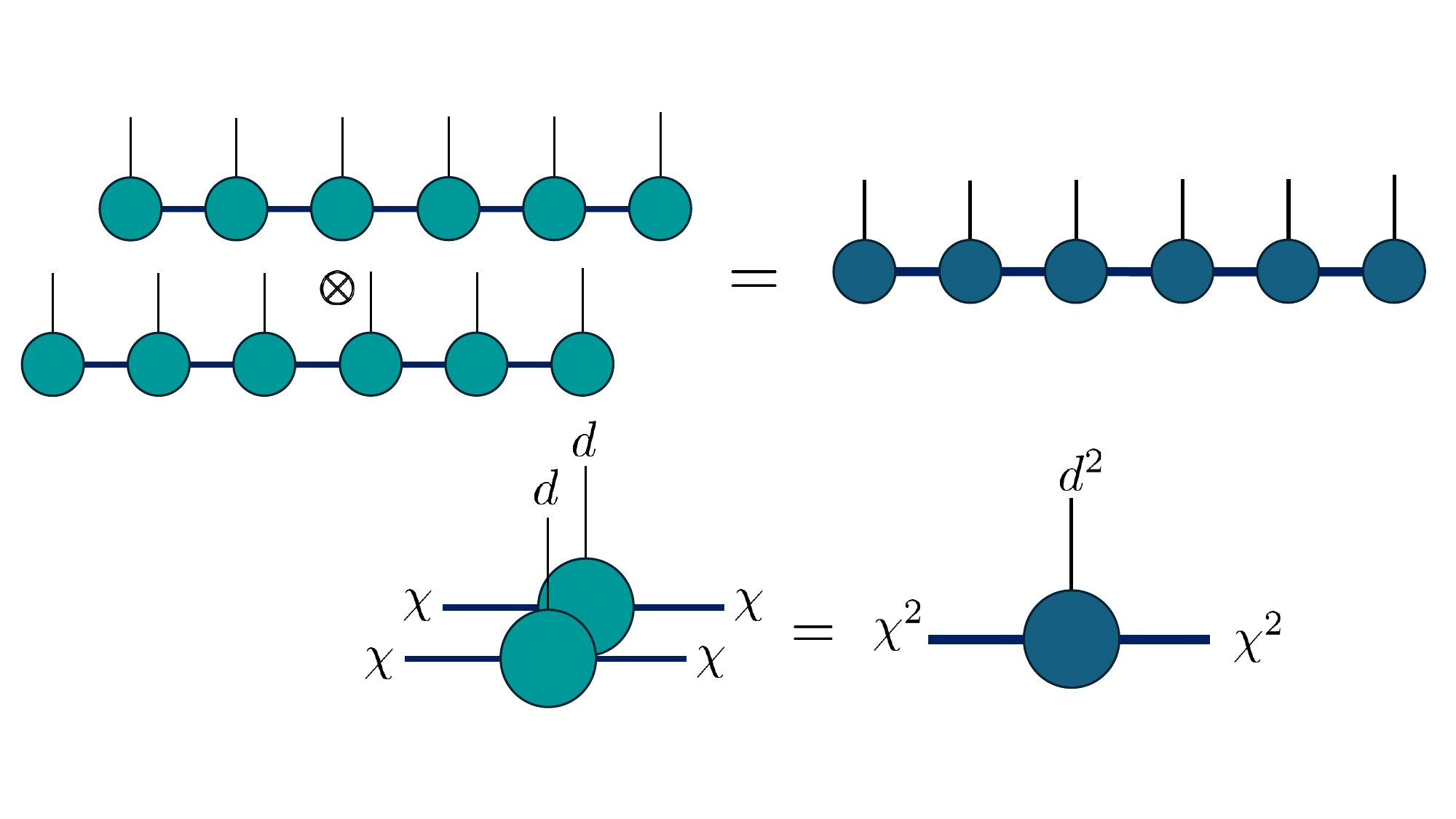}
\end{equation*}
For the evaluation of the SRE, it is technically convenient to rewrite $n$ of the $2n$ real-valued expectation values by taking their complex conjugates, i.e\ $\expval{\hat{\pmb{\sigma}}^*}{\psi^*}$. In this way, we can write the argument of the $\log$ in the SRE definition as
\begin{equation}
    \frac{1}{2^N} \sum_{\pmb{\sigma}}  \big( \langle \psi, \psi^* | \big)^{\otimes n} (\hat{\pmb{\sigma}} \otimes \hat{\pmb{\sigma}}^*)^{\otimes n} \big( | \psi, \psi^* \rangle \big)^{\otimes n} \, ,
\end{equation}
Which is graphically represented as follows
\begin{equation*}
\includegraphics[width=.8\linewidth]{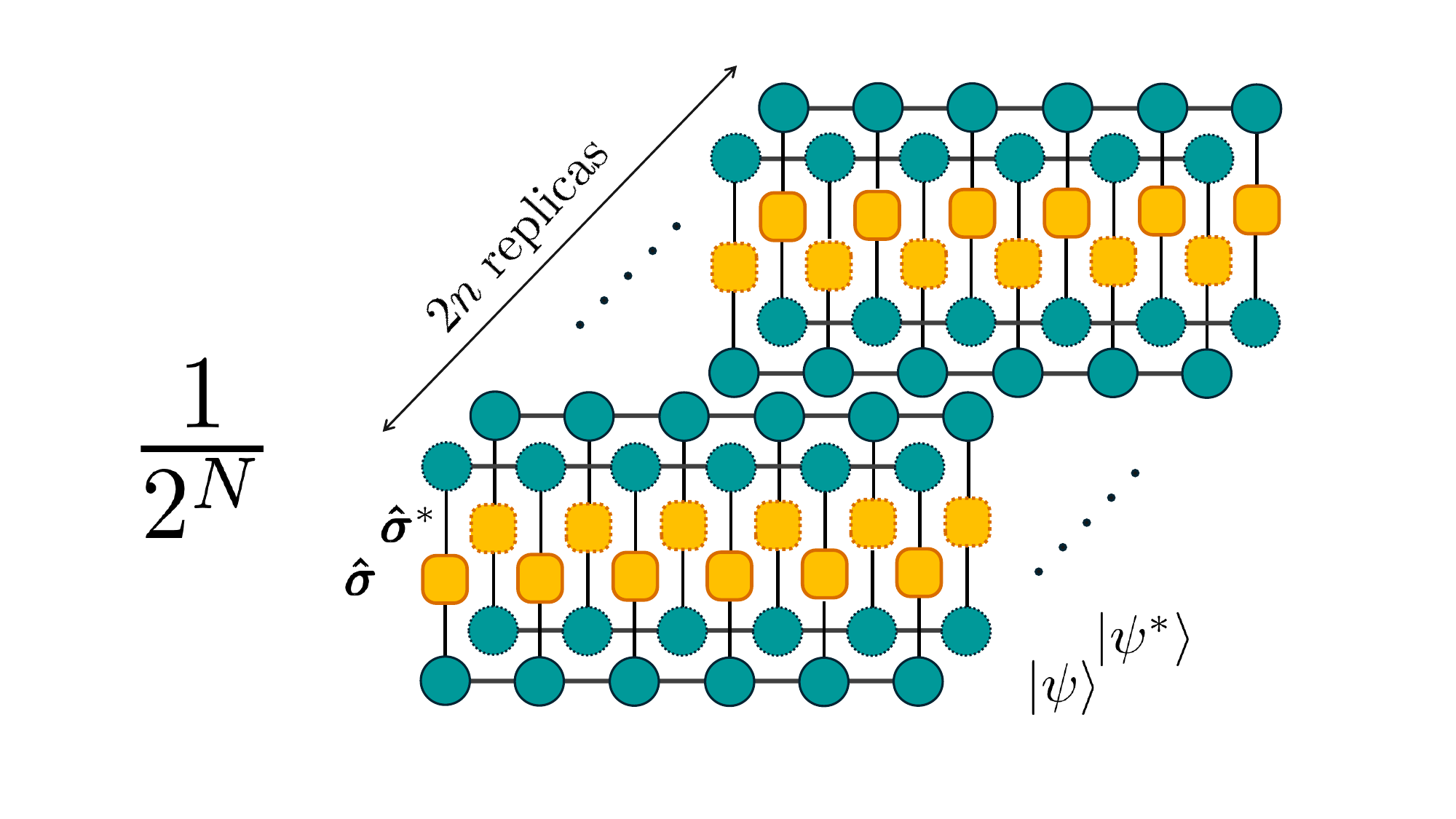}
\end{equation*}
where we used dotted, lighter shapes for the complex conjugate tensors.
It is now convenient to introduce the following object
\begin{equation}
    \Lambda^{(n)}_j = \frac{1}{2} \sum_{\mu=0}^3 (\hat{\sigma}^{\mu}_j \otimes \hat{\sigma}^{\mu, *}_j)^{\otimes n} \, ,
\end{equation}
which, if thought as a matrix, has shape $2^{2n} \times 2^{2n}$. One can easily show that $\Lambda^{(n)}$ is a positive matrix with rank $2^{2n -2}$. Therefore, one can decompose $\Lambda^{(n)}$~as
\begin{equation}
    \Lambda^{(n)} = \Gamma^{\dag} \Gamma \, ,
\end{equation}
where $\Gamma$ is a rectangular matrix of shape $2^{2n-2} \times 2^{2n}$.
These steps are illustrated below for $n=2$:
\begin{equation*}
\includegraphics[width=.8\linewidth,valign=c]{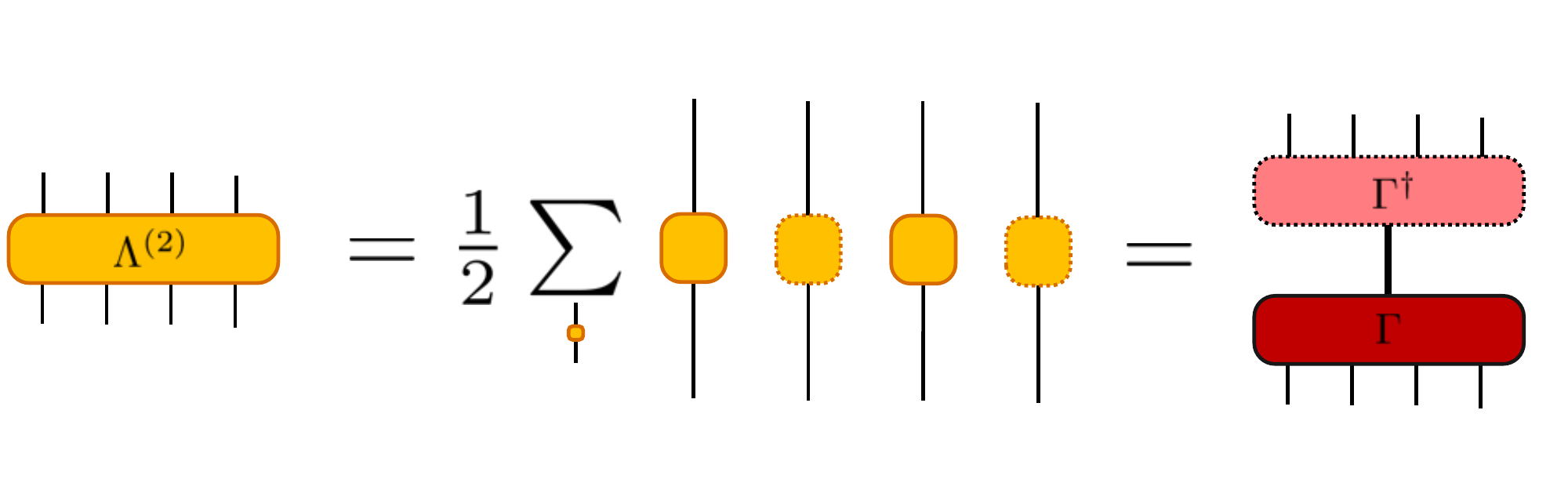} \, ,
\end{equation*}
Now the $\Gamma_j$ matrices can be incorporated in the MPS $\big( | \psi, \psi^* \rangle \big)^{\otimes n}$ by letting them act locally on the (replicated) physical spaces. The result is a new state $\ket{\Phi^{(n)}} = \left( \bigotimes_{j=1}^N \Gamma_j \right) \big( | \psi, \psi^* \rangle \big)^{\otimes n}$. Finally, the SRE can be computed by taking the state norm of $| \Phi^{(n)} \rangle$, namely\index{Stabilizer Rényi Entropies!replica method}
\begin{equation}\label{eq:sre_replica_final}
    M_{n} ( \ket{\psi} ) = \frac{1}{1 - n} \log \bigg( \langle \Phi^{(n)} | \Phi^{(n)} \rangle \bigg) \, .
\end{equation}
Last steps can be graphically understood as follows
\begin{equation*}
\includegraphics[width=.7\linewidth,valign=c]{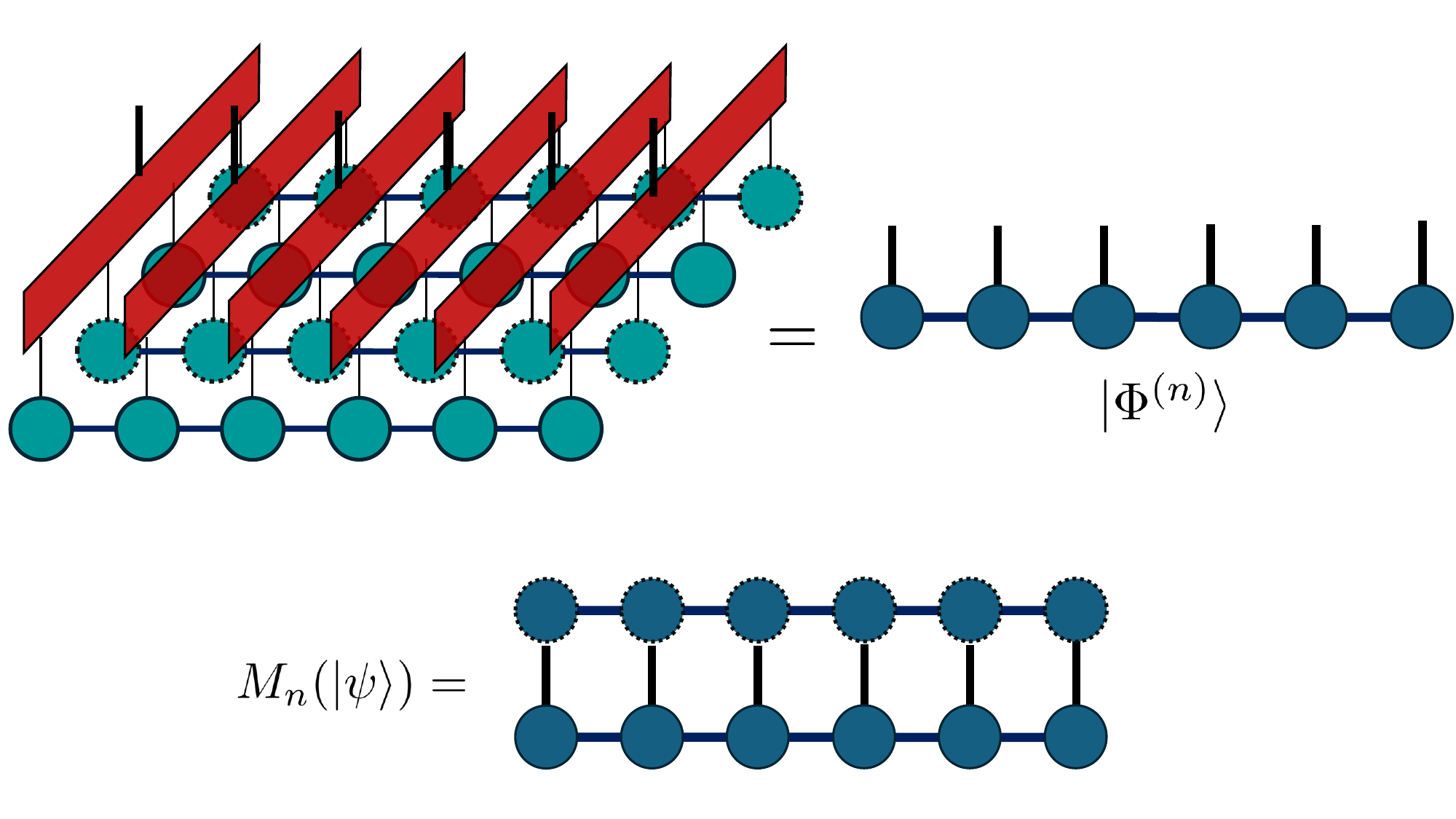} \, ,
\end{equation*}
The norm of the MPS $| \Phi^{(n)} \rangle$ in Eq.~\eqref{eq:sre_replica_final} can be computed exactly at a cost $\order{N \chi'^3 d'}=\order{N \chi^{6n} d^{2n-2}}$. This cost is often prohibitively high, but may be manageable for $n=2,3$ and small bond dimension $\chi$. In particular, for $n=2$ one can exploit a certain symmetry of $\Gamma$ to reduce the computational cost (see~\cite{Haug_2023} for details).

\begin{example}{Replica Pauli-MPS}{replica_pauli_mps}
To efficiently calculate Stabilizer Rényi Entropies (SREs) using tensor networks, we can leverage the representation of quantum states $\hat\rho = \ket{\psi}\bra{\psi}$ in the Pauli basis (using the density-matrix formalism of Chapter~\ref{chap5}).\index{Stabilizer Rényi Entropies!Pauli MPS} This allows us to store Pauli expectation values compactly and compute them efficiently~\cite{PhysRevLett.133.010601}.

If a state is originally represented by a Matrix Product State (MPS) with bond dimension $\chi$, its equivalent representation in the Pauli basis as in Eq.~\eqref{chapt5_eq:rho_pauli}, will have a bond dimension of $\chi^2$:
\begin{equation}
\includegraphics[width=0.8\textwidth,valign=c]{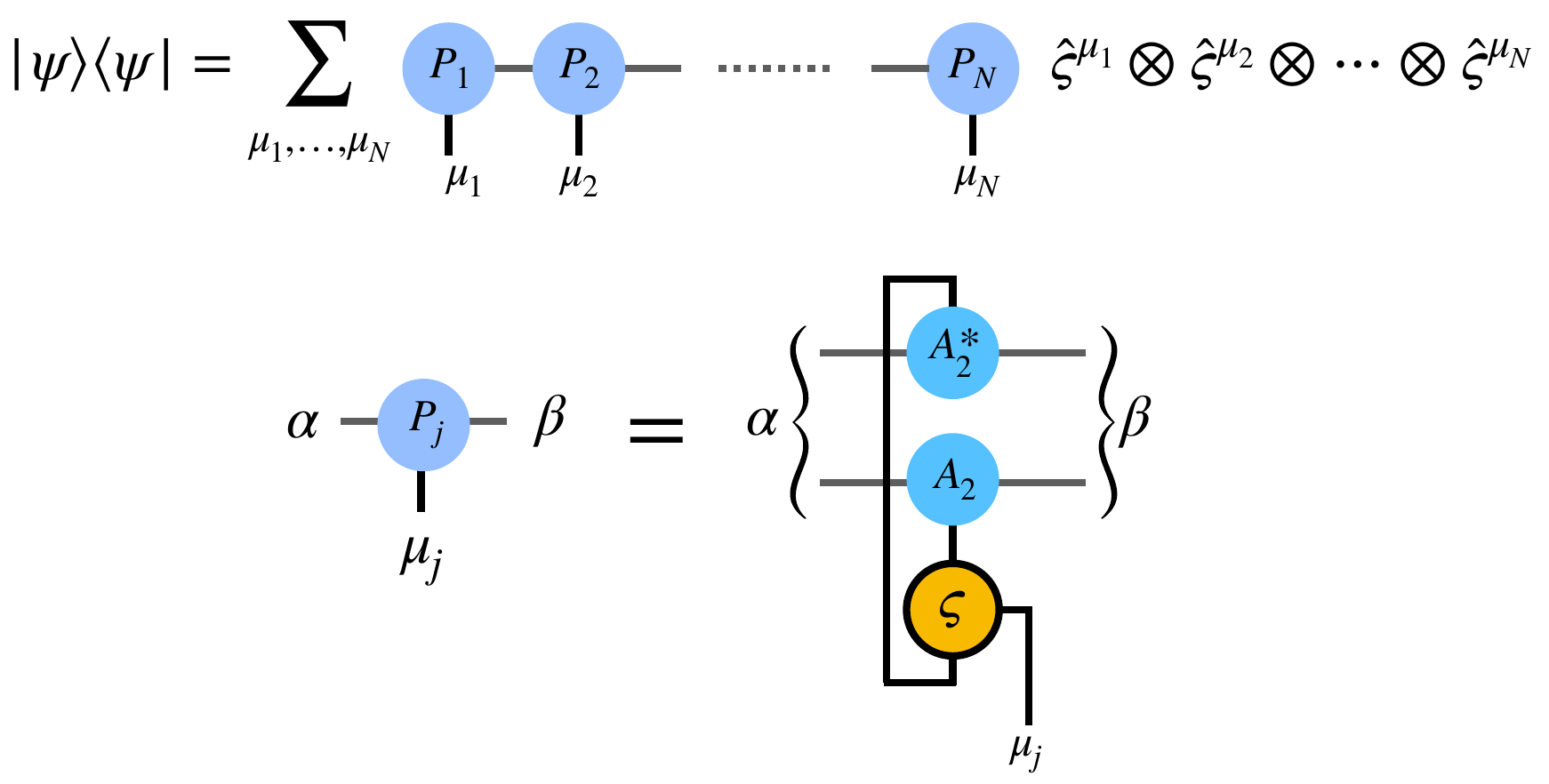}
\end{equation}

This representation facilitates the computation of SREs, where the SRE of index $n$ is derived from the contraction of $2n$ replicas of the Pauli-MPS:
\begin{equation}
\includegraphics[width=0.8\textwidth,valign=c]{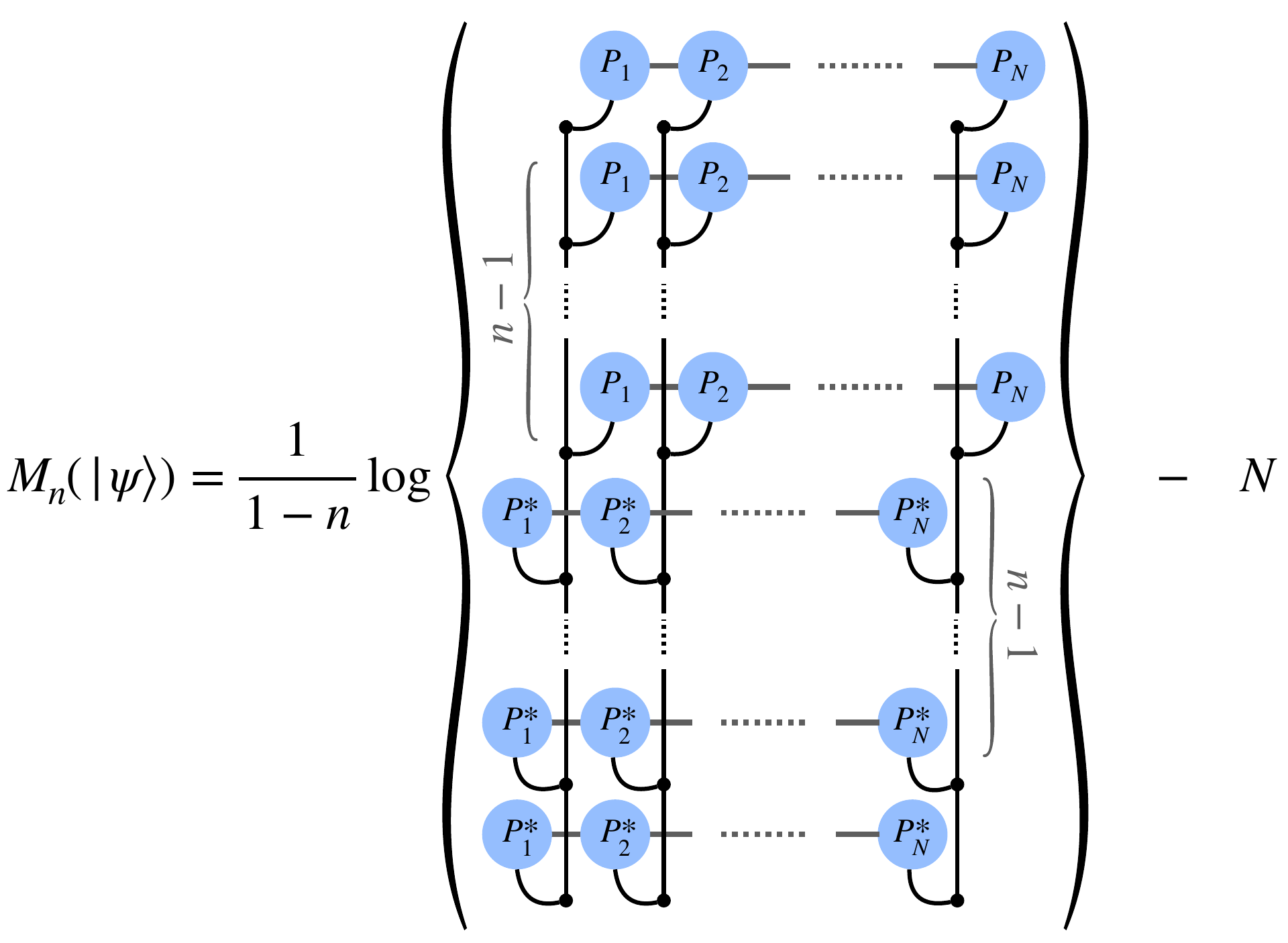}
\end{equation}

This approach offers significant advantages compared to earlier methods like the original replica trick presented in Section~\ref{chapt6_sec:replica_methods}. One key benefit is the constant physical dimension $d^2$ for the intermediate MPS needed to compute the SRE, as opposed to the exponentially growing physical dimension $d^{2(n-1)}$ in the original method. While both methods maintain a bond dimension of $\chi^{2n}$, the computational cost for exact contraction with the Pauli-MPS method scales as $\mathcal{O}(N d^2 \chi^{6n})$, a notable improvement over the $\mathcal{O}(N d^{2(n-1)} \chi^{6n})$ cost of the replica trick.

Another advantage lies in the approximation scheme allowed by the Pauli-MPS method. As the contraction involves repeated MPO-MPS multiplications, it becomes possible to sequentially compress the intermediate MPS at each step using standard tensor network techniques. This controlled approximation, though not exact, is highly practical as it allows monitoring of truncation errors, thus enabling the method to be used beyond very small bond dimensions that exact contractions would require. The approximate contraction, assuming bond dimensions of order $\chi$, scales as $\mathcal{O}(N d^2 \chi^4)$, offering significantly better efficiency compared to the exact approach.
\end{example}

\subsection{Sampling approach}
Stabilizer Rényi Entropies can be efficiently estimated using an MPS sampling algorithm~\cite{PhysRevLett.131.180401,PhysRevLett.133.010602,Haug2023stabilizerentropies}, closely resembling the method described in Section~\ref{sec:Measurements} to sample from the computational basis. The starting point is the definition
\begin{equation}
    M_{n} ( \ket{\psi} ) = \frac{1}{1 - n} \log \bigg( \sum_{\pmb{\sigma}} \big(\Pi_{\rho}(\hat{\pmb{\sigma}}) \big)^{n} \bigg) - N \log 2 \, ,
\end{equation}
from which we see evaluating the SRE essentially requires computing\linebreak
$\sum_{\pmb{\sigma}} \Pi_{\rho}(\hat{\pmb{\sigma}})^n$, which is a sum over the $4^N$ possible strings of Pauli operators. This sum can be expressed as an average over the probability distribution $\Pi_{\rho}(\hat{\pmb{\sigma}})$ itself. In fact:
\begin{equation}
    \sum_{\pmb{\sigma}} \Pi_{\rho}(\hat{\pmb{\sigma}})^n = \sum_{\pmb{\sigma}} \Pi_{\rho}(\hat{\pmb{\sigma}}) \cdot \Pi_{\rho}(\hat{\pmb{\sigma}})^{n-1} = \mathbb{E}_{\pmb{\sigma} \sim \Pi_{\rho}}[\Pi_{\rho}(\hat{\pmb{\sigma}})^{n-1}] \, ,
\end{equation}
where $\mathbb{E}_{\pmb{\sigma} \sim \Pi_{\rho}}[\ldots ]$ denotes the expected value over the probability distribution $\Pi_{\rho}$.
Thus, computing the SRE reduces to efficiently sampling Pauli string operators $\hat{\pmb{\sigma}}$ according to the probability distribution $\Pi_{\rho}(\hat{\pmb{\sigma}})$, and using these samples to estimate $\mathbb{E}_{\pmb{\sigma} \sim \Pi_{\rho}}[\Pi_{\rho}(\hat{\pmb{\sigma}})^{n-1}]$. One can simply use the sample average as the estimator. Specifically, if $\{\pmb{\sigma}_{\mu}\}_{\mu=1}^{N_{\text{samples}}}$ are the samples, the entropy can be estimated as follows:
\begin{equation}
    \sum_{\pmb{\sigma}} \Pi_{\rho}(\hat{\pmb{\sigma}})^n \simeq  \frac{1}{N_{\text{samples}}} \sum_{\mu=1}^{N_{\text{samples}}} \Pi_{\rho}(\hat{\pmb{\sigma}}_{\mu})^{n-1} \, .
\end{equation}
Now, how to sample Pauli strings $\pmb{\sigma}$ exactly with probability $\Pi_{\rho}(\pmb{\sigma})$? The idea is to exploit the following decomposition $\Pi_{\rho}$ in terms of conditional and prior (or marginal) probabilities
\begin{equation}
     \Pi_{\rho}(\pmb{\sigma}) =  \pi(\sigma_1)  \pi(\sigma_2|\sigma_1) \pi(\sigma_3|\sigma_1, \sigma_2) \ldots  \pi(\sigma_N|\sigma_1, \sigma_2 \ldots  \sigma_{N-1}) \, ,
\end{equation}
where
\begin{equation}
\pi_{\rho}(\sigma_j|\sigma_{1}\cdots \sigma_{j-1}) =
\frac{\pi_{\rho}(\sigma_1\cdots \sigma_j)}
{\pi_{\rho}(\sigma_1\cdots \sigma_{j-1})}
\end{equation}
is the probability that the Pauli matrix $\sigma_{j}$ occurs at position $j$ given that the string $\sigma_1\cdots\sigma_{j-1}$ has already occurred at positions $1 \dots j-1$, no matter the occurrences in the rest of the system (i.e.\ marginalising over all possible Pauli strings for the reaming qubits $j+1 \dots N$). Specifically, one has
\begin{equation}
\pi_{\rho}(\sigma_1\cdots \sigma_j)
= \sum_{\pmb\sigma \in \mathcal{P}_{N-j}}
 \frac{1}{2^N} \Trace[\rho \, \sigma_{1}\cdots\sigma_{j}\pmb{\sigma}]^{2} \, .
\end{equation}
In other terms, the conditional probability at the step $j$, i.e.\ $\pi_{\rho}(\sigma_j|\sigma_{1}\cdots \sigma_{j-1})$,
can be thought as the probability $\pi_{\rho_{j-1}}(\sigma_{j})$ of getting $\sigma_{j}$ in the partially projected state
\begin{equation}\label{eq:rho_j}
\rho_{j-1}  \equiv
\frac{\rho|_{\sigma_1\cdots\sigma_{j-1}}}{\pi_{\rho}(\sigma_1\cdots \sigma_{j-1})^{1/2}}
\end{equation}
where we have defined the state
\begin{equation}
\rho|_{\sigma_1\cdots\sigma_{j-1}}
\equiv \frac{1}{2^{N}} \sum_{\pmb{\sigma}\in\mathcal{P}_{N-j+1}}
\,\sigma_{1}\cdots\sigma_{j-1}\pmb{\sigma} \, \Trace[\rho \, \sigma_{1}\cdots \sigma_{j-1} \pmb{\sigma}]
\end{equation}
where, in the Pauli matrices decomposition of $\rho$, we are only keeping the contribution with fixed $\sigma_1\cdots\sigma_{j-1}$. Notice that such state is not normalized, however $\Trace[\rho_{j-1}^2]=1$, and the probability that the remaining string $\pmb{\sigma}\in\mathcal{P}_{N-j+1}$ occurs is exactly given by $\pi_{\rho}(\pmb{\sigma}|\sigma_{1}\cdots\sigma_{j-1})$.
From the definition in Eq.~(\ref{eq:rho_j}), we can easily get the recursive relation $\rho_{j} = \pi_{\rho_{j-1}}(\sigma_{j})^{-1/2} \rho_{j-1}|_{\sigma_{j}}$. Thanks to that, we can generate the outcomes (and the probabilities of that outcomes) by iterating over each single qubits, and sampling each local Pauli matrix according to the conditional probabilities. Once a local outcome occurs, the state is updated accordingly, and the iteration proceeds until all qubits are sampled. At the end of this procedure, as a result of Eq.~(\ref{eq:chain_prob}), we generated configurations $\pmb\sigma$ with probability $\Pi_{\rho}(\pmb{\sigma})$.\index{Stabilizer Rényi Entropies!Pauli perfect sampling}

Following this prescription, we start from the first term of the expansion in Eq.~(\ref{eq:chain_prob}). This can be written as
\begin{align}\label{eq:pi1}
    \begin{split}
    \pi_{\rho}(\sigma_1) &= \frac{1}{2^N} \sum_{\pmb{\sigma} \in \mathcal{P}_{N-1}} \langle \psi| \sigma_1  \pmb{\sigma} |\psi \rangle \langle \psi^*| \sigma_1^* \pmb{\sigma}^* |\psi^* \rangle  \, ,
    \end{split}
\end{align}
where we used the fact that the Pauli matrices are hermitian. Graphically this equation is represented as
\begin{equation*}
\includegraphics[width=.85\linewidth,valign=c]{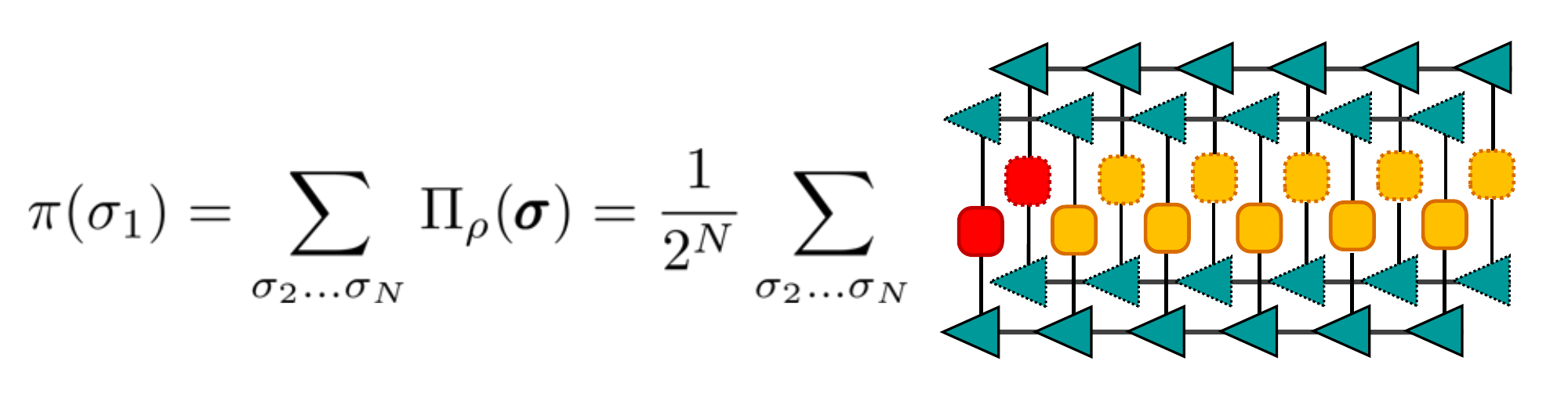} \, ,
\end{equation*}
In terms of the operators $\Lambda_{\sigma_i} = \frac{1}{2} \sigma_i \otimes \sigma_i^*$ and $\Lambda_i = \frac{1}{2} \sum_{\sigma_i} \big( \sigma_i \otimes \sigma_i^* \big)$, each acting on the local Hilbert space given by a spin and its replica, the previous
equation reads
%\begin{equation}
$\pi_{\rho}(\sigma_1) =
\big[ \bra{\psi} \otimes \bra{\psi^*} \big] \Lambda_{\sigma_1}  \Lambda_{2}\cdots
\Lambda_{N}  \big[ \ket{\psi} \otimes \ket{\psi^*} \big]$.
%\end{equation}
%that is an expectation value computed on the two-replicas state $\ket{\psi} \otimes \ket{\psi^*}$.
Now, the following property can easily be proven
\begin{equation}
\big[ \bra{s_i'}\otimes\bra{r_i'} \big] \Lambda_{i} \big[\ket{s_i}\otimes\ket{r_i}\big] = \delta_{s_i',r_i'} \delta_{s_i,r_i} \, ,
\end{equation}
meaning that $\Lambda_{i}$ is just two copies of the identity operator connecting the spin $|s_i\rangle$ and its replica (whose local computational basis is now indicated as $|r_{i}\rangle \in \{|0\rangle, |1\rangle\}$). 
Graphically this identity looks as follows (see Definition~\ref{def:paulitensor})\index{Pauli matrices!completeness}
\begin{equation}\label{eq:Lambda}
\includegraphics[width=.4\linewidth]{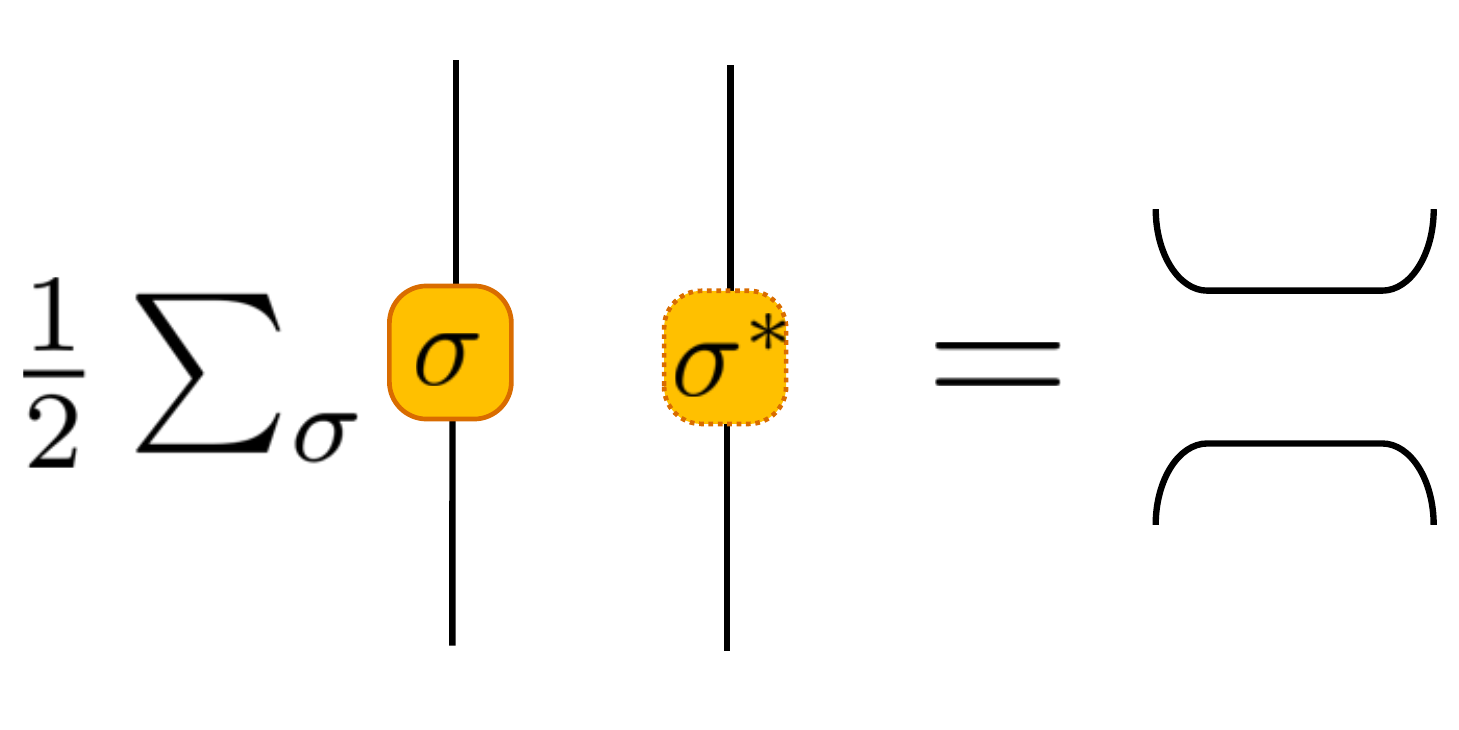}
\end{equation}
Using Eq.~\eqref{eq:Lambda} we get
\begin{equation*}
\includegraphics[width=.7\linewidth,valign=c]{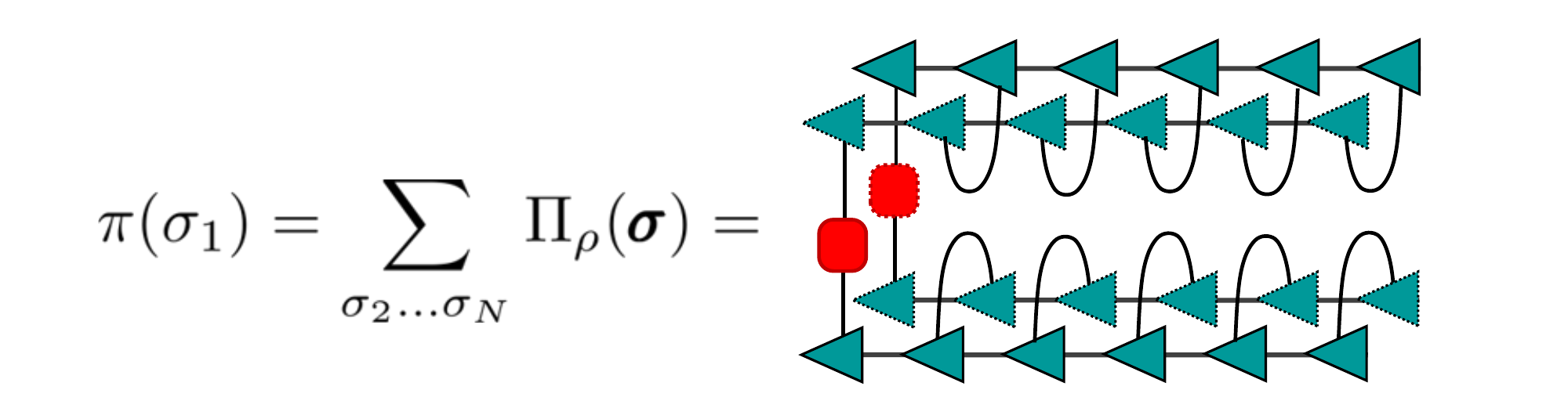} \, .
\end{equation*}
By applying the right-normalization condition of the MPS tensors, we can further simplify the calculation to a local tensor contraction as follows
\begin{equation}\label{eq:prob_1}
\includegraphics[width=.55\linewidth,valign=c]{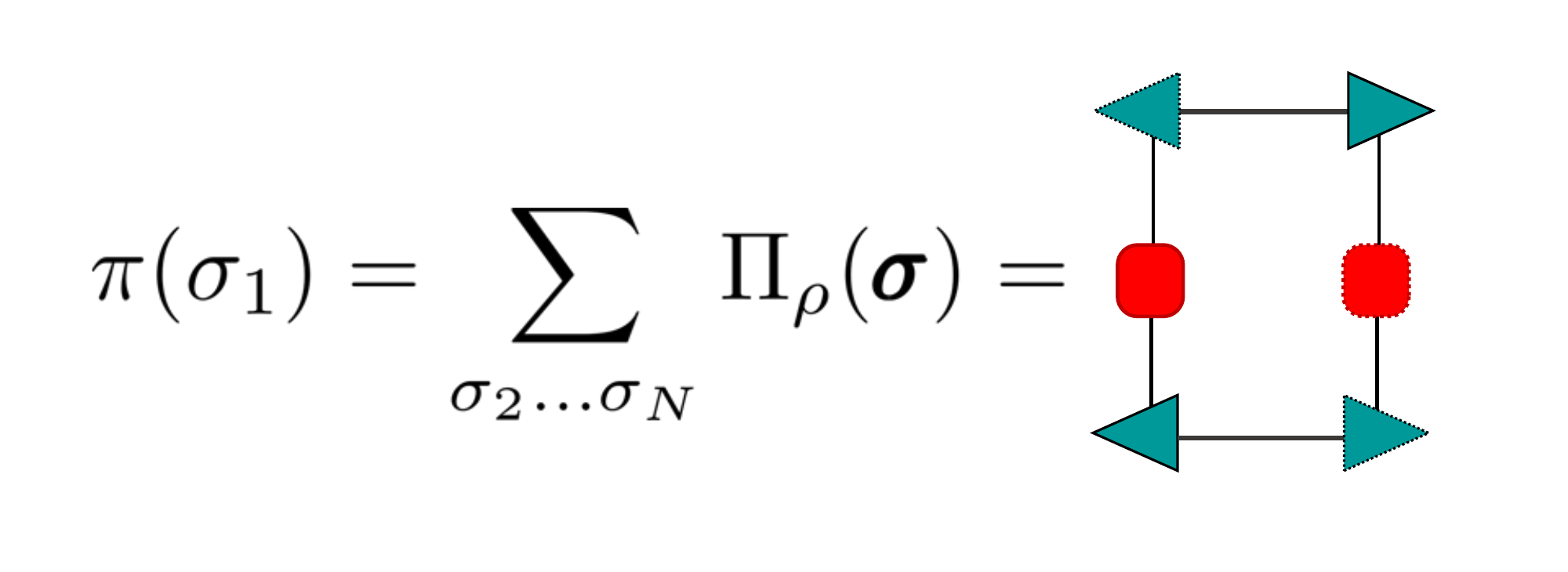} \, ,
\end{equation}
which as an equation reads:
\begin{equation*}
\pi_{\rho}(\sigma_1) =
\frac{1}{2}
\sum_{{s_1,s_1',r_1,r_1'}}
(\mathbb{A}^{s_1'}_{1})^* \mathbb{A}^{r_1'}_{1}
(\sigma_1)_{s_1' s_1} (\sigma_1^*)_{r_1' r_1} \mathbb{A}^{s_1}_{1} (\mathbb{A}^{r_1}_{1})^* \, .
\end{equation*}
After evaluating $\pi_{\rho}(\sigma_1)$ for $\sigma_1 \in \{ \sigma^0, \sigma^1, \sigma^2, \sigma^3 \}$, one can extract a sample from this distribution, obtaining the first element of the string. The information about the partially projected state
Eq.~(\ref{eq:rho_j}) is encoded in an effective environment matrix $\mathbb{L}$ initialized as follows:
\begin{equation*}
\includegraphics[width=.85\linewidth,valign=c]{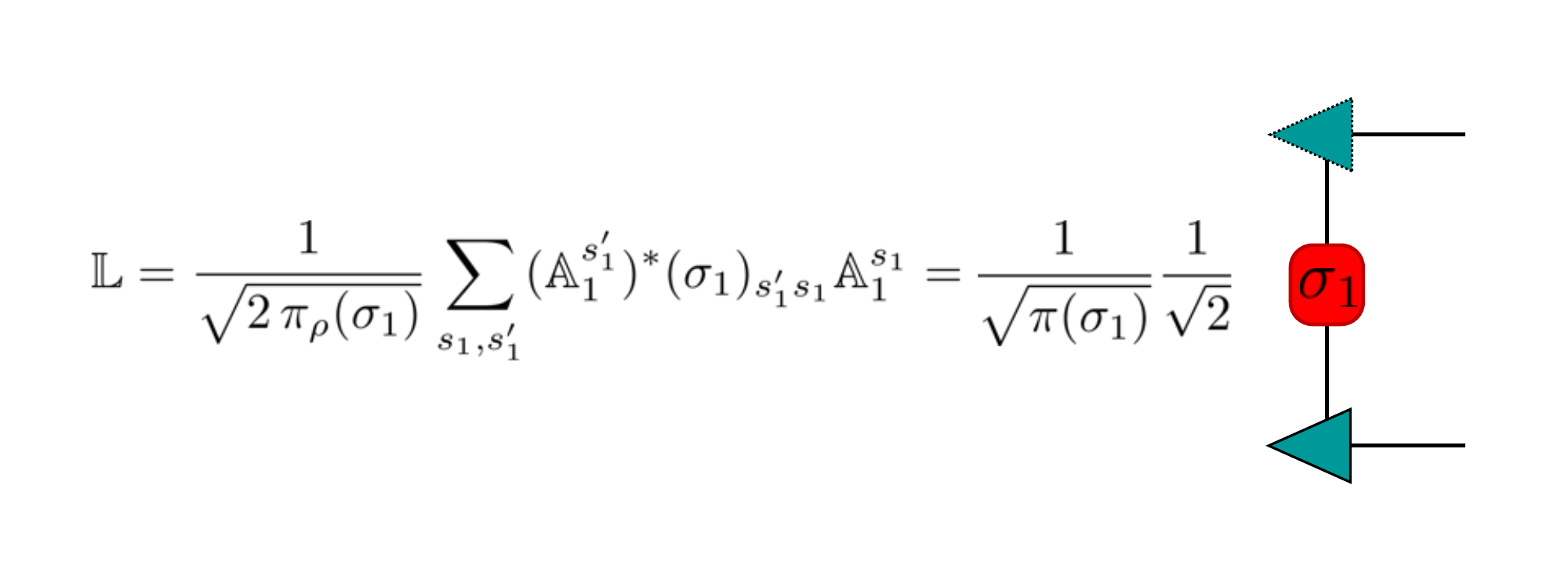} \, .
\end{equation*}
The calculation of the next terms of Eq.~(\ref{eq:chain_prob})
and the extraction of the remaining $\sigma_i$ proceeds following the same line. In fact, it is not difficult to show that at a generic site $i$ Eq.~\eqref{eq:prob_1} becomes
\begin{equation}\label{eq:prob_3}
\includegraphics[width=.55\linewidth,valign=c]{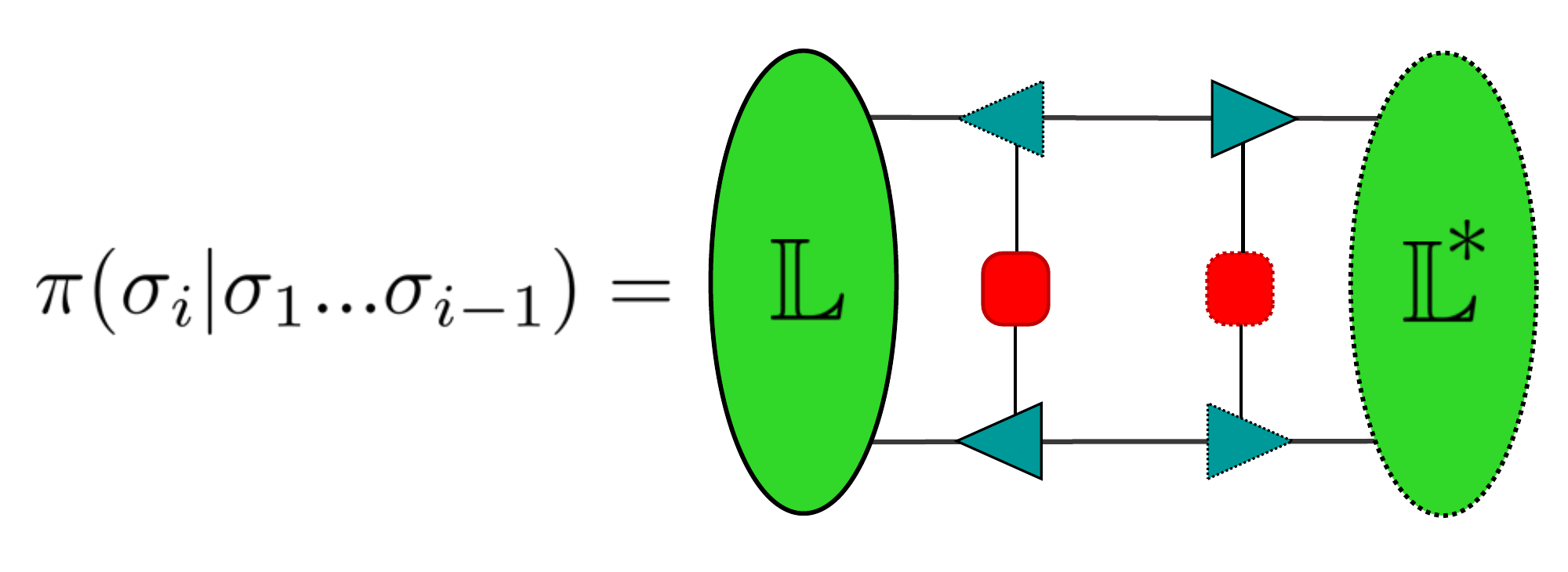} \, .
\end{equation}
After the extraction of $\sigma_i$ the environment $\mathbb{L}$ has to be updated as
\begin{equation}\label{eq:prob_2}
\includegraphics[width=.55\linewidth,valign=c]{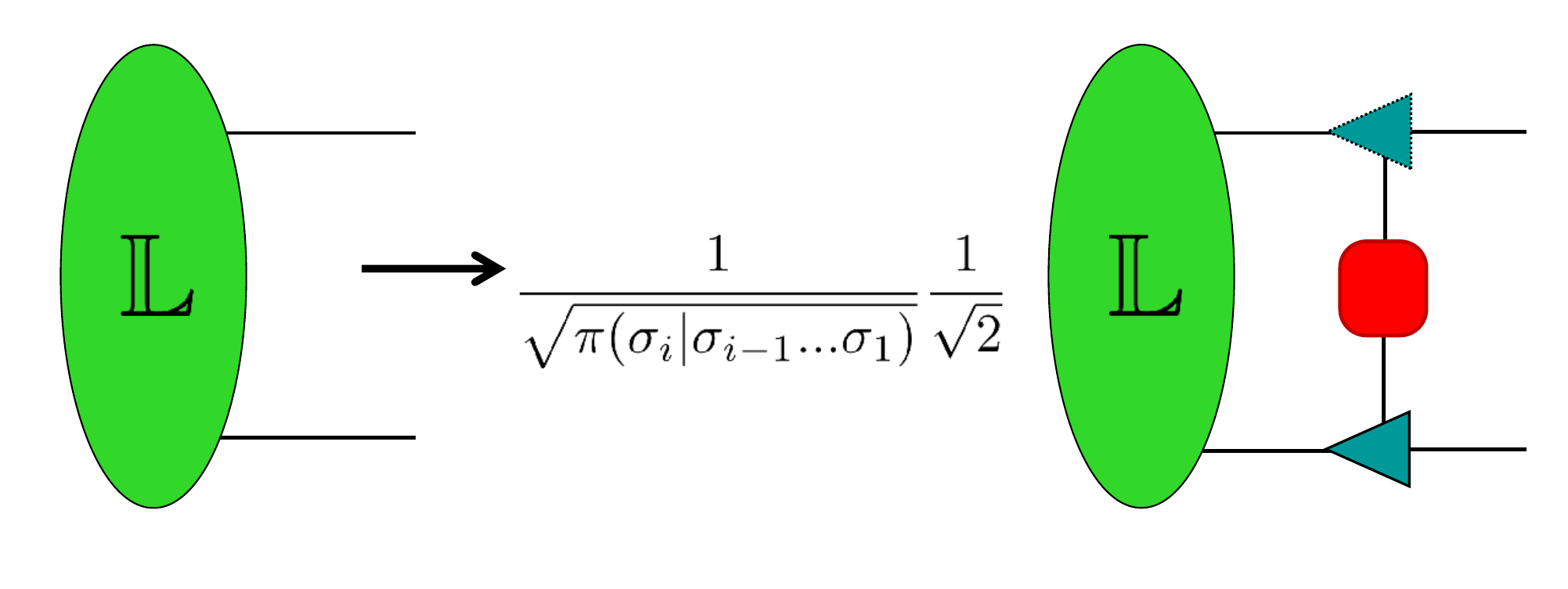} \, ,
\end{equation}
The full sampling recipe is summarized in the Algorithm~\ref{alg:QA}\index{MPS!Pauli perfect sampling}.
\begin{algorithm}[H]
\caption{Pauli sampling from MPS.}\label{alg:QA}
\hspace*{\algorithmicindent} \textbf{Input}: an MPS $\ket{\psi}$ of size $N$
\begin{algorithmic}[1]
\State Put the MPS in right-normalized form.
\State Initialize $\mathbb{L}=(1)$ and $\Pi = 1$
\For{($i=1$, $i=N$, $i++$)}
     \State Compute the probabilities
     $\pi(\alpha)=\pi_{\rho}(\sigma^{\alpha}|\sigma_1\cdots\sigma_{i-1})$ for $\alpha \in \{ 0,1,2,3 \}$ \noindent as in Eq.~\eqref{eq:prob_3}.
     \State  Generate a random value of $\alpha$ according to $\pi(\alpha)$
     \State Set $\sigma_i = \sigma^{\alpha}$, update $\Pi \rightarrow \Pi \cdot \pi(\alpha)$
     \State Update $\mathbb{L}$  as in Eq.~\eqref{eq:prob_2}.
\EndFor
\end{algorithmic}
\hspace*{\algorithmicindent} \textbf{Output}: a Pauli string $\pmb{\sigma}$ and the probability $\Pi(\pmb{\sigma})$

\end{algorithm}

\begin{example}{Stabiliser group of an MPS}{stab_mps}
One question which may be addressed using the approaches outlined previously is to identify the stabilizer group $\mathcal{S}(\ket{\psi})$ of an MPS.

\paragraph{Pauli-mps approach ---}
This can be done for example exploiting  the fact that the stabiliser nullity $\nu = N-k$, where $|\mathcal{S}| = 2^{k}$, is given by $\nu = \lim_{n\to\infty} (n-1)M_n$.
Using the Pauli basis algorithm in ref.~\cite{PhysRevLett.133.010601}, on can find the fixed point
of the power-series of the Pauli basis MPS, namely
$$
\includegraphics[width=0.6\textwidth,valign=c]{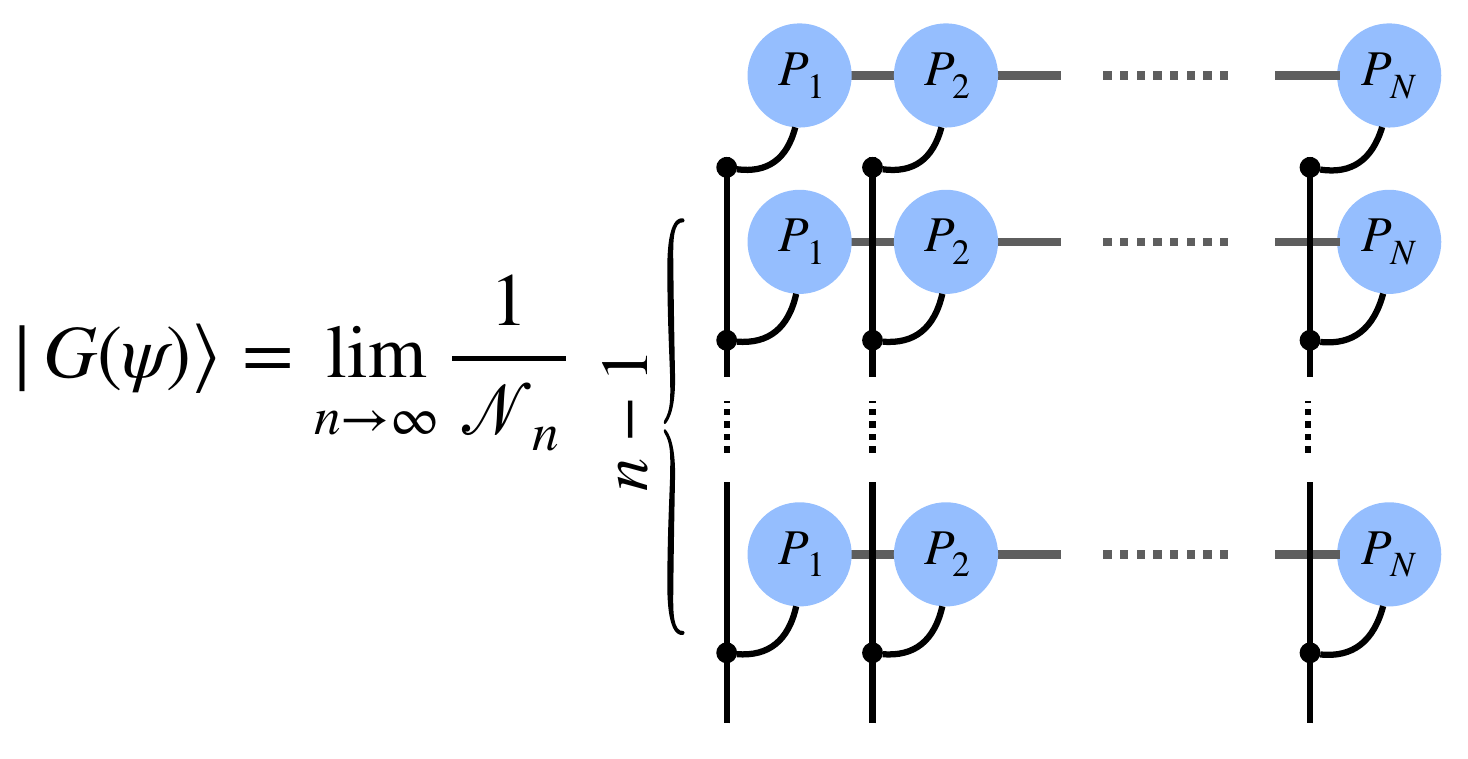}
$$
which is the normalized Pauli vector (hence the factor $1/\mathcal{N}_n$), compressed accordingly to the computation resources (usually $\chi_G \sim \chi^2$), and which can be used to extract the information about the stabiliser group of $|\psi\rangle$, since
$\langle \boldsymbol{\mu}|G(\psi)\rangle = 2^{k/2}$, if and only if  $\hat\sigma^{\mu_1}\cdots\hat\sigma^{\mu_N}|\psi\rangle = \pm |\psi\rangle$, otherwise is zero. This can be achieved by sampling $|G(\psi)\rangle$ in the computational basis, as described in Chapter~\ref{chap3} and Chapter~\ref{chap5}.
The overall computational scaling of this approach is still subjected to the compression cost of of the Pauli MPS, i.e.\ $O(N \chi^4)$.

\paragraph{Biased Pauli-sampling approach ---}
Another possibility is based on a clever biasing of the Pauli sampling techniques~\cite{PhysRevLett.133.010602}.

In this method, at each generic step $i$ of the sweep, a set of $K$ sub-strings $\{ \pmb{\sigma}_{[1,i]}^{\mu} \}_{\mu=1}^K$ is stored, where $\pmb{\sigma}_{[1,i]}$ is shorthand for the string of Pauli operators $(\sigma_1 \dots \sigma_i)$. Along with these sub-strings, their associated partial probabilities $\{\pi_{\rho}(\pmb{\sigma}_{[1,i]}^{\mu})\}_{\mu=1}^K$ are also retained. The partial probability $\pi_{\rho}(\sigma_1 \dots \sigma_i)$ is given by:
\begin{equation}
\pi_{\rho}(\sigma_1 \cdots \sigma_i) = \sum_{\pmb\sigma \in \mathcal{P}_{N-i}} \frac{1}{2^N} \text{Tr}\left[\rho \, \sigma_{1} \cdots \sigma_{i}\pmb{\sigma}\right]^2 \, ,
\end{equation}
where the sum runs over all possible Pauli strings acting on the remaining $N-i$ sites of the system. The ultimate goal is to track these sub-strings until the last step $i=N$, at which point one aims to identify those sub-strings that meet the stabilizer condition $\Pi_{\rho}(\pmb{\sigma}^{\mu}) = \pi_{\rho}(\pmb{\sigma}_{[1,N]}^{\mu}) = 1/2^N$.

However, storing all possible sub-strings quickly becomes computationally infeasible, as their number grows as $4^i$. To manage this growth, efficient strategies for discarding certain sub-strings while maintaining a maximum allowable number $\mathcal{N}$ of stored sub-strings have been developed. The challenge is to retain only those sub-strings that are more likely to form stabilizer strings at the end of the sweep. Two key strategies are used for this purpose:
\begin{enumerate}
\item
\textbf{Threshold-based filtering:}
It is observed that for any stabilizer string $\pmb{\sigma} \in \mathcal{S}(\ket{\psi})$, the partial probability at step $i$ satisfies a lower bound: $\pi_{\rho}(\pmb{\sigma}_{[1,i]}) \geq \frac{1}{2^i \chi_i}$, where $\chi_i$ represents the bond dimension at site $i$ of the MPS representing $|\psi\rangle$. Thus, sub-strings for which $2^i \chi_i \pi_{\rho}(\pmb{\sigma}_{[1,i]}) < 1$ can be safely discarded.

\textbf{Selective pruning:}
When the number $K$ of stored sub-strings exceeds the limit of resources $\mathcal{N}$, the sub-strings are sorted based on their partial probabilities $\pi_{\rho}(\pmb{\sigma}_{[1,i]}^{\mu})$ in descending order. Only the top $\mathcal{N}$ sub-strings with the highest probabilities are retained. These sub-strings have the greatest potential to maximize the final stabilizer probability $\Pi_{\rho}(\pmb{\sigma})$ at the end of the sweep. This approach ensures that the computational resources are focused on the most promising candidates.
\end{enumerate}

These strategies collectively allow for efficient management of sub-string storage, ensuring that the computation remains feasible even for large systems while retaining the most relevant sub-strings for stabilizer identification.
Let us mention that, in most scenarios a sinlge sweep is sufficient to extract the entire set of stabiliser generators, leading to a computational cost $O(N\chi^3)$; however further refined strategies have been proposed in ref.~\cite{PhysRevLett.133.010602} to allow the algorithm to target unsampled regions of $\mathcal{S}(|\psi\rangle)$ with minor increasing of the computational effort.
\end{example}

\section{Clifford enhanced matrix product states ($\mathcal{C}$MPS)}

Here, we show the possibility of improving tensor network algorithms by introducing a class of quantum states described by an MPS of bond dimension $\chi$, where a Clifford unitary is applied to the state.\index{Stabiliser tensor network} We call this set of states Clifford enhanced matrix product states $\mathcal{C}{\rm MPS}$\cite{masotllima2024stabtn,lami2024cmps}
\begin{equation}
    \mathcal{C}{\rm MPS} =  \{ U_\mathcal{C} \ket{\psi_\chi},\quad \forall U_\mathcal{C} \in \mathcal{C}_N, \quad \ket{\psi_\chi} \in {\rm MPS\ bond\ dimension\ }\chi\}.
\end{equation}
Note that, in general, $\ket{\phi} = U_\mathcal{C}\ket{\psi_\chi}$ may describe a quantum state with higher entanglement, highlighting the potential of applying Clifford unitaries to MPS in order to represent more entangled quantum states.

Moreover, by combining stabilizer and tensor network methods, it is easy to figure out how to compute the expectation value of a Pauli string $\Sigma^{\boldsymbol{\mu}} = \sigma^{\mu_1}\otimes \ldots  \otimes \sigma^{\mu_N}$ over a $\mathcal{C}MPS$ indeed\begin{equation}
    \expval{\Sigma^{\boldsymbol{\mu}}}{\phi} = \expval{U_\mathcal{C}^\dagger \Sigma^{\boldsymbol{\mu}} U_\mathcal{C}}{\psi_\chi} = \expval{\Sigma^{\boldsymbol{\mu'}}}{\psi_\chi}
\end{equation}
where $\Sigma^{\boldsymbol{\mu'}}$ is the new Pauli string under the action of the Clifford unitary $U_\mathcal{C}$. Therefore, even though $\ket{\phi}$ can be highly entangled, the complexity of computing the expectation value of a Pauli string over $\ket{\phi}$ remains the same as computing it over $\ket{\psi_\chi}$.

The use of such hybrid tensor networks, which blend classical stabilizer elements with quantum resources, represents a cutting-edge strategy for simulating quantum systems that are otherwise too complex to handle using purely classical techniques. This approach underscores the importance of quantum magic in understanding the full computational potential of quantum systems and their simulation via tensor networks.

\subsection{Hybrid stabilizer MPO}

In general, any unitary operation can be decomposed into a set of Clifford unitaries and a magic quantum gate. In order to be as general as possible, we will consider local rotations as the set of non-Clifford operations that introduce non-stabilizerness to a quantum circuit. We define $ R_l^\mu(\theta) $ as a local rotation gate acting on qubit $ l $, rotating it by an angle $ \theta $ around the $ \mu $-axis. This is given by the expression
\begin{equation}
    R_l^\mu(\theta) = e^{-i \theta/2 \sigma_l^\mu}.
\end{equation}
A unitary acting on $N$ qubits has the following representation\begin{equation}\label{chpt6:eq:Udecomposition}
    U =\prod_{j} R_{{l_j}}^{{\mu_j}}(\theta_j) C_j
\end{equation}
where $R_{{l_j}}^{{\mu_j}}(\theta_j)$ are local rotations and $C_j \in \mathcal{C}_N$ Clifford unitaries.
\begin{equation}
\includegraphics[width=10cm,valign=c]{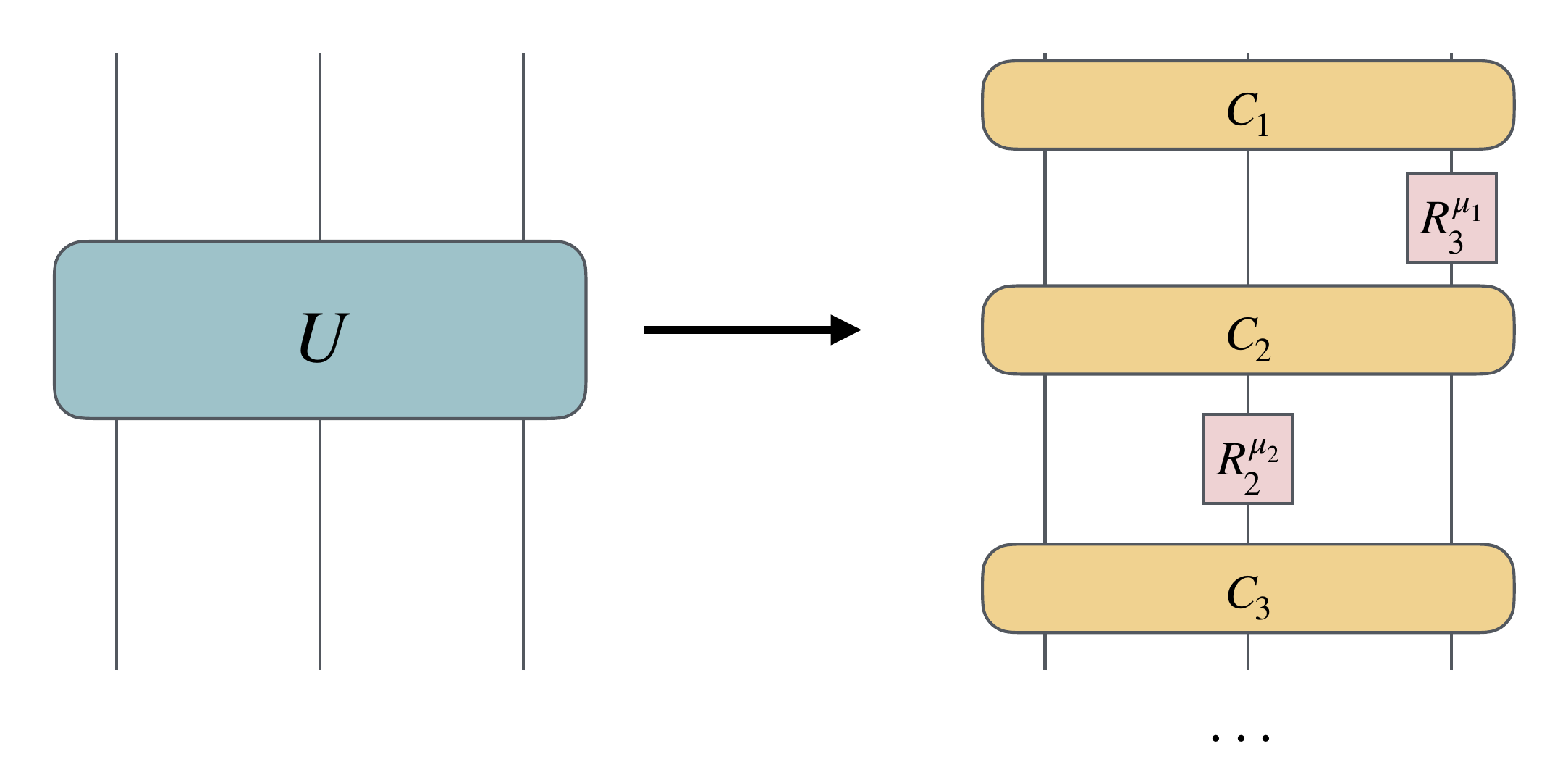}
\end{equation}
We want to compute the action of $U$ on a quantum state $\ket{\psi_0}$. In the following, we will provide a procedure to construct a $\mathcal{C}MPS$ from $U\ket{\psi_0} = C \ket{\phi}$, such that the entanglement entropy of $\ket{\phi}$ for any bipartition is less than that of $U\ket{\psi_0}$. In particular, we notice that\begin{equation}
    C^\dagger R_{l}^\mu(\theta) C= \cos(\theta/2) \mathbb{I}\mp i \sin(\theta/2) \Sigma^{\boldsymbol{\mu}}
\end{equation}
where we used the fact that $C^\dagger \sigma_l^\mu C = \pm \Sigma^{\boldsymbol{\mu}}$.

Before providing the full expression for the decomposition of  U  in\linebreak Eq.~\eqref{chpt6:eq:Udecomposition}, it is instructive to first examine how the transformation of  U  works for a small number of layers, as shown in the following example.

\begin{example}{Implementation of the stabilizer MPO}{Two_unitaries}
Here, we show an explicit implementation of the algorithm for two layers of $U$, that is\begin{equation}
    U = R_{m}^\mu(\theta_m) C_2  R_{n}^\nu(\theta_n) C_1
\end{equation}
first of all we insert the identity $C_1C^\dagger_1$
\begin{equation}
     U = R_{m}^\mu(\theta_m) C_2 C_1 C^\dagger_1 R_{n}^\nu(\theta_n) C_1
\end{equation}
we define the since Clifford unitaries form a group we can define the composite Clifford $C_{12} = C_2C_1$. We then insert the identity $C_{12}C^\dagger_{12}$ in order to transform the second local rotation\begin{equation}
      U =C_{12}C_{12}^\dagger R_{m}^\mu(\theta_m) C_{12} C^\dagger_1 R_{n}^\nu(\theta_n) C_1
\end{equation}
now we can transform the local rotations and obtain\begin{equation}
    U = C_{12} \left[\cos(\theta_m/2)\mathbb{I} \mp i \sin(\theta_m/2)\Sigma^{\mu}\right]\left[\cos(\theta_n/2)\mathbb{I} \mp i \sin(\theta_n/2)\Sigma^{\nu}\right]
\end{equation}
\end{example}

To write the decomposition of $U$ we define $C_{1\ldots j} \equiv C_jC_{j-1}\ldots C_1$. By sequentially inserting the identities $C_{1\ldots j}C^\dagger_{1\ldots j}$ we obtain the following
\begin{equation}
    U= C_{1\ldots N} \prod_j^N {C}_{1\ldots j}^\dagger R^{\mu_j}_{l_j}(\theta_j) {C}_{1\ldots j} = C_{1\ldots N}  \prod_j^N (\cos( \theta_j/2) \mathbb{I} \mp i \sin(\theta_j/2) \Sigma^{\boldsymbol{\mu}_j} ).
\end{equation}
We can now focus on constructing an MPO description of the terms\begin{equation}
    \cos( \theta/2) \mathbb{I} \mp i \sin(\theta/2) \Sigma^{\boldsymbol{\mu}}  = \mathbb{T}_1 \ldots  \mathbb{T}_N
\end{equation}
Since we are summing two operators, i.e.\ $\mathbb{I}$ and $\Sigma^{\boldsymbol{\mu}}$, whose MPO representation has an auxiliary dimension $D=1$, it is straightforward to see that their sum results in an MPO with an auxiliary dimension  $D=2$, such that\begin{equation}
    \mathbb{T}_j = \begin{pmatrix}
        \cos(\theta/2)^{1/N} \mathbb{I}_j && 0\\
        0 && (\mp i \sin(\theta/2))^{1/N} \sigma^{\mu_j}_j
    \end{pmatrix}
\end{equation}
for $j=2,\ldots ,N-1$, and with boundaries
\begin{equation}
    \mathbb{T}_1 = \begin{pmatrix}
        \cos(\theta/2)^{1/N} \mathbb{I}_1 & (\mp i \sin(\theta/2))^{1/N} \sigma^{\mu_1}_1
    \end{pmatrix},\    \mathbb{T}_N = \begin{pmatrix}
        \cos(\theta/2)^{1/N} \mathbb{I}_N \\ (\mp i \sin(\theta/2))^{1/N} \sigma^{\mu_N}_N
    \end{pmatrix}.
\end{equation}
We call this MPO with bond dimension $D=2$, \textbf{stabilizer MPO}~\cite{PhysRevLett.133.150604}.\index{Stabiliser tensor network!stabiliser MPO}

As an example, we can illustrate the expression $\expval{U^\dagger \Sigma^{\boldsymbol{\gamma}} U}{\psi_0}$ for a two-layer unitary $U$ by providing both its representation as a quantum circuit and as a tensor network
\begin{equation}
\includegraphics[width=.95\linewidth,valign=c]{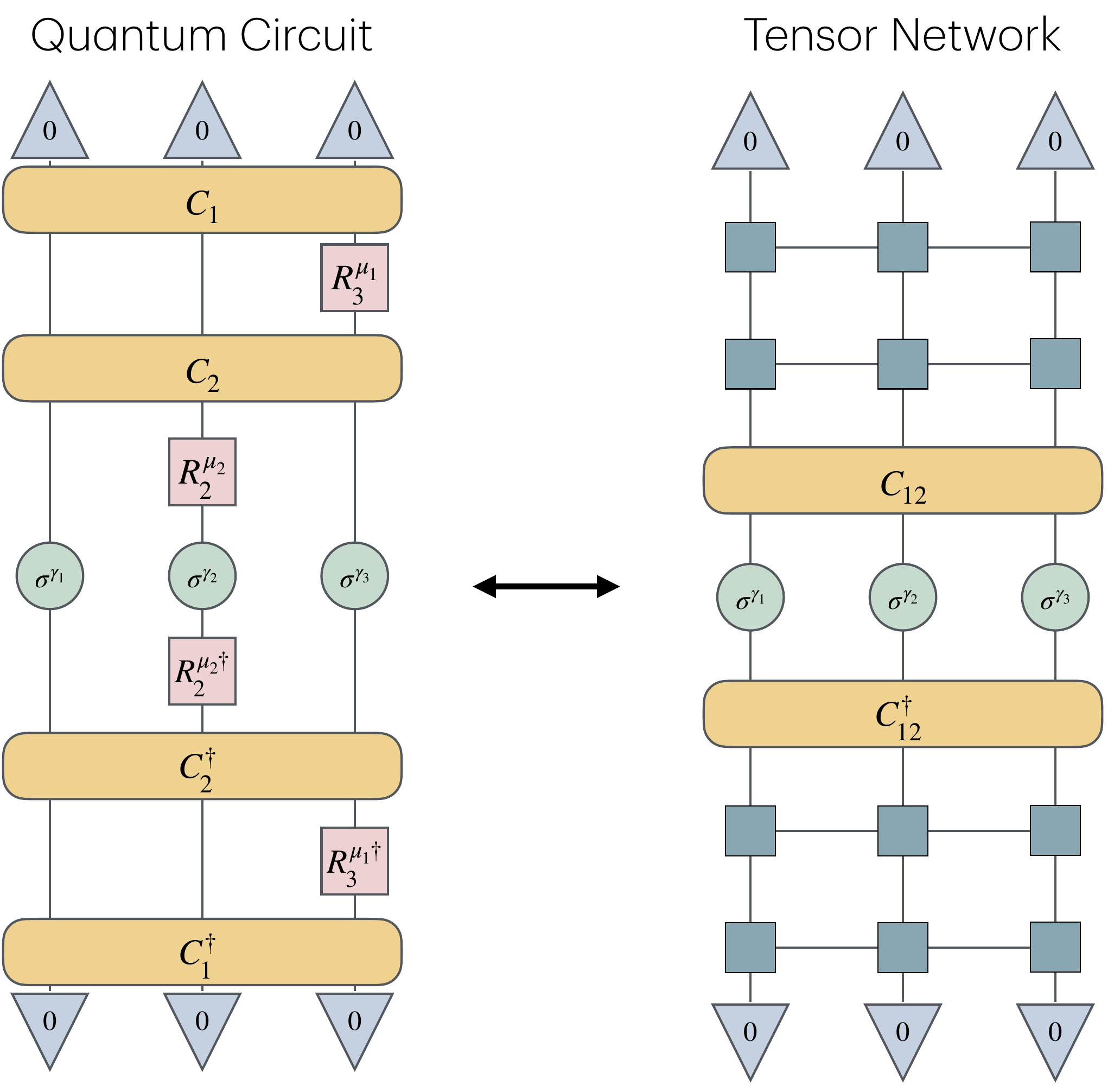}
\end{equation}

Notice that the entangling Clifford $C_{12}$ does not increase the complexity of the circuit, as it can be absorbed by conjugating the Pauli product $C_{12}^\dagger \Sigma^{\boldsymbol{\gamma}} C_{12} = \Sigma^{\boldsymbol{\tilde{\gamma}}}$, which, since it is a MPO with auxiliary dimension $1$, does not increase the bond dimensions of the network, as shown in the diagram
\begin{equation}
\includegraphics[width=.88\linewidth,valign=c]{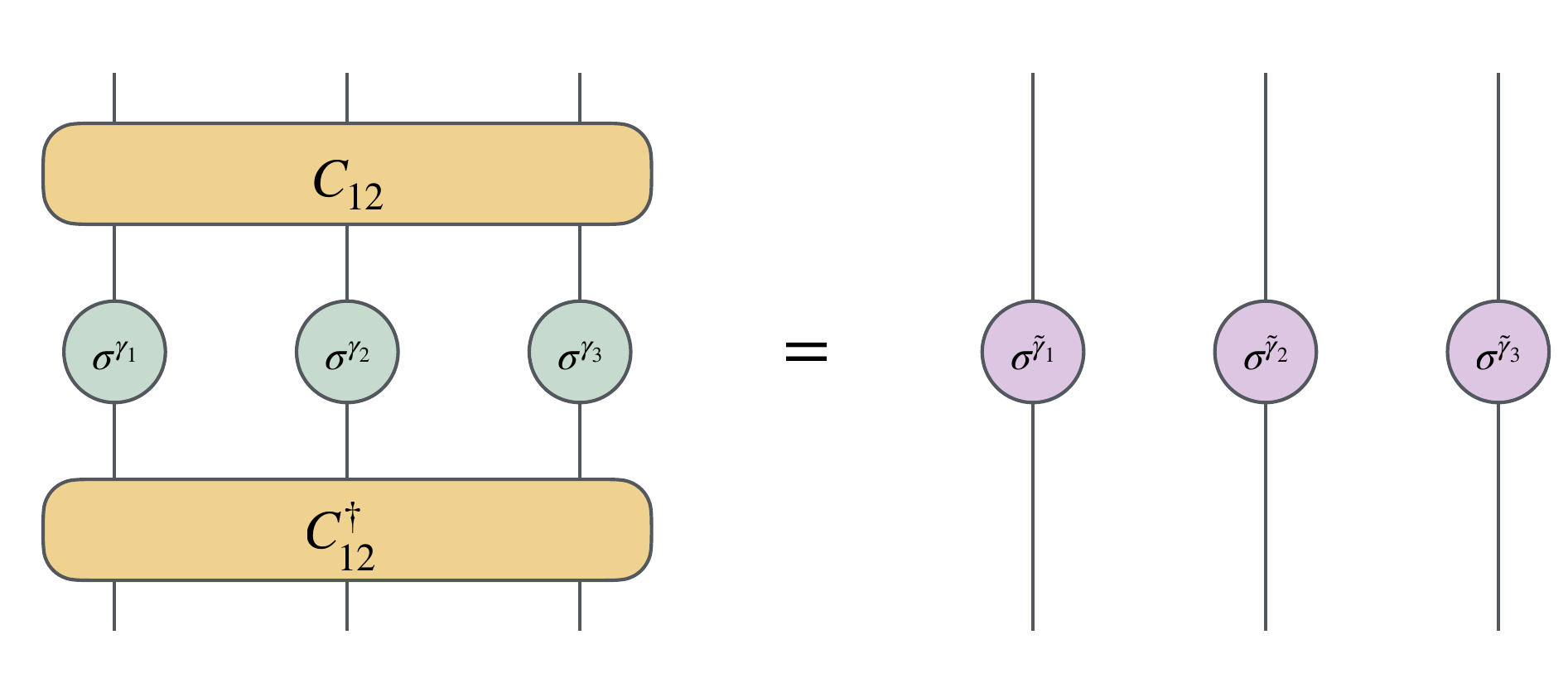}
\end{equation}

\paragraph{Auxiliary Spin Picture ---}
The stabilizer MPO with an auxiliary dimension of 2 is depicted in the following diagram
\begin{equation}   \includegraphics[width=.75\linewidth,valign=c]{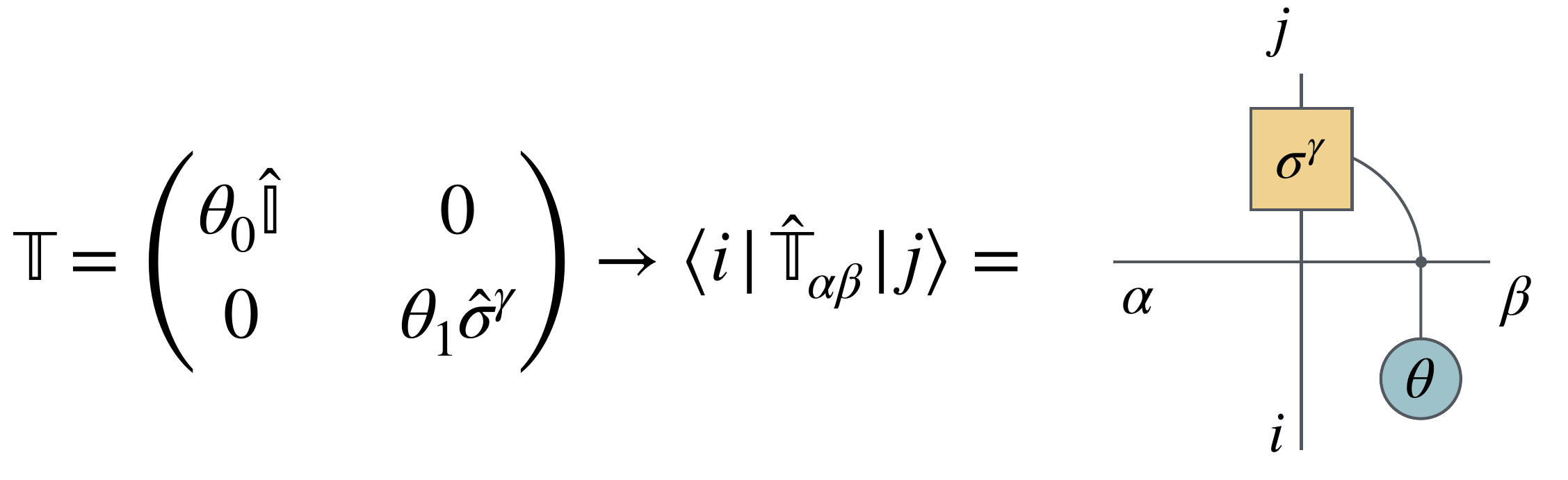}
\end{equation}
In this diagram, the stabilizer MPO can be interpreted as a Control-Pauli operator with complex weights $\theta_0$ and $\theta_1$, where $|\theta_0^N|^2$ + $|\theta_1^N|^2 = 1$.  As shown, we can absorb these coefficients into an auxiliary qubit state that evolves along the temporal dimension
\begin{equation}
    \ket{\theta} = \theta_0^N \ket{0} + \theta_1^N\ket{1}=\includegraphics[width=.66\linewidth,valign=c]{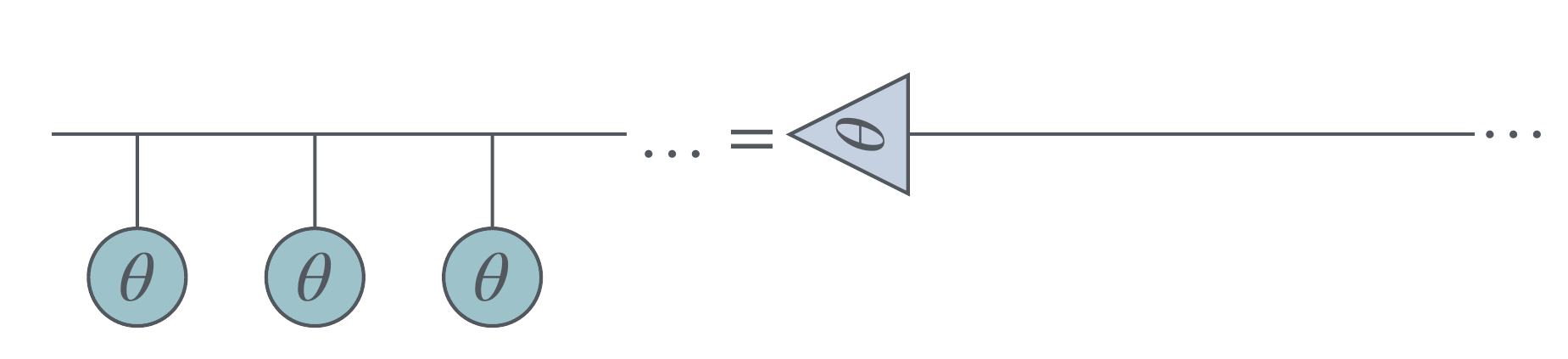}
\end{equation}
We define also the following not-normalized auxiliary qubit state which will serve as a boundary vector
\begin{equation}
    \ket{X} = \ket{0}+\ket{1}
\end{equation}
This formulation allows us to represent the tensor network in terms of controlled Pauli gates only, simplifying the overall structure. The corresponding operator for each site $j=1,\ldots ,N$ is given by
\begin{equation}
    \mathbb{K}_j = \begin{pmatrix}
         \mathbb{I}_j && 0\\
        0 &&  \sigma^{\mu_j}_j
    \end{pmatrix}
\end{equation}
where $\mathbb{I}_j$ represents the identity and $\sigma^{\mu_j}_j$ is a Pauli operator acting on site $j$. Along the spatial dimension, the tensor network reduces to the following product of blocks\begin{equation}
    \mathbb{T}_1 \ldots  \mathbb{T}_N = \mel{\theta}{\mathbb{K}_1 \ldots  \mathbb{K}_N}{X} =\includegraphics[width=.5\linewidth,valign=c]{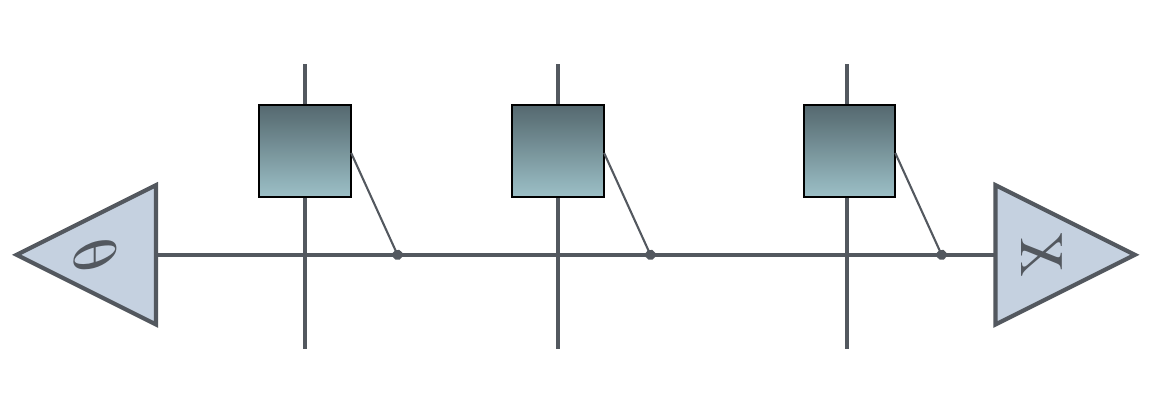}
\end{equation}
This relation enables us to express the evolution of the operator $U$ as\begin{equation}
    U = C_{1\ldots N} \prod_{m=1}^M \mel{\theta_m}{\mathbb{K}_1^m \ldots  \mathbb{K}_N^m}{X}
\end{equation}
where $C_{1\ldots N}=C_N\ldots C_1$. Therefore we obtain the following\begin{equation}
    \expval{U^\dagger \Sigma^{\boldsymbol{\gamma}} U}{\psi_0} = \expval{ {\prod_{m=1}^N} \mel{\theta_m^*}{\mathbb{K}_1^{m\dagger}\ldots \mathbb{K}_N^{m\dagger}}{X} \Sigma^{\boldsymbol{\tilde{\gamma}}} {\prod_{m=1}^N} \mel{\theta_m}{\mathbb{K}_1^{m}\ldots \mathbb{K}_N^{m}}{X}}{\psi_0}
\end{equation}
This expression can be represented as a tensor network, which is shown below, and can be contracted efficiently in any direction
\begin{equation}   \includegraphics[width=\linewidth,valign=c]{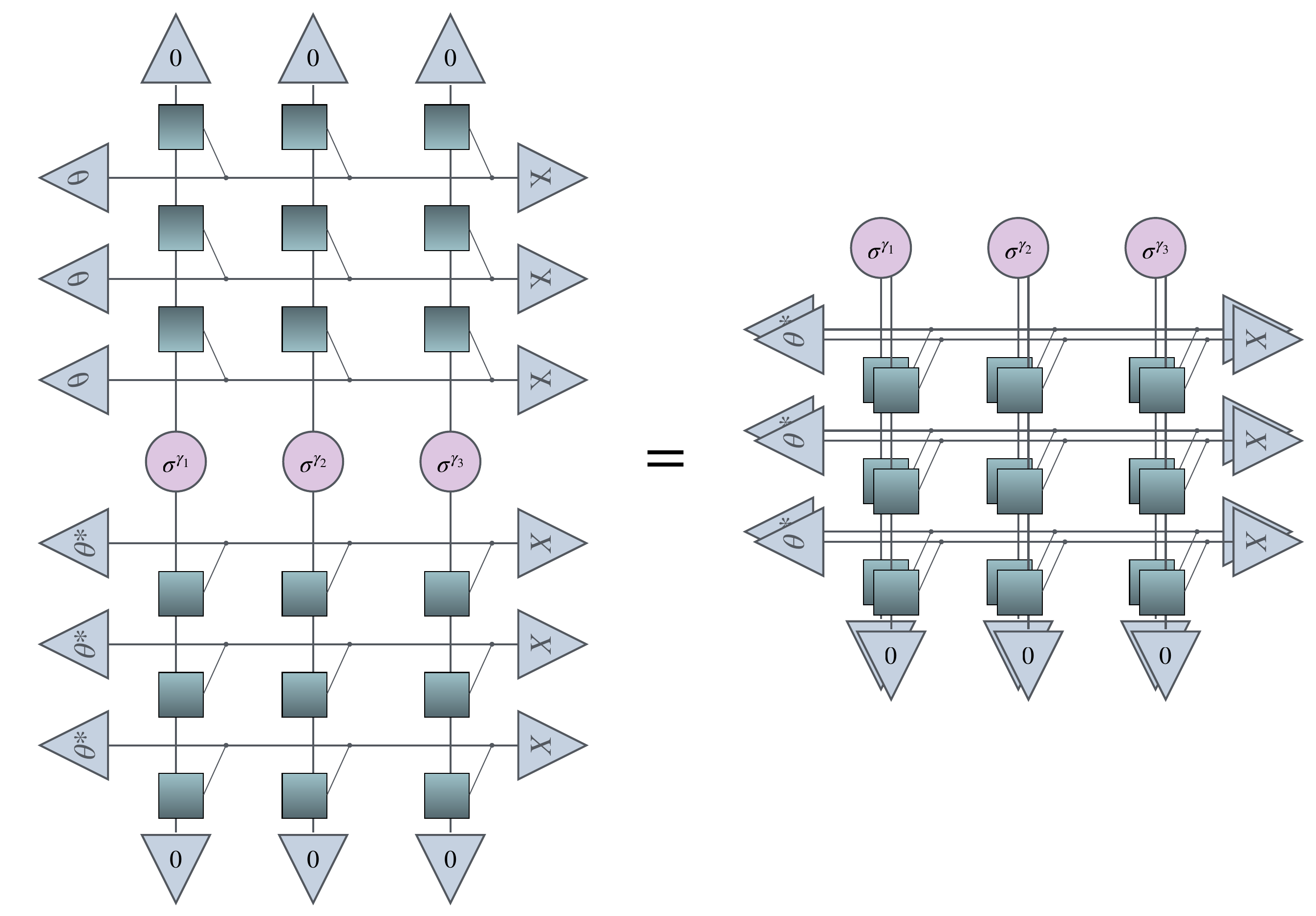}
\end{equation}
Furthermore, by folding the tensor network, as shown in the diagram, we can compute the evolution of Pauli strings in the Heisenberg picture.

%\subsection{Clifford dressed DMRG}
\subsection{Clifford-Dressed TDVP for Hamiltonian Dynamics}

Another typical scenario which can be consider involves the dynamics, from a short-range correlated initial state (e.g., a product state), generated by the Hamiltonian
$\hat H_0 = \sum_{k}^{O(N)} J_k \hat\Sigma^{\boldsymbol{\mu}_k}$, which in graphical notation looks like
\begin{equation}
\includegraphics[width=0.7\linewidth,valign=c]{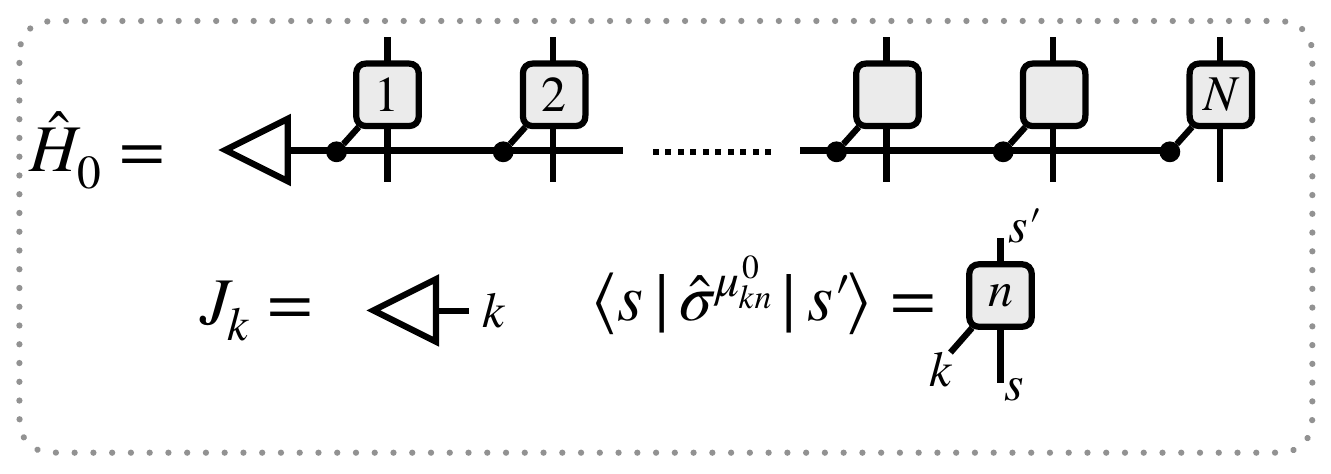}
\end{equation}
which is a diagonal MPO with bond dimension equal to the number $O(N)$ of independent Pauli strings entering in the sum.
Here the subscript zero is indicating that the Hamiltonian is in its ``bare'' form, meaning it has not yet been modified by any Clifford disentangler.

The algorithm that we are going to describe in this section has been recently proposed in ref.~\cite{mello2024clifforddressed, qian2024cliffordcircuits}.
A similar approach for ground-state search has been presented in ref.~\cite{qian2024augmentingdensitymatrix}.

We are primarily interested in evolving the state and monitoring local observables that are experimentally relevant. This includes computing the expectation value of Pauli strings:
\begin{equation}
    \bra{\psi(0)} e^{i \hat{H}_0 t} \hat{\Sigma}^{\boldsymbol{\mu}} e^{-i \hat{H}_0 t} \ket{\psi(0)}.
\end{equation}
Typically, the time evolution is discretized into small intervals of length $dt$. The state, which is approximated as an MPS with bond dimension $\chi$, is evolved using the single-site TDVP scheme. In this symplectic time-evolution integrator, the bond dimension $\chi$ is fixed from the beginning of the simulation.

However, the entanglement entropy typically grows rapidly, making the MPS representation inadequate after a relatively short time. This is where Clifford disentangling strategies, or ``entanglement cooling'' become essential.

At each time step $m \in \{1, \dots\}$, corresponding to time $t_m = m \, dt$, we apply a suitable Clifford transformation $\hat{C}_m$ to the state $\ket{\psi(t_m)}$ to reduce the entanglement of the new state:
\begin{equation}
\includegraphics[width=0.6\linewidth,valign=c]{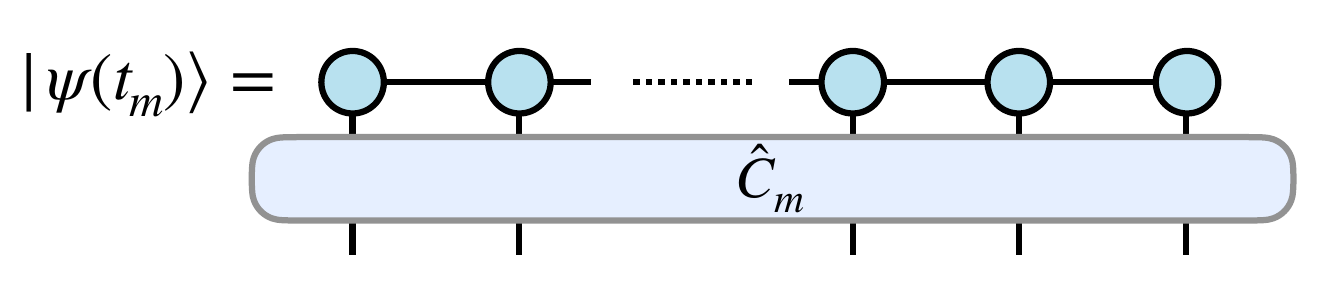}
    %\ket{\tilde{\psi}(t_m)} = \hat{C}_m %\ket{\psi(t_m)}.
\end{equation}
This induces a transformation of the Hamiltonian as well:
\begin{equation}
  \includegraphics[width=\linewidth,valign=c]{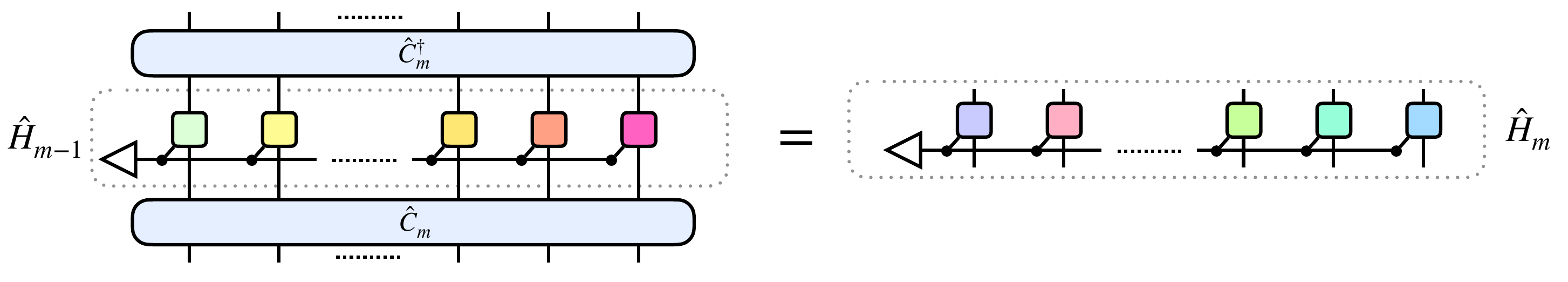}
\end{equation}
Which can be efficiently performed using the stabilizer tableau formalism \cite{Gottesman_1997}, and does not increase the number of Pauli strings involved in its definition. In practice, as depicted in the previous figure, the Clifford transformation preserves the diagonal structure of the MPO and only modifies the Pauli strings composing the operator (hence the different colors in the pictorial representation of the MPO).

In practice, we are recasting theoriginal time evolution,
\begin{equation}
    \prod_{m=0}^{\lfloor t/dt \rfloor - 1} e^{-i \hat{H}_0 dt} \ket{\psi(0)},
\end{equation}
into a so called \textbf{Clifford-dressed} evolution\index{Stabiliser tensor network!clifford-dressed TDVP}
\begin{equation} \label{chapt6_eq:clifford_dressed}
    \prod_{m=0}^{\lfloor t/dt \rfloor - 1} e^{-i \hat{H}_m dt} \hat{C}_m \ket{\psi(0)},
\end{equation}
where $\hat{C}_0 = \hat{I}$ is the identity operator. The Clifford-dressed evolution iteratively constructs the final Clifford-enhanced MPS ($\mathcal{C}$MPS), yielding:
\begin{equation}
   \includegraphics[width=0.6\linewidth,valign=c]{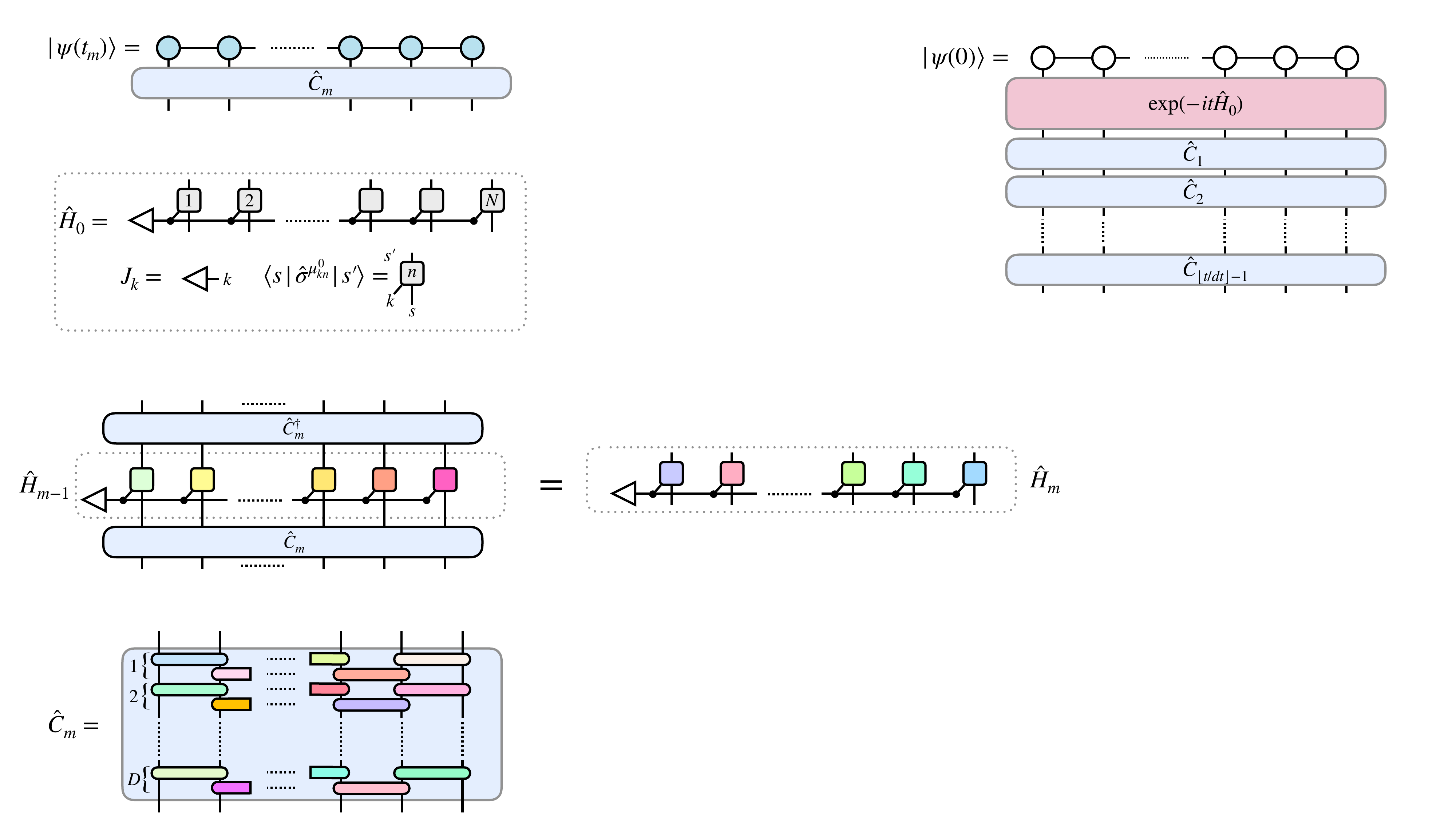}
\end{equation}
which however has been constructed iterativelly from Eq.~(\ref{chapt6_eq:clifford_dressed}) by optimising the Clifford transformations $\{\hat C_1, \hat C_2, \dots\}$ at intermediate time steps and not just at the end of the full bare time evolution.
Additionally, for any Pauli string $\hat{\Sigma}^{\boldsymbol{\mu}}$, its time evolution must be transformed as
\begin{equation}
    \hat{C}_{m} \cdots \hat{C}_1 \hat{\Sigma}^{\boldsymbol{\mu}} \hat{C}_1^{\dagger} \cdots \hat{C}_m^{\dagger},
\end{equation}
which thanks to the Clifford properties, it is a trivial transformation.

While one can generally apply the Clifford disentangler at every step of the TDVP evolution, this is not strictly required. One can space out the disentangling steps, applying the routine every
$k$ iterations instead. This approach helps reduce computational costs while still controlling the growth of entanglement over time.

The structure of the Clifford disentanglers has not been defined yet, and, in principle, they could belong to the generic $N$-qubit Clifford group. However, this leads to two main difficulties: (1) the number of possible global Clifford operations exceeds any reasonable limit for performing a systematic search for the best disentangler; (2) even computing the action of a generic $N$-qubit Clifford unitary on an MPS to test whether it is disentangling or not is computationally infeasible for efficient tensor-network manipulation.

A concrete computationally feasible disentangling strategy involves instead iteratively constructing an optimal Clifford transformation by performing sweeps over two-qubit Clifford gates between adjacent sites in a checkerboard pattern. The number of sweeps is denoted by $D$. Basically a possibility is thus to enforce the following structure to the Clifford unitaries
\begin{equation}
\includegraphics[width=0.6\linewidth,valign=c]{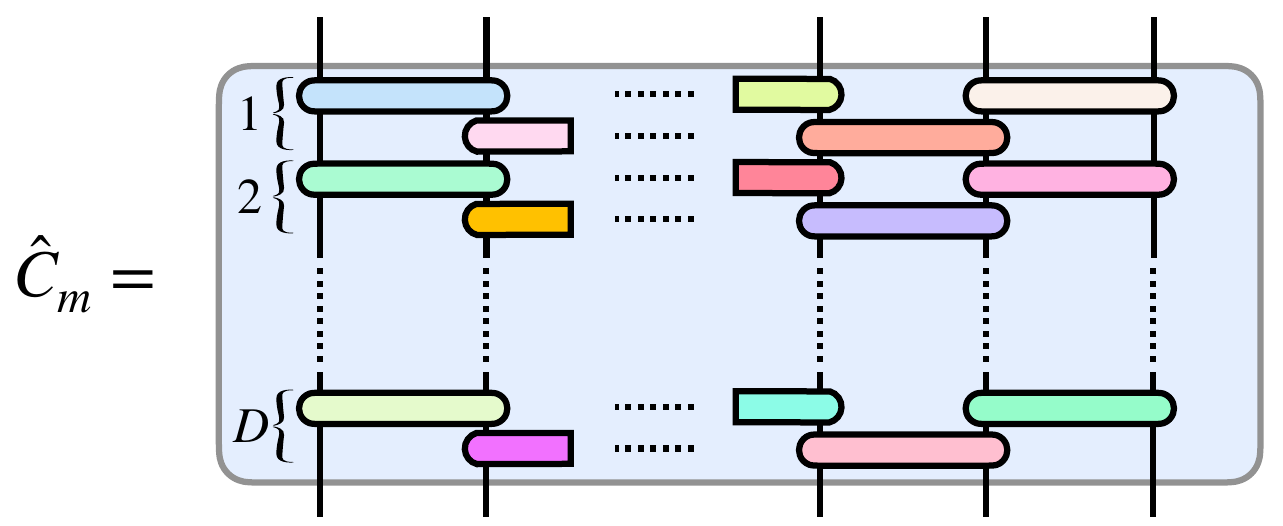}
\end{equation}

Sweeping over the layers, each optimal Clifford gate can be selected from a suitable subset of two-qubit Clifford transformations which allows to minimise the von Neumann entropy.
In practice we follow the following diagramatic routine:
\begin{equation}
\includegraphics[width=0.55\linewidth,valign=c]{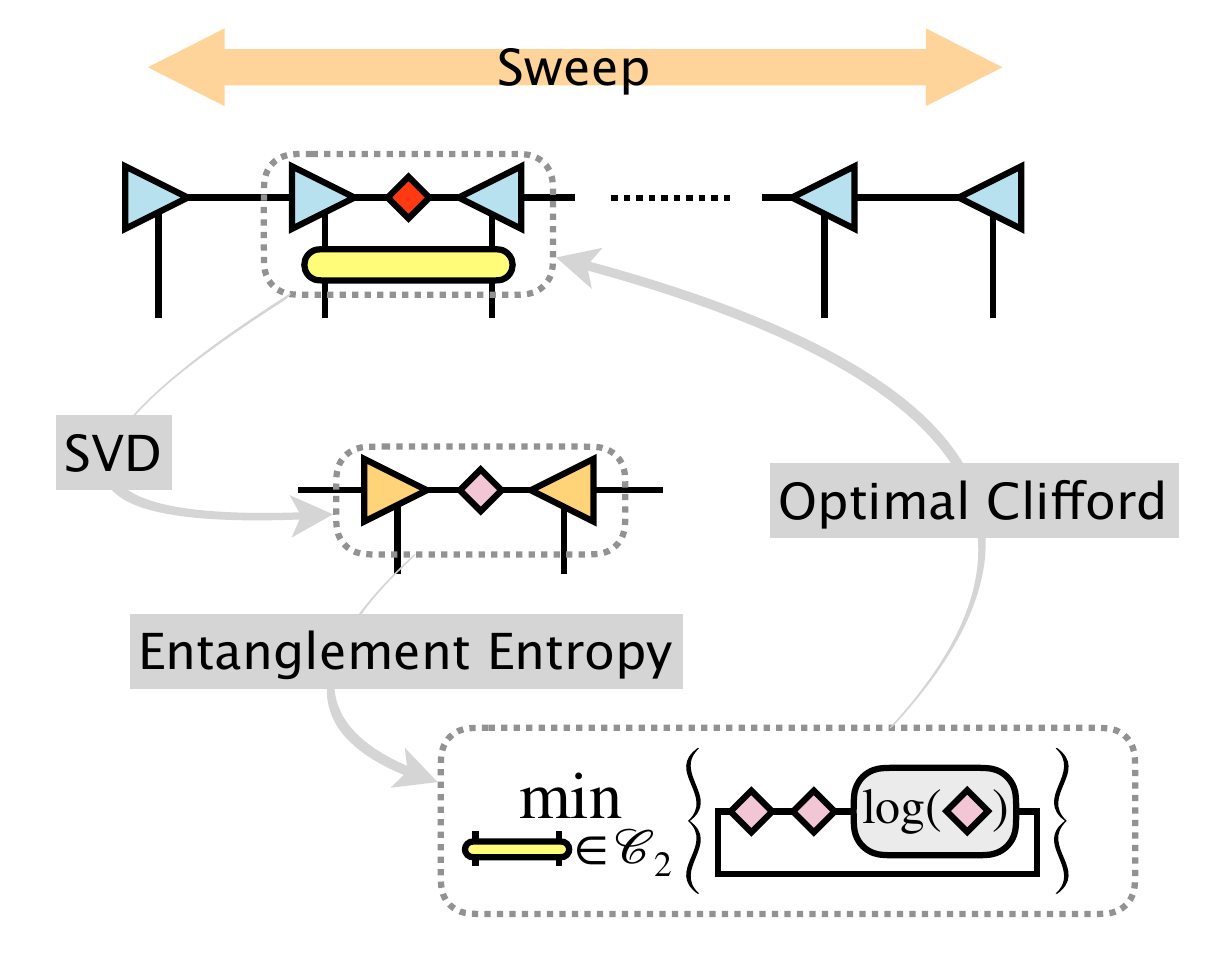}
\end{equation}

In this approach, the time evolution is push forward using the single-site Time-Dependent Variational Principle (1-TDVP) scheme\index{TDVP}, which, due to its symplectic nature, is ensuring that the ``Clifford-dressed'' energy is preserved during the system's time evolution. At each time step, for a particular local tensor evolution, the dressed Hamiltonian $\hat{H}_m$ is projected onto the Matrix Product State (MPS) tensors of the Clifford-enhanced state
$$
    %|\tilde{\psi}(t_{m})\rangle = \mathbb{A}^{s_1}_L \cdots \mathbb{A}^{s_n}_C \cdots \mathbb{A}^{s_N}_R \ket{s_1, \dots, s_N},
\includegraphics[width=0.7\linewidth,valign=c]{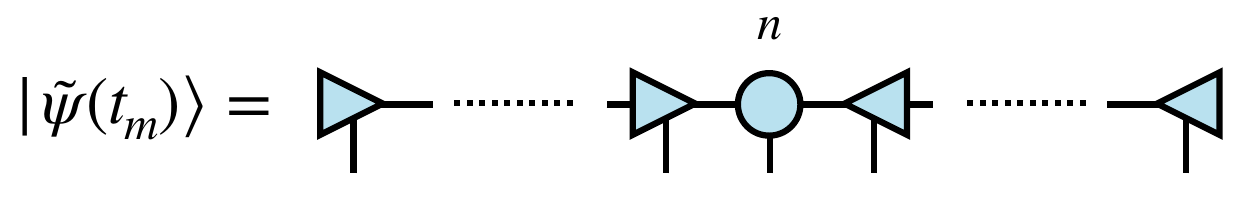}
$$
where the MPS is represented in mixed canonical form with respect to the central site $n$. The effective Hamiltonian at site $n$ is expressed as:
\begin{equation}
    %H^{\text{eff}}_m(n) = \sum_{k}^{O(N)} J_k \, \mathbb{L}^{m}_k(n-1) \otimes \hat{\sigma}_n^{\mu^m_k} \otimes \mathbb{R}^{m}_k(n+1),
    \includegraphics[width=0.9\linewidth,valign=c]{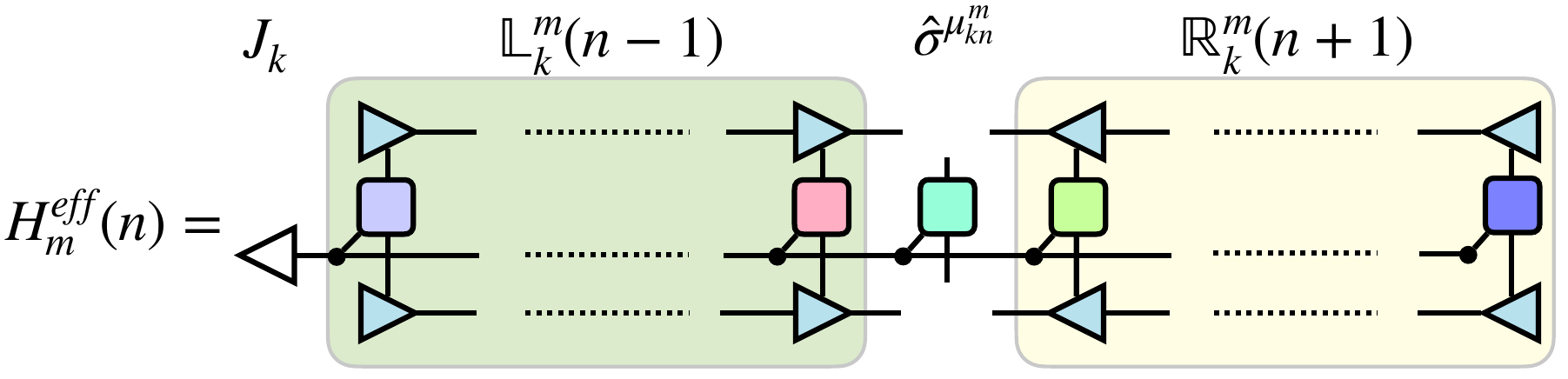}
\end{equation}
where $J_k$ denotes the couplings associated with each Pauli string, $\mathbb{L}^{m}_k(n-1)$ and $\mathbb{R}^{m}_k(n+1)$ represent the left and right block projected Hamiltonians, and $\hat{\sigma}^{\mu^m_{kn}}$ is the Pauli operator acting on site $n$. Notice that the effective Hamiltonian remains diagonal in the auxiliary index, preserving the structure of the dressed Hamiltonian during the evolution process.

A similar transformation is performed for the operator
\begin{equation}
\includegraphics[width=0.9\linewidth,valign=c]{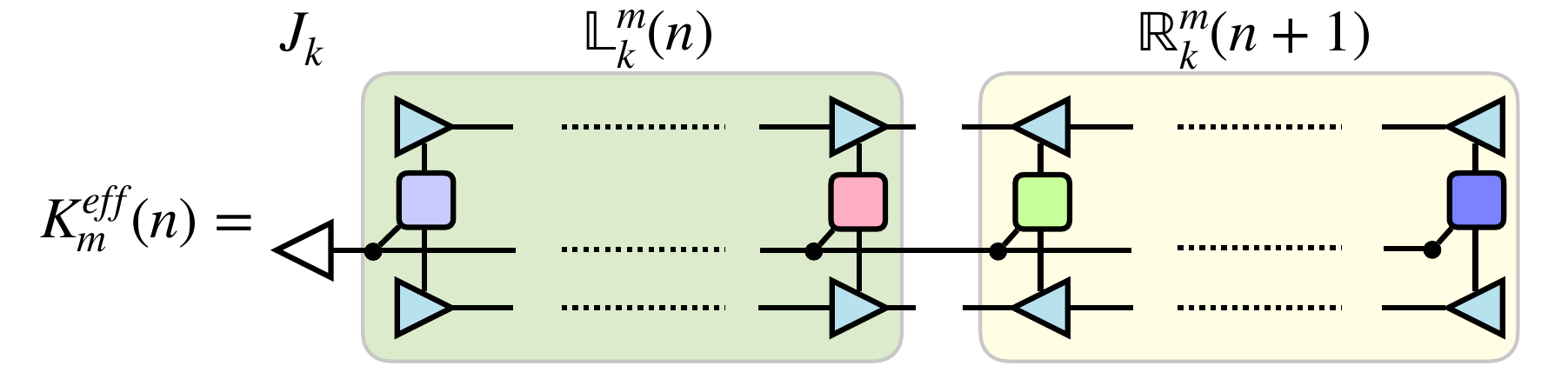}
\end{equation}
which is nothing more than the effective Hamiltonian projected into the MPS $\ket{\tilde\psi(t_m)}$ in its central bond representation
$$
\includegraphics[width=0.7\linewidth,valign=c]{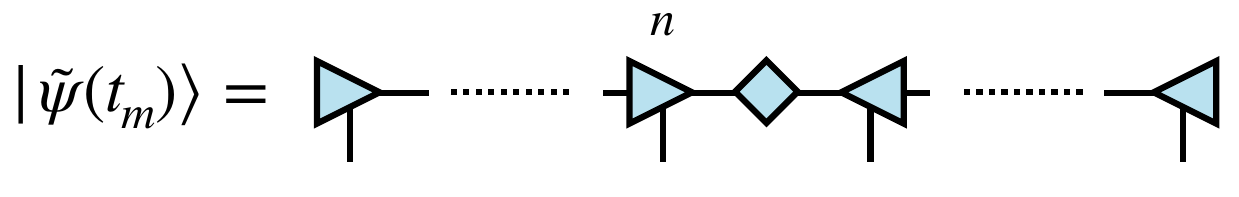}.
$$
The operator $\hat K_{m}^{\mathrm{eff}}(n)$ need to be used to perform a backward time-step evolution for the bond tensor, as described in details in ref.~\cite{PhysRevB.94.165116}.

This method ensures an efficient computation of the time-evolved states while maintaining the underlying structure of Clifford-enhanced Matrix Product States.

\paragraph{Two-site TDVP scheme ---}
The Time-Dependent Variational Principle (TDVP) algorithm in the Matrix Product State (MPS) formalism traditionally supports a two-site integration scheme (see Chapter~\ref{chap4}).\index{TDVP} While this modification abandons the symplectic nature of single-site integration, it compensates by allowing the bond dimension to dynamically adapt to the evolving entanglement structure. In this approach, the effective Hamiltonian acts on a two-site central block, updating the corresponding local MPS tensors. After the time evolution, a Singular Value Decomposition (SVD) is performed to extract the single-site tensors, with the rightmost tensor adapted through backward evolution.

Building upon this, a new two-site TDVP scheme enhanced by Clifford disentanglers has been proposed in ref.~\cite{qian2024cliffordcircuits}.
Inspired by the disentangling techniques used in ref.~\cite{qian2024augmentingdensitymatrix} for the Density Matrix Renormalization Group (DMRG), the 2-site method applies a local two-qubit Clifford gate before the SVD step. This ``Clifford dressing'' aims to further reduce entanglement by locally optimizing the two-site MPS tensors. After finding the optimal disentangler, the effective Hamiltonian is transformed under conjugation by the Clifford operator. This transformation remains localized, only impacting the Pauli strings corresponding to the two lattice sites currently being evolved.

The computational cost of this additional disentangling step should be modest, as it only involves optimizing local gates. However, a potential drawback of this approach is its propensity to get trapped in local entanglement minima. Without a global sweeping mechanism, the disentanglers may fail to optimally reduce global entanglement across the entire wave function. Nonetheless, this approach could offer a computationally efficient way to mitigate entanglement growth in systems where bond dimension scaling poses a challenge.

\newpage

\biblio

%%%%%%%%%%%%%%%%%%%%%%%%%%%%%%%%%%%%%%%%%%%%%%%%%%%%%%%%%%%%%

%\part{Supplementaries}

\addcontentsline{toc}{chapter}{\indexname}
\printindex

\end{document}